\documentclass[a4paper,11pt,twoside,openright]{book}
\usepackage{epsfig} 
\usepackage{amssymb}
\usepackage{amsfonts}
\usepackage{amsmath}
\usepackage{mathrsfs}
\usepackage{nicefrac}
\usepackage{amsthm}
\usepackage{latexsym,graphicx,color,bbm,subfigure}
\usepackage{pdfsync}   
\usepackage{bbm}
\usepackage{eso-pic}
\usepackage[T1]{fontenc}  
\usepackage{textcomp}
\usepackage{fancyhdr}
\usepackage{emptypage}

\DeclareMathOperator{\deter}{det}
\DeclareMathOperator{\im}{Im}

\DeclareMathOperator{\diag}{diag}

\newcommand{\be}{\begin{equation}}
\newcommand{\ee}{\end{equation}}
\newcommand{\beq}{\begin{eqnarray}}
\newcommand{\eeq}{\end{eqnarray}}
\newcommand{\non}{\nonumber\\}

\newcommand{\p}{\partial}
\newcommand{\Tr}{{\rm Tr}}

\newcommand{\bea}{\begin{eqnarray}}
\newcommand{\eea}{\end{eqnarray}}
\def\Tr{ \hbox{\rm Tr}}
\def\tr{ \hbox{\rm tr}}
\def\de{\partial}
\def\im{\hbox{\rm Im}}

\newcommand{\bqa}{\begin{eqnarray}}
\newcommand{\beas}{\begin{eqnarray*}}
\newcommand{\eeas}{\end{eqnarray*}}

\newcommand{\bquo}{\begin{quote}}
\newcommand{\enqu}{\end{quote}}
\newcommand{\De}{\partial}
\newcommand{\te}{\theta}

\newcommand{\exl}{\left<}
\newcommand{\exr}{\right>}

\newcommand{\Pf}{\mathrm{Pf}}
\newcommand{\m}{\mathfrak{m}}
\newcommand{\W}{\mathcal{W}}
\newcommand{\V}{\mathrm{v}}

\def\p{{}^{\prime}}

\newcommand{\bt}{\bar{\theta}}
\newcommand{\da}{\dot{\alpha}}

\newcommand{\sm}{\sigma^{\mu}}

\newcommand{\dg}{\dagger}

\newcommand{\F}{\mathcal{F}}
\newcommand{\LL}{\mathcal{L}}
\newcommand{\wt}{\widetilde}
\newcommand{\MM}{\mathcal{M}}

\def\de{\partial}
\def\Tr{ \hbox{\rm Tr}}
\def\tr{ \hbox{\rm tr}}

\def\im{\hbox{\rm Im}}

\def\Re{\hbox {\rm Re}}

\def\diag{\hbox{\rm diag}}

\begin{document}
\frontmatter

\begin{titlepage}
\begin{center}
\AddToShipoutPicture*{%
  \AtTextCenter{%
    \makebox(0,0)[c]{\resizebox{\textwidth}{!}{%
      \includegraphics[width=\textwidth]{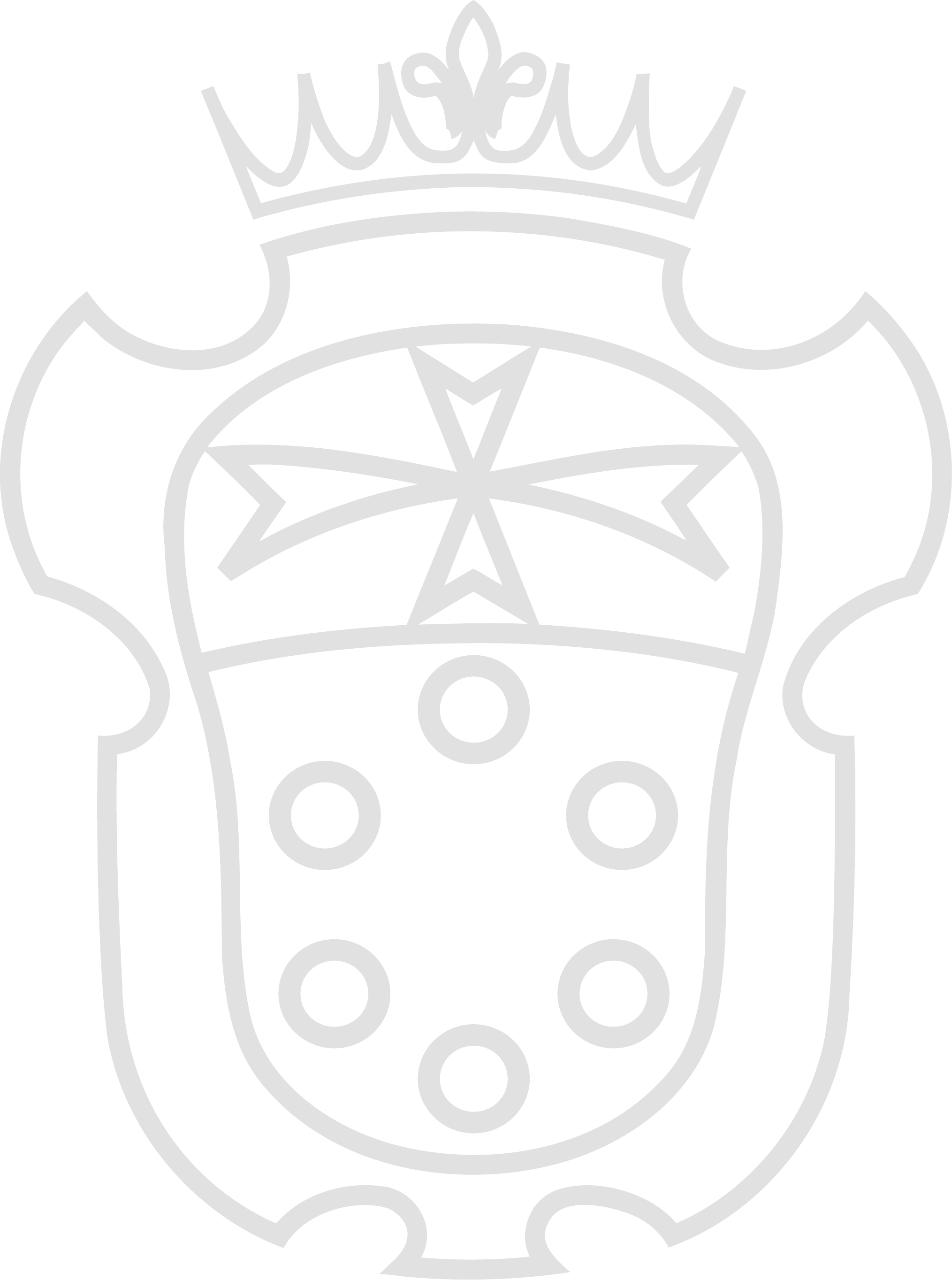}}}}}%
      
{\Huge\textbf{\textsc{Scuola Normale Superiore}}}

\vspace{0.5cm}
\textbf{\Large Classe di Scienze}\\

\vspace{0.5cm}
\makebox[0pt][c]{\textbf{\textsc{\LARGE {Corso di Perfezionamento in Fisica}}}}\\
\rule{0.5\linewidth}{0.2 mm}\\

\vspace{3cm}
\begin{minipage}{1.0\textwidth}{\begin{center}\Huge
{\textsc{\textbf{Confinement and duality in supersymmetric gauge theories}}}
\end{center}}\end{minipage}

\vspace{1.5cm}
\textbf{\Large Tesi di Perfezionamento}\\

\vspace{2.5cm}
\LARGE
\textbf{\textbf{Simone \textsc{GIACOMELLI}}}

\end{center}

\vspace{2cm}
\begin{center}
\Large					
\textbf{\textbf{\emph{Relatore:}}}	\textbf{{Prof.~Kenichi \textsc{KONISHI}}}

\end{center}

\vspace*{\fill}
\begin{center}
\rule[2mm]{0.8\linewidth}{0.2 mm}\\
\Large
\textbf{\textsc{Anno Accademico 2012-2013}}
\normalsize
\end{center}
\end{titlepage}

\thispagestyle{empty}
\
\newpage

\chapter*{Acknowledgements}

I would like to thank first of all my advisor Kenichi Konishi for all his support during these years, starting from my 
undergraduate course, and for having introduced me to this beautiful field of research. He constantly encouraged me and actively 
contributed in the development of this project. He will always be of great inspiration for me and I really owe him a lot. 

\noindent This thesis 
includes most of my work as a graduate student at Scuola Normale. It is a pleasure for me to acknowledge this beautiful 
institution that gave me so much during the past eight years, providing a stimulating environment in which my passion for science 
has grown constantly. I would also like to thank professor Augusto Sagnotti for many precious comments and suggestions. 

\noindent I am grateful to all my collaborators and to various colleagues, in particular Alessandro Tanzini, Giulio Bonelli, Michele Del 
Zotto and Sergio Cecotti. I really benefited from many illuminating discussions with them. I would also like to thank Yuji Tachikawa 
for many helpful comments on my work.

\noindent I acknowledge the Center for the Fundamental Laws of Nature of Harvard University, The Scuola Internazionale Superiore di Studi 
Avanzati and IPMU, where part of this work was done, for hospitality.

\noindent I would also like to thank my family and my friends for all their support and for encouraging me. A special thank 
goes to my girlfriend Anna.

\newpage 
\thispagestyle{empty}
\
\newpage

\tableofcontents
\mainmatter
\chapter*{Introduction}
\addcontentsline{toc}{chapter}{Introduction}

Gauge symmetry is ubiquitous in physics and is one of the key ingredients underlying the dynamics of elementary particles. The Standard Model, the theory which unifies
the present-day particle physics is described by a lagrangian invariant under $SU(3)\times SU(2)\times U(1)$ local transformations,
where $SU(2)\times U(1)$ is the gauge group associated to electroweak forces and $SU(3)$ is the gauge group associated to strong interactions (the color group)

From a theoretical point of view, the model for strong interactions (QCD) remains very challenging, despite the fact that it has been formulated 
almost fourty years ago. One of the most remarkable properties of the model is asymptotic freedom: the theory is weakly coupled
at very high energy, making it tractable with standard perturbative techniques in that regime, whereas it becomes strongly interacting
at low energies. Since there is no general technique that allows to approach a strongly coupled field theory, it has so far been impossible
to follow analytically the flow from high to low energies (RG flow) and understand in a precise way the properties of the theory 
in the strongly coupled regime.

Both from an experimental and theoretical (and numerical) point of view, there are strong indications that QCD in the infrared limit is
characterized by confinement and chiral symmetry breaking and that these two phenomena are deeply related, being originated by the
same mechanism. 

Confinement means that the elementary fields in QCD charged under the color group (the quarks) tend to form bound states at low energies
since their mutual interaction becomes stronger and stronger and these bound states (hadrons) become the effective dynamical
variables in the infrared. Chiral invariance refers instead to the symmetry that QCD acquires in the limit of massless quarks.
This is a very reasonable approximation as long as just the two lightest quarks are considered. Furthermore, observations suggest that this approximate
symmetry of nature is spontaneously broken by a non vanishing condensate $\langle\bar{\psi}\psi\rangle\neq0$. 
Neither of these properties can be explained in the framework of standard perturbation theory and thus must be related
to some still unknown nonperturbative effect. Understanding precisely the infrared dynamics of QCD is thus one of the most challenging open problems
in field theory.

Due to the complexity of the theory of strong interactions it is important to approach the problem starting from simpler models that
can help understanding the main features and the mechanism underlying confinement and symmetry breaking. In this respect supersymmetric
non-abelian gauge theories play a leading role and have been for decades a source of ideas for the exploration of the strongly-coupled gauge dynamics.

There is a distinguished subclass of supersymmetric gauge theories, namely those with extended supersymmetry ($\mathcal{N}=2$ theories) in which
it is almost always possible to find a dual weakly coupled description (electric-magnetic duality), allowing thus to determine the structure of the low energy
effective action and consequently the infrared dynamics of the theory (Seiberg-Witten solution) \cite{SWI,SWII}. One of the most remarkable byproducts
of this construction is the presence of massless monopoles and dyons which are best described as solitons in the original description of the theory
(the description used to analyze the theory in the UV). The theory in the infrared can be essentially understood in terms of these
distinguished particles.

The above mentioned models are just distant relatives of QCD and many key properties are different; for instance $\mathcal{N}=2$ theories are not confining,
as opposed to QCD. Nevertheless, just breaking softly extended supersymmetry to $\mathcal{N}=1$, one can still use many of the properties derived
for the undeformed theory but at the same time flows to a model in which confinement is expected. The outcome is that the massless
particles of the $\mathcal{N}=2$ theory condense, leading to confinement via the 't Hooft-Mandelstam mechanism \cite{TP}-\cite{NM2}: as the condensation
of electrically charged particles in a superconductor confines magnetically charged objects \cite{abri}, the condensation of magnetic monopoles
in these theories form a sort of dual superconductor in which electrically charged particles are confined. Furthermore, when quark fields
are introduced in the theory these massless particles acquire flavor charges via the Jackiw-Rebbi effect \cite{JR}-\cite{JRII}, thus explaining what is the relation (at least for this
class of theories) between confinement and chiral symmetry breaking.

We can see from the above discussion that the study of supersymmetric models is extremely helpful in addressing problems essentially
related to the details of the dynamics since we have no tools to address them directly in QCD, where all the constraints coming from
supersymmetry are not available. The idea is thus to learn as much as possible about strong dynamics in the supersymmetric case
and then to figure out how analogous mechanisms can be at work in the QCD case \cite{stra,straI} (see also \cite{EHsu}-\cite{LAZ2}).

Let us discuss now the picture that has emerged so far from the study of supersymmetric theories (we will discuss mainly the $SU(N)$
case) with softly broken $\mathcal{N}=2$ supersymmetry. In the pure gauge case (in the absence of quarks) it has been shown that the theory dynamically abelianizes: despite the 
lagrangian is invariant under a non-abelian group, the effective theory in the infrared is invariant just under its maximal abelian
subgroup due to quantum corrections \cite{DouglasShenker} (see also \cite{auz1}). The study of these vacua is by now standard since a weakly coupled description can always be
found making use of electric-magnetic duality, which is well understood in the abelian case.

More interesting is the situation in the theory with flavors, where vacua which do not abelianize in the infrared indeed exist 
\cite{APS,CKM}. Also in this case confinement is realized via the 't Hooft-Mandelstam mechanism, but this time it is less clear how
the electric-magnetic duality works: in the abelian case it just amounts to a change of variables in the path integral whereas in the non-abelian
case this procedure cannot be carried out. 

The most established examples of such a duality are Seiberg duality in $\mathcal{N}=1$
SQCD \cite{SD} and its analogue for the models we have been discussing so far, called Kutasov duality \cite{KD}. 
In both cases the duality requires the introduction of a new set of dynamical variables, which include magnetic variables of 
a non-abelian kind, and involves a change in the gauge group. The validity of these dualities is
still a conjecture but they have anyway remarkably passed numerous checks and led to many significant results (see e.g. 
\cite{ind1}-\cite{ISS} and \cite{srev} for a review). 

Many aspects of these theories are still unclear and deserve further investigations, especially due to the fact that the observed 
degeneracies in the hadron spectrum suggest that the right answer for QCD is not abelianization (see e.g. 
\cite{Konishi:2005qt}-\cite{Kenn} and references therein). Motivated by these considerations, in this thesis we try to better 
understand the properties of softly broken $N=2$ SQCD and the possible mechanisms leading to confinement and chiral symmetry 
breaking. 

This thesis is based on \cite{noi}-\cite{CMS} and is organized as follows: the first two chapters contain some review material 
which is used in the main body of the thesis (part of chapter 1 appears also in \cite{noi2}). Chapter 1 contains some basic 
material, including a brief review 
of supersymmetry representations and supersymmetric field theories in four dimensions, which is the case of interest for us. 
We then analyze in some detail the properties of $\mathcal{N}=2$ theories and describe the original argument given by Seiberg 
and Witten in \cite{SWI} for the determination of the infrared effective action for $SU(2)$ SYM. Chapter 2 collects more 
recent and advanced results about $\mathcal{N}=2$ theories which are relevant for the present work. We discuss in particular the brane 
realization of these models in type IIA/M-theory and the more recent six-dimensional construction.
Chapters 3-5 are based on my original work and summarize my results. Every chapter contains an introductory section in which I 
explain in detail the setup. 

Chapter 3 is based on \cite{noi,noi2} and is devoted to the study of $N=2$ SQCD, softly broken by a mass term for the chiral field in the adjoint representation of the gauge group.
The basic techniques that have been used in earlier works on the subject are essentially the classical analysis using the equations
of motion and the nonperturbative one using Seiberg-Witten theory. Looking at the details of the theory, it turns out that the first
approach is reliable just for very large masses of the flavors whereas the second, altough in principle always applicable, can be used
just in the small mass limit, due to technical complications that make the method almost intractable in the general case. 

We approach this problem using the generalized Konishi anomaly and the Dijkgraaf-Vafa superpotential. This method has been applied to similar models
and some of our results are very similar. Anyway, what has not been emphasized in previous works is that this 
technique allows to extract more information than other methods about the generic mass case, making it possible to follow the vacua
from weak to strong coupling. As a Byproduct, we are able to shed new light on the relation between semiclassical and quantum vacua, 
which involves an infrared duality very close to Seiberg duality in $\mathcal{N}=1$ SQCD \cite{BK}.

Another interesting result is that for a particular value of the mass some of the vacua merge in a superconformal point, signalling
the transition between Higgs and confining phase. Understanding the low energy dynamics at this point is a nontrivial task, since 
the theory is strongly interacting and the Seiberg-Witten curve becomes singular, making it difficult to extract  any precise 
information. A very similar class of singular points has been considered from a different perspective in \cite{Bolo}. The content of this chapter can be seen as a preliminary work that allows to 
understand the following steps of my analysis. From this work one can extract a number of nontrivial results such as the pattern 
of dynamical flavor symmetry breaking.

Chapter 4 is based on \cite{Simone} which is devoted to the analysis of singular points (where the Seiberg-Witten curve degenerates) in the moduli space of 
$\mathcal{N}=2$ SQCD, with particular attention to the points which are not lifted by the $\mathcal{N}=1$ perturbation and are thus 
relevant for the study of confinement in these models. As I mentioned before, it is difficult to understand the infrared dynamics 
in these cases, since the SW curve does not lead directly to a weakly coupled action and the theory is often intrinsically interacting. 
In order to shed new light on this problem it is necessary to make use of the most recent results on $\mathcal{N}=2$ 
superconformal theories \cite{G,DT}. 

For our purposes it is particularly important the analysis performed in
\cite{GST}, in which the authors argued that the low-energy physics at the above mentioned singular points in $SU(N)$ SQCD 
involves two (possibly free) superconformal sectors. In \cite{Simone} I essentially generalized to $SO$ and $USp$ SQCD this 
analysis, recovering an analogous structure. As we will see, it turns out that this observation is fundamental in order to get 
new insight about the mechanism underlying confinement in $SO$ and $USp$ theories, in which typically the relevant points in the 
moduli space are singular. 

The second part of the chapter contains a revised version of my contribution to \cite{CMS} and is devoted to explain how it is possible to make use of the results presented in 
\cite{CM}, in which the authors define a broad class of $\mathcal{N}=2$ SCFTs exploiting their type IIB superstring realization, 
to derive information about the BPS spectrum of IR fixed points in $\mathcal{N}=2$ SQCD. This is done by carefully matching the 
SW curves associated to the theories constructed in the above mentioned paper. In the last section I discuss the properties of 
some of the infrared fixed points in quiver gauge theories, always exploiting the construction presented in \cite{CMS}. 

In chapter 5 we come to our final payoff: the results of chapter 4 allow us to explicitly study the mechanism underlying 
confinement in $USp$ and $SO$ theories (in some cases), where the standard 't Hooft-Mandelstam mechanism does not seem to work. Indeed we will 
see that, although confinement is realized as expected by the condensation of magnetically charged objects, the details of the 
confining mechanism are rather unusual, making it possible to reproduce all the semiclassical expectations which might seem quite 
hard to combine at first looking \cite{Konishi:2005qt}. 

This can be also important for the study of confinement in the context of QCD, where the assumption that the underlying mechanism is given 
by the 't Hooft-Mandelstam scenario is known to lead to various difficulties (doubling of the meson spectrum, excess in the number 
of Goldstone bosons...) \cite{Kenn}. This chapter is based on \cite{noi2,Ion} 
and is devoted to the study of the low-energy dynamics at singular points in $USp(2N)$ and $SU(N)$ SQCD with four flavors, 
$SO(2N)$ SQCD with two flavors and $SO(2N+1)$ SQCD with one flavor. 

We are still not able to approach the general case and the 
restriction on the number of flavors comes from the fact that in these special cases we are able, exploiting the analysis performed 
in chapter 4, either to identify a weakly coupled dual description or to find a description in terms of a strongly interacting theory 
that we know well enough to extract the information we are interested in (such as chiral condensates). In all the cases our 
results are perfectly consistent with the semiclassical results obtained using the techniques presented in chapter 3 (pattern of 
dynamical flavor symmetry breaking, number of vacua...).

At the initial stage of my PhD I also tried to apply the ideas emerging from supersymmetric gauge theories to nonsupersymmetric Yang-Mills theory; 
in particular I focused on the Faddeev-Niemi decomposition \cite{FN}. In this paper the authors claim that $SU(2)$
Yang-Mills theory is on-shell equivalent to an abelian theory with a unit three vector and a complex scalar as matter fields. 
This reformulation aims at rewriting the theory in terms of a set of variables suited to investigate the scenario of dynamical abelianization in Yang-Mills theory, 
as is the case in supersymmetric theories, and has received much attention in the past years especially for lattice simulations. 

However, this reformulation does not allow to recover the non abelian Gauss' constraints of the original theory. At a closer inspection I
found out in collaboration with Jarah Evslin that their claim is actually wrong and some solutions of the Faddeev-Niemi equations do not solve the above mentioned constraints
and thus are not solutions of Yang-Mills equations \cite{FNNoi}. I then formulated in collaboration with Jarah Evslin, Kenichi Konishi and Alberto Michelini a similar 
decomposition for $SU(3)$, suited for investigation of
the non-abelian scenario (we rewrite the theory in terms of $SU(2)\times U(1)$ variables) \cite{FNnoi2}. 

In order to avoid the problems we found in the original
proposal by Faddeev and Niemi, we constructed a decomposition such that only a partial gauge fixing is implied, without affecting 
the Gauss' constraints.
Our parametrization includes all the solutions of Yang-Mills equations of physical interest such as monopoles, Witten's generalized
instantons \cite{Witsol} and merons. Furthermore, since only a partial gauge fixing is implied, our formula can be used to perform path integral
computations. Finally, we reproduce the no-go theorem on non-Abelian monopoles (the so-called topological obstruction) in the pure Yang-Mills 
theory \cite{CDI}-\cite{CDIV}.  Also, we show that the knot-solitons discussed by Faddeev and Niemi in \cite{nodi1,nodi2} do not exist if the system does not 
dynamically Abelianize. These topics have not been included in the present thesis.

\chapter{Supersymmetric field theories and the Seiberg-Witten solution}

In this chapter we briefly review the basic concepts of supersymmetric field theories and the Seiberg-Witten solution of $\mathcal{N}=2$ 
gauge theories. This is not a comprehensive review of supersymmetry and for further details the reader is referred to the many good 
reviews available in the literature. Those which are closer to my selection of topics are \cite{G-H,Bilal}. In the first two 
sections we introduce the supersymmetry algebra and describe how to build supersymmetric lagrangians. Most of the material is 
taken from \cite{Weinberg,W-B}. This is the starting point for the analysis performed by Seiberg and Witten in \cite{SWI} which 
is the central topic of section three. In section four we review how the Seiberg-Witten solution allows to shed light on 
confinement and summarize the available results in the literature on this topic. Since this is the central theme of the present 
work, the material in this section represents the starting point for our analysis.

\section{Supersymmetric theories in four dimensions}

\subsection{Notation and conventions}

We will make use of the notation adopted in \cite{W-B}. Our convention for the Minkowsky metric is $\eta_{\mu\nu}=\text{diag}(1,-1,-1,-1)$ 
and we will indicate left and right spinors with dotted and undotted indices respectively.
In order to raise and lower indices we will use the antisymmetric tensor $\varepsilon$, $$\varepsilon^{\alpha\beta}=\varepsilon^{\dot{\alpha}\dot{\beta}}= \left(\begin{array}{ll}
0 & 1 \\
-1 & 0 \\
\end{array}\right)=\imath \sigma_{2}.$$ Let us now define $$(\sigma^{\mu})_{\alpha\dot{\alpha}}\equiv(1,\vec{\sigma}),$$ 
$$(\bar{\sigma}^{\mu})^{\dot{\alpha}\alpha}=-(\sigma^{\mu})^{\alpha\dot{\alpha}}=\varepsilon^{\dot{\alpha}\dot{\beta}}
\varepsilon^{\alpha\beta} (\sigma^{\mu})_ {\beta\dot{\beta}}=(1,-\vec{\sigma}).$$ 
With these conventions, Lorentz transformations are generated by 
\[\begin{array}{l}
\displaystyle (\sigma^{\mu\nu})_{\alpha}^{\beta}=\frac{1}{4}\left[\sigma^{\mu}_{\alpha\dot{\beta}}\bar{\sigma}^{\nu\dot{\beta} \beta}-(\mu\leftrightarrow\nu)\right],\\
\displaystyle (\bar{\sigma}^{\mu\nu})^{\dot{\alpha}}_{\dot{\beta}}=\frac{1}{4}\left[\bar{\sigma}^{\mu\dot{\alpha}\beta}\sigma^{\nu}_{\beta\dot{\beta} }-(\mu\leftrightarrow\nu)\right].\\   
\end{array}\]
For the product of spinors we will use the following conventions: \[\begin{aligned}
\psi\chi &= \psi^{\alpha}\chi_{\alpha}=-\psi_{\alpha}\chi^{\alpha}=\chi^{\alpha}\psi_{\alpha}=\chi\psi,\\
\bar{\psi}\bar{\chi}&=\bar{\psi}_{\dot{\alpha}}\bar{\chi}^{\dot{\alpha}}=\bar{\chi}\bar{\psi} ,\\
(\psi\chi)^{\dagger}&=\bar{\chi}_{\dot{\alpha}}\bar{\psi}^{\dot{\alpha}}=\bar{\chi}\bar{\psi}=\bar{\psi}\bar{\chi} .\\
\end{aligned}\]
In this notation $\gamma$ matrices, Dirac and Majorana spinors can be written as follows: \[\gamma^{\mu}=\left(\begin{array}{ll}
0 & \sigma^{\mu}\\\bar{\sigma}^{\mu}\\                                                                                                  \end{array}\right),\qquad \psi_{D}=\left(\begin{array}{l}
\psi_{\alpha}\\
\bar{\chi}^{\dot{\alpha}}\\
\end{array}\right),\qquad \psi_{M}=\left(\begin{array}{l}
\psi_{\alpha}\\
\bar{\psi}^{\dot{\alpha}}\\
\end{array}\right).\] 
The following identities hold: \[\begin{aligned}
&\chi\sigma^{\mu}\bar{\sigma}^{\nu}\psi =\psi\sigma^{\nu}\bar{\sigma}^{\mu}\chi,\quad
(\chi\sigma^{\mu}\bar{\sigma}^{\nu}\psi)^{\dagger}= \bar{\psi}\bar{\sigma}^{\nu}\sigma^{\mu}\bar{\chi},\quad\\
&\chi\sigma^{\mu}\bar{\psi}=-\bar{\psi}\bar{\sigma}^{\mu}\chi ,\quad
(\chi\sigma^{\mu}\bar{\psi})^{\dagger}=\psi\sigma^{\mu}\bar{\chi},\quad
\end{aligned}\]

\subsection{Supersymmetry algebra}

The SUSY algebra can be written as follows \cite{HLS}:
\begin{equation}\label{cao}
\begin{array}{l}
\displaystyle{\{Q_{\alpha}^A , \bar{Q}_{\dot\beta B}\} = 2\sigma_{\alpha\dot\beta}^\mu P_\mu \delta_B^A},\\
\displaystyle{\{Q_{\alpha}^A , {Q}_{\beta}^B\} = 2\sqrt{2}\varepsilon_{\alpha\beta}Z^{AB}} \vphantom{\sigma_{\alpha\dot\beta}^\mu},\\
\displaystyle{\{\bar{Q}_{\dot\alpha A} , \bar{Q}_{\dot\beta B}\} = 2\sqrt{2}\varepsilon_{\dot\alpha\dot\beta}Z^{*}_{AB}}.
\end{array}
\end{equation}
$Q$ and $\bar{Q}$ are the generators of supersymmetry transformations (supercharges) and transform as 
operators of spin $1/2$ under the Lorentz group. Indices $A$ and $B$ run from 1 to N, where N is the number of supersymmetries. 
The supersymmetry charges commute with $P^{2}$, so all the states in a given representation of the SUSY algebra have the same mass. 
The operators $Z$ e $Z^{*}$, antisymmetric in the $A,B$ indices, are called central charges. Clearly they can be nonzero only 
if we have at least two supersymmetries. Their presence will be relevant below, when we will discuss theories with eight supercharges.
A field theory will be supersymmetric if the set of its states fall in representations of the algebra (\ref{cao}). 
We will now briefly review the properties of the SUSY algebra representations, specializing to the four-dimensional case.

\subsubsection{Massive irreducible representations}

Let us discuss first the case without central charges. In the massive case we can go to the rest frame, in which $P_{\mu}=(M,0,0,0)$ and define
$$a^{A}_{\alpha}=Q^{A}_{\alpha}/\sqrt{2M},\qquad (a^{A}_{\alpha})^{\dagger}=\bar{Q}_{\dot{\alpha}A}/\sqrt{2M}.$$ The SUSY algebra then becomes 
$$\{a^{A}_{1},(a^{B}_{1})^{\dagger}\}=\delta^{AB},\qquad \{a^{A}_{2},(a^{B}_{2})^{\dagger}\}= \delta^{AB},$$ 
The vacuum $\vert\Omega\rangle$ is defined by the relation $a^{A}_{\alpha}\vert\Omega\rangle=0$ and the representation can be constructed 
applying the operators $(a^{A}_{\alpha})^{\dagger}$ to the vacuum. 
If $\vert\Omega\rangle$ has spin zero, it is easy to see that we can construct $\binom{2N}{m}$ distinct states applying $m$ 
raising operators. The representation thus has dimension \[ \sum^{2N}_{m=0}\binom{2N}{m}=2^{2N}.\] and the maximum spin in the multiplet is $N/2$. 
For example, an $\mathcal{N}=1$ multiplet describes $2^{2}=4$ states, $$\vert\Omega\rangle,\qquad a^{\dagger}_{\alpha}\vert\Omega\rangle,\qquad \frac{1}{\sqrt{2}}\varepsilon^{\alpha\beta}a_{\alpha}^{\dagger}a_{\beta}^{\dagger}\vert\Omega\rangle.$$ 
The spin content is $(0)\oplus(1/2)\oplus(0)$. If the vacuum $\vert\Omega_{s}\rangle$ has instead spin $s$ the representation will include $(2s+1)2^{2N}$ states. 
For instance, in the $\mathcal{N}=1$ case the SUSY multiplet includes particles with spin $(s)\oplus (s+1/2)\oplus (s-1/2)\oplus(s)$.
In any case the number of bosonic and fermionic degrees of freedom is the same.

\subsubsection{Massless irreducible representations}

In the massless case we can go to a reference frame in which $P_{\mu}=E(1,0,0,1)$ and reduce the SUSY algebra to: \[ \{ Q^{A}_{\alpha},\bar{Q}_{\dot{\alpha B}}\}= 
\left(\begin{array}{ll}
0 & 0\\
0 & 4E\\\end{array}\right)\delta^{A}_{B}.\] Since the anticommutator of $Q^{A}_{\alpha}$ and $\bar{Q}_{\dot{\alpha} A}$ 
is a positive operator, from the relation $\{ Q_{1}^{A}, \bar{Q}_{\dot{1} B}\}= 0$ we can conclude that 
both $Q_{1}^{A}$ and $\bar{Q}_{\dot{1}A}$ annihilate all physical states. The algebra thus contains only 
$\mathcal{N}$ nontrivial supercharges. We can now define as before the raising and lowering operators $$a^{A}=\frac{1}{2\sqrt{M}}
Q_{2}^{A},\qquad (a^{A})^{\dagger}=\frac{1}{2\sqrt{M}}\bar{Q}^{A}_{\dot{2}}.$$ 
The SUSY algebra can thus be rewritten in the form: $$\{a^{A},(a^{B})^{\dagger}\}=\delta^{AB},\qquad 
\{a^{A},a^{B}\}=0,\qquad\{(a^{A})^{\dagger},(a^{B})^{\dagger}\}=0.$$ This is a Clifford algebra with $2\mathcal{N}$ generators 
and the representation will have dimension $2^{N}$. The operators $a^{A}$ increase the helicity 
of massless states by $1/2$ whereas the operators $(a^{A})^{\dagger}$ decrease it by the same quantity. We define as before 
the vacuum as the state $\vert\Omega\rangle$ annihilated by all the operators $a^{A}$. We are free to assume that it is an eigenstate 
of $J_{z}$. The other states in the multiplet are generated applying the operators $(a^{A})^{\dagger}$ to $\vert\Omega\rangle$. 
In this way we can build $\binom{N}{m}$ states applying $m$ lowering operators, all with helicity $J_{z}\vert\Omega\rangle-m/2$.
Notice that the representations obtained in this way are not necessarily CPT invariant so, if we want to build a physical theory 
we must add the CPT conjugate states. 

Indicating with $\lambda$ the helicity of the vacuum, let us notice that for $\mathcal{N}=1$ a massless representation 
contains a Majorana spinor and a complex scalar if $\lambda=1/2$; if $\lambda=1$ we get a 
Majorana spinor and a massless vector. For $\mathcal{N}=2$ and $\lambda=1/2$ the multiplet includes two Majorana spinors and two complex 
scalars, i.e. two copies of the corresponding $\mathcal{N}=1$ multiplet. If $\lambda=1$ we get a massless vector, two majorana 
spinors and a complex scalar, which is the content of the $\mathcal{N}=1$ multiplets with $\lambda=1$ and $\lambda=1/2$. 
For $\mathcal{N}=4$ and $\lambda=1$ the representation is CPT selfconjugate and contains a massless vector, four spinors and three complex scalars. This 
is the field content of the $\mathcal{N}=2$ multiplets with $\lambda=1$ and $\lambda=1/2$ together. We cannot have more than 16 
supercharges without introducing gravity.

\subsubsection{Central charges and BPS states}

Let us focus on the $N$ even case: modulo a unitary transformation we can assume that the central charge matrices are of the form 
$Z=\varepsilon\otimes D$ with $D$ a diagonal matrix. We can thus focus on the $\mathcal{N}=2$ case and defining now 
$$a_{\alpha}=\frac{1}{2}\left(Q^{1}_{\alpha}+\varepsilon_{\alpha\beta}(Q^{2}_{\beta})^{\dagger}\right),\qquad 
b_{\alpha}=\frac{1}{2}\left(Q^{1}_{\alpha}-\varepsilon_{\alpha\beta}(Q^{2}_{\beta})^{\dagger}\right),$$ the algebra (\ref{cao}) 
reduces to
\begin{equation}\label{z}
\{a_{\alpha},a_{\beta}^{\dagger}\}=\delta_{\alpha\beta}(M+\sqrt{2}Z),\qquad\{b_{\alpha},b_{\beta}^{\dagger}\}=\delta_{\alpha\beta}(M-\sqrt{2}Z). 
\end{equation}
We can thus immediately deduce the relation $M\geq\sqrt{2}\vert Z\vert$ (in particular all massless states have $Z=0$). 
When $M=\sqrt{2}\vert Z\vert$, either $\{a,a^{\dagger}\}$ or $\{b,b^{\dagger}\}$ in (\ref{z}) are zero and the dimension 
of the representation decreases, since the states in the multiplet are annihilated by a subset of the supercharges. In the 
case $N=2$, a reduced (or BPS) massive multiplet has the same dimension as a massless one.

\subsection{Superspace and superfields}

In order to write down the lagrangian for supersymmetric theories it is convenient to adopt the superspace and superfield formalism. 
We will now briefly review these topics.

\subsubsection{Superspace}

The superspace is defined adding anticommuting Grassmann variables $\theta^{\alpha},\bar{\theta}_{\dot{\alpha}}$ to the spacetime 
coordinates $x^{\mu}$ and can thus be identified with the set of triples $(x,\te,\bt)$ (we will from now on focus on the 
$\mathcal{N}=1$ superspace in four dimensions). We indicate with $\te\te$ e $\bt\bt$ the expressions $\te^{\alpha}\te_{\alpha}=-2\te^{1}\te^{2}$ and $\bt_{\da}\bt^{\da}=2\bt_{\dot{1}}\bt_{\dot{2}}$. 
Introducing auxiliary anticommuting parameters $\xi$ e $\bar{\xi}$ we can rewrite (\ref{cao}) in terms of $\xi Q$ e $\bar{\xi}\bar{Q}$, obtaining an algebra which involves commutation relations only. We can now write a SUSY transformation with parameters 
$\xi$ and $\bar{\xi}$ just exponentiating: $$G(x,\te,\bt)=e^{\imath(-x^{\mu}P_{\mu}+\te Q+\bt\bar{Q})}.$$ We thus obtain 
$$G(0,\xi,\bar{\xi})G(x,\te,\bt)=G(x^{\mu}+\imath\te\sm\bar{\xi}-\imath\xi\sm\bt,\te+\xi,\bt+\bar{\xi}),$$ so under a SUSY 
transformation $\xi,\bar{\xi}$ the superspace coordinates transform as: \begin{equation}\label{x}
\begin{aligned}
x^{\mu}&\rightarrow x^{\mu}+\imath\te\sm\bar{\xi}-\imath\xi\sm\bt,\\
\te &\rightarrow \te+\xi,\\
\bt &\rightarrow \bt+\bar{\xi}.\\
\end{aligned}\end{equation}
We can now easily write the supercharges as differential operators on the superspace:
\begin{equation}
Q_{\alpha}=\frac{\de}{\de\te^{\alpha}}-\imath\sm_{\alpha\da}\bt^{\da}\de_{\mu},\qquad \bar{Q}_{\da}=-\frac{\de}{\de\bt^{\da}}+\imath\te^{\alpha}\sm_{\alpha\da}\de_{\mu}. 
\end{equation}
Let us finally introduce the super derivatives
\begin{equation}
D_{\alpha}=\frac{\de}{\de\te^{\alpha}}+\imath\sm_{\alpha\da}\bt^{\da}\de_{\mu},\qquad \bar{D}_{\da}= -\frac{\de}{\de\bt^{\da}}-\imath\sm_{\alpha\da}\te^{\alpha}\de_{\mu}.
\end{equation}
which satisfy the relation $\{D_{\alpha},\bar{D}_{\da}\}=-2\imath\sm_{\alpha\da}\de_{\mu}$. We introduce them because they
anticommute with $Q$ and $\bar{Q}$ and this will make it easier to write down supersymmetric lagrangians.

\subsubsection{Superfields}

A superfield is a function defined on the superspace $F(x,\te,\bt)$ (defined in terms of its expansion in powers of the anticommuting variables) \[
F(x,\te,\bt)=f(x) + \theta \varphi(x) + \bar\theta\chi(x) + \theta\theta m(x) +  \bar\theta\bar\theta n(x) +  \theta\sigma^{\mu}\bar\theta v_{\mu}(x)\]\[ + \theta\theta\bar\theta\bar\lambda(x) + \bar\theta\bar\theta\theta \psi(x) + \theta\theta\bar\theta\bar\theta d(x).\] 
A supersymmetry transformation acts on super fields as $\delta F=(\xi Q+\bar{\xi}\bar{Q})F$. From this we can read the 
transformation properties of the various components. Notice that the variation of the highest component 
of a superfield is always a total derivative. The spacetime integral of this component will then be invariant under supersymmetry 
transformations. This is basically how supersymmetric lagrangians are constructed, as we will now see.
 
Clearly every function of superfields is a superfield and since the algebra of superfields is closed under SUSY transformations 
they give a representation of the SUSY algebra. However, this will be reducible and in order to obtain irreducible representations 
we have to impose some constraints. The most common (and the only ones we will use) are:
\begin{itemize}
\item $\bar{D}_{\da}\Phi=0$ which defines \textbf{Chiral Superfields}:\\
It is convenient to define $y^{\mu}= x^{\mu}+\imath \te\sm\bt$, so that the above condition becomes $$\frac{\de}{\de\bt^{\da}}\Phi=0.$$
This simply tells that a chiral superfield is a function of $y$ and $\te$ only and thus is of the form 
$$\Phi = A(y) + \sqrt{2}\theta\psi(y) + \theta\theta F(y),$$ or in terms of the original coordinates
\begin{equation}\label{cc}
\begin{array}{l}
\begin{aligned}
 \Phi(x,\te,\bt)=&A(x)+\imath\te\sm\bt\de_{\mu}A-\frac{1}{4}\te^{2}\bt^{2}\Box A
+\sqrt{2}\te\psi(x)\\&-\frac{i}{\sqrt{2}}\te\te\de_{\mu}\psi\sm\bt
+\te\te F(x).
\end{aligned}
\end{array}
\end{equation} This superfield describes the matter content of a $\mathcal{N}=1$ scalar multiplet. Obviously the product 
of chiral superfields defines a chiral superfield. This fact will allow to include in the lagrangian arbitrary polynomial 
functions of chiral superfields and introduce the so-called superpotential. The only general constraint is that the superpotential should 
be holomorphic.

\item $V=V^{\dagger}$ which gives the so-called \textbf{vector superfields}:
They are used to describe $\mathcal{N}=1$ gauge multiplets. The supersymmetric counterpart of gauge transformations is given (in the abelian case) by 
$V\rightarrow V+\Lambda+\Lambda^{\dg}$, where $\Lambda$ is a chiral superfield. Notice that $V\vert_{\te\sm\bt}=A_{\mu}\rightarrow 
A_{\mu}+\imath\de_{\mu}(A-A^{\dg})$ (where $A$ is the lowest component of $\Lambda$). Some components of our superfield can be set with a suitable choice of $\Lambda$ to zero. 
In particular we can impose the Wess-Zumino gauge in which the vector superfield becomes 
$$V=-\te\sm\bt A_{\mu}+\imath\te^{2}\bt\bar{\lambda}-\imath\bt^{2}\te\lambda+\frac{1}{2}\te^{2}\bt^{2}D.$$ Notice that $V^3=0$ 
in this gauge. The field strength is then defined by  
\begin{equation}
W_{\alpha}=-\frac{1}{4}\bar{D}^{2}D_{\alpha}V,\qquad \bar{W}_{\da}=-\frac{1}{4}D^{2}\bar{D}_{\da}V,
\end{equation} which is a gauge invariant chiral superfield.

The generalization to the non-abelian case is straightforward: we will have $V=V_{a}T^{a}$ ($T^a$'s are the group generators), the 
gauge transformations are given by $$e^{-2V}\rightarrow e^{-\imath\Lambda^{\dg}}e^{-2V}e^{\imath\Lambda},$$ where 
$\Lambda=\Lambda_aT^a$ and the field strength can be written as 
\begin{equation}
W_{\alpha}=\frac{1}{8}\bar{D}^{2}e^{2V}D_{\alpha}e^{-2V}. 
\end{equation} It transforms in the expected way: $W_{\alpha}\rightarrow e^{-\imath \Lambda}W_{\alpha}e^{\imath \Lambda}$ 
\end{itemize}

\subsection{Lagrangian $\mathcal{N}=1$ field theories}

I this section we will explain how to build lagrangians for $\mathcal{N}=1$ theories describing chiral and vector multiplets.  
These constitute the building blocks for gauge theories lagrangians, in particular those with $\mathcal{N}=2$ supersymmetry, 
which represent the starting point of the Seiberg-Witten analysis. The basic idea is to consider the highest component of a 
suitable superfield.

\subsubsection{Chiral multiplets}

The kinetic terms for the fields in the SUSY multiplet can be introduced considering the highest component of $\Phi^{\dg}_{i}\Phi_{j}$.
Neglecting total derivatives and summing over $i=j$, we get the lagrangian
\[
\mathcal{L}=\Phi^{\dg}_{i}\Phi_{i}\vert_{\te^{2}\bt^{2}}= \de_{\mu}A^{\dg}_{i}\de^{\mu}A_{i}+ F^{\dg}_{i}F_{i}- \imath \bar{\psi}_{i}\bar{\sigma}^{\mu}\de_{\mu}\psi_{i}. 
\]
This lagrangian describes as expected a scalar and a spinor massless fields. It also describes an auxiliary field $F$, which can 
be eliminated by means of the equations of motion. 

Mass and interaction terms can be added introducing a superpotential. The most general superpotential compatible with renormalizability is 
\begin{equation}
\mathscr{L} = \Phi_{i}^{\dagger}\Phi_{i}|_{\theta\theta\bar\theta\bar\theta} + \left[\left(\frac{1}{2}m_{ij}\Phi_i\Phi_j + \frac{1}{3}g_{ijk}\Phi_i\Phi_j\Phi_k + \lambda_i\Phi_i\right)\bigg|_{\theta\theta} + h.c.\right].
\end{equation}
We can rewrite everything in terms of an integral over superspace 
\[
\mathcal{L}=\int d^{4}\te\Phi^{\dg}_{i}\Phi_{i}+ \int d^{2}\te\mathcal{W}(\Phi)+\int d^{2}\bt\bar{\mathcal{W}}(\Phi^{\dg}).
\]

This is not the most general lagrangian compatible with supersymmetry. We can also consider the highest component of 
$K(\Phi,\Phi^{\dg})$ where $K$ is called \textbf{K\"{a}hler potential} and should satisfy the constraint $\bar{K}(z_{i},\bar{z}_{j})=K(\bar{z}_{i},z_{j})$. 

At the classical level any supersymmetric theory has a $U(1)$ simmetry called R-simmetry, which acts on chiral superfields as
\[\begin{aligned}
R\Phi(x,\te)&=e^{2\imath n\alpha}\Phi(x,e^{-\imath\alpha}\te),\\
R\Phi^{\dg}(x,\bt)&=e^{-2\imath n\alpha}\Phi^{\dg}(x,e^{\imath\alpha}\bt). 
\end{aligned}\]
$n$ is called R-charge. Since $d^{2}\te\rightarrow e^{-2\imath\alpha}d^{2}\te$, the R-charge of the superpotential $\mathcal{W}$ should be one.

\subsubsection{Supersymmetric gauge theories}

The standard kinetic term for the abelian multiplet is  
\begin{equation}
\mathcal{L}=\frac{1}{4}\left(\int d^{2}\te W^{\alpha}W_{\alpha}+\int d^{2}\bt\bar{W}_{\da}\bar{W}^{\da}\right). 
\end{equation}
Similarly, the SUSY analog of Yang-Mills (SYM) theory is described by the lagrangian
\begin{equation}
\mathcal{L}=\frac{1}{4g^{2}}\left(\int d^{2}\te W^{\alpha}W_{\alpha}+\int d^{2}\bt\bar{W}_{\da}\bar{W}^{\da}\right), 
\end{equation}
where $g$ is the coupling constant. If we want to include the $\theta$ term we can simply consider
$$\mathcal{L}=\frac{1}{8\pi}Im\left(\tau Tr\int d^{2}\te W^{\alpha}W_{\alpha}\right),$$
where $\tau=\te/2\pi+4\pi\imath/g^{2}$. 

If we want to include matter fields the minimal coupling can be implemented introducing the term 
$$ \int d^{4}\te \Phi^{\dg} e^{-2V}\Phi$$ for any matter field. Interaction terms can be simply included using the superpotential 
as before.

\subsection{$\mathcal{N}=2$ gauge theories}

\subsubsection{SYM theory}

As we have explained above, the field content of a $N=2$ vectormultiplet $(A_{\mu},\lambda,\psi,\phi)$ is equivalent to that of 
two $N=1$ multiplets, one vector multiplet $(A_{\mu},\lambda)$ e one scalar multiplet $(\psi,\phi)$. The $\mathcal{N}=2$ pure gauge 
theory is thus equivalent to a $\mathcal{N}=1$ gauge theory with a chiral multiplet in the adjoint representation. The relative 
normalization is fixed by the requirement that the two fermions should enter symmetrically in the lagrangian. Using the receipt 
given previously, we can immediately write down the lagrangian for $\mathcal{N}=2$ SYM:
\begin{equation}\label{sym}
\mathcal{L}=\frac{1}{8\pi}\im\Tr\left[\tau\left(\int d^{2}\te W^{\alpha}W_{\alpha}\right)+2\int d^{4}\te\Phi^{\dg} e^{-2V}\Phi\right]. 
\end{equation}
If we rewrite it in terms of the component fields and eliminate the two auxiliary fields using their equations of motion, we get
\begin{multline}
\mathcal{L}=\frac{1}{g^{2}}\Tr\left(-\frac{1}{4}F_{\mu\nu}F^{\mu\nu}+g^{2}\frac{\te}{32\pi^{2}}F_{\mu\nu}\widetilde{F}^{\mu\nu}+(D_{\mu}\phi)^{\dg}(D^{\mu}\phi)-\frac{1}{2}[\phi^{\dg},\phi]^{2}\right.\\
 \left.-\imath\lambda\sm D_{\mu}\bar{\lambda}-\imath\bar{\psi} \bar{\sigma}^{\mu}D_{\mu}\psi-\imath\sqrt{2}[\lambda,\psi]\phi^{\dg} -\imath\sqrt{2}[\bar{\lambda},\bar{\psi}]\phi\right). 
\end{multline}
The scalar potential of the theory is thus
\begin{equation}\label{V}
\mathcal{V}=-\frac{1}{2g^{2}}\Tr\left([\phi^{\dg},\phi]^{2}\right). 
\end{equation} 
At the classical level the vacuum configurations can be found minimizing the potential (\ref{V}). We can immediately see from 
the above formula one of the characterizing properties of field theories with extended supersymmetry: the presence of flat directions. 
The potential attains its minimum when the field $\phi$ commutes with its hermitian conjugate and can thus be diagonalized. The set 
of gauge inequivalent vacua can be parametrized by the eigenvalues of $\phi$ (or more precisely their gauge invariant combinations 
$\Tr\phi^{k}$) and is called moduli space.

\subsubsection{Including matter fields}

Matter multiplets coupled to the $\mathcal{N}=2$ gauge multiplet can also be included. They are described by the so called 
hypermultiplets, which can be constructed using two $\mathcal{N}=1$ chiral multiplets $Q$ and $\widetilde{Q}^{\dagger}$ in the same 
representation of the gauge group. We can for instance describe $\mathcal{N}=2$ SQCD adding $N_f$ hypermultiplets in the 
fundamental representation. The lagrangian can be written down simply adding to (\ref{sym}) the terms (for $SU(N)$ gauge theories)
\begin{equation}\nonumber
\mathcal{L}=\int d^{4}\te\left(Q_{i}^{\dg}e^{-2V}Q_{i}+\widetilde{Q}_{i}e^{2V}\widetilde{Q}^{\dg}_{i}\right)+ 
\int d^{2}\te \left(\sqrt{2}\widetilde{Q}_{i}\Phi Q_{i}+m_{i}\widetilde{Q}_{i}Q_{i}\right)+\text{h.c.} 
\end{equation}
The term $\sqrt{2}\widetilde{Q}_{i}\Phi Q_{i}$ is linked through $\mathcal{N}=2$ SUSY to the coupling of the hypermultiplet with 
the vector superfield $V$.

\subsubsection{Central charges in $\mathcal{N}=2$ gauge theories}

We have seen that gauge theories with extended supersymmetry always include a scalar field in the adjoint representation. Its 
vacuum expectation value generically breaks the gauge group down to the maximally abelian subgroup. All the theories with this 
property contain in their spectrum magnetic monopoles and dyons \cite{'t Hooft}-\cite{J-Z}. As shown by Olive and Witten in \cite{O-W}, the central charge 
of SUSY representations in these theories is proportional to the electric and magnetic charges. Their computation leads to the 
following result for $SU(2)$ SYM theory $$Z=a(n_{e}+\tau n_{m}),\qquad \tau=\frac{\te}{2\pi}+\frac{4\pi\imath}{g^{2}},$$ where 
$a$ is the vev of $\phi$. If we include matter fields in various representations this formula should be modified adding the 
flavor charges $S_i$: $$Z=a(n_{e}+\tau n_{m})+\frac{1}{\sqrt{2}}\sum m_{i}S_{i},$$ where $m_i$ are the masses of the matter 
fields. This result holds for theories with gauge group $SU(2)$, which has rank one. In the general case we will have one electric 
and one magnetic charges for each Cartan generator.

\section{$\mathcal{N}=2$ SYM and low energy effective action}

As we have seen the classical moduli space (Coulomb branch) of SYM theory can be parametrized by the vev of the scalar field $\phi$ in the vector multiplet, 
which can be supposed to be diagonal, more precisely by the gauge invariant combinations of its eigenvalues $\Tr\phi^{k}$ 
with $k=1,\dots,\text{rank G}$. If we introduce matter fields in the theory the Coulomb branch is just a submanifold of the whole 
moduli space, which also includes the so-called Higgs branch. It is parametrized by the vev of the scalar fields in the matter 
hypermultiplets, whereas $\langle\phi\rangle=0$. On the Higgs branch the gauge group is generically completely broken. The Higgs and Coulomb branches can also intersect along the so called mixed 
branches, on which the vev of $\phi$ is different from zero. A key property of the Higgs branch (dictated by extended supersymmetry) 
is that it is a hyperk\"{a}hler manifold and is not modified by quantum corrections \cite{APS}. This is not true for the Coulomb branch. 
The purpose of this section is to understand how the Coulomb branch is modified by quantum corrections. For definiteness we will restrict 
our discussion to $SU(N)$ gauge theories.

\subsection{Breaking of the R-symmetry}

We will now discuss how quantum corrections break the classical R-symmetry of the theory. This will play an important role in the 
other chapters.

We can rewrite the lagrangian for $\mathcal{N}=2$ SYM theory with gauge group $SU(N_{c})$ in terms of the Dirac spinor $\psi_{D}=\left(\begin{array}{l}
\lambda\\
\bar{\psi}
\end{array}
\right)$ constructed using the spinors in the vector multiplet
$$\begin{aligned}
\LL=&\frac{1}{g^{2}}\Tr \left(-\frac{1}{4}F_{\mu\nu}F^{\mu\nu}+g^{2}\frac{\te}{32\pi^{2}}F_{\mu\nu}\wt{F}^{\mu\nu}+(D_{\mu}\phi)^{\dg}(D^{\mu}\phi)-\frac{1}{2}[\phi^{\dg},\phi]^{2}\right.\nonumber\\
&\left.+\imath\bar{\psi}_{D}\gamma^{\mu}D_{\mu}\psi_{D}+\imath\sqrt{2}[\bar{\psi}_{D},\frac{1+\gamma_{5}}{2}\psi_{D}]\phi^{\dg}-\imath\sqrt{2}[\bar{\psi}_{D},\frac{1-\gamma_{5}}{2}\psi_{D}]\phi\right).
\end{aligned}$$
One can then easily check that the following transformation 
\[
\begin{array}{llll}
U(1)_{\mathcal{R}}:&\phi&\rightarrow&\phi'=e^{2\imath\alpha}\phi,\\
 &\psi_{D}&\rightarrow&\psi'_{D}=e^{\imath\alpha\gamma_{5}},\\
 &\bar{\psi}_{D}&\rightarrow&\bar{\psi}'_{D}=\bar{\psi}_{D}e^{\imath\alpha\gamma_{5}},\\ 
\end{array} 
\]
is a symmetry of the classical theory but is broken at the quantum level by the chiral anomaly: $\de_{\mu}J^{\mu}_{5}=-\frac{N_{c}}
{8\pi^{2}}F_{\mu\nu}\wt{F}^{\mu\nu}$. From the path integral representation we find infact
\[
\int[d\phi'][d\bar{\psi}'_{D}][d\psi'_{D}]\exp \{\imath S[\phi',\bar{\psi}'_{D},\psi'_{D},A_{\mu}]\}  
\]
\begin{equation}\label{pippo}
=\int[d\phi][d\bar{\psi}_{D}][d\psi_{D}]\exp \{\imath S[\phi,\bar{\psi}_{D},\psi_{D},A_{\mu}]\}\exp \{-\imath\alpha4N_{c}\nu\}, 
\end{equation}
where $\nu=\frac{1}{32\pi^{2}}\int d^{4}xF_{\mu\nu}\wt{F}^{\mu\nu}$ is the instanton number. The transformations $U(1)_{\mathcal{R}}$ 
which leave the path integral invariant satisfy the relation $\alpha=\frac{2\pi n}{4N_{c}}$ where $n=1,\dots, 4N_{c}$. 
The residual symmetry group is then $\mathbb{Z}_{2N_{c}}$, which is further broken in each vacuum by the vev of $\phi$. If we 
include matter hypermultiplets the corresponding spinors contribute to the anomaly. For SQCD with $N_c$ colors and $N_f$ flavors, 
which is the case of interest for us, the unbroken subgroup is $\mathbb{Z}_{2N_{c}-N_f}$.

\subsection{Low energy effective action}

We have seen that semiclassically the theory abelianizes on the Coulomb branch due to the vev $\langle\phi\rangle=\diag(a_1,
\dots,a_N)$, with $\sum_{i}a_i=0$. If $a_i-a_j\gg\Lambda$ for all i and j, this breaking occurs at very high energy where the theory 
is weakly coupled. Below that scale the theory becomes abelian and the coupling constant decreases at lower energies. The theory 
is thus weakly coupled at all scales and the semiclassical picture is reliable. The gauge multiplets associated to broken gauge 
generators acquire mass of order $\langle\phi\rangle$ by the Higgs mechanism and at low energy the dynamics will be encoded in 
an abelian effective action, written in terms of the vector multiplets associated with the Cartan generators. We expect this 
picture to remain valid at the quantum level however, a perturbative computation will not be reliable in the inner region of the 
moduli space, where quantum corrections become important. The question is then how one can determine the effective action encoding 
the infrared dynamics.

As a preliminary step, one can notice that extended supersymmetry imposes strong constraints on the structure of this effective 
action, which must have the following form, as found in \cite{2SS} using the $\mathcal{N}=2$ superspace formalism:
\begin{align}\label{U1eff}
\mathcal{L}=\frac{1}{8\pi}\im\left(\int d^{2}\te\F_{ab}(\Phi)W^{a\alpha}W^{b}_{\alpha}+2\int d^{4}\te(\Phi^{\dg}e^{-2V})^a
\F_{a}(\Phi)\right).
\end{align}
In the previous formula $\F_{a}(\Phi)=\de\F/\de\Phi^{a},\quad\F_{ab}(\Phi)=\de^{2}\F/\de\Phi^{a}\de\Phi^{b}$ and $\F$ is a 
holomorphic function called \textbf{prepotential}. The Seiberg-Witten solutions allows to determine it exactly.

Let us now focus on the case $N_{c}=2$, the generalization being straightforward. We will then have $\langle\phi\rangle=a\tau_{3}$.
At the classical level the prepotential assumes the form $\F=\frac{1}{2}\tau a^{2}$ and it receives perturbative corrections
only at the one loop level. This can be determined using R-symmetry \cite{S88}: as we have seen the group $U(1)_{\mathcal{R}}$ is broken by the  
chiral anomaly and under its action the lagrangian changes as 
\[
\delta\LL=-\frac{\alpha}{4\pi^{2}}F_{\mu\nu}\wt{F}^{\mu\nu}. 
\]
We can rewrite this result in terms of the prepotential applying the R-symmetry transformation to the effective action (\ref{U1eff}), obtaining
\[
\delta\LL=-\frac{\alpha}{8\pi}\im\left[a\F'''(a)(\wt{F}_{\mu\nu}F^{\mu\nu}+\imath F_{\mu\nu}F^{\mu\nu})\right]. 
\]
Comparing the above equations we find $\F'''(a)=\frac{2\imath}{a\pi}$ and integrating we get the one loop correction to the prepotential
\begin{equation}\label{1loop}
\F_{one-loop}(a)=\frac{\imath}{2\pi}a^{2}\ln\frac{a^{2}}{\Lambda^{2}}, 
\end{equation}
Where $\Lambda$ is the dynamical scale of the theory.

This description of the effective action in terms of the superfields $\Phi$ and $W^{\alpha}$ is appropriate in the semiclassical 
region (for $u=\langle\Tr\phi^2\rangle\gg\Lambda^2$) but cannot be valid globally on the moduli space. In order to see this one can notice 
that unitarity requires the kinetic term to be positive definite, 
which in turn implies that $$\im\tau\equiv\im\frac{\de^{2}\F}{\de a\de \bar{a}}=\frac{\te}{2\pi}+\imath\frac{4\pi}{g^{2}}$$ should be positive. 
On the other hand, since $\F$ is holomorphic, its imaginary part is harmonic and consequently cannot be positive everywhere on the complex plane. 
The solution to this problem is that our choice of coordinates is valid only on a region of the moduli space and when we approach 
a point where $\im\tau(a)=0$ we should adopt a different set of coordinates $\tilde{a}$ such that $\im\tilde{\tau}(\tilde{a})>0$ 
in a neighbourhood of that point. Let us now study how to implement this change of coordinates.

\subsection{Duality transformations}

Let us define $\Phi_{D}$ to be the field dual to $\Phi$ and the function $\F_{D}(\Phi_{D})$ to be the dual of the prepotential $\F(\Phi)$ through
the equations
\begin{equation}\label{duale}
\Phi_{D}=\F'(\Phi),\qquad \F'_{D}(\Phi_{D})=-\Phi. 
\end{equation}
Using these relations in (\ref{U1eff}) we can rewrite the second term as follows
\begin{align}
\im\int d^{4}xd^{2}\te d^{2}\bt\Phi^{\dg}\F'(\Phi)&=\im\int d^{4}xd^{2}\te d^{2}\bt(-\F'_{D}(\Phi_{D}))^{\dagger}\Phi_{D}\nonumber\\\nonumber
&=\im\int d^{4}xd^{2}\te d^{2}\bt\Phi^{\dg}_{D}\F'_{D}(\Phi_{D}).
\end{align}
We thus see that this term is invariant under the duality transformation (\ref{duale}). Let us now analyze the term 
$\F''(\Phi)W^{\alpha}W_{\alpha}$. Remember that $W_{\alpha}$ contains the abelian field-strength $F_{\mu\nu}=\de_{\mu}A_{\nu}-\de_{\nu}A_{\mu}$ 
for some $A_{\mu}$. This constraint can be implemented imposing the Bianchi identity $\de_{\nu}\wt{F}^{\mu\nu}=0$; the corresponding 
constraint in the superspace formalism is $\im D_{\alpha}W^{\alpha}=0$. 

Inserting this in the path integral we can now integrate  
either with respect to $V$ or $W^{\alpha}$, imposing the condition $\im D_{\alpha}W^{\alpha}=0$ by means of an auxiliary vector 
superfield $V_{D}$ which plays the role of a Lagrange multiplier:
\begin{align}
\int[dV]&\exp\left[\frac{\imath}{8\pi}\im\int d^{4}xd^{2}\te\F''(\Phi)W^{\alpha}W_{\alpha}\right]
\simeq\int[dW][dV_{D}]\nonumber\\\nonumber
&\exp\left[\frac{\imath}{8\pi}
\im\int d^{4}x\left(\int d^{2}\te\F''(\Phi)W^{\alpha}W_{\alpha}
+\frac{1}{2}\int d^{2}\te d^{2}\bt V_{D}D_{\alpha}W^{\alpha}\right)\right]. 
\end{align}
The second term can now be rewritten as
\begin{align}
\int d^{2}\te d^{2}\bt V_{D}D_{\alpha}W^{\alpha}=&-\int d^{2}\te d^{2}\bt D_{\alpha}V_{D}W^{\alpha}=\int d^{2}\te \bar{D}^{2}(D_{\alpha}V_{D}W^{\alpha})\nonumber\\\nonumber
=&\int d^{2}\te(\bar{D}^{2}D_{\alpha}V_{D})W^{\alpha}=-\frac{1}{4}\int d^{2}\te (W_{D})_{\alpha}W^{\alpha},
\end{align}
where we have defined the dual of $W$ by means of the relation $(W_{D})_{\alpha}=-\frac{1}{4}\bar{D}^{2}D_{\alpha}V_{D}$ and 
we have exploited the chirality of the field strength $\bar{D}_{\dot{\beta}}W^{\alpha}=0$. Integrating explicitly with respect to $W$ 
we finally get
\[
\int[dV_{D}]\exp\left[\frac{\imath}{8\pi}\im\int d^{4}xd^{2}\te\left(-\frac{1}{\F''(\Phi)}W^{\alpha}_{D}W_{D\alpha}\right)\right]. 
\]
We have thus rewritten the lagrangian (\ref{U1eff}) in terms of dual variables. The generalized coupling $\tau(a)$ is replaced 
in these variables by $-\frac{1}{\tau(a)}$. This transformation is the SUSY counterpart of electric-magnetic duality: the 
transformation $W\rightarrow W_{D}$ implies $F_{\mu\nu}\rightarrow\wt{F}_{\mu\nu}$. Notice that 
$$\F''_{D}(\Phi_{D})=-\frac{d\Phi}{d\Phi_{D}}=-\frac{1}{\F''(\Phi)},$$ and this  
implies $$\tau_{D}(a_{D})=-\frac{1}{\tau(a)},$$ where we have defined $a_{D}=\langle\Phi_{D}\rangle$. 
Substituting in the lagrangian we finally get
\begin{equation}
\frac{1}{8\pi}\im\int d^{4}x\left[\int d^{2}\te\F''_{D}W^{\alpha}_{D}W_{D\alpha}+\int d^{2}\te d^{2}\bt\Phi^{\dg}_{D}\F'_{D}(\Phi_{D})\right]. 
\end{equation}
Notice that the metric induced on the moduli space
\begin{equation}\label{DS}
ds^{2}=\im\tau dad\bar{a}=\im(da_{D}d\bar{a})= \frac{\imath}{2}(dad\bar{a_{D}}-da_{D}d\bar{a}), 
\end{equation}
is left invariant by the duality transformation (\ref{duale}). We will now determine the group of transformations which leave 
the metric (\ref{DS}) invariant and which map our theory into an equivalent one, thus determining all the possible parametrizations 
of the moduli space.

\subsubsection{The duality group}

In order to analize the duality group it is convenient to write the effective action (\ref{U1eff}) in the form
\begin{equation}\label{leffe}
\LL=\frac{1}{8\pi}\im\int d^{2}\te\frac{d\Phi_{D}}{d\Phi}W^{\alpha}W_{\alpha}+\frac{1}{16\imath\pi}\int d^{2}\te d^{2}\bt(\Phi^{\dg}\Phi_{D}-\Phi^{\dg}_{D}\Phi). 
\end{equation}
We have just seen that the transformation
\begin{equation}
\left(\begin{array}{l}
\Phi_{D}\\
\Phi 
\end{array}\right)\rightarrow
\left(\begin{array}{ll}
0 & 1\\
-1 & 0 
\end{array}\right)\left(\begin{array}{l}
\Phi_{D}\\
\Phi 
\end{array}\right) 
\end{equation}
is in the duality group and from (\ref{DS}) it is easy to see that also the following transformation leaves the metric invariant:
\begin{equation}\label{theta}
\left(\begin{array}{l}
\Phi_{D}\\
\Phi 
\end{array}\right)\rightarrow
\left(\begin{array}{ll}
1 & a\\
0 & 1 
\end{array}\right)\left(\begin{array}{l}
\Phi_{D}\\
\Phi 
\end{array}\right),\qquad a\in \mathbb{R}.  
\end{equation}
Since $a$ is real, this transformation leaves the second term of the lagrangian (\ref{leffe}) invariant whereas the first one is 
modified as follows:
\[
\frac{a}{8\pi}\im\int d^{4}xd^{2}\te W^{\alpha}W_{\alpha}=-\frac{a}{16\pi}\int d^{4}xF_{\mu\nu}\wt{F}^{\mu\nu}=-2\pi a\nu 
\]
Since the functional integral is invariant if the action changes by an integer multiple of $2\pi$, we can conclude that the 
transformations (\ref{theta}) are in the duality group if
$a\in\mathbb{Z}$. Their effect is to shift the $\te$ angle by $2\pi a$.
Equations (\ref{duale}) and (\ref{theta}) together generate the duality group $SL(2,\mathbb{Z})$. It is easy to verify that the 
action of this group on the generalized coupling constant is $$\tau\rightarrow\frac{a\tau+b}{c\tau+d},$$ where $ad-bc=1$ and 
$a,b,c,d\in\mathbb{Z}$. For $N_{c}>2$ the duality group which leaves the metric invariant is $Sp(2N_{c}-2,\mathbb{R})$.

\subsection{Masses of monopoles and dyons}

We have seen that in this class of models monopoles and dyons arise in a very natural way and their mass is bounded below by 
$M\geq\sqrt{2}\vert Z\vert$, where $Z$ is the central charge. States saturating this bound are called BPS (see \cite{Bog,BPS}) and are organized in short multiplets of the $\mathcal{N}=2$ 
SUSY algebra. Semiclassically $Z=a(n_{e}+\tau n_{m})$ and the purpose of this section is to achieve an exact formula, written in 
terms of the low energy effective quantities.
Let us suppose that the effective theory contains a matter hypermultiplet described by the chiral superfields $M,\wt{M}$. 
When the vev $a$ of the field $\phi$ is different from zero the multiplet becomes massive . If its electric charge is 
$n_{e}$ (and its magnetic charge zero), the interaction term assumes necessarily the form $$\sqrt{2}n_{e}M\Phi\wt{M}.$$ 
by supersymmetry. We can thus conclude that its central charge is $Z=n_{e}a$. If we have instead a magnetic monopole with 
charge $n_{m}$, we can reduce ourselves to study a system equivalent to the previous one by means of a duality transformation
and conclude that the central charge in this case is $Z=a_{D}n_{m}$. This argument implies that for a generic dyon with charges 
$(n_{e},n_{m})$ we will have the formula
\begin{equation}\label{Z}
Z=an_{e}+a_{D}n_{m}. 
\end{equation}
Comparing the two equations we find that semiclassically $a_{D}=\tau a$. The generalization of this formula to a rank $r$ 
theory is simply given by
\begin{equation}
Z=a^{i}n_{e,i}+a_{D}^{i}n_{m,i}, \qquad i=1,\dots,r 
\end{equation}
where $a^{i}$ are local coordinates on the moduli space and $n_{e,i},n_{m,i}$ are the charges with respect to the various $U(1)$ 
subgroups.
We conclude remarking that, since the central charge determines the mass of the various particles in the spectrum, the duality 
transformations discussed previously should not modify the value of the central charge. If we organize the parameters $a_{D}$ and 
$a$ in a vector $v$, under the action of a duality transformation $v\rightarrow Mv$. 
In order to leave $Z$ invariant we should then impose the transformation rule $w\rightarrow wM^{-1}$ where $w=(n_{m},n_{e})$. 

\subsection{$\beta$ function for the effective $U(1)$ theory}

In the next section we will make use of the one-loop expressions for $a$ and $a_D$ to compute the monodromy matrices. We will now 
see how we can determine them starting from the beta function of this theory.
If we have Weyl fermions with charges $Q_{f}$ and complex scalars with charges $Q_{s}$ (with respect to the $U(1)$ gauge group), 
their contribution to the beta function of the theory is:
\[
\beta(g)\equiv\mu\frac{dg}{d\mu}=\frac{g^{3}}{16\pi^{2}}\left(\sum_{f}\frac{2}{3}Q_{f}^{2}+\sum_{s}\frac{1}{3}Q_{s}^{2}\right). 
\]
Since the beta function is positive the theory is infrared free and at very low energies we expect the one-loop approximation to 
be reliable.
If we indicate with $b$ the coefficient of the term $g^{3}$ and set $\alpha=g^{2}/4\pi$, we can write $$\mu\frac{d}{d\mu}\left(\frac{1}{\alpha}\right)=-8\pi b.$$ 
Since in a hypermultiplet we have two Weyl fermions and two complex scalars with the same charge $Q$, its contribution will be
\[
b=\frac{1}{16\pi^{2}}Q^{2}\left(2\cdot\frac{2}{3}+2\cdot\frac{1}{3}\right)=\frac{1}{8\pi^{2}}Q^{2}. 
\]
Setting now $\tau=\imath/\alpha$ we obtain $$\mu\frac{d\tau}{d\mu}=-\frac{1}{\pi}Q^{2}.$$ If we now identify the energy scale $\mu$ 
with $a$ and set $Q=1$ we are led to the equation
\[
\tau\approx-\frac{\imath}{\pi}\ln\frac{a}{\Lambda}. 
\]

If the hypermultiplet we are considering becomes massless at let's say $u_{0}$, we will have $\lim_{u\rightarrow u_{0}}a(u)=0$. 
We are now free to choose the parametrization in such a way that $a\approx c(u-u_{0})$ with $c\in\mathbb{C}$. 
Since $\tau=\frac{da_{D}}{da}$ we obtain integrating
\begin{equation}\label{ne=1}
\begin{aligned}
a(u)&\approx c(u-u_{0}),\\
a_{D}(u)&\approx a_{D}(u_{0})-\frac{\imath}{\pi}c(u-u_{0})\ln\frac{u-u_{0}}{\Lambda}.
\end{aligned}
\end{equation}
If instead the hypermultiplet describes a monopole becoming massless at $u_{0}$, we have $\lim_{u\rightarrow u_{0}}a_{D}(u)=0$. 
Performing the above computation in terms of magnetic variables we find
\[
\tau_{D}\approx-\frac{\imath}{\pi}\ln\frac{a_{D}}{\Lambda}. 
\]
Recalling now that $\tau_{D}=-\frac{da}{da_{D}}$ we find
\begin{equation}\label{nm=1}
\begin{aligned}
a_{D}(u)&\approx c(u-u_{0}),\\
a(u)&\approx a(u_{0})+\frac{\imath}{\pi}c(u-u_{0})\ln\frac{u-u_{0}}{\Lambda}.
\end{aligned} 
\end{equation}
Notice that the dual coupling constant tends to zero when the monopole becomes massless, i.e. for $u\rightarrow u_{0}$. The 
theory can thus be analyzed using standard techniques once we adopt the magnetic description.

\section{The Seiberg-Witten solution}

\subsection{Monodromies and singularities of the moduli space}

In the previous section we have seen that we can parametrize the moduli space with the coordinate $u=\langle\tr\phi^{2}\rangle$ 
and for large $u$ the theory can be described using the parameter $a$ defined by $\langle\phi\rangle=\frac{1}{2}a\sigma_{3}$. 
However, this description is not globally valid and at strong coupling we should perform a duality transformation. 
It is thus convenient to introduce the vector $v=(a_{D}(u),a(u))^{T}$.
One of the key ingredients of the Seiberg-Witten solution is the presence of singular points: 
the functions $a_{D}(u)$ and $a(u)$ are not single-valued and if we loop around a singular point 
they undergo a nontrivial transformation which can be conveniently described in terms of a matrix (monodromy matrix) acting 
on the $v$ vector. We will now explore this point.

\subsubsection{Monodromy at infinity}

Let us start from the semiclassical region, where the theory can be analyzed by means of a standard one-loop computation. 
Using the formula for the prepotential (\ref{1loop}), and recalling that $a_{D}=\de\F/\de a$ we get
\begin{equation}
a_{D}(u)=\frac{\imath}{\pi}a\left(1+\ln\frac{a^{2}}{\Lambda^{2}}\right),\qquad u\rightarrow\infty. 
\end{equation}
Looping now around the point at infinity in the moduli space ($u\rightarrow e^{2\pi\imath}u$) we find
\begin{equation}
\begin{aligned}
a&\rightarrow-a,\\
a_{D}&\rightarrow\frac{\imath}{\pi}(-a)\left(1+\ln\frac{e^{2\pi\imath}a^{2}}{\Lambda^{2}}\right)=-a_{D}+2a.
\end{aligned}
\end{equation}
The monodromy matrix at infinity is then
\begin{equation}
\left(\begin{array}{l}
a_{D}\\
a 
\end{array}\right)\rightarrow
M_{\infty}\left(\begin{array}{l}
a_{D}\\
a 
\end{array}\right),\qquad M_{\infty}=\left(\begin{array}{cc}
-1 & 2\\
 0 & -1 
\end{array}\right). 
\end{equation}

\subsubsection{Structure of the moduli space}

Since there is a singularity at infinity we necessarily have at least another one in the $u$ plane (otherwise every loop would 
be contractible and we would not have any monodromy) and the associated monodromy should be in $SL(2,\mathbb{Z})$ as we have seen before. 
We cannot actually have only two singular points in the moduli space, as shown in \cite{SWI}. If this were the case the monodromy 
associated to the second singular point would be equal to $M_{\infty}$. On the other hand this transformation acts trivially on 
the metric (\ref{DS}), which would then be globally expressible as a harmonic function $\im\tau(a)$ on the moduli space. 
As we have seen this is in contradiction with the requirement of positivity. 
We can thus conclude that we need at least three singular points. In \cite{SWI} Seiberg and Witten assume that this ``minimal'' 
choice is indeed correct. We will now see how this assumption leads to a self-consistent picture and allows to determine 
explicitly the prepotential.

Let us suppose that we have exactly two singular points (plus the singular point at infinity studied before). First of all, due to the $\mathbb{Z}/2\mathbb{Z}$ symmetry of the theory 
which acts on the moduli space sending $u$ in $-u$, we can immediately say that these two singularities should be located at 
opposite points in the moduli space, which we can assume to be $u=\pm1$. What is then their physical interpretation?
The key point is that at the singularities the effective description in terms of the lagrangian 
(\ref{U1eff}) is not adequate. Since we are dealing with a Wilsonian effective action, in which all massive fields (whose mass 
depends on $u$) are integrated out (see \cite{WILS,WILSI}), the most natural explanation of the singularities is that some of the massive fields  
actually become massless at these points due to strong quantum corrections. 
As argued in \cite{SWI}, these massless multiplets cannot be identified with gauge vectormultiplets; they must then be 
BPS hypermultiplets and the only such multiplets in the spectrum of our theory are the solitonic monopoles and dyons.
Following this interpretation we can now safely assume (with a suitable choice of conventions) that one of the singularities 
is due to a massless monopole. 

We are essentially proposing that the infrared dynamics at the singularity is described by a SQED, in which the massless electron 
is actually a magnetic monopole in terms of the original UV variables of the theory. Let us now see how we can determine the 
monodromy matrix at the singularity.

\subsubsection{Monodromies in the strongly coupled region}

Let us analyze the monodromies at finite $u$. Assuming that the singularity at $u=1$ is due to a massless monopole,
we can conclude from (\ref{Z}) that $a_{D}$ must vanish there and, turning to the dual description as explained in 
the previous section, we find equation (\ref{nm=1}). Looping around the singularity counterclockwise in the moduli space
$u-1\rightarrow e^{2\pi\imath}(u-1)$ we then find the transformation law:
\begin{equation}\label{M_1}
\left(\begin{array}{l}
a_{D}\\
a 
\end{array}\right)\rightarrow
M_{1}\left(\begin{array}{l}
a_{D}\\
a 
\end{array}\right)=\left(\begin{array}{c}
a_{D}\\
a-2a_{D}
\end{array}\right),\qquad M_{1}=\left(\begin{array}{cc}
1 & 0\\
 -2 & 1 
\end{array}\right),
\end{equation}
where $M_{1}$ is our monodromy matrix.

In order to determine the monodromy matrix at the second singularity $u=-1$, it is enough to observe that a loop around the 
point at infinity is equivalent to the composition of a loop around the point $u=-1$ and another one around $u=1$. 
This tells us that $M_{\infty}$ is equivalent to the composition of $M_{1}$ and $M_{-1}$. We can thus conclude that the following 
relation holds:
\begin{equation}\label{fattore}
M_{\infty}=M_{1}M_{-1}. 
\end{equation}
We then immediately get
\begin{equation}\label{M_-1}
M_{-1}=\left(\begin{array}{cc}
-1 & 2\\
-2 & 3
\end{array}\right). 
\end{equation}
In order to determine the charges of the particle responsible for this singularity we can proceed as follows \cite{G-H}: 
The monodromy due to a dyon with charges $(n_{m},n_{e})$ can be computed determining first of all the duality transformation 
which turns it into an electron $(0,1)$. 

Under an arbitrary $SL(2,\mathbb{Z})$ transformation \begin{footnotesize}$\left(\begin{array}{cc} 
\alpha & \beta\\
\gamma & \delta\end{array}\right)$\end{footnotesize}
we have the following relations
\[
\left(\begin{array}{l}
a_{D}\\
a 
\end{array}\right)\rightarrow\left(\begin{array}{l}
\alpha a_{D}+\beta a\\
\gamma a_{D}+\delta a
\end{array}\right),\qquad\left(\begin{array}{l}
n_{m}\\
n_{e} 
\end{array}\right)\rightarrow\left(\begin{array}{l}
\delta n_{m}-\gamma n_{e}\\
-\beta n_{m}+\alpha n_{e} 
\end{array}\right). 
\]
Imposing that the resulting dyon has charges $(0,1)$, we can determine the duality transformation. In these variables the monodromy 
associated to the dyon is equal to that of an electron, which we can easily determine from (\ref{ne=1}):
\[
\left(\begin{array}{cc}
1 & 2\\
0 & 1 
\end{array}\right). 
\]
Going back to the original variables we find the monodromy matrix
\begin{equation}\label{dione}
\left(\begin{array}{cc}
1+2n_{m}n_{e} & 2n_{e}^{2}\\
-2n_{m}^{2} & 1-2n_{m}n_{e} 
\end{array}\right). 
\end{equation}
Comparing now with (\ref{M_-1}) we conclude that the dyon becoming massless at $u=-1$ has charges $(1,-1)$. 

\subsection{Solution of the model}

We will now see how the structure of the quantum moduli space described above allows us to determine explicitly
$a(u)$ and $a_{D}(u)$ interpreting them as periods of a suitable elliptic curve. 

\subsubsection{Moduli space and elliptic curves}

We have seen that the moduli space $\mathcal{M}$ of $SU(2)$ SYM is the complex plane with singularities at 1,$-1$ and $\infty$. 
We can parametrize it using the coordinate $u$ equal to (semiclassically) $\langle\tr\Phi^{2}\rangle$ and it is characterized 
by a $\mathbb{Z}_{2}$ symmetry that acts as $u\rightarrow-u$.
The quantities of interest for us $a$ and $a_{D}$ can be expressed as (many-valued) functions of $u$. 

The first key observation is that the duality we have studied before implies that we can construct a flat $SL(2,\mathbb{Z})$ 
bundle $V$ over $\mathcal{M}$ and the pair $(a_D(u),a(u))$ can be interpreted as a holomorphic section of $V$. We have the following 
monodromies around 1,$-1$ and $\infty$:
\[
M_{\infty}=\left(\begin{array}{cc}
-1 & 2\\
 0 & -1 
\end{array}\right),\qquad M_{1}=\left(\begin{array}{cc}
1 & 0\\
 -2 & 1 
\end{array}\right),\qquad M_{-1}=\left(\begin{array}{cc}
-1 & 2\\
-2 & 3
\end{array}\right). 
\] 
We can notice that the monodromy matrices generate the group $\Gamma(2)$ of matrices in $SL(2,\mathbb{Z})$ congruent to the  
identity modulo 2, and that $\mathbb{C}\setminus{-1,1}$ coincides with the quotient of the upper half plane 
$H$ by $\Gamma(2)$, where the group action is defined by
\begin{equation}\label{gamma}
\tau\rightarrow\frac{a\tau+b}{c\tau+d},\qquad\tau\in H,\quad\left(\begin{array}{cc}
a & b\\
 c & d 
\end{array}\right)\in \Gamma(2). 
\end{equation}
We can now establish a link with the theory of algebraic curves noticing that the space $H/\Gamma(2)$ also parametrizes 
the family of elliptic curves \cite{clemens}
\begin{equation}\label{curvass}
y^{2}=(x-1)(x+1)(x-u). 
\end{equation}
The idea is then to associate to every point $u$ in the moduli space $\MM$ a genus one Riemann surface $E_{u}$ determined 
by the above equation. The curve becomes singular whenever two of the branch points in the $x$ plane coincide and this precisely 
happens for $u=1,-1,\infty$. Let us notice that (\ref{curvass}) has a $\mathbb{Z}_{4}$ symmetry which acts as 
$u\rightarrow-u, x\rightarrow-x, y\rightarrow\pm\imath y$. However, only a $\mathbb{Z}_{2}$ subgroup acts nontrivially on the 
$u$ plane. These are precisely the properties characterizing our theory that we discussed before. 

The first de Rham cohomology group $V_{u}=H^{1}(E_{u},\mathbb{C})$ of any torus $E_{u}$ has dimension 2 and can be thought to as 
the space of meromorphic (1,0)-forms with vanishing residues on $E_{u}$. We can then construct a vector bundle having as base space 
$\mathbb{C}\setminus{\{-1,1\}}$ and as fibers $V_{u}$; it can be locally trivialized choosing two continuously varying cycles 
$\gamma_{1},\gamma_{2}$ on $E_u$, in such a way that their intersection number is one, and integrating over them a representative of the 
equivalence classes in $H^{1}(E_{u},\mathbb{C})$. The crucial point is that this bundle can be identified with $V$. 
Our section $(a_{D},a)$ can then be written as $\omega=a_{1}(u)\omega_{1}+a_{2}(u)\omega_{2}$, where $\omega_{1}$ and $\omega_{2}$ 
are two independent elements in $H^{1}(E_{u},\mathbb{C})$. In order to extract $a_{D}$ and $a$ it is enough to choose two 
cycles on $E_{u}$ $\gamma_{1},\gamma_{2}$ as before and set
\begin{equation}\label{periodi}
a_{D}=\oint_{\gamma_{1}}\omega,\qquad a=\oint_{\gamma_{2}}\omega. 
\end{equation}
Furthermore, if we identify the periods $p_{1},p_{2}$ of the torus $E_{u}$, defined as
\[
p_{i}=\oint_{\gamma_{i}}\frac{dx}{y}\qquad i=1,2 
\]
with $da_{D}/du$ and $da/du$ respectively, we find
\begin{equation}\label{tau}
\tau(u)=\frac{da_{D}}{da}=\frac{da_{D}/du}{da/du}=\frac{p_{1}}{p_{2}}. 
\end{equation}
The positivity condition $\im\tau(u)>0$ follows now from the second Riemann relation. This identification allows us to find a 
solution which satisfies all the physical constraints described above. Conversely, assuming to have found a solution $\tau(u)$, 
we can determine for every value of $u$ the associated elliptic curve $E_{\tau}$, and consequently its periods. Since $(a_{D},a)$ 
and $(da_{D}/du,da/du)$ transform in the same way under the action of the group $SL(2,\mathbb{Z})$,  
the family of elliptic curves determined by $\tau(u)$ has the same monodromies as (\ref{curvass}). We can thus conclude that they 
coincide and that the given $\tau(u)$ function coincides with the one provided by the Seiberg-Witten solution.

From equation (\ref{periodi}) and from the definition of periods we can see that (\ref{tau}) is automatically satisfied if we impose 
the relation
\[
\frac{d\omega}{du}=f(u)\frac{dx}{y}. 
\]
All we need to do now is to determine $f(u)$, matching the asymptotic expansion of our solution with the behaviour of $a_{D}$ and 
$a$ in a neighborhood of the points 1,$-1$ e $\infty$. Expanding at first order as in \cite{SWI} we find that $f=-\sqrt{2}/4\pi$ does 
the job. Integrating in $u$ we can determine $\omega$ and thus the fundamental relations (with a suitable choice of periods $\gamma_{1},\gamma_{2}$)
\begin{align}\label{solsw}
a&=\frac{\sqrt{2}}{\pi}\int_{-1}^{1}\frac{\sqrt{x-u}}{\sqrt{x^{2}-1}}dx,\\
\nonumber\\
a_{D}&=\frac{\sqrt{2}}{\pi}\int_{1}^{u}\frac{\sqrt{x-u}}{\sqrt{x^{2}-1}}dx. 
\end{align}

\subsection{Solution for SQCD with classical gauge groups}

\subsubsection{The SW curves for $\mathcal{N}=2$ SQCD}

The idea of encoding the infrared effective action in an auxiliary family of algebraic curves
can be applied to a wide class of $\mathcal{N}=2$ models, including SYM theory with any gauge group and any matter content.
We will now list in Table \ref{ddd} the curves for SQCD with classical gauge groups (this is the class of models we will be concerned about in this 
thesis) which were found in \cite{AF}-\cite{HananyI} using arguments similar to the one we have just reviewed. We refer to these papers for the detailed 
derivation.

\begin{table}[!ht]
\begin{center}
\vskip .3cm \footnotesize
\begin{tabular}{|c|c|}
\hline
\text{Gauge group} & \text{SW curve} \\
\hline
 $SU(N)$ & $y^2=P_{N}^2(x)-4\Lambda^{2N-N_f}\prod_{i=1}^{N_f}(x+m_i)$\\
\hline
 $USp(2N)$ & $xy^2=[xP_{N}(x)+2\Lambda^{2N-N_f+2}\prod_{i=1}^{N_f}m_i]^{2}-4\Lambda^{4N-2N_f+4}\prod_{i=1}^{N_f}(x-m_{i}^{2})$\\
\hline
 $SO(2N)$ & $y^2=xP_{N}^{2}(x)-4\Lambda^{4N-2N_f-4}x^3\prod_{i=1}^{N_f}(x-m_{i}^{2})$\\
\hline
 $SO(2N+1)$ & $y^2=xP_{N}^{2}(x)-4\Lambda^{4N-2N_f-4}x^2\prod_{i=1}^{N_f}(x-m_{i}^{2})$ \\
\hline
\end{tabular}
\caption{\emph{The SW curves for $\mathcal{N}=2$ SQCD with $N_f$ flavors. $P_N(x)$ is a monic poynomial of 
degree $N$. In the $SU(N)$ case the coefficient of the term $x^{N-1}$ is set to zero. Turning it on gives the curve for $U(N)$ 
SQCD. There is no such constraint in the other cases.}}
\label{ddd}
\end{center}
\end{table}

\subsubsection{Singular points in the moduli space}

We have seen that there are (complex) codimension one singular submanifolds of the Coulob branch where some BPS states become 
massless. The Coulomb branch has dimension equal to the rank of the gauge group so, when the group is different from $SU(2)$, 
it is possible that two (or more) such singular submanifolds intersect, leading to singular points in which two (or more) different states become 
massless at the same time. If the Dirac product between these states is zero (i.e. they are relatively local), the low energy 
dynamics can be described in terms of a local lagrangian and it is possible to find a duality frame in which all the 
states have zero magnetic charges. We thus have in the infrared an abelian theory with a bunch of electrons.

If the states are relatively nonlocal such a duality frame does not exist and the low energy theory cannot be described by a 
local lagrangian. Such singularities usually signal the presence of an interacting IR fixed point and are ubiquitous 
in $\mathcal{N}=2$ theories. They are usually referred to as Argyres-Douglas (AD) points, since Argyres and Douglas described 
for the first time such singularities in \cite{AD}. Their analysis focuses on $SU(3)$ SYM theory, whose SW curve is 
$$y^2=(x^3-ux-v)^2-4\Lambda^{6}.$$ They observed that setting $u=0$ and $v=\pm2\Lambda^{3}$, the curve degenerates as 
$y^2=x^3(x^3\pm2\Lambda^3)$, and in a neighbourhood of $x=0$ can be approximated as $y^2\approx x^3$. They checked explicitly that two 
relatively nonlocal states simultaneously become massless at these points, signalling the presence of an infrared fixed point. This analysis has been soon extended to $SU(2)$ 
SQCD \cite{ADSW} and then to more general models \cite{EH,EHIY}. 

An important point I would like to stress is that it is not always enough to check that in a neighbourhood of the singular point 
there are relatively nonlocal states becoming lighter and lighter to infer that the low energy theory at the singular point is 
nonlocal: typically, when we include flavors, there are points in the Coulomb branch where the non-abelian gauge symmetry is 
partly restored. Formally the corresponding monopoles and dyons, which indeed are included in the spectrum in a neighbourhood 
of the singular point, become massless. This has been checked explicitly for the singular points in $SU(3)$ and $USp(4)$ SQCD 
with 4 flavors \cite{kauz,auz2}. This however does not imply that the theory becomes nonlocal. We will infact propose a local 
lagrangian description for the singular point of $USp(2N)$ SQCD with four flavors in chapter 5. These singular points will play an important role later.

\section{Confinement in softly broken $\mathcal{N}=2$ gauge theories}

One of the most important outcomes of the Seiberg-Witten solution is that it allows to understand the phenomenon of confinement 
in a subclass of four dimensional gauge theories, as explained by the authors in \cite{SWI}. Since this will be a central topic 
in this thesis, let us review the argument.

\subsection{Confinement in $SU(2)$ SYM theory}

As we have remarked in the introduction, $\mathcal{N}=2$ gauge theories are just distant relatives of QCD and many key properties indeed differ. For instance, these  
models do not exhibit confinement. However, we can achieve a confining theory just making the adjoint chiral multiplet massive, 
thus breaking extended supersymmetry. As long as the mass is small we can understand these models as perturbations of the 
$\mathcal{N}=2$ theory, whose behaviour in the IR is explicitly under control thanks to the SW solution.

Focusing on the by now familiar $SU(2)$ SYM case, if we turn on the superpotential term $\mu\Tr\Phi^2$ the degeneracy of vacua 
characterizing the $\mathcal{N}=2$ theory disappears and we are left with two vacua, as can be seen e.g. using Witten's index \cite{Windex}. 
We can take into account the effect of this perturbation in the IR adding to the effective lagrangian the superpotential $\mu U$ 
(it can be actually shown that this superpotential is exact also for large $\mu$ \cite{SWI}). 

At a generic point in the moduli space, where the only multiplet appearing in the effective action is the abelian vectormultiplet, 
this superpotential has no minimum and the corresponding vacuum is lifted by the $\mathcal{N}=1$ perturbation. The only candidate 
vacua are thus the two singular points and the vacuum counting indeed suggests that they are not lifted. As we have seen, the low 
energy effective action contains in this case a charged hypermultiplet $M$, $\tilde{M}$. Let us analyze e.g. the monopole vacuum; the other case 
is analogous. The effective superpotential now is $$\sqrt{2}\tilde{M}A_DM+\mu U.$$ The F-term equations are $$A_DM=0;\quad 
\sqrt{2}\tilde{M}M+\mu\frac{\de U}{\de A_D}=0.$$ At the monopole point $A_D$ has to vanish, since the monopole is massless, 
consistently with the first equation. From the SW solution one can verify that the derivative of $U$ with respect to $A_D$ is 
nonzero at $A_D=0$, so we get \begin{equation}\label{conf}\langle\tilde{M}M\rangle=\frac{\mu}{\sqrt{2}}\frac{\de U}{\de A_D}
\neq 0.\end{equation} We thus get that the magnetic monopole condenses! This breaks the $U(1)$ gauge symmetry, making the 
vectormultiplet massive via the (dual) Higgs mechanism. This implies that the theory (as expected) develops a mass gap and becomes confining, thanks 
to the dual superconductor mechanism for confinement proposed by 't Hooft and Mandelstam. What we thus get is basically a 
supersymmetric version of the abelian Higgs model, which is known to include topological vortex-like configurations labelled 
by an element of the fundamental group of the gauge group, in this case $\Pi_{1}(U(1))\simeq\mathbb{Z}$ \cite{NIEL}. These abelian vortices 
play a role analogous to the confining string of QCD. This represents the first example of four-dimensional theory in which the 
confinement mechanism is explicitly under control.

\subsection{Softly broken ${\cal N}=2$ $SU(2)$ SQCD and chiral symmetry breaking} 

We have seen that the prepotential for $SU(2)$ SYM is encoded in an auxiliary family of tori called SW curve. 
In a second paper Seiberg and Witten analyzed along the same lines $SU(2)$ SQCD \cite{SWII}. The associated curves can be easily 
deduced from table (\ref{ddd}). As the study of SYM theory taught us something about confinement, the analysis of these models 
led to new insights about chiral symmetry breaking.

In order to introduce some of the issues discussed in later chapters, let us briefly review the physics of the $N_{f}=2$ theory.   
There are four singularities in the $u=
\langle\Tr \Phi^{2} \rangle$ plane where some hypermultiplets become massless. The effective theory at these points is Abelian, dual $U(1)$ gauge theory. 
For small, nearly equal bare quark masses,  $m_{1}\simeq m_{2} \ll \Lambda$,     the singularities group into two pairs of nearby  singularities. 
The massless hypermultiplets in these two singularities are the Abelian monopoles in one or the other of the spinor representations 
\beq   ({\underline 2}, {\underline 1})\qquad   {\rm  or} \qquad   ({\underline 1}, {\underline 2})   \label{JR} \eeq 
of the flavor symmetry group $SO(4)\sim SU(2)\times SU(2)$.  

When the perturbation  $\mu \, \Tr \Phi^{2}$  ($\mu \ll \Lambda$)   is added in the system,  the monopole, say in  $({\underline 2}, {\underline 1})$, condenses, 
\beq   \langle M_{1} \rangle \sim \sqrt{\mu \Lambda}\;,     \label{mpcondense}
\eeq
 the dual $U_{D}(1)$ gauge group is Higgsed, and the system is in a confinement phase. 
An interesting feature of this case is that the confinement order parameter at the same time breaks the global symmetry as  
\beq    SU(2)\times SU(2) \to SU(2)\;. \label{symbrk}
\eeq
 
The effective action of Seiberg-Witten correctly describes the low-energy excitations:  the exactly massless Nambu-Golstone bosons 
of the symmetry breaking  (\ref{symbrk})  and their superpartners.  Unlike the light flavored standard  QCD,  the massless 
Nambu-Goldstone bosons do not carry  the quantum numbers of the remaining unbroken $SU(2)$.
There are also light but massive dual photon and dual photino of the order of $\sqrt{\mu \Lambda}$, which arise as a result of 
the dual Higgs mechanism.  

 All these light particles are gauge invariant states (they are asymptotic states); the presence of the original quarks degrees 
 of freedom can be detected in the flavor quantum numbers \cite{JR}.  
 
The low energy system is a dual Abelian $U(1)$ gauge theory broken by the magnetic monopole condensation (\ref{mpcondense}).   The 
ANO vortex of this system, with tension $\sim \mu \Lambda$,  carries the (Abelianized) chromoelectric flux.  
The fact that the underlying  $SU(2)$ theory is simply connected, means that such a vortex must end:  the endpoints are the 
quarks (and squarks) of the underlying theory.  Quarks are confined. 

An important point we want to stress is the fact that the particles becoming massless at each abelian singularity of the Seiberg-Witten 
curve  are  pure magnetic monopoles even though they carry distinct labels  $\{n_{m\,i}, n_{e\, i}\}$ ($i=1,2, \ldots, N-1$)   and coupled to different ``magnetic duals'',  $n_{m} A_{D \, \mu} + n_{e} A_{\mu}$. 
In the case of $SU(2)$ theory where  there is only a single $U(1)$ gauge group at low energies, this fact is easily seen \cite{KT,CKT}.
 At a singularity of the quantum moduli space
where the  $(n_m, n_e)$ dyon   becomes massless  
\be n_m   a_D  +  n_e   a =0, 
\ee
the exact SW  solution tells us that 
\beq 
\frac{n_m (d a_D / du) +  n_e (d a / du) }{ (d a / du)}  =0, 
\label{theta0}
\eeq
due to a logarithmic singularity in the denominator.  Thus
\beq 
\theta_{eff} = \Re \frac{d a_D }{ da}  \, \pi = - \frac{n_e }{ n_m}\pi
\label{theeff}
\eeq 
and  the electric charge of  a $(n_m, n_e)$ ``dyon''  is   \cite{Witten}
 \beq
\frac{2 }{ g}Q_e = n_e + \frac{\theta_{eff} }{ \pi} n_m =0.
\label{dyonismonop}
\eeq

In the case of the $SU(2)$ theory with $N_{f}=2$ with small masses, the massless dyons (which condense upon $\mu \Phi^{2}$ perturbation)  carry $(n_{m},n_{e})$
charges  
\be  (n_{m},n_{e}) = (1,0), \qquad  \theta_{eff} =0\;, 
\ee
in one doublet of singularities,  and 
\be  (n_{m},n_{e}) = (1,1), \qquad  \theta_{eff} = -  \pi\;,  
\ee
in the other.   Thus in all vacua  the quarks (with charges $(n_{m}^{1},n_{e}^{1})= (0,1)$) are confined, carrying a relative nonzero  Dirac unit 
\be   D = n_{m}^{1}  n_{e}^{2} -     n_{m}^{2}  n_{e}^{1} \qquad Mod \,\,{2},  
\ee
with respect to the condensed fields, $(n_{m}^{2}, n_{e}^{2})$.  In the $m \to 0$  limit, a ${\mathbbm Z}_{2}$ symmetry ensures that the physics at the two vacua look identical,  even though the light monopoles (dyons) are coupled locally to two 
different magnetic duals.   

As was shown in \cite{SWII}, all massless ``dyons'' in $SU(2)$ theory with various $N_{f}$  carry  $n_{m}=1$.  Their condensation upon the $\mu \Phi^{2}$  perturbation leads to quark confinement. The only 
exception occurs \cite{SWII}  in one of the vacua  of $N_{f}=3$ theory,  where  massless $(2,1)$ dyons appear as the infrared degrees of freedom.  This vacuum  (where $\theta_{eff}= - 1/2$)  survives the $\mu \Phi^{2}$ deformation,  the $(2,1)$ dyons condense,  but  quarks are unconfined:  it is in  't Hooft's  oblique confinement phase \cite{TP}.  The phase  of the  pure (non supersymmetric)  $SU(2)$ Yang-Mills theory with $\theta = -\pi$ is believed to be in such a phase, where the composite of the $(1,0)$
monopole and the $(1,1)$ dyon with charges $\mp\tfrac{1}{2}$ condense.

\subsection{Confinement in $SU(N)$ SQCD \label{quantumr}}

The classical and quantum moduli space of the vacua of the ${\cal N}=2$ supersymmetric $SU(N)$ QCD has been first studied 
systematically by Argyres, Plesser and Seiberg and then by others \cite{APS,CKM,APSII,HOO}.  
 Of particular interest are the $r$-vacua characterized by an effective low-energy $SU(r)\times U(1)^{N-r}$ gauge symmetry, with massless monopoles carrying the 
 charges shown in  the Table  taken from ~\cite{APS} (we will discuss further their properties later on). When the adjoint scalar mass  $\mu \Tr\Phi^{2}$  term, which breaks supersymmetry to  ${\cal N}=1$,
is added the massless Abelian $M_{k}$ and non-Abelian monopoles $\cal M$ (see \cite{GNO}-\cite{EWW}) all condense, bringing the system to a confinement phase.
\begin{table}[b]
\begin{center}
\vskip .3cm
\begin{tabular}{|ccccccc|}
\hline
&   $SU(r)  $     &     $U(1)_0$    &      $ U(1)_1$
&     $\ldots $      &   $U(1)_{N-r-1}$    &  $ U(1)_B  $  \\
\hline
$n_f \times  {\cal M}$     &    ${\underline {\bf r}} $    &     $1$
&     $0$
&      $\ldots$      &     $0$             &    $0$      \\ \hline
$M_1$                 & ${\underline {\bf 1} } $       &    0
&
1      & \ldots             &  $0$                   &  $0$  \\ \hline
$\vdots $  &    $\vdots   $         &   $\vdots   $        &    $\vdots   $
&             $\ddots $     &     $\vdots   $        &     $\vdots   $
\\ \hline
$M_{N-r-1} $    &  ${\underline {\bf 1}} $    & 0                     & 0
&      $ \ldots  $            & 1                 &  0 \\ \hline
\end{tabular}
\caption{The massless non-Abelian and Abelian monopoles  and their charges  at the $r$ vacua
at the root of a ``non-baryonic'' $r$-th   Higgs branch.  }
\label{tabnonb}
\end{center}
\end{table}
The form of the effective action describing these light degrees of freedom is dictated by the ${\cal N}=2$ supersymmetry and the gauge and flavor symmetries. 
The effective superpotential has the form ~\cite{APS,CKM}
   \bea
W_{r-vacua} &=& \sqrt2 \,  \Tr ({\cal M} \phi {\tilde {\cal M}}) + \sqrt2  \, a_{D 0} \Tr ({\cal M}
{\tilde {\cal M}}) + \sqrt 2 \sum_{k=1}^{N-r-1} a_{D k} M_k {\tilde M}_k + \nonumber \\
&+& \mu \left(\Lambda \sum_{k=0}^{N-r-1} \,c_k  a_{D  k} + \frac{1}{ 2} \Tr \phi^2\right),
\label{nonbaryonic}
\eea
where $\phi$ and $a_{D\, 0}$ are the adjoint scalar fields in the  ${\cal N}=2$    $SU(r)\times U(1)$ vector multiplet, $a_{D\,  k}$, $k=1,2,\ldots, N-r-1$  are the adjoint scalars of the Abelian $U(1)^{N-r-1}$ gauge multiplets. $M_{k}$'s are  the Abelian monopoles, each carrying one of the magnetic $U(1)$ charges, whereas ${\cal M}$ (with $r$ color components and in the fundamental representation of the flavor $SU(N_{f})$ group)  are the non-Abelian monopoles.   The terms linear in $\mu$ is generated by the microscopic   ${\cal N}=1$ perturbation $\mu \, \Tr \Phi^{2}$, written in terms of the infrared degrees of freedom  $a_{D k}$ and $\phi$, and $c_{k}$ are some dimensionless  constants of order of unity.  


These  quantum $r$-vacua are known to exist only for $r \le  \left[ \frac{N_{f}}{2} \right]$. 
 
When small, generic bare quark mass terms 
\beq     W_{masses} =  m_{i} \, Q_{i} {\tilde Q}_{i}
\eeq
are added in the microscopic theory,
 the infrared theory gets modified further by the addition 
\beq
\Delta W_{masses} = m_i  \, {\cal M}^{i} {\tilde {\cal M}}_i + \sum_{k=1}^{N-r-1} S_k^j\,
m_j  M_k {\tilde M}_k ,\qquad  i,j =1,\dots, N_{f}
\label{masses1}\eeq
where $S_{k}^{j}$ are the $j$-th quark number carried by the $k$-th monopole.  
Supersymmetric vacua are found by minimizing the potential following from Eq.~(\ref{nonbaryonic}) with Eq.~(\ref{masses1}),  
and by vanishing of the $D$-term potential.  

   The part of the decoupled $U(1)^{N-r-1}$ theory involving  Abelian monopoles  is trivial and gives the VEV's 
   \beq   a_{D k}  \sim O(m_i); \quad    M_k= {\tilde M}_k \sim \sqrt{\mu
\Lambda}\;,   \label{asinsu2} \eeq
as in the $SU(2)$  theories.

   The  equations for the $SU(r)\times U(1)$ sector   (see Eq.~(\ref{effD1})-Eq.~(\ref{effF4})) are less trivial.  The equations look rather similar to the  
semiclassical equations of the microscopic $SU(N)$ theory,  which are valid for $\mu \gg \Lambda$,  $|m_{i}| \gg \Lambda$,  
reported in the Appendix,  but there are a few crucial differences. 

One is that the effective gauge group $SU(r)\times U(1)$ is not simply connected and the low-energy system generates vortex solutions, while the microscopic theory cannot possess
such solitons.   Secondly,  the massless hypermultiplets in the system describe magnetically charged particles,  in contrast to those in the  
original ultraviolet Lagrangian.  Finally,  the range of validity of the effective theory is limited to the excitations of energies much less than the dynamical scale $\Lambda$, as the particles of masses of the order of $\Lambda$ or larger have been integrated out in obtaining it.

This last fact makes  the identification of the correct solutions of  Eq.~(\ref{effD1})-Eq.~(\ref{effF4}) somewhat a subtle task
(i.e., fake solutions involving VEVs of the order of $\Lambda$ must be disregarded):
the solutions are given by \cite{CKM}:  
\beq   \phi =\frac{1 }{ \sqrt{2}} \, \left(\begin{array}{ccc}-m_1- \sqrt{2}  \psi_0 &  &  \\ & \ddots &  \\ &  & -m_r- \sqrt{2} \psi_0\end{array}\right)\;, 
    \label{phiveveff}   \eeq
\beq  {\cal M}_a^i  =
\left(\begin{array}{cccc}d_1 &  &  &  \\    & \ddots &    &  \\  &  & d_r &      \\   &  &    & {\mathbf 0}\end{array}\right)\;, \qquad {\tilde {\cal M}}_i^a  =
\left(\begin{array}{cccc}{\tilde d}_1 &  &  &  \\    & \ddots &    &  \\  &  & {\tilde d}_r &      \\   &  &    & {\mathbf 0}\end{array}\right)\;, 
\eeq
where   $d_i$, ${\tilde d}_i$'s   and $\psi_{0}$  are given by
\beq  \psi_0= -\frac{  1 }{ \sqrt{2} \,   r}    \sum_i m_i , \label{corrvacua}  \eeq
\beq    d_i  {\tilde d}_i  = \mu \left( m_i - \frac{1}{r}\sum_j^r
m_j\right) - \frac{\mu
\Lambda}{\sqrt{2}\,  r}\;.  \eeq
In the limit $m_{i}\to 0$   the monopole VEV's tend to 
\beq     \langle  {\cal M}_{a}^{i} \rangle =\delta_{a}^{i} \, \sqrt{\frac{\mu
\Lambda}{\sqrt{2}\,  r}}, \quad i=1,\ldots, r; \quad    \langle  {\cal M}_{a}^{i}\rangle  =0,  \quad  i = r+1, \ldots, N_{f}\;.   \label{cflock}
\eeq
The system is in a color-flavor locked phase of the dual $SU(r)$ gauge theory. The flavor $SU(N_{f})\times U(1)$  symmetry of the underlying SQCD is 
dynamically broken as
\beq       SU(N_{f})\times U(1) \to  U(r) \times U(N_{f}-r)\;.  \label{symbr}
\eeq
The fact that $ \langle  {\cal M}_{a}^{i}\rangle $  is nonvanishing in the limit $m_{i}\to 0$ means that the symmetry breaking is dynamical, 
and this property distinguishes the $r$ vacua appearing at these nonbaryonic roots from the vacua at the baryonic root\footnote{ The vacua 
at the baryonic root, present only for $N_{f}> N$,  are interesting as they are characterized \cite{APS} by the low-energy 
effective $SU({\tilde N})$ gauge group, ${\tilde N}\equiv N_{f}-N$. 
Thus it was argued in \cite{APS} that these might be relevant for the  understanding of the Seiberg duality in the ${\cal N}=1$ 
SQCD, and some further observations on this point were made recently \cite{SYI}-\cite{SYIV}.  These vacua at the baryonic root 
are however nonconfining \cite{CKM} in the limit $m \to 0$, $\mu \ne 0$, and this is the reason why we focus on the $r$ vacua. 
}. 

Until now we have discussed only the case of degenerate (or slightly unequal) bare masses $m_i$ for the flavors. An important observation is the 
fact that, when the $m_i$'s are generic, each $r$ vacuum splits in $\binom{N_f}{r}$ abelian vacua. Taking into account the Witten effect as in section 2, 
it is easy to generalize the argument we gave for $SU(2)$ to $SU(N)$ and conclude, as we will now see, that the particles becoming massless in each one of these vacua 
are magnetic monopoles and not dyons.
Consider a Cartan basis for $SU(N)$: 
 \begin{equation}\begin{aligned}   &[H_i, H_k]=0, \qquad (i,k=1,2,\ldots, r);  
    \qquad   [H_i,  E_{\alpha}] = \alpha_i \, E_{\alpha}; \\
    &[E_{\alpha}, E_{-\alpha}]= \alpha^i  \, H_i;\qquad 
 [E_{\alpha}, E_{\beta} ]=  N_{\alpha \beta}\,E_{\alpha + \beta} \qquad (\alpha+ \beta \ne 0).  
  \end{aligned}\end{equation}
where $\alpha$'s are the root vectors.    $3(N-1)$  generators can be grouped into  $SU(2)$ subsets of  generators,  
\beq \qquad   [H_i,  E_{\alpha}] = \alpha_i \, E_{\alpha}; \qquad
    [E_{\alpha}, E_{-\alpha}]= \alpha^i  \, H_i,
    \eeq
containing $N-1$ diagonal $U(1)$ generators. 

Assuming  Abelianization  the magnetic monopoles are the 't Hooft-Polyakov monopoles living in these broken $SU(2)$ groups.  Each of the $SU(2)$ group acquires a $\theta$ term,
\beq   \frac{\theta}{32\pi^{2}}    \sum_{j=1}^{3}  F_{\mu \nu}^{j} \tilde F^{j \, \mu \nu}  
=  \frac{\theta}{8\pi^{2}}      \sum_{j=1}^{3}   {\bf E}^{j} \cdot   {\bf B}^{j},
\eeq
The $i$-th magnetic monopole contributes to the electromagnetic static energy
\beq   \frac{\theta}{8\pi^{2}}      \sum_{j=1}^{3}   {\bf E}^{j} \cdot   {\bf B}^{j}=\
\frac{\theta}{8\pi^{2}}    (-\nabla \phi) \cdot  \nabla \frac{  g_{m}}{r} =-\frac{\theta}{2\pi} g_{m} \phi\, \delta^{3}({\bf r})\;. 
\eeq
thus carries the $i$-th ``electric''  $U(1)$ charge,  $-\frac{\theta}{2\pi} g_{m}$.  Of course,  the monopole of the $i$-th  $SU(2)/U(1)$ sector  ($i=1,2,\ldots N-1$) is neutral with respect to all other $U(1)$'s. 

Under the dynamical hypothesis of Abelianization,  thus each $U(1)$ factor has its own Witten effect.  The argument made in the $SU(2)\to U(1)$ theories   works here too. 

If we now take the equal mass limit, we recover $r$ vacua with their non-abelian sector. The particle becoming massless in each abelian vacuum now 
combine into multiplets in the fundamental representation of the $SU(r)$ gauge group. We can thus conclude that these are truly non-abelian monopoles.

\subsubsection{Low-energy excitations and non-Abelian chromoelectric vortices}

The obvious low-energy excitations of this system are the massless and light particles described by the effective Lagrangian described above. These can be found by expanding around the vacua (\ref{phiveveff})-(\ref{cflock}).  They contain massless Nambu-Goldstone bosons of the breaking (\ref{symbr}) and their superpartners, as well as light pseudo Nambu-Goldstone particles of the $SU_{R}(2)$  breaking. Also, the dual $SU(r)\times U(1)^{N-r}$ gauge bosons and gauginos form light massive multiplets. 

What is perhaps not so well known (however, see \cite{SYI}-\cite{SYIV} for related remarks)  is the fact that, apart from these elementary excitations, the system described by Eqs.(\ref{phiveveff}) and (\ref{masses1})  has low-energy non-Abelian excitations of a different sort.  As $\Pi_{1}(U(r) \times U(1)^{N-r-1}) = {\mathbbm Z}^{N-r}$, the low-energy system possesses soliton vortices. In the vacuum (\ref{cflock})  the minimum vortex configuration  (see Eq.~(\ref{minivort1})) breaks the color-flavor diagonal symmetry to $SU(r-1)\times U(1)$: it is a non-Abelian vortex \cite{HT}-\cite{GJK}.  The fluctuation of the orientational modes of 
\be  CP^{r-1}= SU(r)/SU(r-1)\times U(1)
\ee
  is described by a vortex worldsheet sigma model, 
\begin{align}
S_{1+1}   & =   2\beta \int dtdz \; \tr\left\{
X^{-1}\p_\alpha B^\dag
Y^{-1}\p_\alpha B
\right\}   \non
&=  2\beta \int dtdz \; \tr\left\{
\left(1 + B^\dag B\right)^{-1}\p_\alpha B^\dag
\left(\mathbf{1}_{r-1} + B B^\dag\right)^{-1}\p_\alpha B
\right\} \ , \label{eq:sigmamodelaction}
\end{align}
where $B$, a $r-1$ component vector,  represents the inhomogeneous  coordinates of $CP^{r-1}$  (see Eq.~(\ref{eq:Umatrix})) and   $\beta$ is a constant. 
The low energy system has also $N-r-1$  distinct Abelian (Abrikosov-Nielsen-Olesen) vortices, as the dual $U(1)^{N-r-1}$  theory is in the Higgs phase (see Eq.~(\ref{asinsu2})). 

The point of crucial importance is the fact that the underlying $SU(N)$ theory, being simply connected, 
does not support a  vortex solution.  It means that  both the non-Abelian vortex (\ref{eq:sigmamodelaction}) and the Abelian vortices of the $U(1)^{N-r-1}$  sectors  must end. 
These vortices in the dual, magnetic theory  carry chromoelectric fluxes.  The endpoints are quarks of the fundamental theory,  which, 
being relatively nonlocal to the low-energy effective degrees of freedom, and also  having dynamical masses of the order of $\Lambda$, 
are not explicitly visible in the low-energy effective action \footnote{Not all effects related to the underlying quarks are invisible at low energies, however. The zero-energy quark fermion modes are indeed  responsible for giving the flavor quantum numbers to the monopoles
as  in Table~\ref{tabnonb}.}.  The quarks are confined\footnote{Of course, as the underlying theory contains scalars in the fundamental representation there are no distinct 
phases between the confinement and Higgs phase in these theories (complementarity).}. 

The system produces more than  one kinds of confining strings as the $SU(N)$ gauge symmetry of the ultraviolet theory is dynamically broken to $U(r)\times U(1)^{N-r-1}$ at low energies, so the mesons appear in various Regge trajectories. The only exception is the case of $SU(3)$ theory,
where  the only nontrivial $r$ vacua ($r=2$ in this case) corresponds to a low energy $U(2)$ theory with 
$  \Pi_{1}(U(2)) = {\mathbbm Z}$: there is a unique universal Regge trajectory.

\section*{Appendix}
\addcontentsline{toc}{section}{Appendix}

\subsection*{Classical vacuum equations}

The superpotential has the form
\beq W= \mu \Tr \, \Phi^2  + \sqrt2 \,{\tilde Q}_i^a \Phi_a^b Q_b^i +
     m_i \, {\tilde Q}_i^a Q_a^i\;.\eeq
The vacuum equations read
\beq   [ \Phi, \Phi^{\dagger}] = 0 \, ;
\label{D1}
\eeq
\beq
\nu \delta_a^b=  Q_a^i (Q^{\dagger})_i^b - ({\tilde Q}^{\dagger})_a^i
{\tilde Q}_i^b \, ;
\label{D2}
\eeq
\beq
Q_a^i {\tilde Q}_i^b - \frac{ 1}{ N} \delta_a^b (Q_c^i {\tilde Q}_i^c) +
\sqrt2  \, \mu \Phi_a^b = 0 \, ;
\label{F1}
\eeq
\beq
Q_a^i m_i + \sqrt2 \,\Phi_a^b Q_b^i = 0  \qquad ( { \hbox {\rm no sum
over}} \, \,i) \, ;
\label{F2}
\eeq
\beq
m_i {\tilde Q}_i^a +  \sqrt2 \, {\tilde Q}_i^b \Phi_a^b = 0 \qquad (
{\hbox{\rm no sum over}} \,\,i) \, .
\label{F3}
\eeq
By gauge rotation $\Phi$ can be taken as
\beq
\Phi = \diag \, (\phi_1, \phi_2, \ldots  \phi_{N}) \, , \qquad \sum
\phi_a = 0 \, .
\label{phivev}
\eeq
  $Q_a^i $ and ${\tilde  Q}_i^b$
are either nontrivial eigenvectors of the matrix $\Phi $ with possible
eigenvalues $m_i$,  or null vectors.    The solution with eigenvalues $m_1, m_2, \ldots, m_r$ is    
\beq
\Phi = \frac{1}{\sqrt{2}}\diag \, (-m_1, \ldots, -m_r, c,
\ldots , c) \, ; \quad c = \frac{1 }{ N-r} \sum_{k=1}^r m_k.
\label{diagphi}
\eeq
\beq   Q_a^i  =
\left(\begin{array}{cccc}   f_1 &  &  &  \\    & \ddots &    &  \\  &  & f_r &      \\   &  &    & {\mathbf 0}\end{array}\right)\;, \qquad {\tilde Q}_i^a  =
\left(\begin{array}{cccc}{\tilde f}_1 &  &  &  \\    & \ddots &    &  \\  &  & {\tilde f}_r &      \\   &  &    & {\mathbf 0}\end{array}\right)\;, 
\label{vevofqti}\eeq
where
\beq   r=0,1, \ldots, {\hbox {\rm min}} \, \{N_f, N-1\},
\eeq
The solution for $f_{i}, {\tilde f}_{i}$ is   (see \cite{CKM} for more details)
\beq
f_i {\tilde f}_i =  \, \mu \,m_i  + \frac{ 1 }{ N -r} \mu \,\sum_{k=1}^r
m_k \;, \qquad  
f_i^2 = | {\tilde f}_i |^2 \,, \qquad 
(f_{i} > 0)\, .
\label{solnd}
\eeq
The number of the quark flavors ``used'' to make solutions define  various classical $r$-vacua. 
As the  solution with a given $r$ leaves a local  $SU(N-r)$ invariance  it counts as a set of $N-r$  solutions (Witten's
index \cite{Windex}).
In all  there are precisely
\beq
{\cal N} = \sum_{r=0}^{{\hbox {\rm min}} \, \{N_f, N-1\}}\, (N-r)
\, \binom{N_f}{r}
\label{nofvac}
\eeq
classical  solutions for generic $m_{i}$'s and $\mu \ne 0$.

\subsection*{Equations determining VEVs in the quantum  $r$ vacua}

The D-tem potential gives 
\beq  0=  [ \phi, \phi^{\dagger}]; \label{effD1}\eeq
\beq \nu \delta_a^b=  q_a^i (q^{\dagger})_i^b - ({\tilde
q}^{\dagger})_a^i
   {\tilde q}_i^b; \label{effD2}\eeq
\beq 0=  q_a^i (q^{\dagger})_i^a - ({\tilde
q}^{\dagger})_a^i
   {\tilde q}_i^a; \label{effD2bis}\eeq
   while the F-term equations are 
\beq q_a^i {\tilde q}_i^b - \frac{ 1}{r} \delta_a^b (q_c^i {\tilde q}_i^c)
+ \sqrt2  \, \mu \phi_a^b =0; \label{effF1} \eeq
\beq   0= \sqrt2 \,\phi_a^b q_b^i + q_a^i  (m_i  + \sqrt2  \, a_{D0})
; \quad ( {\hbox{\rm no
sum
over}} \,\,i,\,\, a) \,
   \label{effF2} \eeq
\beq  0=  \sqrt2 \, {\tilde q}_i^b
\phi_b^a +   (m_i  + \sqrt2  \, a_{D 0}) \,  {\tilde q}_i^b
\quad ( {\hbox{\rm no sum over}}\,\,i, \,\, a).  \, \label{effF3} \eeq
\beq  \sqrt2 \, \Tr ( q {\tilde q})  + \mu \Lambda =0.  \label{effF4} \eeq
The $SU(r)$  adjoint  scalars can be diagonalized  by color rotations,
\beq   \diag \, \phi = ( \phi_1, \phi_2, \ldots  \phi_{r}), \quad
\sum \phi_a=0. \label{effphivev} \eeq

\subsection*{Non-Abelian vortex in the $r$ vacua}

\begin{align}
{\cal M} &=
\begin{pmatrix}
e^{i\theta}\phi_1(\rho)    1 & 0 \\
0 & \phi_2(\rho)\mathbf{1}_{r-1}
\end{pmatrix}
= \frac{e^{i\theta}\phi_1(\rho)+\phi_2(\rho)}{2}\mathbf{1}_{r}
+ \frac{e^{i\theta}\phi_1(\rho)-\phi_2(\rho)}{2} T \;,   \nonumber   \\
A_i &=
\frac{1}{2}\epsilon_{ij}\frac{x^j}{\rho^2}\left[
\left(1-f(\rho)\right) \mathbf{1}_{r}+\left(1-f_{\rm NA}(\rho)\right) T
\right]  \ ,     \label{minivort1}
\end{align}
which is oriented to a specific direction.  In (\ref{minivort1})
\beq T \equiv \diag\left(1,-\mathbf{1}_{r-1}\right)\ , \eeq
and $z, \rho,\theta $ are cylindrical coordinates.  
The profile functions $\phi_{1,2}(\rho), \, f(\rho)$ and $f_{\rm NA}(\rho)$ satisfy the boundary conditions 
\begin{align}
&\phi_{1,2}(\infty) = \frac{v}{\sqrt{2N}} \ , \quad
f(\infty)=f_{\rm NA}(\infty) = 0, \\
&\phi_1(0) = 0 \ , \quad
\partial_r\phi_2(0) = 0 \ , \quad
f(0) = f_{\rm NA}(0) = 1 \ .
\label{eq:bcs}
\end{align}
The vortex oriented in a generic direction in color-flavor space can be written as 
\begin{align}
{\cal M} &= U
\begin{pmatrix}
\phi_1(\rho)  1 & 0 \\
0 & \phi_2(\rho)\mathbf{1}_{r-1}
\end{pmatrix} U^{-1},\\
A_i &= -\frac{1}{2}\epsilon_{ij}\frac{x^j}{r^2} \left[
f(\rho)\, \mathbf{1}_{r} + f_{\rm NA}(\rho) \, U T U^{-1} \right].\label{genericorient}
\end{align}
The matrix $U$ represents the coset  
\be     SU(r) / SU(r-1)\times U(1) \sim CP^{r-1},
\ee 
and is expressed in terms of an $r-1$ dimensional complex vector $B$ as 
$$\begin{aligned}
U =
\begin{pmatrix}
1 & - B^\dag \\
0 & \mathbf{1}_{r-1}
\end{pmatrix}
\begin{pmatrix}
X^{-\frac{1}{2}} & 0 \\
0 & Y^{-\frac{1}{2}}
\end{pmatrix}
\begin{pmatrix}
1& 0 \\
B & \mathbf{1}_{r-1}  
\end{pmatrix}
=
\begin{pmatrix}
X^{-\frac{1}{2}} & - B^\dag Y^{-\frac{1}{2}} \\
B X^{-\frac{1}{2}} & Y^{-\frac{1}{2}}
\end{pmatrix}
\label{eq:Umatrix} \ ,
\end{aligned}$$
where the matrices $X$ and $Y$ are defined by
\beq
X\equiv   1+ B^\dag B \ , \quad
Y\equiv\mathbf{1}_{r-1} + B B^\dag \ ,
\eeq
This form of the unitary $SU(r)$ matrices containing only the  coset coordinates $B$ is known as the reducing matrix.

\chapter{Four dimensional $\mathcal{N}=2$ theories from six dimensions}

The recent progress in the understanding of supersymmetric field theories is mostly due to their realization in string/M-theory, 
which allow to rephrase most of the physical content of these models in geometric terms, leading eventually to completely new 
insights \cite{HW,branea}. These constructions will indeed play a role in our analysis and this chapter is devoted to reviewing them. We will 
first outline the basic properties of the brane realization of $\mathcal{N}=2$ theories proposed by Witten and its lift to 
M-theory \cite{branea} and we will then review the fundamental work by Seiberg and Argyres \cite{AS} on the S-dual description of superconformal 
theories. These results constitute the starting point of the recent six-dimensional construction proposed by Gaiotto, which 
will play a fundamental role in chapters 4 and 5 and represents the central topic of this chapter. We will then briefly discuss 
the BPS quiver technique developed by Vafa and Cecotti, which allows to extract several nontrivial information about the BPS 
spectrum of a large class of $\mathcal{N}=2$ theories.

\section{Type IIA string theory and Witten's construction}

\begin{figure}
\centering{\includegraphics[width=.8\textwidth]{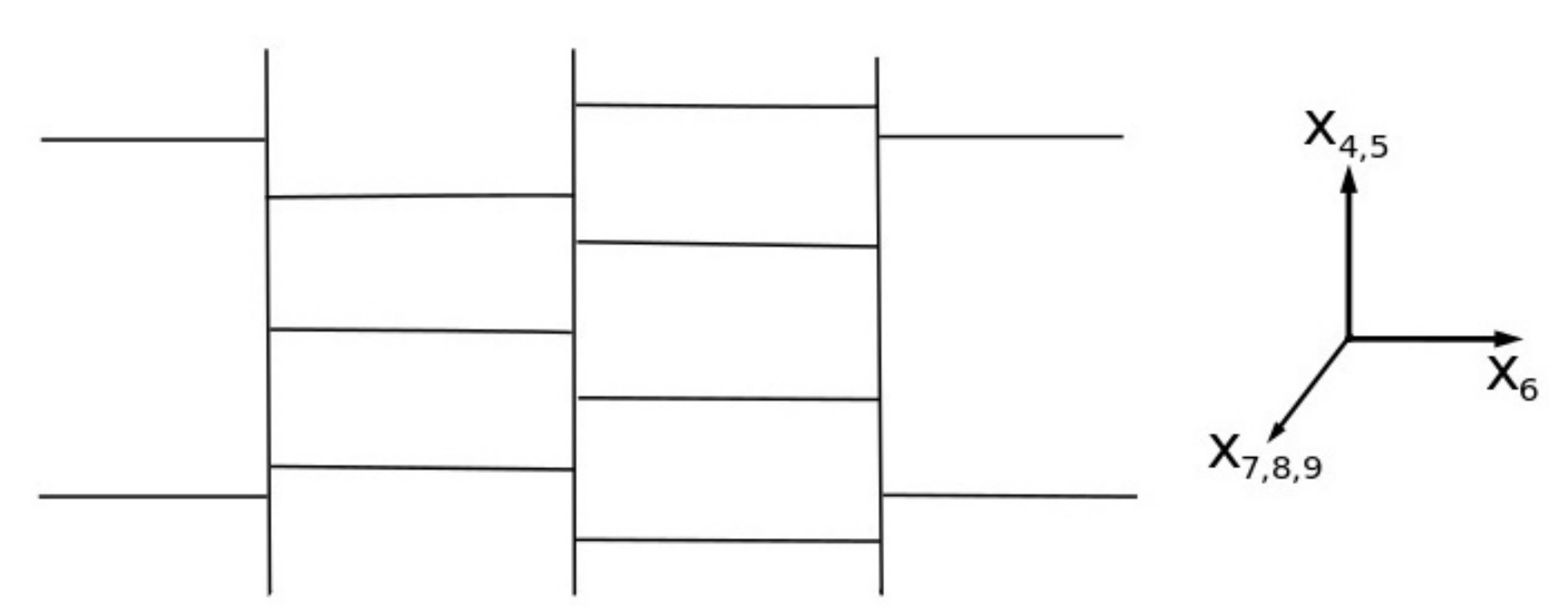}} 
\caption{\label{brane}\emph{The brane diagram associated to a $\mathcal{N}=2$ $SU(3)\times SU(4)$ gauge theory. The matter content 
is given by a hypermultiplet in the bifundamental, two in the fundamental of $SU(3)$ and other two in the fundamental of $SU(4)$. 
The $D4$ branes extend along the $x_6$ direction and are represented by horizontal lines. Vertical lines describe the $NS5$ 
branes extended along the $x_4$,$x_5$ coordinates.}}
\end{figure}

In \cite{branea} Witten showed that the $\mathcal{N}=2$ linear quivers of unitary groups can be constructed using a system of branes 
in Type IIA string theory. The construction involves inserting $NS5$ branes located at $x_{7,8,9}=0$ and classically at a fixed 
position in the $x_6$ direction. We also introduce parallel $D4$ branes extended along the $x_{0,1,2,3}$ and $x_6$ directions. They 
are suspended between consecutive $NS5$ branes or terminate on a fivebrane at one end and go to infinity at the other end. Classically every 
fourbrane is located at a fixed position in the ($x_4,x_5$) plane (see Figure \ref{brane}). It will be convenient to introduce 
the complex variable $v=x_4+ix_5$.

Actually the fivebranes are not really sitting at a definite position in $x_6$ as the classical picture suggests: the $D4$ branes 
pull them creating a dimple and the $x_6$ coordinate is actually a function of $v$ which can be determined by minimizing the 
worldvolume of the $NS5$ branes. Asymptotically (for very large $\vert v\vert$) we have \cite{branea} \be\label{asymp} x_6=k\sum_i\ln\vert v-a_i\vert-
k\sum_j\ln\vert v-b_j\vert+\text{const.}\ee where $k$ is some constant and $a_i$, $b_j$ are the $v$ positions of the $D4$ branes ending on the given fivebrane 
respectively from the left and from the right. The fivebrane has asymptotically a well defined position in the $x_6$ direction 
if the forces exerted by the fourbranes from the left and from the right exactly balance. We will see that this condition of 
mechanical equilibrium for the brane system corresponds to the vanishing of the beta function for the four-dimensional theory.

Since the $D4$ branes are finitely extended in the $x_6$ direction, their worldvolume theory is macroscopically a four-dimensional 
field theory in $x_{0,1,2,3}$ with eight supercharges. Open strings stretching between two $D4$ branes suspended between the 
same fivebranes give rise to a $SU(N)$ vectormultiplet (where $N$ is the total number of fourbranes), whereas strings connecting 
two $D4$ branes ending on the same fivebrane from the left and from the right give a hypermultiplet in the bifundamental of the 
two gauge groups. We thus end up with a linear quiver of unitary gauge groups. The infinite fourbranes at both ends of the brane 
system give hypermultiplets in the fundamental of the first and last gauge group respectively. See Figure \ref{brane} for an example.

The motion of the $D4$ branes in the $v$ direction give a contribution to the fivebrane kinetic energy, which is proportional to 
the integral on its worldvolume of $\vert\partial_{\mu}x_6\partial^{\mu}x_6\vert$. The integral converges only if 
\be\sum_ia_i-\sum_jb_j=\text{const.}\ee and this constant determines the mass parameter of the corresponding hypermultiplet 
in the bifundamental, which is more precisely given by the difference of the average position of the branes on the left and on 
the right. The branes at the ends of the brane system are an exception, since they are infinitely massive and cannot move. The 
difference between their position and the average position of the neighbouring brane system can be identified with the mass of 
the corresponding multiplet in the fundamental. Such a costraint implies that the $U(1)$ factor of each gauge group is ``frozen'', and 
the corresponding vectormultiplet is thus missing from the spectrum. This is the reason why the gauge groups are $SU(N)$ rather 
than $U(N)$. 

Apart from this restriction the $a_i$'s and $b_j$'s are free to vary (with the exception of the semi-infinite branes) and their position 
depends on the choice of vacuum of the four-dimensional theory. We thus recover the defining property of $\mathcal{N}=2$ gauge 
theories of the previous chapter: the model is characterized by a continuum of vacua, the moduli space. As we have seen 
semiclassically this can be parametrized by the eigenvalues of the scalar field in the vectormultiplet. In the present context 
these are naturally identified with the positions of the $D4$ branes in the $v$ complex plane.

Denoting with $x_6^{\alpha}$ and $x_6^{\alpha+1}$ the ``position'' of two consecutive $NS5$ branes, the coupling constant of the 
corresponding $SU(N_{\alpha})$ gauge group is given by the relation ($\lambda$ is the string coupling constant) $$\frac{1}{g^2}=
\frac{x_6^{\alpha}-x_6^{\alpha-1}}{\lambda}.$$ Actually, the above relation between $x_6$ and $v$ forces us to interpret this 
ratio as a function of $v$: $$\frac{1}{g^2(v)}=\frac{x_6^{\alpha}(v)-x_6^{\alpha-1}(v)}{\lambda}.$$ If we now interpret $v$ as 
setting the mass scale we obtain from (\ref{asymp}) that for large $v$ (or equivalently at high energies, where the one-loop approximation is 
reliable) $$\frac{v}{g^3}\frac{\de g}{\de v}\propto -(2N_{\alpha}-N_{\alpha-1}-N_{\alpha+1}).$$ Since $N_{\alpha-1}+N_{\alpha+1}$ 
is precisely the number of flavors coupled to the $SU(N_{\alpha})$ gauge group, we recover precisely the perturbative beta 
function of  $\mathcal{N}=2$ theories.

\subsection{The lift to M-theory}

We have seen that the brane system described above captures the semiclassical properties of the four-dimensional gauge theory. 
In order to see the effect of quantum corrections, we will now lift the system to M-theory. This is convenient because in the type 
IIA description we have to deal with the ending of a fourbrane on a fivebrane, which is hard to understand in detail. If we 
consider the M-theory setup instead, all the relevant information is encoded in the low energy limit of the theory, which is 
accessible. We then have to understand how our brane system is embedded in $\mathbb{R}^{1,9}\times S^1$. The $NS5$ branes arise from $M5$ branes which are sitting at a point on the M-theory circle, whose coordinate we 
denote with $x_{10}$. The $D4$ branes correspond instead to $M5$ branes wrapping it. The theta angles of the various gauge groups 
are encoded in the difference between the $x_{10}$ coordinate of the corresponding $NS5$ branes. If we define 
$s=\frac{x_6+ix_{10}}{R}$, where $R$ is the radius of the circle and clearly $x_{10}$ is periodic with period $2\pi R$, we have the 
relation $$i\tau_{\alpha}=s_{\alpha-1}(v)-s_{\alpha}(v),$$ where $\tau$ is the generalized coupling costant of the theory introduced in 
chapter 1 and $s_{\alpha}(v)$ is the ($v$ dependent) position of the fivebrane. Equation (\ref{asymp}) now becomes 
\be\label{MM}s=\sum_i\ln(v-a_i)-\sum_j\ln(v-b_j)+\text{const.}\ee from which the previous relation clearly follows. The 
holomorphic dependence of $s$ on $v$ is a consequence of supersymmetry.

The type IIA brane system is actually lifted to a single smooth $M5$ brane wrapping a surface $\Sigma$ and extended in $x_{0,1,2,3}$. The 
$s$ and $v$ coordinates parametrize the complex manifold $Q\simeq\mathbb{R}^3\times S^1$. Supersymmetry demands that $\Sigma$ is 
a complex Riemann surface in $Q$. The low-energy effective theory can now be understood in terms of the worldvolume theory of 
the $M5$ brane which is free; all the information about the dynamics of the four-dimensional field theory is encoded in the 
geometric properties of $\Sigma$, which turns out to coincide with the SW curve describing the theory. It is known that the 
worldvolume theory of the $M5$ brane contains a rank-two tensor field whose field-strength is self-dual (remember that we are in 
six-dimensions). The compactification on $\Sigma$ will then lead to a four-dimensional abelian gauge theory, in which the number 
of $U(1)$ vectormultiplets is equal to (half) the dimension of $H^1(\Sigma)$, which is equal to the genus of the Riemann surface 
or more precisely of its closure (see \cite{branea}). This is exactly the relation between the genus of the SW curve and the rank of the low-energy effective 
theory. We thus see that the $M5$ picture captures directly the low-energy dynamics of our field theory. The vevs of the scalar fields 
in the vectormultiplets specify the embedding of $\Sigma$ in $Q$.

It is now convenient to introduce the variable $t=e^{-s}$, since $s$ itself is not single-valued. The surface $\Sigma$ will then be 
described by a polynomial equation $F(t,v)=0$, which can be determined by analyzing the boundary conditions at large $v$ and $t$ 
and matching the number of solutions at fixed $t$ or $v$ with the number of fourbranes and fivebranes respectively. 
The detailed analysis of the moduli space of the theory which is the starting point for the SW solution as I presented it in 
chapter 1 is replaced in the present framework by the analysis of the asymptotic behaviour of the $M5$ brane.
This can in turn be determined using the equations given above. The SW differential emerges from the holomorphic two-form 
defined on $Q$ and can be written as $\lambda_{SW}=\frac{v}{t}dt$, independently on the specific model considered. $F(t,v)$ was 
explicitly determined in \cite{branea} and assumes the form
\be t^{n}P_{k_0}(v)+\sum_{i=1}^{n-1}t^{n-i}P_{k_i}(v)+P_{k_n}(v).\ee where the $P_i$'s are polynomials in $v$ with This is Witten's proposal for the SW curve associated to a 
linear quiver with $n-1$ gauge groups. The i-th gauge group is $SU(k_i)$, there are hypermultiplets in the bifundamental of 
$SU(k_i)\times SU(k_{i+1})$ for $i=1,\dots,n-1$ and there are $k_0$ hypermultiplets in the fundamental of the first gauge group 
and $k_n$ in the fundamental of the last one. If we have only one gauge group (SQCD with $N$ colors and $N_f$ flavors) the above 
formula reduces to $$t^2P_n(v)+tP_N(v)+P_{N_f-n}(v)=0;\quad\lambda_{SW}=\frac{v}{t}dt.$$ Since this is not the formula given in chapter 1, 
let us see explicitly how to recover it from the above expression: we simply have to multiply everything by $4P_n(v)$ and define 
$y=2tP_n(v)+P_N(v)$. The curve becomes then $y^2=P_N(v)^2-4Q_{N_f}(v)$. If we replace $v$ with $x$ this becomes precisely the curve 
given in chapter 1 if we identify the coefficient of the leading term in $Q_{N-f}(v)$ with $\Lambda^{2N-N_f}$. It is easy to 
check that we recover the expected SW differential as well. It is also possible to include other flavors adding $D6$ branes 
extended along $x_{0,1,2,3}$ and $x_{7,8,9}$ to the brane system. Their effect when we lift the brane system to M-theory is to 
replace $Q\simeq\mathbb{R}^3\times S^1$ with TAUB-NUT space \cite{branea}. 

In \cite{branea} Witten also showed that if the number of suspended $D4$ branes between any pair of $NS5$ branes is the same, it 
is possible to compactify the $x_6$ direction and achieve a brane description of elliptic quivers. From the field theory point of 
view these can be obtained taking a linear quiver with $SU(N)$ gauge groups with $N$ flavors in the fundamental for the first 
and last gauge groups (always the same $N$) and gauging the diagonal combination of the two $SU(N)$ flavor symmetries. The most 
notable example is the case with one gauge group (or equivalently a single $NS5$ brane), which corresponds to $\mathcal{N}=2^*$. 
This brane system can then be lifted to M-theory 
and from this we can recover the SW curve associated to the theory, which will now be embedded in an affine budle on the torus. In 
this case one $U(1)$ factor is not frozen by the constraint discussed before and the gauge group is $SU(k)^n\times U(1)$. I 
refer the reader to the original paper for further details about this.

The M-theory construction can be generalized in some cases to $\mathcal{N}=1$ models as in \cite{HOO,nm1,nm2,nm3}. Adding e.g. a 
superpotential which breaks extended supersymmetry, it was found that the curve wrapped by the $M5$ brane is embedded in a 
Calabi-Yau threefold rather than a twofold. However, it is no longer 
true that this construction encodes in a simple way the infrared dynamics of the theory. This picture is very effective in 
analyzing protected quantities in the theory of interest (such as chiral ring operators...). It is a special property of 
$\mathcal{N}=2$ theories that the prepotential, which encodes all the information about the IR dynamics, is included in the list. 
Generically, the less supersymmetry we have the less we can learn from the knowledge of holomorphic quantities. Maybe it is 
possible to identify the M-theory description for Yang-Mills theory as well. The point is that it is not clear at the moment what 
we can learn from this.

\subsubsection{Orthogonal and simplectic groups}

What we have done so far can be repeated for linear quivers of alternating SO-USp gauge groups and half-hypermultiplets in the 
bifundamental representation of the ``neighbouring'' gauge groups. This can be done considering the brane system described above 
and inserting an $O4$ orientifold plane along the directions $x_{0,1,2,3,6}$. An important point is that there are two kind of 
$O4$ planes that we will denote as $O4^+$ and $O4^-$, depending on the $D4$ charge they carry. When an $O4^+$ plane crosses an 
$NS5$ brane it becomes an $O4^-$ plane and vice versa. If we have an $O4^+$ plane between two fivebranes the corresponding gauge 
group in four dimensions is $USp(N)$, where $N$ (which must be even) is the number of $D4$ branes suspended between the two 
$NS5$ branes. If it is instead an $O4^-$ plane the group will be $SO(N)$. 

Also in this case we can consider elliptic quivers 
obtained compactifying the $x_6$ coordinate. This requires the insertion of the same number of $D4$ branes everywhere as for the 
unitary case. There is the further constraint that the number of $NS5$ branes should be even (in order to gauge the diagonal 
combination of the two flavor groups, they obviously must be both USp or both SO). As in the previous case the whole brane system can be 
lifted to M-theory where it is described by a single $M5$ brane wrapping the SW curve as before. In the M-theory setting the 
orientifold plane is replaced by a $\mathbb{Z}_2$ orbifold (involving the coordinates $x_{4,5,7,8,9}$). This setup has been 
considered in \cite{SOP,SO1,SO2} (see also \cite{SO3,SO4,SO5} for a detailed discussion on the orientifold procedure).

The SW curves describing these linear alternating quivers was found in \cite{SOP}. The result is similar to the previous case:
\be t^{n}P_{k_0}(v^2)+\sum_{i=1}^{n-1}t^{n-i}P_{k_i}(v^2)+P_{k_n}(v^2).\ee
In these case we have only even powers of $v$. The degree $2k$ polynomial $P_k(v^2)$ describes either an $SO(2k)$ or a $USp(2k-2)$ 
gauge group. In the second case we have no constant term and the polynomial is always divisible by $v^2$. The first and last 
polynomials describe the flavors in the fundamental of the extremal gauge groups and have degree respectively $v^{2N_f}$ or $v^{2N_f+2}$ 
depending on whether the corresponding gauge group is USp or SO ($N_f$ counts the number of hypermultiplets).

\section{Argyres-Seiberg duality}

In this section we will be mainly intersted in $\mathcal{N}=2$ SQCD with zero beta function. These models are conformal and possess 
a marginal coupling (the generalized coupling of chapter 1) $\tau=\frac{\theta}{2\pi}+\frac{4\pi i}{g^2}$. These models are 
characterized by a remarkable S-duality: there are infinitely many descriptions of the same theory and going from one S-dual description 
to another the coupling constant is changed, thus allowing sometimes to trade a strong-coupling limit for a weak-coupling one.

As we have seen in chapter 1, unitarity requires that $\tau$ is valued in upper-half complex plane. Since we have at our disposal 
S-duality, a different value of $\tau$ does not imply that the theory is different: given a certain value of the marginal coupling, 
all other values which can be obtained by acting with an S-duality transformation actually correspond to the same theory, just 
written in a different set of variables. The set of inequivalent values of $\tau$ in this sense is the quotient of the upper-half 
plane by the action of the S-duality group and I will refer to this space as the fundamental domain.

\begin{figure}
\centering{\includegraphics[width=.5\textwidth]{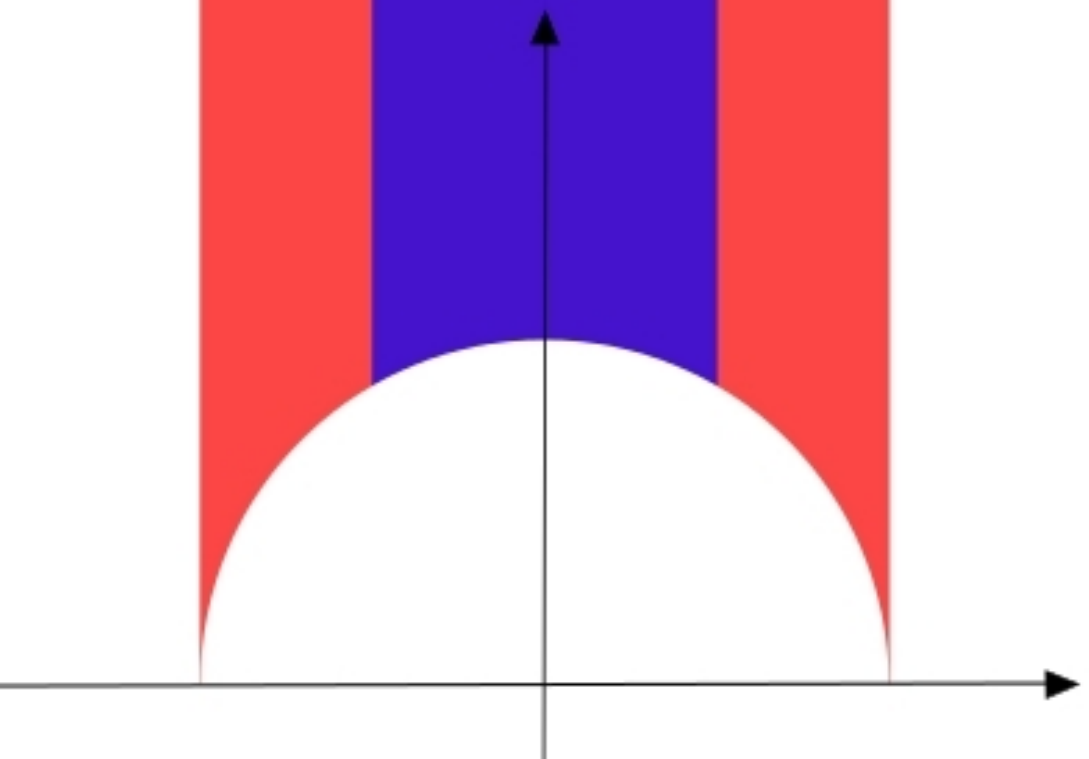}} 
\caption{\label{dominio}\emph{Picture of the upper-half complex plane. The fundamental domain for the marginal coupling $\tau$ 
of $SU(2)$ SQCD is the blue region. For the parameter $\tilde{\tau}$ of the $SU(3)$ theory with six flavors it includes also 
the red region. It is not bounded away from the real axis, signaling the presence of the infinite coupling point.}}
\end{figure}

In this class the $SU(2)$ theory with four flavors is special because the S-duality group is $SL(2,\mathbb{Z})$, which acts on 
$\tau$ as $$\tau\rightarrow\frac{a\tau+b}{c\tau+d};\quad \left(\begin{array}{cc} a & b \\ c & d \end{array}\right)\in SL(2,\mathbb{Z}).$$ 
Its fundamental domain is the blue region in Figure \ref{dominio}, which is bounded away from the real axis (which corresponds 
to the infinite-coupling limit). This reflects the fact that the apparently infinite-coupling limits $\tau\rightarrow 0,1$ 
actually correspond to a weak-coupling limit but in a different S-duality frame.

For higher rank theories the situation is different: the S-duality group is in general only a subgroup of $SL(2,\mathbb{Z})$ and 
accordingly the fundamental domain is larger. For $SU(3)$ for example, the S-duality group is $\Gamma_0(2)$, which is generated by 
the transformations 
$$\tilde{\tau}\rightarrow\tilde{\tau}+2;\quad\tilde{\tau}\rightarrow-\frac{1}{\tilde{\tau}};\quad\tilde{\tau}=2\tau.$$ Its modular 
domain is depicted in Figure \ref{dominio} and includes the red region, which is not bounded awy from the real axis.
Whereas the limit $\tau\rightarrow0$ is equivalent to the weak-coupling limit in some S-dual description in this case as well, 
the limit $\tilde{\tau}\rightarrow1$ is really an infinite-coupling limit. The question is then: how can one characterize the 
physics in this limit? The answer to this question was given by Argyres and Seiberg in \cite{AS}, who found that at the  
infinite-coupling cusp two superconformal sectors emerge, coupled through a weakly-coupled $SU(2)$ vectormultiplet. This does not correspond 
to a weakly-coupled description, since one of the sectors is intrinsically strongly-coupled, but removes the problematic 
infinite-coupling limit. Their analysis is based on a detailed study of the SW curves describing these theories.

\subsection{$SU(3)$ SQCD with six flavors}

Let us review their argument for the $SU(3)$ theory with six flavors. In this case the proposed dual description involves a rank one 
scale invariant theory without marginal coupling (isolated SCFT) which has $E_6$ flavor symmetry discovered in \cite{MNI}. This theory is described 
by the SW curve (setting to zero the six mass parameters) $$y^2=x^3-u^4;\quad \frac{\de\lambda_{SW}}{\de u}=\frac{dx}{y}.$$ The 
Coulomb branch coordinate $u$ has scaling dimension three (requiring that each term in the curve has the same dimension one can 
easily fix the scaling dimensions of $x$ and $y$). The $E_6$ group has an $SU(2)\times SU(6)$ subgroup; in our context the 
$SU(2)$ factor is gauged and we couple to it a hypermultiplet. The claim is then that $SU(3)$ $\mathcal{N}=2$ SQCD with six 
flavors can be described by an $SU(2)$ vectormultiplet coupled to two ``matter'' sectors: one is just a doublet of $SU(2)$ and 
the other is this complicated interacting theory with $E_6$ flavor symmetry $$\boxed{1}-SU(2)-\boxed{E_6\;\; \text{Minahan-Nemeschansky theory}}.$$
We will now provide some checks that substantiate the above conjecture. First of all we can notice that the rank of the two theories 
agree: in both cases the Coulomb branch is described by two coordinates of dimension two and three. The flavor symmetry coincides 
as well, since in the proposed dual description we have an $SU(6)$ global symmetry coming from the strongly coupled sector and 
a further $U(1)$ is associated to the doublet of $SU(2)$, precisely matching the $U(6)$ symmetry of the $SU(3)$ theory. The 
number of marginal couplings is the same only if the $SU(2)$ gauge group is scale invariant, which is true provided that the 
contribution to the $SU(2)$ beta function from the $E_6$ theory is the same as three doublets. We will now see that the SW curve 
confirms this prediction.

As we have seen in chapter 1 the curve describing the $SU(3)$ theory with six flavors is \be\label{ssu3} y^2=(x^3-ux-v)^2-f(\tau)x^6,\ee where 
$f(\tau)$ tends to one in the infinite coupling limit. In order to study this limit it is convenient to factorize the curve as 
follows $$y^2=[(1-\sqrt{f})x^3-ux-v][(1+\sqrt{f})x^3-ux-v].$$ As $f\rightarrow1$ the curve clearly degenerates to a genus one 
curve. The differential $xdx/y=\de\lambda_{SW}/\de u$ develops a pair of poles at infinity whereas $dx/y=\de\lambda_{SW}/\de v$ remains holomorphic. This corresponds to 
the fact that only one cycle vanishes in this degeneration limit. Setting $u$ to zero the residue vanishes, suggesting that it 
is the mass parameter associated to a flavor symmetry. This symmetry is actually gauged in the original theory and the limit 
$f\rightarrow1$ corresponds to turning off the gauge interaction. In this limit a new flavor symmetry emerges and the Coulomb 
branch coordinates associated to the decoupled gauge multiplet (in this case $u$) can be interpreted as mass parameters. We 
thus see the emergence of the ``hidden'' $SU(2)$ gauge group. 

As the curve degenerates we are then left with a rank one theory whose Coulomb branch coordinate is $v$. The SW differential 
is just inherited from that of SQCD. According to our proposed duality, this can be identified with the $E_6$ MN theory. The 
curve indeed confirms this claim: setting $u=0$ the $SU(3)$ curve becomes $$y^2=-v(2x^3-v);\quad \frac{\de\lambda}{\de v}=
\frac{dx}{y}.$$ With the redefinition $$x\rightarrow-i\frac{x}{v};\quad y\rightarrow2\frac{y}{v};\quad v\rightarrow2iv,$$ we
finally get $$y^2=x^3-v^4;\quad \frac{\de\lambda}{\de v}=\frac{dx}{y},$$ which is precisely the curve for the $E_6$ theory \cite{MNI}.

If we instead set to zero $v$ we can recover the curve associated to the emergent scale invariant $SU(2)$ group: the curve (\ref{ssu3}) 
becomes $$y^2=x^2[(x^2-u)^2-fx^4].$$ The factor $x^2$ corresponds to a pinched cycle at $x=0$ and we can scale to the remaining 
genus one curve by redefining $y\rightarrow y/x$. The curve then becomes $$y^2=(x^2-u)^2-fx^4;\quad\frac{\de\lambda_{SW}}
{\de u}=\frac{dx}{y},$$ which is precisely the curve describing $SU(2)$ SQCD with four flavors. This confirms that the $SU(2)$ 
group entering in our dual description is scale invariant, thus providing the marginal coupling we were looking for. As we already 
said, in the limit $f\rightarrow1$, which corresponds to the infinite coupling limit of the original theory, this $SU(2)$ theory 
becomes again weakly coupled but in a different S-duality frame (in which the new $f$ encoding the marginal coupling tends to 0 
\cite{AS}). This argument confirms that the contribution to the $SU(2)$ beta function coming from the $E_6$ theory is three times 
that of a hypermultiplet.

As is well known, the beta function can be determined computing in the background field formalism the two point function for the 
background gauge bosons. The contribution from matter fields is in turn proportional to the two point function of the gauge 
current $J_{\mu}^{a}T_a$. We have the general formula $$J_{\mu}^{a}(x)J_{\nu}^{b}(0)=\frac{3k}{4\pi^4}\frac{x^2g_{\mu\nu}-
2x_{\mu}x_{\nu}}{x^8}+\frac{2}{\pi^2}f^{abc}\frac{x_{\mu}x_{\nu}x^{\rho}J_{\rho c}(0)}{x^6}+\dots,$$ where the dots indicate less 
singular terms and $f^{abc}$ are the structure constants. $k$ is called flavor central charge and in this normalization is exactly 
twice the contribution to the beta function \cite{AS}.

The scale invariance of the $SU(2)$ theory then implies that $k_{SU(2)\subset E_6}=6$. It was shown in \cite{AS} that 
if we gauge a subgroup $H$ of the global symmetry group $G$ of a theory, the corresponding central charges satisfy the relation 
$$k_{H\subset G}=\mathcal{I}_{H\hookrightarrow G}k_G,$$ where $\mathcal{I}$ is the embedding index, which is defined as follows: 
consider a representation ${\bf r}$ of $G$ (the result is indeed independent on the representation chosen) and see how it 
decomposes as $\bigoplus_i{\bf r_i}$, where ${\bf r_i}$ are representations of $H$. The index is given by the formula 
$$\mathcal{I}_{H\hookrightarrow G}\equiv\frac{\sum_iT({\bf r_i})}{T({\bf r})}.$$ This ratio turns 
out to be one both for $SU(2)$ and $SU(6)$ (see Appendix C of \cite{AS}), and from this we can easily conclude that the $E_6$ 
central charge is 6. This remarkably allows to determine exactly the leading term of the two point function of the current and at 
the same time provides a check of our duality: If we gauge the $SU(6)$ flavor symmetry we can easily determine the contribution 
to the beta function on the SQCD side (since the gauge group is $SU(3)$ it is equal to that of three multiplets in the ${\bf 6}$). 
This tells us that the $SU(6)$ central charge is equal to 6 and the dual description should allow to recover this result. This is 
clearly true from the above discussion.

Another interesting observation comes from the evaluation of the $U(1)$ flavor central charge. On the $SU(3)$ side we find 
$k_{U(1)}=36$ and on the dual side, since the only field charged under $U(1)$ is the doublet of $SU(2)$, we get $k_{U(1)}=4q^2$ 
where $q$ is the charge of the doublet. Matching the two we find $q=3$, which suggests that this doublet is an $SU(3)$ magnetic 
monopole, since 3 is precisely the $U(1)$ charge of a monopole neutral under $SU(6)$ as discussed in \cite{AS}. This implies 
that the ``emergent'' $SU(2)$ is not a subgroup of $SU(3)$. 

So far this duality has remarkably passed several other checks than those just dicussed, such as the matching of the a and c 
central charges, which can be computed independently for the $E_6$ theory exploiting its holographic dual \cite{AT} or using the 
field theory argument proposed in \cite{ST} (see also \cite{TIII}), and the matching of the Higgs branch \cite{GNT}.

\subsection{$USp(4)$ SQCD with six flavors}

This is the second example discussed by Argyres and Seiberg and will play an important role in chapter 4. In this case the duality involves a strongly coupled 
theory with $E_7$ flavor symmetry introduced in \cite{MN}. The proposed S-dual description is $$SU(2)-\boxed{E_7\;\; \text{Minahan-Nemeschansky theory}}.$$ By this we 
mean that we gauge an $SU(2)$ subgroup of $E_7$, whose commutant is 
$SO(12)$. The commutant is just the flavor symmetry of the resulting theory as in the previous case, matching the global symmetry 
of $USp(4)$ scale invariant SQCD.

The curve describing the massless $USp(4)$ theory with six hypermultiplets is (see chapter 1) $y^2=x(x^2-ux-v)-fx^5.$ We can 
rewrite it as $$y^2=x[(1-\sqrt{f})x^2-ux-v][(1+\sqrt{f})x^2-ux-v].$$ In this case $x$ has scaling dimension two and the infinite 
coupling limit is $f\rightarrow1$ as before. We can take as a basis for the holomorphic differentials 
$\omega_u=xdx/y$ and $\omega_v=dx/y$. The first develops a pair of poles at infinity in the limit $f\rightarrow1$, whereas the 
second remains holomorphic. The analysis we have done for $SU(3)$ then applies also in this case. 

Setting $v=0$ and rescaling $y$ as before we identify the scale invariant $SU(2)$ theory: 
$$y^2=x[(x-u)^2-fx^2];\quad\omega_u=\frac{dx}{y}.$$ 
The equivalence of this curve with the one written in the previous section is discussed in \cite{ASI}. Setting instead to zero 
$u$ we find the curve describing the isolated rank one SCFT $$y^2=-v(2x^3-vx);\quad\omega_v=\frac{dx}{y}.$$ Making now the change of 
variables $$y=-\frac{\tilde{y}}{2u};\quad x=-\frac{\tilde{x}}{2u},$$ we recover precisely the curve
for the $E_7$ SCFT of Minahan and Nemeschansky \cite{MN}: $$\tilde{y}^2=\tilde{x}^3-2u^3\tilde{x};\quad\omega_v=\frac{d\tilde{x}}{\tilde{y}}.$$ 
We can now repeat the checks performed for $SU(3)$: we have on both sides of the duality one marginal parameter, the scale invariance 
of the $SU(2)$ theory requires that the flavor central charge is 8. If we gauge the $SO(12)$ flavor symmetry on the SQCD side 
we find a $USp(4)\times SO(12)$ gauge theory with one half-hypermultiplet in the bifundamental, so the contribution to the $SO(12)$ 
beta function is four, or equivalently its flavor central charge is 8. Combining this with the previous result we find that necessarily 
the embedding indices of the two groups in $E_7$ should be the same and the direct evaluation shows that they are both equal to one. 
This implies in particular that the $E_7$ flavor central charge is 8. 

Also in this case we can test the duality matching the value of the a and c central charges on both sides using the techniques 
of \cite{AT,ST}. These results have been generalized to a variety of cases in \cite{AW}, using arguments similar to those 
proposed in this section. We will now see how the setting introduced by Gaiotto allows to easily identify all possible dualities 
of this kind in a very broad class of $\mathcal{N}=2$ theories.

\section{Six-dimensional SCFTs and $M5$ branes}

One of the basic aspects of M-theory is the presence of extended objects of dimension two and five called $M2$ and $M5$ branes. 
Understanding their properties is very hard and in particular the determination of their worldvolume theories is a challenging 
problem (see \cite{Wcft} and for a review \cite{revm}). The worldvolume theory on $M2$ branes was found in \cite{ABJM} (see also 
\cite{BLG,BLGII,ABJ}). The analogous problem for $M5$ branes is still open and even in the case of a single brane, whose worldvolume 
theory is known to be free, determining the action is a nontrivial task due to the self-duality condition for the three-form field 
strength \cite{self} \footnote{This problem is similar to the one arising in type IIB supergravity (see \cite{Wm5}), where the RR four-form has a selfdual 
five-form field strength. It can be solved using e.g. the technique proposed in \cite{PST}.}. This should be a six-dimensional 
SCFT with $\mathcal{N}=(2,0)$ supersymmetry, whose dimensional reduction to 5d gives maximally supersymmetric YM theory. Basically, 
we know only the susy multiplet, which includes a two-form (with selfdual field strength), two spinors and five scalars. Since 
an action is lacking we don't have a path integral definition for the theory, which remains mysterious. 

However, as we will see these basic properties of the 6d SCFT are enough to learn a lot about its compactified version and the 
resulting lower dimensional theories. These ideas are particularly important in the study of $\mathcal{N}=2$ theories in four 
dimensions, as found by Gaiotto in \cite{G}: the 6d constructions are suited to analyze in a systematic way all possible Argyres-Seiberg 
like dualities, considerably enlarge the landscape of $\mathcal{N}=2$ SCFTs, allow to identify a connection between the instanton 
partition function of $\mathcal{N}=2$ theories and conformal blocks of 2d CFTs (AGT) \cite{AGT,AGTI} and provide an algorithm that allows to identify 
the SW curve for any theory which can be constructed in this way (usually called class $\mathcal{S}$ theories). This section is 
devoted to review the basic aspects of these recent developments.

\subsection{Rank one theories}

\subsubsection{The Gaiotto curve}

Let us start from the simplest example of $\mathcal{N}=2$ superconformal theory: $SU(2)$ SQCD with four flavors. As we have seen 
the theory has a marginal coupling $\tau$ which lives in the upper-half complex plane and, modulo the action of the 
S-duality group $SL(2,\mathbb{Z})$, we can restrict to consider the modular domain. The first step is to recognize that this 
coincides with the complex structure moduli space of a sphere with four (equivalent) marked points. If we call $z$ the coordinate parametrizing 
the sphere, we can assume that three of them are located at $z=0,1,\infty$. We cannot fix the position of the fourth marked point, 
which will be located at a generic point $z=q$. What is the relation between our field theory and this auxiliary four-punctured 
sphere?

In order to answer this question, let us manipulate a little bit the SW curve of the theory. As we have seen, the parametrization 
one obtains from Witten's construction is \be\label{www}z^2(v-m_1)(v-m_2)+c_1z(v^2-u)+c_2(v-m_3)(v-m_4)=0,\ee
and the corresponding SW differential is $\lambda_{SW}=\frac{v}{z}dz$.
Let us set to zero the mass parameters for the moment. Collecting all terms with the same power of $v$ 
and with a simple redefinition of $z$ we can bring it to the form $$(z-1)(z-q)z^2x^2=uz\quad\lambda_{SW}=xdz,$$ where we have 
defined $x=v/z$. Dividing now by $z^2(z-1)(z-q)$ we get \be\label{qcd2}x^2=\frac{u}{z(z-1)(z-q)}\quad \text{or}\quad \lambda^2= 
\frac{uz}{(z-1)(z-q)}\left(\frac{dz}{z}\right)^2\equiv\phi_2(z),\ee where we have defined the quadratic differential $\phi_2(z)$ 
and exploited the particular form of the SW differential. We can now see the connection with the four-punctured sphere introduced 
previously: in the formula above we have written the SW curve as a double cover of the sphere, and the quadratic differential 
$\phi_2(z)$ defined on it has simple poles precisely at four points ($0,1,q,\infty$). The information about the marginal coupling 
is encoded in $q$. This fact suggests that we can actually identify the base of the fibration with our auxiliary four-punctured 
sphere, which is usually referred to as the {\bf Gaiotto curve}. 

This correspondence can be easily generalized: let us consider e.g. the conformal quiver with $SU(2)^2$ gauge symmetry 
$$\boxed{2}-SU(2)-SU(2)-\boxed{2},$$ with a bifundamental hypermultiplet and two doublets charged under each $SU(2)$ factor (that 
we have indicated with a 2 inside the box). This theory has two marginal couplings and their fundamental domain coincides with the 
moduli space of a theory with five identical marked points. Its SW curve and differential can be written as 
$$z^3y^2+c_1z^2(y^2-u_1)+c_2z(y^2-u_2)+c_3z^2;\quad\lambda_{SW}=\frac{y}{z}dz.$$ With the same manipulations described above we 
can bring it to the form $$x^2=\frac{u_1z+u_2}{z(z-1)(z-a)(z-b)}\quad\lambda_{SW}=xdz.$$ We recover the same structure 
encountered before: the SW curve is the double cover of a sphere and the corresponding quadratic differential 
$$\phi_2=\frac{u_1z^2+u_2z}{(z-1)(z-a)(z-b)}\left(\frac{dz}{z}\right)^2$$ has five simple poles. We can indeed repeat this procedure 
for a longer quiver with $n$ $SU(2)$ gauge groups and write the SW curve in the universal form \be\label{su22}x^2=F(z),\quad\lambda_{SW}=xdz 
\quad\phi_2(z)=F(z)(dz)^2,\ee where the quadratic differential has simple poles at $n+3$ punctures. 

We have seen that the SW curve describing a linear quiver of $SU(2)$ groups can always be written as a double cover of a sphere with a 
certain number of punctures (which correspond to the poles of the quadratic differential). What happens if we turn on the mass parameters 
for the doublets and the bifundamentals? The answer is that the simple poles of the quadratic differential become double poles. 
This can be easily checked in the $SU(2)$ $N_f=4$ theory going back to the curve (\ref{www}) and repeating the above manipulations. 
This result should not be surprising: as we have seen in chapter 1 the SW differential can have simple poles and the corresponding 
residues are just proportional to the mass parameters of the matter fields. From the relation $\lambda^2=\phi_2(z)$ we can easily 
see that a double pole for $\phi_2$ corresponds to a simple pole for $\lambda$. As we turn off the mass parameters the double pole 
is replaced by a simple one and the corresponding residue of the SW differential vanishes as it should. 

With a further fractional 
linear redefinition of $z$ and a redefinition of mass parameters described in detail in \cite{G}, we can also rewrite the SW curve 
in such a way that the behaviour of $x$ near the punctures is $x\sim\pm m/z$ (+ on one sheet and - on the other), for some mass 
parameter. This fact is not automatic in Witten's parametrization and tells us that in a neighbourhood of the puncture we can 
approximate the SW curve as $x^2-m^2=\det(xI_2-M_2)=0$, where $M_2=\frac{1}{z}\text{diag}(m,-m)$. Indeed, we can think of $M_2$ as the Cartan 
element of an $SU(2)$ symmetry group of the theory, and $m$ is the corresponding mass parameter. Having $n+3$ punctures (as is the 
case for the $SU(2)^n$ theory) thus tells us that the flavor symmetry of the theory is $SU(2)^{n+3}$, which is indeed the right 
result: every bifundamental contributes an $USp(2)=SU(2)$ factor (as the bifundamental of $SU(2)$ is real) and each group of two 
doublets at the ends of the quiver contributes a further $SO(4)\simeq SU(2)\times SU(2)$. The case $n=4$ is special since the 
flavor symmetry enhances to $SO(8)$: in this setting we are not taking into account the possibility of ``mixing'' the two pairs of doublets, 
which is indeed possible in this particular case. This discussion taught us that the flavor symmetry of the theory is encoded 
in the punctures of the corresponding sphere.

\subsubsection{$SU(2)$ generalized quivers}

Let us now go back to the $SU(2)$ theory with four flavors. We have seen that the SW curve is just the double cover of a sphere with 
four punctures, whose complex structure moduli space can be in turn identified with the fundamental domain of the marginal coupling of the theory. 
However, we did not discuss a key point underlying this fact: the singularity of the fundamental domain, which corresponds to the 
weak coupling limit, has a transparent geometric interpretation in this setting, since it corresponds to a degeneration limit 
of the four-punctured sphere. At the level of the SW curve (\ref{qcd2}) this amounts to sending $q$ to infinity, which 
corresponds as expected to taking the weak coupling limit. Clearly, we can also send $q$ to 0 or 1: the sphere degenerates 
also in these cases. However, from the field theory analysis we know that these limits are actually equivalent to the process we 
have just described: we can always find an S-dual description of the theory in which the $SU(2)$ gauge group becomes weakly-coupled. 
This fact has a nice geometric counterpart in our setting: the four punctures are all on the same footing and the three collision 
processes mentioned above are perfectly equivalent to each other.

We can understand the decoupling of the gauge multiplet as follows: 
when $q$ is sent to infinity (\ref{qcd2}) becomes $$\lambda^2=\frac{uz}{z-1}\left(\frac{dz}{z}\right)^2.$$ The quadratic 
differential has now simple poles at 0 and 1 and a double pole at infinity, which corresponds to a pole for the SW differential. 
The residue is proportional to $u$, the Casimir of the $SU(2)$ gauge group. When we turn off the gauge coupling our theory degenerates to two decoupled 
sectors, each one having an $SU(2)$ flavor symmetry. From the SW curve this can be seen noticing that the Coulomb branch coordinate 
$u$ in the degenerated theory can be interpreted as the mass parameter associated to the new $SU(2)$ flavor symmetry (it is the 
residue of the pole of the SW differential).

It is not harder to analyze this phenomenon in the general case: a sphere with many punctures will have many marginal couplings 
with an intricate fundamental domain. All its singular points correspond to degeneration limits of the sphere, which are 
described by the collision of a certain number of punctures. There is always an S-duality frame in which this corresponds to the 
weak-coupling limit of some $SU(2)$ groups and this is captured by the way in which the sphere degenerates. We have thus found an 
easy graphical rule which encodes all possible S-dual descriptions, even for very complicated theories. 

\begin{figure}
\centering{\includegraphics[width=\textwidth]{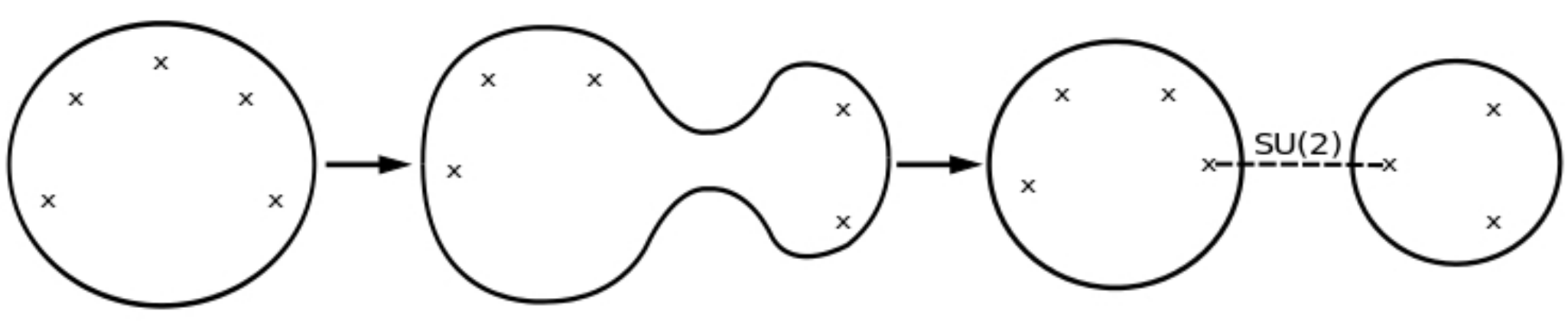}} 
\caption{\label{su2}\emph{A sphere with more than three punctures describes a linear quiver of $SU(2)$ gauge groups. All possible 
degeneration limits correspond to a weak-coupling limit in a suitable S-duality frame. On the left we have a sphere with five punctures 
describing a theory with $SU(2)^2$ gauge symmetry. The collision of two punctures is equivalent to the weak-coupling limit of one of the 
two $SU(2)$ factors (on the right).}}
\end{figure}

In figure \ref{su2} 
we have depicted this for the sphere with five punctures: the collision of the right-most two punctures correspond to decoupling 
one of the two $SU(2)$ factors. We thus expect to end up with $SU(2)$ SQCD with four flavors and a decoupled free sector. This is 
indeed confirmed by the curve: after the collision we are left with a four-punctured sphere, which describes precisely the 
scale-invariant $SU(2)$ SQCD, and a three punctured sphere. What does it correspond to? The answer is easy: it must describe the 
decoupled free sector, i.e. two doublets of $SU(2)$, or equivalently an half-hypermultiplet in the trifundamental of $SU(2)^3$. 
We can now go on and degenerate the sphere with four punctures thus decoupling the second $SU(2)$ group. We then find four 
doublets of $SU(2)$, which is twice the matter content of the three-punctured sphere. This is confirmed graphically: the degeneration 
of the four punctured-sphere produces precisely two spheres with three punctures. 

So far we have discussed the decoupling process: any sphere with an arbitrary number of punctures can be degenerated to a collection 
of spheres with three punctures. Of course we can also do the opposite and recover the original theory just gauging the diagonal 
combination of the $SU(2)$ flavor symmetries associated to two puncures. Graphically, this amounts to connecting the two punctures 
with a tube, thus gluing together the two spheres. 

We can actually construct several new theories using this procedure: starting from a collection of three-punctured spheres we can 
connect them in many ways and build Riemann surfaces of arbitrary genus with any number of punctures. As we have seen, on the 
field theory this amounts to gauging the diagonal combination of the various $SU(2)$ global symmetries associated to 
the punctures. In this way we construct the so-called generalized quivers. Only the theories associated to surfaces of genus 0 
and 1 can be engineered using a brane system in type IIA. This approach thus allow to greatly extend Witten's construction. 

Conversely, starting from 
a given punctured surface we can decompose it into a collection of three punctured spheres. Indeed, the same surface can be 
decomposed in many different ways. Any such decomposition corresponds to a singular point of the fundamental domain, in which the 
theory becomes weakly coupled in a certain S-duality frame. All degenerations limits are connected to each other by an S-duality 
transformation.

We have seen that for genus 0 the curve can 
always be written in the form $$\lambda^2=\phi_2(z);\quad\lambda=xdz.$$ In \cite{G} it is shown that also the curve proposed by 
Witten for elliptic quivers can be recast in the above universal form, where now the Gaiotto cuve is a punctured torus and $\phi_2(z)$ is 
a quadratic differential defined on it, with poles at all the punctures. The simplest example in this class is the 
$\mathcal{N}=2^*$ theory. We can obtain it in the following way: consider a three punctured sphere and connect two of its punctures 
with a handle. The Gaiotto curve now is a punctured torus. As we have seen this corresponds to gauging the diagonal combination 
of the corresponding $SU(2)$ flavor symmetries. The matter fields described by it transform in the ${\bf 2}\otimes{\bf 2}= 
{\bf 3}\oplus{\bf 1}$ of the $SU(2)$ gauge group. The neutral multiplet decouples and we are left with a hypermultiplet in the 
adjoint, which is precisely the matter content of $\mathcal{N}=2^*$ theory. The $SU(2)$ flavor symmetry and mass parameter of the 
adjoint hypermultiplet are encoded in the puncture. Setting to zero the mass, supersymmetry enhances to $\mathcal{N}=4$.

One of the great virtues of the present setting is that we can straightforwardly identify the SW curves associated to all these 
exotic theories: they are simply given by the above curve, where $\phi_2(z)$ is a differential defined on the genus g Riemann 
surface with poles at all the punctures. Another important point that I would like to stress is the following: let us consider 
the following SW curve $$y^2=x^3+3\tau u^2x+2u^3.$$ We can now ask what is the $\mathcal{N}=2$ theory associated to it. The answer 
is not unique \cite{AWII}: both $SU(2)$ SQCD with four massless flavors and $\mathcal{N}=4$ $SU(2)$ SYM are described by it. The SW curve 
may be insufficient to distinguish various theories between each other. We can instead immediately distinguish these two models 
focusing on the Gaiotto curve: in the first case it is a sphere with four punctures and in the second a torus with one puncture. 
The point is that the SW curve is a two-sheeted covering of the Gaiotto curve. Focusing just on it we have control only on the total 
space of the fibration whereas, if we consider the Gaiotto curve, which is the base space, we achieve control on the whole 
fibration and this allows to extract more information. In our example the total space of the covering is in both cases a torus, 
which is basically the information encoded in the SW curve written above. One can now wonder whether this framework has the same 
problem or not: identifying the Gaiotto curve is always enough to distinguish two theories? The six dimensional origin of these 
theories that we will now discuss suggests that the uniqueness of the theory is guaranteed.
 
\subsubsection{The six-dimensional construction} 

We have seen that linear and elliptic quivers can be constructed considering a system of branes and the rank of each gauge group 
is specified by the number of $D4$ branes suspended between two $NS5$. This system can in turn be lifted to 
a single smooth $M5$ brane, which captures the infrared dynamics of the theory.

We can also adopt a different viewpoint, as explained in detail in \cite{G,GNM}, and interpret each $D4$ brane as an $M5$ brane 
wrapping another surface, a sphere in the case of linear quivers and a torus in the elliptic case. We thus have a system of 
parallel $M5$ branes, as many as the number of $D4$ branes (two in the case of $SU(2)$ quivers), wrapping a surface which turns 
out to be precisely the Gaiotto curve. Whereas in Witten's setup the single $M5$ brane is smooth, in this case the intersection 
of the $NS5$ branes with our brane system will produce singularities, leading to a punctured Riemann surface. These precisely 
correspond to the punctures discussed before. In the case of linear quivers, we have other two punctures at zero and infinity which 
correspond to the semi-infinite $D4$ branes at the ends of the quiver. This fits well with our analysis: a quiver with $n$ gauge 
groups is constructed inserting $n+1$ $NS5$ branes, taking into account the singularities at zero and infinity we find precisely 
$n+3$ punctures which is the expected result. 

The worldvolume theory of two coincident $M5$ branes is the mysterious $A_1$ six-dimensional SCFT discussed previously. 
The above discussion teaches us that we can obtain the four-dimensional $\mathcal{N}=2$ linear and elliptic quivers by 
compactifying it on the corresponding Gaiotto curve. Actually, there is no reason to restrict to these two cases: we can 
compactify the $A_1$ theory on an arbitrary punctured Riemann surface and, with a suitable twist described in detail in 
\cite{G}, the compactification can be done preserving eight supercharges, which form the $\mathcal{N}=2$ superalgebra of the 
four-dimensional theory. 

The $A_1$ theory has a protected operator of R-charge two, which descends to a quadratic differential defined on the compactifying 
surface. This is precisely the $\phi_2$ entering in the SW curve of the theory. In this framework the punctures correspond to 
codimension-two defects of the six-dimensional theory sitting at a point in the Riemann surface and extending along the flat 
four-dimensional spacetime. The existence of these defects is in a sense a prediction of this construction and their properties 
are still poorly understood. For our purposes we can limit ourselves to say that their effect in the compactified theory is to 
encode the flavor symmetry of the four-dimensional theory.
The generalized quivers studied before can alternatively be defined in this way. This construction ensures the uniqueness of 
the theory, as we mentioned in the previous section. At low energies, the $M5$ branes recombine into a single smooth object, thus 
recovering Witten's construction. This is why the gaiotto curve is sometimes referred to as the UV curve.

\subsection{Rank two theories}

What we have said so far for rank one theories can be extended to the rank N case. Let us start from the simplest case of rank two. 
Starting from Witten's curve for a quiver of $SU(3)$ gauge groups we can bring it to a canonical form, rather as for rank one 
theories. The required manipulations are similar: we have to collect the same powers of $v$ and define $x=v/z$. This brings the 
SW differential to the by now familiar form $\lambda=xdz$. The curve will become instead $$\lambda^3=\lambda\phi_2(z)+\phi_3(z),$$ 
where $\phi_2$ and $\phi_3$ are respectively quadratic and cubic meromorphic differentials on the sphere. As in the rank one case, 
these theories can be constructed compactifying on a certain punctured sphere the rank two 6d $(2,0)$ theory. This theory has 
protected operators of R-charge two and three, corresponding to the Casimirs of the $A_2$ algebra. Once we have compactified the theory, 
they become precisely the $\phi_i$ differentials appearing in the SW curve. 

\begin{figure}
\centering{\includegraphics[width=.9\textwidth]{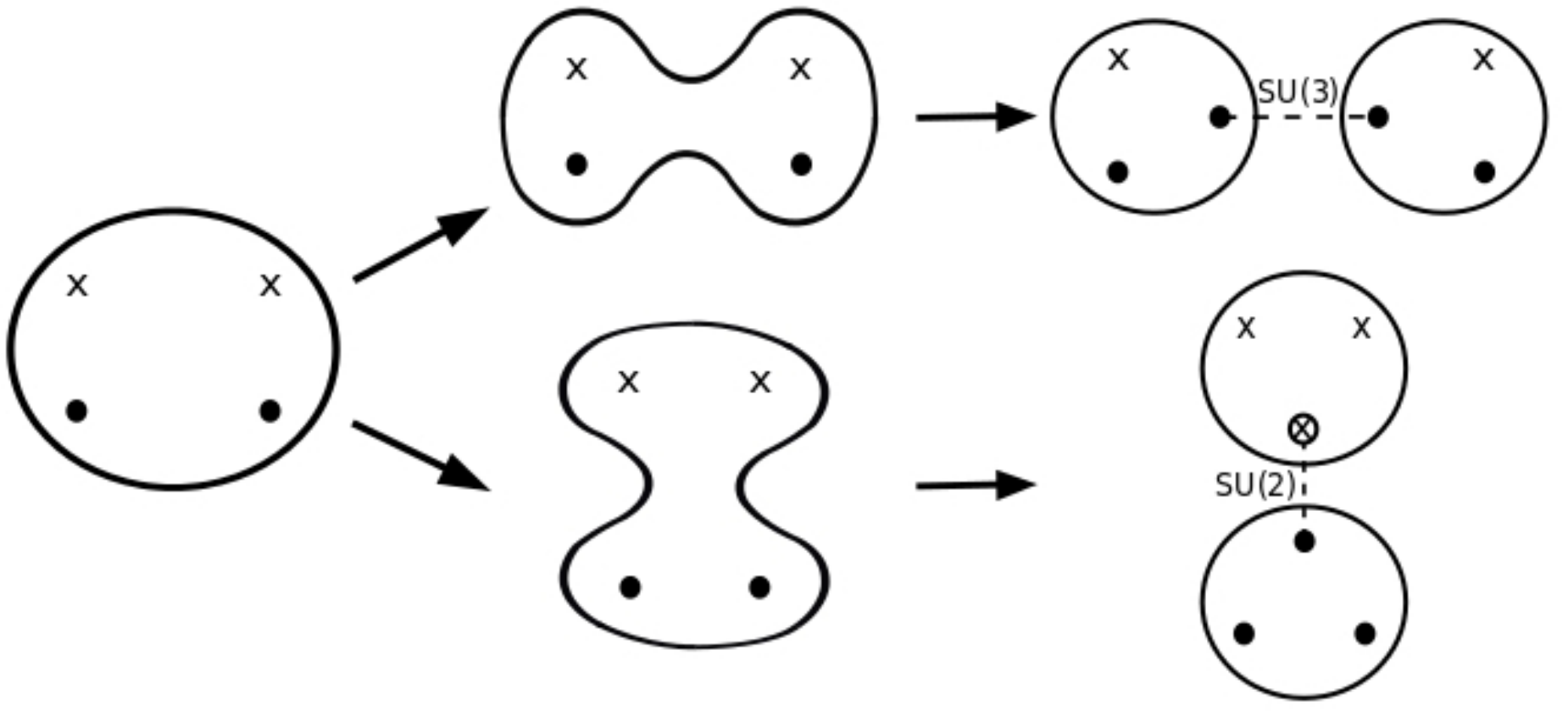}} 
\caption{\label{su3AD}\emph{Possible degeneration limits of the Gaiotto curve describing $SU(3)$ SQCD with six flavors. The collision 
of a maximal and a simple punctures corresponds to the weak-coupling limit (above). The collision of the two simple punctures 
(below) corresponds to the infinite coupling limit, in which the $SU(2)$ gauge group entering the Argyres-Seiberg 
description becomes weakly-coupled.}}
\end{figure}

We will now work out the main properties starting from the simplest example: $SU(3)$ SQCD. We have seen that the fundamental domain 
for its marginal coupling is the quotient of the upper-half plane under the action of $\Gamma_0(2)$. This can in turn be 
interpreted as the complex structure moduli space of a sphere with two couples of equivalent punctures
as in Figure \ref{su3AD} on the left. Comparing with the rank one case, we expect 
this auxiliary sphere to be related to the SW curve. Indeed, this will be manifest once we have manipulated the SW curve as above. 
Let us do that explicitly: Witten's curve (for the massless theory) is $$v^3z^2+c_1z(v^3-u_2v-u_3)+c_2v^3=0.$$ Collecting powers of 
$v$ and rescaling $z$ it becomes $$(z-1)(z-q)v^3=vu_2z+u_3z.$$ Introducing now $x=v/z$ we can bring it to the desired form 
\be\label{333} x^3=x\frac{u_2}{z(z-1)(z-q)}+\frac{u_3}{z^2(z-1)(z-q)};\quad\lambda=xdz.\ee We can easily see that the quadratic 
differential has a simple pole at $0,1,q,\infty$, whereas the cubic differential has a simple pole at $1,q$ and a pole of degree two 
at 0 and $\infty$. This is precisely what we expected: the punctures at 1 and q are identical, those at 0 and $\infty$ are equal 
too, but the two pairs of punctures are different, since the cubic differential has poles of different degree. For reason that will become 
clear later on, I will call the former minimal and the latter maximal.

As in the $SU(2)$ case, the parameter $q$ can be ``identified'' with the marginal coupling and the degeneration limits, i.e. when two 
punctures collide, correspond to the singular points of the fundamental domain or equivalently to the weak and infinite coupling 
limits. From the field theory perspective we expect the limits $q\rightarrow0,\infty$ to correspond both to weak coupling in 
different S-dual descriptions, whereas the infinite coupling point $q\rightarrow1$ has a different structure. This fits perfectly 
with the geometric picture (see Figure \ref{su3AD}): the first two cases correspond to the collision of a maximal puncture with a minimal one. The third 
process is different, being associated to the collision of two minimal punctures. We will now see how Argyres-Seiberg duality is 
automatically encoded in this formalism.

Let us see why the limit $q\rightarrow0$ ($q\rightarrow\infty$ is analogous) truly corresponds to weak coupling: taking this limit 
in the curve (\ref{333}) the pole structure at 1 and $\infty$ remain unchanged, whereas the puncture at zero is now characterized 
by a double pole for the quadratic differential and a cubic one for the cubic differential. This means that the SW differential 
has now a simple pole at zero, whose residue depends on two parameters ($u_2$ and $u_3$). As we have seen before, these should be 
interpreted as mass parameters associated to an ``emergent'' flavor symmetry of rank two and whose Casimirs have dimensions two and 
three, so an $SU(3)$ group. In the original theory this symmetry is gauged and in our limit the interaction becomes weaker and 
weaker. At $t=0$ the $SU(3)$ gauge multiplet is completely decoupled and we are left with a collection of hypermultiplets. On 
the geometric side the Gaiotto curve degenerates and we end up with two identical three-punctured spheres, each describing 
$3\times3=9$ hypermultiplets.

In the $q\rightarrow1$ limit we find a different situation (see Figure \ref{su3AD}): the punctures at zero and infinity remain maximal whereas the puncture 
at $z=1$ has now quadratic pole both for the quadratic and cubic differentials. This time the residue of the SW curve is proportional 
only to $u_2$, thus suggesting that the gauge group which decouples in this limit is an emergent $SU(2)$. This is precisely what 
happens in the Argyres-Seiberg dual description! Can we recover the $E_6$ theory as well? The answer is yes and it can be seen as 
follows: we have seen that $u_2$ is related to the $SU(2)$ mass parameter in the degeneration limit. In order to isolate 
the $E_6$ sector, let us then turn it off (notice that this precisely parallels the analysis of Argyres-Seiberg duality in the 
previous section). The $SU(3)$ curve (\ref{333}) becomes $$x^3=\frac{u_3}{z^2(z-1)^2};\quad\lambda=xdz.$$ First of all notice that 
this curve describes a sphere with three full punctures. We thus learn that the collision of two minimal punctures produces a full 
one. The second observation is that this theory has a Coulomb branch of dimension one, parametrized by the coordinate $u_3$ which 
has dimension three (exactly as the $E_6$ theory). Let us now define $y=xz$ and rewrite the curve as $$(z-1)^2=\frac{u_3z}{y^3}.$$ 
We can now rescale the Coulomb branch coordinate $u_3=2v$ and introduce $t=z-1-\frac{v}{y^3}$, bringing the curve to 
$$t^2=\frac{v}{y^3}\left(\frac{v}{y^3}+2\right).$$ If we now multiply everything by $y^6$ and redefine $w=ty^3$ we finally get 
$$w^2=v(2y^3+v);\quad\frac{\de\lambda}{\de v}=-\frac{dy}{w}.$$ With the redefinition $v\rightarrow-v$, this becomes exactly the 
curve describing the $E_6$ theory we have encountered while discussing Argyres-Seiberg duality!

As discussed previously, in the rank one case all punctures are related to an $SU(2)$ flavor symmetry. I would now like to discuss 
how this generalizes to the rank two case. If we turn on mass parameters in the M-theory curve and repeat the above steps to 
bring it to the canonical form, we can easily see that the order of the poles increases and the SW differential now has simple 
poles at the punctures with residue proportional to the mass parameters. It can be seen that modulo a fractional linear redefinition of $z$ discussed in \cite{G} the curve 
can be approximated near the punctures as $\det(xI_3-M_3)=0$, where $M_3=\text{diag}(m,m,-2m)$ for a minimal puncture and 
$M_3=\text{diag}(m_1,m_2,-m_1-m_2)$ for a full one. These matrices can be thought of as the cartans of the corresponding flavor 
group, which is $U(1)$ for a minimal puncture and $SU(3)$ for a full one.

We can repeat the above analysis for linear quivers of $n$ $SU(3)$ gauge groups. The final result is that the corresponding Gaiotto 
curve is a sphere with $n+3$ punctures, two maximal and the other minimal. With the procedure described before we can bring the 
SW curve to the canonical form, in which it is manifestly a three-sheeted covering of the Gaiotto curve. This reflects the fact 
that all these models can be realized compactifying the 6d $A_2$ theory on the Gaiotto curve. The simple punctures reflect the 
presence of the $NS5$ branes and the full punctures are associated to the three semi-infinite $D4$ branes at the ends of the 
brane system. At low energies the three $M5$ branes recombine into a single smooth object, recovering Witten's description. 

According to the above rule this corresponds to a 
$SU(3)^2\times U(1)^{n+1}$ flavor symmetry, which matches precisely the flavor symmetry of the theory. The only exception is the 
case $n=1$: in this case the symmetry enhances to $U(6)$ but in this description only a $U(3)^2$ symmetry is manifest. We have 
already encountered another case in which the symmetry enhances\footnote{this raises the question of how one can recover the full 
flavor symmetry of the theory in nonlagrangian cases. This can be done exploiting the fact that the mirror dual of the $\mathcal{N}=4$ 
3d theory obtained compactifying the theory on $S^1$ is lagrangian \cite{TX,NX1}. The algorithm which determines the full flavor 
symmetry of the theory is given explicitly in \cite{CD}}: in this framework the $E_6$ Minahan-Nemeschansky theory is 
described by a three-punctured sphere, whose naive flavor symmetry is $SU(3)^3$, which is just a subgroup of $E_6$. 

We can now play with these theories as in the rank one case: our sphere can be degenerated to a collection of three-punctured 
spheres in many different ways, all these decompositions are linked by repeated S-dualities or Argyres-Seiberg dualities. 
We can now use these spheres to build other theories: on the field theory side this can be done by gauging the diagonal combination 
of the global symmetries associated to the punctures of two different spheres. The geometric counterpart of this operation is to 
connect the corresponding punctures with a tube. In this way we can form surfaces of arbitrary genus with any prescribed number 
of punctures. In the rank one case we had only one basic building block. In this case we have two: the $E_6$ theory (three full 
punctures) and the sphere with two maximal and one minimal punctures. This arises from the weak-coupling limit of the $SU(3)$ 
theory analyzed before and as we have seen describes a free theory of nine hypermultiplets. 

Among the theories associated to surfaces 
of nonzero genus, which are almost all new, a distinguished class is represented by those obtained connecting together all the 
full punctures of a collection of $n$ spheres with two maximal and one minimal punctures: this corresponds to a genus one 
Gaiotto curve with $n$ simple punctures. These represent the elliptic quivers discussed by Witten in \cite{branea}. Indeed, 
starting from its curve one can bring it to the canonical form and show that it corresponds to a degree three covering of the 
Gaiotto curve as expected. Again the simple punctures correspond to transverse $NS5$ branes. The simplest example is the case 
$n=1$ which describes $\mathcal{N}=4$ $SU(3)$ SYM.

We are anyway facing a further difficulty with respect to the rank one case: one of the basic building blocks has no lagrangian description, 
so it is not clear from the field theory perspective what kind of theory we get using it. These models are best described exploiting the six-dimensional construction.  
An important feature of this framework is that the SW curve can be easily obtained for all these theories: we know a priori that we are 
looking for something of the form $\lambda^3=\lambda\phi_2+\phi_3;\quad\lambda=xdz$, where $\phi_i$ is a degree $i$ differential, 
with poles of prescribed order at the punctures. The set of differentials satisfying this constraint is a vector space $V_i$, whose 
dimension can be immediately found using Riemann-Roch theorem: \be\label{RR}dim V_i=\sum_k p_{k,i}+(g-1)(2i-1),\ee where the sum runs over the 
punctures, $p_{k,i}$ is the order of the pole at the k-th puncture and $g$ is the genus of the surface. Once we have found 
$dim V_i$ independent differentials with this property we can simply take a generic linear combination and plug it into the above 
formula for the curve. The coefficients of the linear combination are just the Coulomb branch coordinates of dimension $i$. We 
thus see that the Coulomb branch of any theory is really a graded vector space. Let us give a simple example: the three-punctured 
sphere with only maximal punctures (we can assume they are located at $z=0,1,\infty$). Just from this information the above formula tells me that $dim V_2=0$ and $dim V_3=1$. So 
there is only one cubic differential (modulo rescaling), which is $$\phi_3=\frac{z}{(z-1)^2}\left(\frac{dz}{z}\right)^3,$$ 
and we get directly from this the curve describing the $E_6$ theory. We have thus found an algorithmic procedure that allows to 
extract the SW curve immediately.

\subsection{Higher rank generalized quivers}

\begin{figure}
\centering{\includegraphics[width=\textwidth]{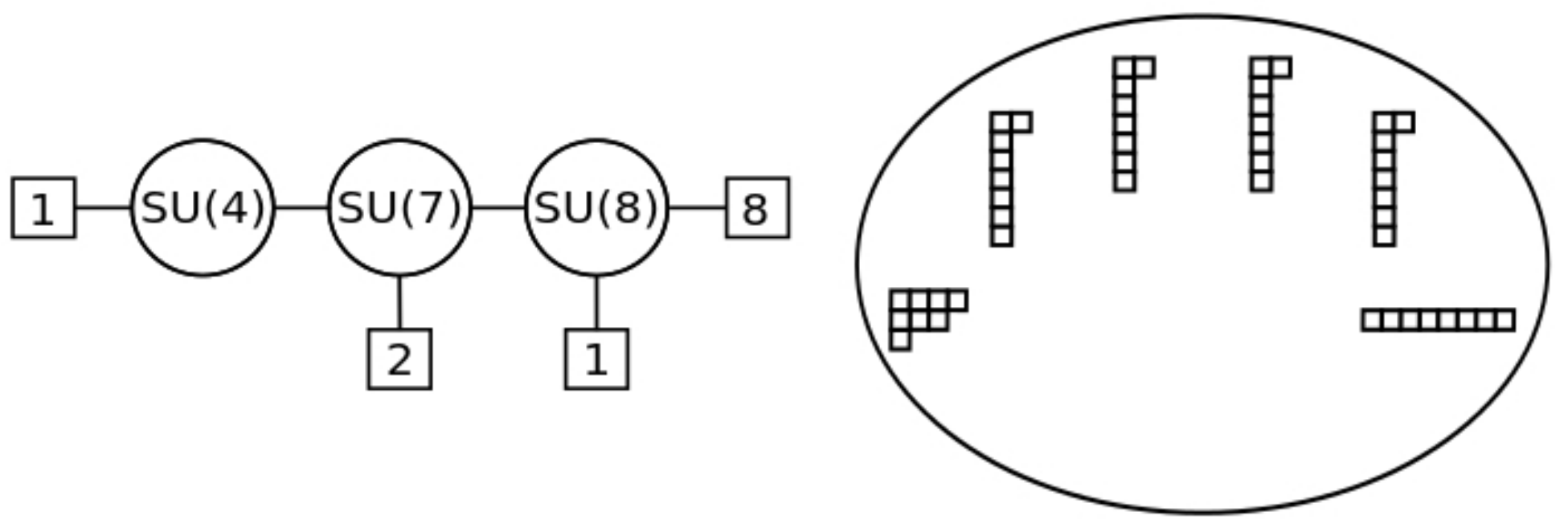}} 
\caption{\label{qq8}\emph{The quiver on the left and the corresponding Gaiotto curve on the right. The leftmost and rightmost 
punctures encode the rank of the gauge groups: the length of the rows give the difference between the ranks of two adiacent 
groups. In this case the first puncture has rows of length 4,3,1 and correspondingly the groups are $SU(4)$, $SU(4+3=7)$ and 
$SU(7+1)$. The last Young tableau has a single row, which tells that the quiver ends with an $SU(8)$ group. In between we can 
have only simple punctures, as many as the number of gauge groups plus one. Notice that the flavor symmetry of the theory is 
perfectly reproduced by adding the contribution of each puncture.}}
\end{figure}

The canonical form for the SW curve in this case is $$\lambda^N=\sum_{k=2}^{N}\lambda^{N-k}\phi_k(z);\quad\lambda=xdz.$$ The basic 
ingredients can be identified analyzing the linear superconformal quivers as before, which correspond to the compactification of 
the six dimensional rank $N-1$ theory on a sphere with punctures, or in the presence of codimension two defects 
(see Figure \ref{qq8}). In this case the 6d theory has protected operators with R-charge $2,\dots,N$, which become differentials 
on the compactifying surface. Their vevs give the $\phi_i$'s entering in the curve. I refer the reader to \cite{G} for the details. 

In this case one finds a proliferation of punctures which can be conveniently described by Young tableaux with $N$ boxes (for rank 
$N-1$ theories), which encode the poles of the various differentials. All possible Young tableaux correspond to some puncture except the totally antisymmetric one (one column). In 
the rank one case we then have only one possibility and in the rank two case there are two, which is the right result. The 
corresponding flavor symmetry is $$S(\Pi_k U(n_k)),$$ where $n_k$ is the number of columns of length k. This can be determined by 
studying the linear quivers and requiring that the global symmetry associated to the punctures matches the flavor symmetry of the 
corresponding theory. All conformal linear quivers are described by a sphere with two generic punctures 
and a collection of minimal punctures (the corresponding Young tableau has one row of length two and the others of length one) 
 (see Figure \ref{qq8}) and elliptic quivers correspond to a torus with minimal punctures only. All other theories do not have 
 a brane construction and are essentially a new outcome of this formalism.

Let us explain how the Young tableaux encode the pole orders of the various differentials at the punctures. This will provide 
an algorithm which allows to extract the SW curve just from these data. The procedure is the following \cite{CD}:
\begin{itemize}
\item Number the boxes of the Young tableau as follows: start with zero in the first box and number the boxes in the
first row with successive integers. When you reach the end of the row, repeat that number in the first box of the following row and continue.
\item Starting from the second box in the first row, the numbers inserted at step 1 are the pole orders of the differentials of 
degree $2,\dots,N$ respectively.
\end{itemize}
A minimal puncture then assigns a pole of order one to all the differentials whereas a maximal one (Young tableau with one row) 
assigns a pole of order $k-1$ to a differential of degree $k$. 

Now that we have learned how the Gaiotto curve and punctures encode the physical data of the corresponding 4d theory, we can 
identify the building blocks of generalized quivers: the three punctured spheres. Starting from a given theory we can degenerate 
the Gaiotto curve letting punctures collide. In this way we will end up with a collection of three punctured spheres and we can 
connect them gauging the diagonal combination of their flavor symmetries. In this way we can build Riemann surfaces with arbitrary 
topology and all of them give a 4d $\mathcal{N}=2$ SCFT when we compactify on them the 6d (2,0) theory of the corresponding rank. 
This analysis has been performed in detail in \cite{CD} (see also \cite{NX2}). I would like to mention that an arbitrary triple of Young tableaux do 
not necessarily correspond to a well defined theory in 4d; there are some restrictions. The basic obstruction is that (\ref{RR}) 
in general assigns negative dimension to some $V_i$'s (consider e.g. a sphere with three minimal punctures, (\ref{RR}) then assigns 
$dim V_3=3-5=-2$). This is in particular true for all spheres with two (or less) punctures, whose interpretation is rather 
subtle \cite{GT}. 

Let us discuss a couple of examples of three-punctured sphere: if we consider $SU(N)$ SQCD with $2N$ flavors we can easily see 
starting from Witten's curve and manipulating it as explained in the previous sections that the corresponding Gaiotto curve is 
a sphere with four punctures: two minimal and two maximal. The collision of two different punctures corresponds to the weak 
coupling limit. As the sphere degenerates we end up with two identical three-punctured spheres with two maximal punctures and 
one minimal. This theory is free and describes $N^2$ hypermultiplets. The collision of two minimal punctures corresponds instead 
to the infinite coupling limit. It can be checked that in the degeneration limit an $SU(2)$ group decouples and we are left with 
two spheres: one describes simply a doublet of $SU(2)$ and the other has two maximal punctures and one described by a Young tableau 
with a row of length three and the others of length one. This is a strongly coupled theory which generalizes the $E_6$ theory 
discussed before (in the rank two case these theories in fact coincide). This is the natural generalization of Argyres-Seiberg 
duality and has been discussed in detail in \cite{CD}. 

Another notable generalization of the $E_6$ theory is the sphere with three 
maximal punctures, which is usually referred to as $T_N$ theory (see \cite{G}). This model and the theories one can build 
connecting several copies of it (both with $\mathcal{N}=2$ and $\mathcal{N}=1$ vectormultiplets) have received much attention 
lately, since one can recover all other three-punctured spheres starting from $T_N$ and moving along the Higgs branch \cite{BBT}. In a sense 
this is the mother of all other three-punctured spheres and the S-dualities involving it encode all possible generalized 
Argyres-Seiberg dualities (an incomplete list of references is \cite{GM}-\cite{KMT}).

\subsubsection{Irregular singularities}

If we want to describe asymptotically free theories we have to introduce the so-called irregular punctures, which are characterized 
by the property that at least one of the k-differentials has a pole of degree greater than k \cite{GNM,NX}. Let us briefly discuss this aspect for 
SYM theory: Witten's curve is $$\Lambda^Nz^2+P_N(v)z+\Lambda^N=0;\quad\lambda=\frac{v}{z}dz.$$ We can easily bring it to the canonical 
form $$x^N=\sum_{k=2}^{N-1}\frac{u_k}{z^k}x^{N-k}+\frac{\Lambda^N+u_Nz+\Lambda^Nz^2}{z^{N+1}};\quad\lambda=xdz.$$ It is easy to 
see from this formula that $\phi_{N}$ has poles of degree $N+1$ both at zero and infinity.

Irregular punctures also emerge in the description of infrared fixed points in AF theories, like AD theories. Usually these are 
described by a sphere with one or two punctures. One of them is always irregular and the other (if any) is instead regular. See 
also \cite{XIE}-\cite{XIE2}. We will discuss a class of such theories at the end of chapter 4.

\subsection{$D_N$ six-dimensional $\mathcal{N}=(2,0)$ theories}

The above analysis has been extended in \cite{DT} to 4d theories obtained from the twisted compactification of the 6d (2,0) 
theories of type $D_N$. These theories can be constructed considering a $\mathbb{Z}_2$ orbifold acting on a five dimensional subspace 
in M-theory (the eleven dimensional spacetime is then $\mathbb{R}^5\times S^1\times(\mathbb{R}^5/\mathbb{Z}_2)$) and placing 
$2N$ coincident $M5$ branes parallel to the subspace invariant under the $\mathbb{Z}_2$ action. The canonical form for the SW 
curve in this case is $$\lambda^{2N}=\sum_{k=1}^{N}\lambda^{2N-2k}\phi_{2k}(z);\quad\lambda=xdz.$$ Rather as in the $SU(N)$ case, 
the basic ingredient of this construction is given by the presence of punctures on the compactifying Riemann surface, which 
encode the flavor symmetry of the theory and at which the k-differentials have punctures. 

The rules determining the degree of  
the poles at the punctures and the flavor symmetry associated to them can be worked out with a technique similar to the one we 
have just reviewed: one tries to bring the curves of the lagrangian subclass to the canonical form and reads out the pole structure. 
In this case the theories obtained compactifying on a punctured sphere correspond to linear alternating quivers of SO/USp groups 
with half-hypermultiplets in the bifundamental. As we have seen in the first section the corresponding SW curves were found in 
\cite{SOP} and using this result it is possible to identify the basic rules of the construction \cite{DT}. We will give the 
algorithm in detail in chapter 4, where this construction plays an important role. The new features is the presence of two different 
classes of punctures, which carry SO or USp flavor symmetry respectively.

There is a new complication with respect to the $A_N$ case: the coefficients one extracts by applying Riemann-Roch theorem are 
not directly the Coulomb branch coordinates: in general they satisfy intricate polynomial relations and only after these are taken 
into account one can extract the true Coulomb branch coordinates. These constraints are discussed in detail in \cite{CDII,CDT} 
(see also \cite{TII,CDT2,RDN}). In principle the construction can be extended to the $E_N$ case. However, there are no known simple 
lagrangian theories in this class that can be used as a starting point (apart from the finite family of SQCD-like theories with 
hypermultiplets in various representations of the gauge group) and moreover finding the SW curves describing these models 
is considerably more complicated with respect to the A,D cases (see \cite{MW} for the determination of the curves describing $E_N$ SYM, 
and \cite{e6} for the relation between the curve for $E_6$ SYM and the stringy realization of the theory). Some results about the reguler punctures have been derived in 
\cite{CDT}.

\section{BPS spectrum and BPS quivers}

The determination of the BPS spectrum of $\mathcal{N}=2$ theories is a notoriously hard problem and the complete answer is known 
only for $SU(2)$ SQCD \cite{B-FI}-\cite{GNM} and Argyres-Douglas theories \cite{SV}. However, the recent progress in the understanding of $\mathcal{N}=2$ theories 
led to several remarkable new insights in this respect. We will now briefly recall the basic aspects of the BPS quiver technique 
(for a different approach using the six-dimensional construction (spectral network) see \cite{GNM}-\cite{GMNN}). 
This method can be applied for theories whose BPS spectrum is finitely generated. By this we mean that there is a finite set of 
states (and we will denote their charge vector with $\gamma_i$) such that all the BPS particles in the 
theory can be written as $$\gamma=\sum_in_i\gamma_i,\quad n_i\in\mathbb{Z}_+\text{or}\; n_i\in\mathbb{Z}_- \forall i.$$ 
This simply states that all the BPS particles in the spectrum are bound states of a finite number of elementary states. 

Many theories respect this property but clearly not all of them: it is easy to see that if the set of phases of the central charges 
of the various BPS states are dense on $S^1$ (the points $\vert z\vert=1$ in the complex plane), the spectrum of the theory cannot 
be finitely generated in the above sense. In particular, S-duality of $\mathcal{N}=4$ implies that its spectrum is not finitely 
generated. However, if we give mass to the adjoint hypermultiplet thus breaking superconformal invariance, it turns out that the 
BPS spectrum can be described using a quiver. Other examples of models that do not have a BPS quiver are class $\mathcal{S}$ 
theories whose Gaiotto curve is a surface without punctures.

\subsection{Defining the BPS quiver}

Let us pick a half-plane in the complex plane (such that the origin lies on its boundary). We define as particles the BPS states whose central 
charges lie in it. The others will be the corresponding antiparticles. Let us consider a certain point in the Coulomb branch and,
supposing to have a set of generators for the BPS spectrum, we can construct an oriented graph called quiver drawing one 
node for each generator and $n$ arrows from node i to node j, where $n$ is the Dirac product between states j and i 
$\langle\gamma_j,\gamma_i\rangle=n$ (if $n$ is negative the arrows will have the opposite orientation). These data can be conveniently 
encoded in a skew-symmetric matrix (called exchange matrix) $B_{ij}\equiv\langle\gamma_i,\gamma_j\rangle$, whose dimension is equal to the number of nodes in 
the quiver and whose kernel is the rank of the flavor symmetry of the theory. Notice that if a finite 
set of generators exists, then it is unique \cite{CV1}, so the quiver is well-defined. A priori it seems that, in order to 
find the quiver, one has to know the BPS spectrum already. Actually, several techniques have been developed to construct it 
for a broad class of $\mathcal{N}=2$ theories \cite{CNV}-\cite{XI2} and they do not require the knowledge of the spectrum in advance. 

A particularly effective technique is based on the 4d/2d correspondence proposed in \cite{CNV}, which states that any theory 
whose BPS spectrum can be described by a quiver is related to a 2d $\mathcal{N}=(2,2)$ model, such that the exchange matrix can 
be written as $$B=S^t-S,$$ where $S$ is the $tt^*$ Stokes matrix of the 2d model \cite{CV92}. When the theory admits geometric 
engineering in type IIB string theory (see e.g. \cite{V11,V12,L1}), the corresponding 2d theory is a Landau-Ginzburg model, whose superpotential can be read 
out from the Calaby-Yau geometry, or equivalently from the SW curve. In this case the nodes of the quiver are in one-to-one 
correspondence with the vacua of the 2d theory and the number of arrows connecting them corresponds to the number of solitons going 
from one vacuum to another. One can then extract the quiver applying the techniques developed in \cite{CV92}. This method can be applied 
e.g. to find the BPS quiver for all SQCD theories with arbitrary gauge group \cite{CNV,CV1,cdelz2}. In particular, in \cite{CNV} 
it was shown that for all theories aring from the compactification of a single $M5$ brane on the SW curve (so all class $\mathcal{S}$ 
theories), the superpotential of the Landau-Ginzburg model is simply given by the SW curve itself. This fact can be exploited to 
determine the BPS quiver for all rank one Gaiotto theories \cite{CVC}. The higher rank case is more complicated and a general 
derivation is still not available. However, a general algorithm has been proposed in \cite{XI1,XI2} (see also \cite{GMNN}).

A key concept is that of quiver representations: we can attach to the node $i$ of the quiver a complex vector space $V_i$ of dimension 
$n_i$ and associate to any arrow connecting node $i$ to node $j$ a linear map $B^a_{ij}:V_i\rightarrow V_j$. These maps satisfy a 
certain set of relations which can be conveniently encoded in a superpotential, in the sense that they coincide with the 
corresponding F-term equations. We refer to the papers listed above for a detailed discussion on the derivation of the superpotential. 

The connection with the BPS spectrum comes from the fact that the above mentioned data define a quantum mechanical problem with 
four supercharges: the gauge group is $$\prod_i U(n_i)$$ and the maps $B^a_{ij}$ are bifundamental fields. The dynamics is encoded 
in the superpotential $\mathcal{W}(B)$ (notice that we can write down a gauge invariant superpotential only if the quiver contains 
closed loops). The quiver representations as defined above (supplemented with a suitable stability condition that I will describe 
later), correspond to the supersymmetric ground states of these quantum mechanical models and can be identified with 
BPS particles in the corresponding four-dimensional $\mathcal{N}=2$ theory \cite{CV1}. The study of quiver representations allows us to 
determine whether the spectrum contains a particle with charge vector $\gamma=\sum_in_i\gamma_i$ or not.

In order to explain the stability condition, we must introduce the notion of quiver subrepresentation: given a representation 
with vector spaces of dimension $n_i$ and maps $B^a_{ij}$, a subrepresentation is another representation with spaces of 
dimension $m_i\leq n_i$ and maps $b^a_{ij}$ such that all diagrams of the following form commute: 
\begin{equation}\begin{array}{ccc}
 \mathbb{C}^{n_i}&{\stackrel {B^{a}_{ij}}{\longrightarrow}}& \mathbb{C}^{n_j}\\
 \text{\textuparrow}& & \text{\textuparrow}\\
 \mathbb{C}^{m_i}&{\stackrel {b^{a}_{ij}}{\longrightarrow}}& \mathbb{C}^{m_j}
  \end{array}
\end{equation}
where the vertical arrows simply indicate an embedding of $\mathbb{C}^{m_i}$ in $\mathbb{C}^{n_i}$. 

As we have explained in the first chapter, the mass of any BPS particle $\gamma$ is given by the norm of its central charge 
$Z(\gamma)$. Given the central charge $Z(\gamma_i)$ of each state associated to the nodes of the quiver, the central charge 
corresponding to our representation $R$ and subrepresentation $S$ will simply be $Z(R)=\sum_in_iZ(\gamma_i)$ and 
$Z(S)=\sum_im_iZ(\gamma_i)$ respectively. The representation $R$ is called stable if all its proper subrepresentations (other 
than $R$ itself and the trivial one) satisfy the relation \be\label{stab}\text{arg}Z(S)<\text{arg}Z(R).\ee
This condition can be related to the Fayet-Illiopoulos terms and D-term constraints in the quantum mechanical problem \cite{CV1}.
The key statement is that stable representations are in one to one correspondence with BPS states of the underlying four-dimensional 
theory \cite{QQ1}-\cite{QQ5}. One can also notice that, in the special case in which all $n_i$'s but two are zero (bound state 
of two objects), the above condition is precisely equivalent to the requirement that the static potential between BPS states 
introduced in \cite{ritzV} is attractive.

The properties of the BPS state (its multiplicity and spin) can be reconstructed by determining the moduli space of vacua of the 
quantum mechanical problem, i.e. the set of solutions of the F-term equations subject to the stability condition, modulo the action 
of the complexified gauge group $\prod_iGL(n_i,\mathbb{C})$ (actually a diagonal $U(1)$ gauge factor is redundant since all bifundamentals 
are uncharged under it. One has to take this into account when taking the quotient). The resulting moduli space $\mathcal{M}$ is by supersymmetry 
a kahler manifold whose complex dimension $d$ encodes the spin of the multiplet (which is $\frac{d+1}{2}$) \cite{CV1}. In practice 
the most frequent cases are: $\mathcal{M}$ is a point, which corresponds to a hypermultiplet and $\mathcal{M}\simeq\mathbb{CP}^1$, 
which corresponds to a vectormultiplet. In higher rank (>1) gauge theories the BPS spectrum is extremely wild and also higher 
spin multiplets can arise. This is not the case for theories whose gauge group is the product of several $SU(2)$ factors (the 
complete theories of \cite{CVC}). 

We can immediately provide a simple application: the state associated to a node is always a hypermultiplet. Its representation 
has only one nonzero $n_i$ which is equal to one. Clearly there are no proper subrepresentations and the stability condition is 
automatically satisfied. Since we have a single nontrivial vector space, all linear maps are necessarily trivial. The moduli space 
of solutions is thus simply given by a point and the above discussion ensures that the BPS state is a hypermultiplet. The BPS 
spectrum of the theory then contains at least as many hypermultiplets as the number of nodes.

\subsection{Quiver mutation and finite chambers}

Once we have found the quiver at a point in the Coulomb branch, the whole BPS spectrum can in principle be derived from the 
study of stable quiver representations. However, this in practice is very hard also for relatively simple quivers and it is 
convenient to adopt a different strategy, which allows to recover the full answer in a simple and algorithmic way at least at 
points in the moduli space where the BPS spectrum is finite: the mutation technique \cite{CVC1,CV1,DX4}. 

Once we have the quiver, we are free to rotate the half-plane containing the particles, until e.g. the left-most BPS particle 
$\gamma$ (which is always associated to a node) exits from it and the corresponding antiparticle $-\gamma$ enters. Now the set of particles 
includes $-\gamma$ and this clearly forces us to change the quiver. Our goal is to understand how. Since we are sitting at a 
point in the moduli space, obviously the spectrum does not change and our new quiver must reproduce this expectation. This can be 
achieved by means of a combinatorial operation called mutation \cite{CV1}. 

\begin{figure}
\centering{\includegraphics[width=\textwidth]{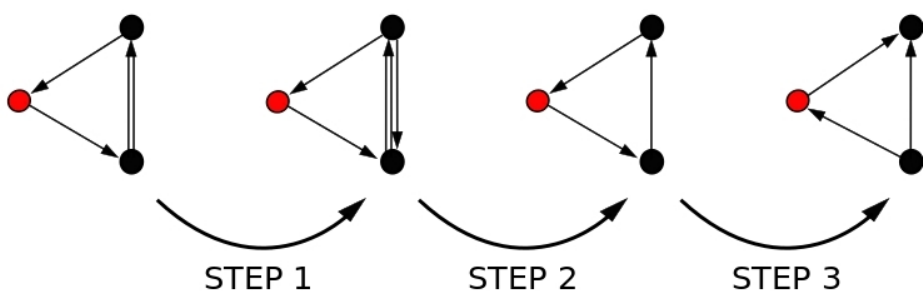}} 
\caption{\label{muta}\emph{The BPS quiver for $SU(2)$ SQCD with one flavor. The mutation at the red node is done following the 
three steps. If before the mutation $\gamma_A$ is the charge vector at the red node, $\gamma_1$ that at the top 
node and $\gamma_2$ that at the lowest node, after the mutation their charge vectors are respectively $-\gamma_A$, $\gamma_1$ and 
$\gamma_2+\gamma_A$.}}
\end{figure}

Mutating at a node (that we will call A) of the quiver corresponds to the following operation\footnote{Actually the algorithm 
also involves a modification of the superpotential. This is important in the proof that the mutation allows to recover the 
same spectrum. Also the cancellation of opposite arrows at the second step can be justified using the superpotential. This is 
discussed in detail in \cite{CV1} and I refer the interested reader to that paper for the details.} (see Figure \ref{muta}): 
\begin{itemize}
\item For any pair of nodes (node 1 and node 2) connected by a path of length two passing through A draw $n_1\times n_2$ arrows 
connecting node 1 to node 2 (pointing in the direction identified by the length two path), where $n_1$ and $n_2$ are respectively 
the number of arrows connecting node 1 and node 2 to A.
\item If nodes 1 and 2 of the previous step are already connected by $n$ arrows pointing in the opposite direction, erase all 
opposite arrows in pairs.
\item Reverse all the arrows that end or start at A.
\end{itemize}
We must then replace the charge vector $\gamma_A$ of the state associated to A with $-\gamma_A$ and change the charge vector of 
any other node from $\gamma$ to $\gamma+n\gamma_A$, where $n$ is the number of arrows pointing now to $A$ from that node (the charge vector 
of the other nodes should not be changed). 

From the above discussion it is clear that as we rotate the half-plane containing the particles the quiver undergoes a sequence of 
mutations: whenever the leftmost BPS state exits we have to mutate on the corresponding node and the corresponding anti particle 
enters in the half-plane. If the spectrum is finite, after a finite sequence of mutations (whose length is clearly equal to the 
number of BPS particles in the spectrum) the half-plane will have made a 
rotation of 180 degrees. At this stage the quiver has come back to his original form and the half-plane 
contains all the antiparticles. In the meanwhile all the particles have been the leftmost state and just keeping track of the 
charge vectors associated to the nodes at which we mutate, we can immediately deduce the full spectrum. This procedure only requires 
the knowledge of the central charges of the BPS states associated to the nodes. Once the spectrum is known at one point in the 
moduli space, in principle the full answer can be deduced using the wall-crossing formula \cite{GM1,wc}. 

We find our first nontrivial prediction: since all the states have become the leftmost state during the rotation, and such a 
state is necessarily associated to a node of the quiver, we can conclude that finite chambers contain only hypermultiplets 
(there are no vectormultiplets or higher spin states).

Clearly, if we start moving around the moduli space the central charges of the BPS states will vary and eventually exit from our 
half-plane. Obviously, also in this case we must mutate at the corresponding node. We therefore learn that when we say that the 
BPS spectrum of a theory is described by a quiver, we actually mean that the 4d theory is associated to the mutation class (or a 
subset of) of that quiver (the set of quivers which are connected by a sequence of mutations). At every point in the moduli space 
the spectrum is captured by an element in the mutation class but each quiver is valid only locally. We will now analyze two 
simple examples to illustrate the technique.

\subsection{$SU(2)$ SYM and Argyres-Douglas theory}

The Gaiotto inspired form for the SW curves describing these two theories are $$x^2=\frac{\Lambda^2}{z}+\frac{u}{z^2}+\frac{\Lambda^2}{z^3},$$ 
for $SU(2)$ SYM and $x^2=z^3+uz+v$ for the AD point of $SU(3)$ SYM \cite{AD}. The SW differential is always $\lambda=xdz$. These 
are both rank one theories and from the discussion of the previous section we know that the superpotentials of the corresponding 
2d theories are simply given by the SW curve itself. It can be simplified further noticing that, since $x$ enters quadratically, 
we can integrate it out and keep only the z-dependent part.

\subsubsection{Argyres-Douglas theory}

\begin{figure}[!h]
\centering{\includegraphics[width=0.6\textwidth]{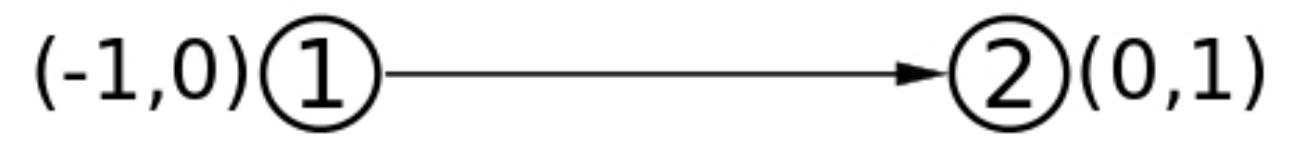}} 
\caption{\label{ADD}\emph{The BPS quiver for the AD theory with the charge vectors determined in \cite{AD} associated to the two nodes.}}
\end{figure}

The 4d/2d correspondence implies that the superpotential for the 2d model is $$\mathcal{W}(X)=X^3+uX+v.$$ This is nothing but the 
$A_2$ minimal model which has two vacua connected by a single soliton. The BPS quiver has therefore two nodes connected by an 
arrow. In Figure \ref{ADD} we draw the quiver and the corresponding charge vectors which were determined in \cite{AD}.

We can determine the BPS spectrum studying the representation theory or applying the mutation technique. Let us consider the 
candidate BPS state $n_1\gamma_1+n_2\gamma_2$, where $\gamma_1$ and $\gamma_2$ are respectively the charge vectors of the states 
1 and 2 in the figure. It is easy to see that $\gamma_2$ is always a subrepresentation. Therefore, if the phase of $Z(\gamma_2)$ 
is larger than that of $Z(\gamma_1)$, it is also larger than that of $Z(n_1\gamma_1+n_2\gamma_2)=n_1Z(\gamma_1)+n_2Z(\gamma_2)$ 
and the stability condition is not satisfied. We thus find that all boundstates are unstable and only the BPS states associated to 
the two nodes exist.

If we move in the moduli space until the phase of $Z(\gamma_1)$ becomes larger the situation is different, since the 
subrepresentation $\gamma_2$ is now stable. On the other hand, if $n_1>n_2$ the linear map has necessarily a nontrivial kernel 
and $\gamma_1$ is a representation if we choose the embedding such that the corresponding one-dimensional vector space is mapped 
in the kernel of the linear map. The representation is then unstable and we can rule out the above mentioned bound states. 
If $n_1<n_2$ it is easy to see that $\gamma_1+\gamma_2$ is a subrepresentation (for a suitable choice of the map $b_{12}$) and it 
violates the stability condition. This argument allows to rule out any bound state apart from $\gamma_1+\gamma_2$. Indeed, the 
only proper subrepresentation $\gamma_2$ satisfies the stability condition and this state is part of the spectrum. Since the 
moduli space is a point, it is a hypermultiplet with charge vector $(-1,1)$.

The same result can be obtained using the mutation technique. If the phase of $Z(\gamma_2)$ is larger it is the first state to 
exit from the half-plane as we start rotating. Therefore we must mutate at node 2 (see the figure) first. Its charge vector becomes 
$-\gamma_2$ whereas the charge vector associated to the first node is unchanged, according to the above rule. As we go on rotating 
we then encounter $\gamma_1$ and when we mutate its charge becomes $-\gamma_1$. The charge at the other node is unaffected. We then 
find the initial quiver with charge labels $-\gamma_1$ and $-\gamma_2$. Tese are the corresponding antiparticles, signaling that we 
have completed our half rotation. Since we have mutated twice the spectrum just contains $\gamma_1$ and $\gamma_2$.

If instead the phase of $Z(\gamma_1)$ is larger, we have to mutate first at $\gamma_1$. The charge of node 2 now is changed to 
$\gamma_1+\gamma_2$ according to the rules of the previous section. This state exits from the half-plane before $\gamma_2$, so we 
have to mutate at the second node. This affects the charge vector at the first node which was $-\gamma_1$ after the first mutation 
and now becomes $-\gamma_1+\gamma_1+\gamma_2=\gamma_2$. If we now mutate again at the first node the quiver is back to its 
original form with labels $-\gamma_1$ and $-\gamma_2$. This mutation sequence tells us that the spectrum contains the three 
states $\gamma_1$, $\gamma_2$ and $\gamma_1+\gamma_2$. If we go on rotating we find the corresponding antiparticles. This agrees 
precisely with the answer from the representation theory and coincides as expected with the result found in $\cite{AD}$.

From this analysis we recover the simplest example of wall-crossing formula: if we consider two hypermultiplets whose Dirac product is one, 
depending only on the ordering of phases of their central charges, the BPS spectrum can include or not their bound state and it is again 
a hypermultiplet. We will see later another famous example.

With analogous arguments we can study the spectrum of any AD point of SYM theory with simply laced gauge group. The quiver is
always the Dynkin diagram associated to the gauge group. These models are the only examples of theory with a finite number of 
BPS states in all BPS chambers (the regions of the moduli space in which the BPS spectrum does not change) \cite{CVC}.

\subsubsection{$SU(2)$ SYM}

If we multiply the curve by $z^2$ and define $z=e^{X}$, we get the superpotential $$\mathcal{W}(X)=\Lambda^2e^X+u+\Lambda^2e^{-X}.$$ 
This describes the 2d $CP^1$ sigma model, which has two vacua connected by a pair of solitons \cite{CV92}. The 4d/2d correspondence then 
tells us that the quiver describing $SU(2)$ SYM has two nodes linked by two arrows (with a similar argument one can deduce the BPS 
quiver for $SU(N)$ SYM theory \cite{CNV}). This is of course the only possible answer: 
as was argued by Bilal and Ferrari in \cite{B-FI}, the strong coupling region of the moduli space contains only two states which 
are those becoming massless at the singularities. If we normalize the charges in such a way that a doublet of $SU(2)$ has electric 
charge one, the charge vectors of these states are $(0,1)$ and $(2,-1)$. Since, as we have seen before, there are at least as 
many states as the number of nodes in the quiver, we find immediately that the BPS quiver describing 
$SU(2)$ SYM (if any) must have exactly two nodes and the corresponding states are precisely the two listed above. Since the 
Dirac pairing between these states is 2, we recover from the definition precisely the quiver predicted by the 4d/2d correspondence.

\begin{figure}
\centering{\includegraphics[width=0.6\textwidth]{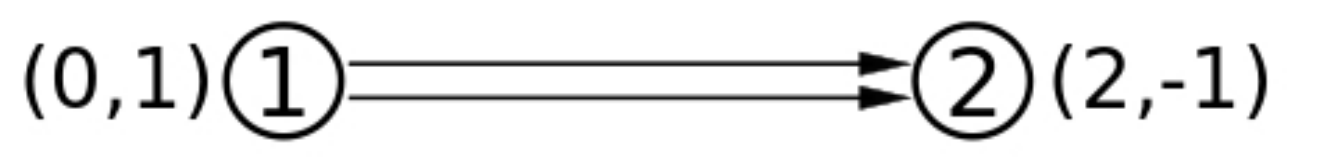}} 
\caption{\label{sim2}\emph{The BPS quiver for $SU(2)$ SYM theory with the charge vectors associated to the two nodes.}}
\end{figure}
We take as half-plane the complex upper half-plane. Using the SW solution we can compute the central charges associated to the 
two nodes. When we are in the strong coupling region the phase of the central charge of $\gamma_2=(2,-1)$ is greater than that of 
$\gamma_1=(0,1)$. When we cross the marginal stability curve the central charges become aligned (wall-crossing) and in the 
weak-coupling chamber the phase of $Z(\gamma_1)$ becomes larger. We will use directly the mutation technique to find the answer.

In the strong coupling region we have to mutate first on node 2. This does not affect the charge label of node 1. As we go on 
rotating we then have to mutate at node 1. At this stage the quiver has come back to its original form and the charge labels 
are $-\gamma_1$ and $-\gamma_2$. We thus find as expected the two BPS states which exist at strong coupling. With an argument 
analogous to that of the previous section, it can be seen that the representation theory confirms this result

In the weak coupling region the phase of $Z(\gamma_1)$ is larger than that of $Z(\gamma_2)$ and we must mutate at node 1 first. 
The charge label of node 2 now is changed to $\gamma_2+2\gamma_1$. This is now the leftmost state and we have to mutate at node 2. 
Then the charge vector associated to node 1 becomes $-\gamma_1+2\gamma_2+4\gamma_1=2\gamma_2+3\gamma_1$. If we go on at the n-th 
mutation we find the state $n\gamma_2+(n+1)\gamma_1=(2n,1)$. We immediately see that this chamber is infinite and contain all the dyons 
of this form, whose central charges accumulate towards that of $\gamma_1+\gamma_2$.

Of course we can also rotate the half-plane in the opposite sense and mutate on the right. The algorithm is slightly different in this 
case and can be determined imposing that a right mutation followed by a left mutation is equivalent to the identity (we rotate 
the plane in one sense and then rotate it back to the initial position). The only difference with the algorithm seen above is 
that the charge labels should be changed when the arrows are pointing to the node at which we are mutating before and not after 
the mutation. 

We must mutate at node 2 first. The charge  label at node 1 then becomes $\gamma_1+2\gamma_2$. When we mutate at node 1 the 
second charge vector becomes $-\gamma_2+2\gamma_1+4\gamma_2=2\gamma_1+3\gamma_2$ and so on. We then get the states 
$(n+1)\gamma_2+n\gamma_1=(2n+2,-1)$. The only state that we could not investigate with this technique is $\gamma_1+\gamma_2$. 
The two towers of dyons we have found accumulate towards this state, which is in a sense infinitely many mutations away. We must 
use the representation theory to find an answer. The two linear maps are $1\times1$ matrices and can be organized in a 
two-dimensional vector. The space of stable representations is thus $\mathbb{C}^2$ minus the origin (if both maps are trivial 
$\gamma_1$ is a subrepresentation and does not satisfy the stability condition). We then have to quotient with respect to the action 
of the complexified gauge group, which means that we consider the above space modulo rescaling by a complex number. This is 
precisely $\mathbb{C}\mathbb{P}^1$ and the state $\gamma_1+\gamma_2=(2,0)$ is a vectormultiplet, the W-boson. This is precisely the 
answer found by Bilal and ferrari in \cite{B-FI} and constitutes the second example of wall-crossing: if we have two states whose 
Dirac product is two, for a specific ordering of phases of the central charges the spectrum includes an infinite tower of dyons 
and a vectormultiplet. One can easily explore the spectrum of $SU(2)$ SQCD with analogous techniques. The details can be found in \cite{CV1}.

\chapter{Chiral condensates in SQCD and the Konishi anomaly}

\section{Introductory remarks}

This chapter is devoted to a careful analysis of vacua in $\mathcal{N}=2$ SQCD with generic bare quark masses, softly broken by a mass term for the chiral superfield 
$\Phi$ in the adjoint representation of the gauge group. As we have seen one of the fundamental aspects of the Seiberg-Witten solution
of $\mathcal{N}=2$ gauge theories is the striking relation between monopoles, confinement and chiral 
symmetry breaking; whether all this is at work in the real world QCD is still an open problem. 

The generalization of their analysis to $SU(N)$ SQCD revealed a very rich structure \cite{APS,CKM,Hanany,HananyI}, in which monopoles of non-abelian kind rather than ordinary 't Hooft-Polyakov monopoles \cite{TP} play a key role \cite{ABEKY,ABEK,ABEKII}. The precise analysis of these vacua (called r vacua) is hard to carry out  just by means 
of the Seiberg-Witten solution and seems to be deeply related to non-abelian duality (see e.g. \cite{SD,KD}). 
In particular the analysis of \cite {CKM} is suited to study the properties of the r vacua for values of the bare quark masses which are very large (semiclassical regime), in which
they behave as Higgs vacua, or very small (nonperturbative regime), in which they are confining; it remains anyway very difficult to make precise predictions
about the ``intermediate range'', leaving some points unclear. 

The aim of this chapter is to approach this problem by making use of the deep connection between these theories and their ``softly broken'' version, 
obtained adding a mass term $\frac{\mu}{2}\Tr\Phi^2$ for the adjoint chiral multiplet (actually the formalism we will use works for more general, 
not necessarily renormalizable superpotentials), lifting the moduli space of the $\mathcal{N}=2$ theory and leaving a finite number of vacua. This strategy has 
been adopted in \cite{GVY} to study the $SU(2)$ theory with one flavor: for $\mu>>\Lambda_{N_f}$ the natural approach is to decouple semiclassically 
the adjoint field. The low-energy theory is $\mathcal{N}=1$ SQCD with a quartic interaction between quarks. Taking into account the corrections due to 
nonperturbative gauge dynamics, the vacua can be found solving the stationarity equations for the quark superpotential. Holomorphicity in the $\mu$ 
parameter ensures that this description gives the same number of supersymmetric vacua and the same pattern of breaking of the global symmetry as 
 those  implied by the SW solution, more suitable for the $\mu<<\Lambda_{N_f}$ limit. This one-to-one correspondence can be established by means 
of an exact relation between chiral condensates derived from the Konishi anomaly \cite{K1,K2}. The analysis of the more interesting $SU(N)$ case along these lines requires the use of 
the Dijkgraaf-Vafa superpotential \cite{DVII,DV,DVI,DVIII}, which allows us to capture all the holomorphic data of the theory by means of a planar calculation 
in a matrix model, and of the generalized Konishi anomaly \cite{FDSW,FSWI,CSW}. These tools allow us also to identify fundamental phase invariants in SYM theories
\cite{FSWI,KO,FF,FFI} and to address nonperturbative investigations in $\mathcal{N}=1$ SQCD (see for instance \cite{AE,BH} and references therein). 

In section 2 we shall review the construction proposed in \cite{GVY} and extend it 
to the other asymptotically free cases (2 and 3 flavors). In the subsequent  sections the general $SU(N)$ case will be analyzed, by  use of the generalized Konishi anomaly relations \cite{FDSW} and the Dijkgraaf-Vafa superpotential. In section 3 we shall derive the anomaly equations and the matrix model superpotential for the theory under consideration (the analysis
is similar to the one proposed in \cite{UN}) and see how it allows to recover the instanton induced superpotentials of \cite{ADS,SS,SI}. In section 4 
the previous results will be used  to derive some general features of the r vacua, in particular regarding the intermediate mass range and the ``transition'' from the Higgs to the pseudo-confining 
phase (as they call it in \cite{CSW}). As a byproduct we find  a clear interpretation of the two to one correspondence noted in \cite{CKM,BK} which 
associates both r and $N_f-r$ semiclassical vacua to the same quantum r vacuum in the theory with degenerate bare masses for
the matter hypermultiplets (for another discussion on the relation between semiclassical and quantum vacua in a slightly different context
see \cite{SYI}-\cite{SYIV}). Section 5 is devoted to a discussion of the results achieved.

\section{The anomaly technique for SU(2) gauge theories}
As a warm-up, in this section we focus on an example with two colors and a renormalizable superpotential.  We consider a version of $\mathcal{N}=1$ SQCD with gauge group SU(2) and $N_f$ flavors, with in addition a chiral superfield $\Phi$ in the adjoint of the gauge group. The superpotential at the tree level is given by:
\begin{equation*}
\W_{tree}=m_i\tilde{Q}_iQ^i + \sqrt{2}h\tilde{Q}_i\Phi Q^i + \mu\Tr \Phi ^2.
\end{equation*}
$i=1,\ldots,N_{f}$ is a flavor index. Setting the Yukawa coupling $h=1$ and the adjoint mass $\mu=0$ we get $\mathcal{N}=2$ SQCD with $N_f$ flavors. For $N_f\leq 3$ this theory is asymptotically free and generates dynamically a scale which we denote by $\Lambda_{N_f}$. 
The case with a single flavor was studied in \cite{GVY}, here we present the generalization to the other asymptotically free cases $N_f = 2, 3$. We show that even the description of the nontrivial flavor structure matches in the two regimes.

\subsection{Classical vacua and symmetries}
As explained in \cite{CKM}, an analysis of the tree level superpotential reveals that for generic values of the quark and adjoint masses there are $N_f + 2$ supersymmetric vacua, and by a Witten index argument the number of vacua must be the same in the full quantum theory.  Flat directions develop when $\mu = 0$ or some of the quark masses coincide.  The 1-(complex) dimensional Coulomb branch of the $\mathcal{N}=2$ theory parametrized by $u=\langle \Tr \Phi ^2 \rangle$ is lifted by the soft breaking to $\mathcal{N} =1$ leaving only this discrete set of points. \newline
\indent The U(1)$_R$ and U(1)$_J \subset$ SU(2)$_R$ simmetries, together with holomorphicity, give constraints on the dependence of chiral condensates on the parameters in the superpotential. The charges are given in the following table:
\begin{center}
\begin{tabular}{|ccc|ccc|ccc|ccc|ccc|ccc|}
\hline
&& && $\Phi$ &&& $Q$ &&& $\tilde{Q}$ &&& $\mu$ &&& $m$ & \\
\hline
& U(1)$_R$ &&& 2 &&& 0 &&& 0 &&& -2 &&& 2 & \\
\hline
& U(1)$_J$ &&& 0 &&& 1 &&& 1 &&& 2 &&& 0 & \\
\hline
\end{tabular}
\end{center}
The U(1)$_R$ is anomalous at the quantum level, thus acting nontrivially on the dynamical scale $\Lambda_{N_f}$. The residual $\mathbb{Z}_{4(4-N_f)}$ symmetry is broken spontaneously to $\mathbb{Z}_4$ by the adjoint VEV leaving a $\mathbb{Z}_{4-N_f}$ acting on the u plane. \newline
\indent Alternatively we can define a modified and nonanomalous U(1)$_{R'}$ symmetry that acts on the Yukawa parameter $h$:
\begin{center}
\begin{tabular}{|c|c|c|c|c|c|c|}
\hline
 & $\Phi$ & $Q$ & $\tilde{Q}$ & $\mu$ & $m$ & $h$  \\
\hline
 $U(1)_{R'}$ & 1 & $\frac{N_f -2}{N_f}$ & $\frac{N_f -2}{N_f}$ & 0 & $\frac{4}{N_f}$ & $\frac{4-N_f}{N_f}$ \\
\hline
\end{tabular}
\end{center}
The combination of the parameters that is neutral under the nonanomalous U(1)'s and adimensional is given by:
\begin{equation}\label{sigma}
 {\sigma_{N_f}}^2 \equiv (\Pf\,\m)^{1-\frac{4}{N_f}}\,h^4\,\Lambda_{N_f}^{4-N_f}
\end{equation}
where the mass matrix has been recast in the form:
\begin{equation*}
\m \equiv  \left( \begin{array}{c|c}
                                  0 & m\\
                                  \hline
                                  -m&0
                                  \end{array}\right), \qquad m=diag(m_1,\ldots , m_{N_f})
\end{equation*}
\indent The limit $\sigma_{N_f}\to 0$ is interesting since it can be interpret in two ways: either as the limit $h \to 0$ in which we recover $\mathcal{N}=1$ SQCD with massive quarks, or as the limit $\m \to \infty$ with $h=1$ in which the quarks decouple and the theory approaches pure $\mathcal{N}=2$ SYM softly broken by the adjoint mass term. Consistently, in both these limits we are left with two discrete supersymmetric vacua, respectively in the Higgs phase and in the confining phase. 
Notice that the Higgs and confining phases are continuosly connected in these theory: it displays complementarity, as expected for a theory with scalar fields in the fundamental.\newline
\indent When the $\mu$ parameter is large with respect to the dynamically generated scale, it is legitimate to study the low-energy theory by integrating out the adjoint field. As a result we get a version of $\mathcal{N}=1$ SQCD modified by a quartic term for the quarks:
\begin{equation*}
\W_{dec} = \frac{h^2}{8\mu}\tr[V^2] - \frac{1}{2}\tr[\m V]
\end{equation*}
where $V$ is the gauge invariant quark bilinear, assembled in a $2N_f \times 2N_f$ antisymmetric matrix, and $\tr$ denotes the trace over flavor indices. At the classical level $V$ is subject to the constraint $\Pf V = 0$. In the massless limit $\m=0$ the O(2$N_f$) flavor symmetry acts on $V$ by conjugation. It is convenient to express the $V$ matrix in term of neutral parameters $\V_1,\dots,\V_{N_f}$ as follows:
\begin{equation}\label{parametrizationV}
 \sigma_{N_f} \frac{\mu}{h^2} \left(\begin{array}{c|c} 
              0 & \begin{array}{ccc}
                      m_1\V_1  &  &                                    \\
                                & \ddots &                                \\
                                &   & m_{N_f}\V_{N_f}               \\
                     \end{array} \\
               \hline
              \begin{array}{ccc}
                        -m_1\V_1  &  &                                \\
                                & \ddots &                                \\
                                &   &   -m_{N_f}\V_{N_f}           \\
                     \end{array} & 0 
             \end{array}\right).
\end{equation}

\subsection{Low-energy effective superpotentials}

The superpotential does not receive any correction in perturbation theory but new terms may appear as a consequence of nonperturbative gauge dynamics. The results for the exact effective superpotentials in SQCD are known. For the theory in consideration it is only necessary to add the quartic term that is reminiscent of the microscopic Yukawa coupling to the $\Phi$. Therefore the general form is:
\begin{equation}\label{s.p.eff}
\W_{eff} = \W_{dec} + \W_{n.p.}
\end{equation}
We list below the results for the different values of $N_f$:
\begin{equation*}
\begin{array}{cl}
N_f=1 & \W_{n.p.} = {\tilde{\Lambda}_1}^{\phantom{1}5}(\mathrm{Pf}V)^{-1} \\
N_f=2 & \W_{n.p.} = X(\mathrm{Pf}V - \tilde{\Lambda}_2^{\phantom{1}4}) \\
N_f=3 & \W_{n.p.} = - \tilde{\Lambda}_3^{\phantom{1}-3}\mathrm{Pf}V.
\end{array}
\end{equation*}
In the second row $X$ is a nondynamical Lagrange multiplier implementing the modified quantum constraint. We denote by $\tilde{\Lambda}$ the dynamical scale in the theory where $\Phi$ has been decoupled. 

\subsection{The Konishi anomaly}
 In order to find where in the Coulomb branch the dicrete vacua are located we have to determine the adjoint field condensate $u=\langle \Tr \Phi ^2 \rangle$. This amounts to exploiting the anomalous Ward identities associated to Konishi anomaly: 
\begin{equation*}
\exl  X \frac{\partial \W}{\partial X} + T(R_X)\frac{\Tr W^2}{8\pi^2} \exr = 0
\end{equation*}
where $X$ is a generic chiral superfield, $T(R_X)$ is the Dynkin index of the representation of the gauge group acting on $X$. Specifyng $X$ to be $Q$, $\tilde{Q}$ or $\Phi$ we end up with:
\begin{equation*}
\left\{\begin{array}{l}
           Q^i, \tilde{Q}^i : \quad \exl \sqrt{2}h\tilde{Q}_{i} \Phi Q^{i} + m_i \tilde{Q}_i Q^i +\frac{1}{16\pi^2} \Tr W^2 \exr =0 \quad i=1,\ldots , N_f\\
           \Phi:\quad \exl 2\mu\Tr\Phi^2 +\sqrt{2}h\tilde{Q}_{i} \Phi Q^{i}+\frac{1}{4\pi^2}\Tr W^2\exr =0 .      
          \end{array}\right.
\end{equation*}
From this equations we eliminate the gaugino condensate  $s\equiv -\frac{1}{16\pi^2}\exl\Tr W^2\exr$ obtaining $s = \exl \frac{1}{2 N_f}\tr\left[\frac{h^2}{2\mu}V^2-\m V\right] \exr $. The remaining equations relates directly the interesting parameter $u$ to the quark VEVs:
\begin{equation}\label{Konishi}
 2 \mu u = \exl \dfrac{4-N_f}{2N_f}\tr\left[\frac{h^2}{2\mu} V^2\right] - \dfrac{2}{N_f}\tr\left[ \m V \right] \exr.
\end{equation}

\subsection{$N_f=1$}
In this section we review the results already obtained in \cite{GVY} with one flavor. Matching the running couplings at the scale $\mu$ we get: $\tilde{\Lambda}_1^{\phantom{1}5} = \mu^2{\Lambda_1}^3$. The stationarity of the superpotential $\eqref{s.p.eff}$ with respect to $V$ gives the equation:
\begin{equation}\label{stationarityNf=1}
\sigma_1\V _1 - 2 + 2 \dfrac{1}{{\V _1}^2} = 0.   
\end{equation}
where $\sigma_1 = h^2{m_1}^{-3/2}\Lambda_1^{3/2}$ and $\V_1 = (h^2\mu\Lambda_1^3)^{-1}m_1^2V_{12}$ are neutral under the nonanomalous U(1)'s.
Once this equation is solved, the corresponding value of $u$ is given by equation $\eqref{Konishi}$:
 \begin{equation}\label{KonishiNf=1}
4 u =(m\Lambda _1 ^3)^{1/2}( - 3\sigma_1{\V _1}^2 + 8 \V _1 )
\end{equation}

\paragraph{$\sigma_1\to 0$ limit:} the equation $\eqref{stationarityNf=1}$ has two solutions for finite values of the VEVs $\V_1 = \pm 1+ \mathcal{O}(\sigma _1)$ and one that goes to infinity $\V _1 = \frac{2}{\sigma _1} +  \mathcal{O}(\sigma _1)$. Correspondingly $\eqref{KonishiNf=1}$ gives:
\begin{equation*}
\left\{\begin{array}{l}
           \V _1 = \pm 1+ \mathcal{O}(\sigma _1) \Rightarrow u = (m\Lambda _1 ^3)^{1/2} (\pm 2 + \mathcal{O}(\sigma _1) ) \\
           \V _1 = \frac{2}{\sigma _1} + \mathcal{O}(\sigma _1) \Rightarrow u = (m\Lambda _1 ^3)^{1/2} (\frac{1}{\sigma_1} + \mathcal{O}(\sigma_1)) \\
          \end{array}\right.
\end{equation*}
From the point of view of the softly broken $\mathcal{N}=2$ theory the two vacua for finite $u$ are the monopole/dyon vacua while the third one is the electric charge vacuum that correctly goes to infinity in this limit.

\paragraph{$m_1\to 0$ limit:} since $\sigma_1 = \mathcal{O}(m^{-\frac{3}{2}})$ and $\V _1 = \mathcal{O}(m^{\frac{1}{2}})$ the equations become:
\begin{equation*}
\left\{\begin{array}{l}
\sigma _1 {\V _1}^3 + 2 = 0  \\
u = 2 (m\Lambda _1 ^3)^{1/2}  \V _1
\end{array}\right.\Rightarrow u = - 2 (m\Lambda _1 ^3)^{1/2} \left(\frac{2}{\sigma _1}\right)^{\frac{1}{3}} e^{\frac{2k\pi i}{3}},\, k=0,\,1,\,2 
\end{equation*}
Notice that we recover the $\mathbb{Z}_3$ symmetry acting on the $u$ plane, again in agreement with softly broken $\mathcal{N}=2$ SQCD with one massless flavor.

\paragraph{Seiberg-Witten curve:} the equivalence with softly broken $\mathcal{N}=2$ SQCD can be established for generic values of the parameters. If we eliminate $\V_1$ from equations $\eqref{stationarityNf=1}$ and $\eqref{KonishiNf=1}$ we get a single equation for the $u$ variable:
\begin{equation*}
u^3 - {m_1}^2 u^2 - \frac{9}{2} m_1 {\Lambda_1}^3 h^2 u + 4 {m_1}^3 {\Lambda_1}^3 + \frac{27}{16} h^4 {\Lambda_1}^6 = 0.
\end{equation*}
Setting $h=1$ these equation is equivalent to the vanishing of the discriminant of the Seiberg-Witten curve for one flavor. The solutions correspond to the vacua of the softly broken theory with $\mu\neq 0$. Therefore the vacua obtained by our analysis are in one to one correspondence to those obtained by the Seiberg-Witten curve.

\subsection{$N_f=2$}
In this case the matching gives ${\tilde{\Lambda}_2}^{\phantom{1}2}=\mu\Lambda_2$. The stationarity of the effective superpotential $\eqref{s.p.eff}$ gives the equations:
\begin{equation}\label{stationarityNf=2}
\left\{\begin{array}{l}
           \sigma_2(\V_1 + \alpha X \V_2) - 2 = 0          \\
           \sigma_2(\V_2 + \alpha^{-1} X \V_1) - 2 = 0  \\
           \V_1\V_2 =  1
          \end{array}\right.
\end{equation}
in which $\sigma_2 = h^2 (m_1\,m_2)^{-1/2}{\Lambda_2}^2$, we have set $\alpha=m_2/m_1$ and a factor of $h^2/\mu$ has been reabsorbed in the definition of $X$. The variables $\V_{1,2}$ were defined in $\eqref{parametrizationV}$. The additional equation derived from the anomaly is:
\begin{equation}\label{KonishiNf=2}
4 u + \Lambda_2 (m_1\, m_2)^{1/2} [ \sigma_2 (\alpha^{-1} {\V_1}^2 + \alpha {\V_2}^2) - 4 (\alpha^{-1} \V_1 + \alpha \V_2) ]= 0.
\end{equation}
As a consistency check one can verify that in the decoupling limit of one of the two flavor, that is $m_2\to \infty$, the equations for the theory with a single flavor are correctly recovered.

\paragraph{$\sigma_2\to 0$ limit:} eliminating $X$, $\eqref{stationarityNf=2}$ has the two solutions $(\V_1,\,\V_2)=\pm(\alpha,\,\alpha^{-1}) + \mathcal{O}(\sigma _2)$ and correspondingly we get $u = \Lambda_2 (m_1\, m_2)^{1/2} (\pm 2 + \mathcal{O}(\sigma_2))$. These two solutions are the monopole/dyon vacua. One can easily see that there are two additional solutions going to infinity in the space of VEVs like $\sigma_2^{-1}$: these are vacua of the softly broken theory in which an electrically charged degree of freedom condensates. As explained above $\sigma_2\to 0$ can be interpreted as the limit in which quarks becomes very massive, and therefore it is correct for these vacua to go to infinity in this limit.

\paragraph{Degenerate masses $m_1=m_2=m$:} in this case $\alpha = 1$. The equations give the following solutions for $u$: 
\begin{eqnarray*}
 u_1 = u_2 =\Lambda_2 m [\sigma_2/2 + {\sigma_2}^{-1}] = \frac{1}{2} {\Lambda_2}^2 + m^2 , \\
  u_{3,4} =  \Lambda_2 m [- \sigma_2/2 \pm 2]= - \frac{1}{2} {\Lambda_2} ^2 \pm 2 m\Lambda_2.
\end{eqnarray*} 
The two coincident values $u_1$ and $u_2$ correspond to two different solutions for $(\V_1, \V_2)$, setting $m_1\neq m_2$ would split the degeneracy on the Coulomb branch and separate the two points. \newline
\indent When $m \to 0$ the full O(4) flavor symmetry is restored and in this limit the four vacua are organized in two pairs of coincident points on the Coulomb branch. This is again in agreement with the interpretation in terms of the softly broken $\mathcal{N}=2$ theory. From this point of view in each vacuum some charged degree of freedom condensates. The magnetic charges fall in spinorial representations of the flavor symmetry group and therefore in representations of Spin(4) $\simeq$ SU(2) $\times$ SU(2). The two pairs of coincident points correspond to two pairs of magnetically charged degrees of freedom that in the $m \to 0$ limit organize in a doublet of one of the two SU(2)s.

\subsection{$N_f=3$}
In this case the matching of the running couplings gives the relation: ${\tilde{\Lambda}_3}^{\phantom{1}3} = \mu^2 \Lambda_3 $. In terms of the $(\V_1, \V_2, \V_3)$ variables of $\eqref{parametrizationV}$ the stationarity conditions of the superpotential is:
\begin{equation}\label{stationarityNf=3}
\left\{ \begin{array}{l}
\sigma_3 \V _1 - 2 + 2{\alpha _1}^{-1} \V _2 \V _3 = 0 \\
\sigma_3 \V _2 -2 + 2{\alpha _2}^{-1} \V _3 \V _1 = 0  \\
\sigma_3 \V _3 -2 + 2{\alpha _3}^{-1} \V _1 \V _2 = 0 
\end{array}\right. 
\end{equation}
where $\sigma _3= h^2 {\tilde{\Lambda} _3}^{\phantom{1}\frac{3}{2}}\mu^{-1}(m_1 m_2 m_3)^{-\frac{1}{6}}$ and we defined $\alpha _1 = {m_1}^{\frac{4}{3}}(m_2 m_3)^{-\frac{2}{3}}$ with cyclic definitions for $\alpha_2$ and $\alpha_3$ such that $\alpha_1\alpha_2\alpha_3=1$. The adjoint field condensate is given by the anomaly equation:
\begin{equation*}
12 u + \sqrt{m_1\,m_2\,m_3 \,\Lambda_3}[\sigma _3 (\alpha _1 {\V _1}^2 + \alpha _2 {\V _2}^2 + \alpha _3 {\V _3}^2) - 8 (\alpha _1 \V _1 + \alpha _2 \V _2 + \alpha _3 \V _3)] = 0
\end{equation*}
Again one can check that in the limit $m_3 \to \infty$ , by imposing a correct scaling of the variables, the $N_f = 2$ equations are recovered, with $\V_3$ formally playing the role of the Lagrange multiplier. 

\paragraph{$\sigma_3\to 0$ limit:} the system $\eqref{stationarityNf=3}$ in this limit has two solutions for finite values of the parameters $(\V _1 ,\, \V _2,\, \V_3) = \pm({\alpha _1}^{-1} ,\, {\alpha _2}^{-1} ,\, {\alpha _3}^{-1}) + \mathcal{O}(\sigma _3)$ which give $u = \pm2\sqrt{m_1\,m_2\,m_3\,\Lambda_3}$. As for $N_f=1,2$ these are the two vacua which correspond to the monopole/dyon vacua. Again, one can find three additional solutions with a runaway $\sigma_3^{-1}$ behavior: these are in correspondence with the three vacua in the semiclassical region associated to the condensation of some electrically charged degree of freedom.

\paragraph{Degenerate masses $m_1=m_2=m_3=m$:} in this case $\alpha_1=\alpha_2=\alpha_3=1$. Solving the equations gives the following values for $u$:
\begin{eqnarray*}
 u_1 = u_2 = u_3 = \sqrt{\Lambda_3 m^3} [\sigma _3^{-1}+ \frac{1}{6} \sigma _3 - \frac{1}{16} {\sigma _3}^3]\\
 u_{4,5} = \sqrt{\Lambda_3 m^3} [-\frac{3}{4}\sigma _3 -\frac{1}{32}{\sigma _3}^3 \pm \frac{1}{4}\left(\frac{{\sigma _3}^2}{4} + 4\right)^{\frac{3}{2}}].
\end{eqnarray*}
Like in the $N_f = 2$ case, the fact that we find coincident points on the Coulomb brach is related to the partial restoration of the flavor symmetry. The full flavor symmetry in this case is O(6) and Spin(6) $\simeq$ SU(4) is explicitily broken by $m$ to SU(3)$\times$U(1). From the point of view of softly broken $\mathcal{N}=2$ there are three vacua associated to the condensation of magnetically charged degrees of freedom: when the masses are switched to have the same value, the three degrees of freedom organize in a fundamental multiplet of SU(3) and the vacua flow in the same point on the $u$ plane. The additional two solutions are singlets of SU(3). \newline
\indent In the limit $m \to 0$ the full O(6) is restored and monopoles/dyons should fall in multiplets of SU(4). This limit is equivalent to $\sigma_3\to\infty$ and indeed we see that the SU(3) triplet solution coincides with one of the two singlets in  $u = -\frac{1}{16}{\sigma _3}^3( 1 + \mathcal{O}({\sigma _3}^{-2}))$ and they form a fundamental multiplet of SU(4), while the other singlets move to $u=0$. Notice that there is no discrete symmetry acting on the $u$ plane when $N_f=3$. The presence of the singlet vacuum indicates that confinement can be realized even without dynamical breaking of the flavor symmetry.

\section{The $SU(N)$ theory}

We now consider the $SU(N)$ case with the same tree level superpotential as before. Decoupling the adjoint field we can express the effective superpotential
in terms of the meson field $M$, obtaining the result
\begin{equation}\label{potM}
\W(M,\Lambda)=-\frac{1}{\mu}\left(\tr M^2-\frac{1}{N}(\tr M)^2\right) + \tr mM + \W_{NP}, 
\end{equation}
where $\W_{NP}$ represents the nonperturbative contribution and its form depends on the range of $N_f$ as before:
\begin{equation}\label{potenziali}
\begin{array}{|c|c|}
\hline
N_f<N_c & \W_{NP}=\left(\frac{\Lambda_{1}^{3N_c-N_f}}{\deter M}\right)^{\frac{1}{N_c-N_f}} \\
\hline
N_f=N_c & \W_{NP}=X\left(\deter M - B\tilde{B} - \Lambda^{2N_c}\right) \\
\hline
N_f=N_c+1 & \W_{NP}=\frac{1}{\Lambda_{1}^{2N_f-3}}\left[\deter M-B^{i}M_{i}^{j}\tilde{B}_{j}\right] \\
\hline
N_f>N_c+1 & \W_{NP}= qM\tilde{q} \\
\hline
\end{array}
\end{equation}
The only new ingredient here is given by the last row, in which the superpotential is written in terms of dual field variables, according to Seiberg duality \cite{SD}.

\subsection{Generalized anomaly equations}

The techniques used so far can be applied in this case as well (see  for instance \cite{CKM} for a detailed calculation). However, we will find more convenient
to use the matrix model superpotential introduced by Dijkgraaf and Vafa. First of all, in order to locate the vacua in the moduli 
space of the $\mathcal{N}=2$ theory we have to determine all the correlators $U_{i}=\frac{1}{i}\langle\Tr\Phi^i\rangle$  and then use the formula given in \cite{CSW} which
relates the SW curve with these quantities:
\begin{equation}\label{curva}
P_{N}(x)=x^{N}e^{-\sum_{i}\frac{U_i}{x^i}}+\Lambda^{2N-N_f}\frac{(x+m)^{N_f}}{x^N}e^{\sum_{i}\frac{U_i}{x^i}}. 
\end{equation}
To extract such information we need to consider the generalised anomaly equations introduced in \cite{FDSW} (their validity has been proven perturbatively
in \cite{FDSW} and nonperturbatively in \cite{NP}). If we consider the following transformations
on the matter fields
$$\begin{aligned}
  \delta\Phi &= \frac{1}{z-\Phi},\\
 \delta\Phi &= \frac{W_{\alpha}W^{\alpha}}{z-\Phi},\\
\delta Q_i &= \frac{1}{z-\Phi}Q_i,
  \end{aligned}
$$
we obtain the Ward identities \cite{SA} (the sum over flavors $i$ is implied)
\begin{equation}\label{angen}
\begin{aligned}
&\left\langle  \Tr\frac{\mu\Phi}{z-\Phi}\right\rangle+\left\langle \sqrt{2}\tilde{Q}^{i}\frac{1}{z-\Phi}Q_{i}\right\rangle =2R(z)T(z),\\
&\left\langle  \Tr\frac{\mu\Phi W_{\alpha}W^{\alpha}}{z-\Phi}\right\rangle = -32\pi^2 R^{2}(z),\\
&\left\langle \tilde{Q}^{i}\frac{\sqrt{2}\Phi+m_i}{z-\Phi}Q_{i}\right\rangle = N_f R(z).
\end{aligned}
\end{equation}
In the previous formula we have used the generating functions of chiral ring correlators:
$$ T(z)=\left\langle\Tr\frac{1}{z-\Phi}\right\rangle,\; R(z)=\frac{-1}{32\pi^2}\left\langle\Tr\frac{W_{\alpha}W^{\alpha}}{z-\Phi}\right\rangle,\; M(z)=\left\langle\tilde{Q}^{i}\frac{1}{z-\Phi}Q_{i}\right\rangle.$$
Once we have determined the above generating functions, we can recover all the correlation functions of operators in the 
chiral ring expanding them about infinity. The $\frac{1}{z^2}$ term of the first equation in (\ref{angen}) and the $\frac{1}{z}$ term of the third
one give the Konishi anomaly, all the others represent various generalizations.

 There is now an important point to consider: the above formulas are valid for $U(N)$ and we have to modify them a little to study the $SU(N)$ theory, since the tracelessness of $\Phi$ is not preserved by
the above listed variations. It is sufficient to modify them by adding a term proportional to the identity such that the condition $\Tr\Phi=0$ is preserved:
$$\begin{aligned}
  \delta\Phi &= \frac{1}{z-\Phi}-\frac{T(z)}{N}I,\\
 \delta\Phi &= \frac{W_{\alpha}W^{\alpha}}{z-\Phi}+\frac{32\pi^2}{N}R(z)I,\\
\delta Q_i &= \frac{1}{z-\Phi}Q_i,
  \end{aligned}
$$
This modification does not affect the anomaly since the identity does not
couple to gluons and the only correction to equations (\ref{angen}) arises due to the presence of the superpotential (the same idea already appeared in \cite{KS},\cite{STA}). 
Making use of the ``modified variations'' we find the Ward identities for the $SU(N)$ theory:
\begin{equation}\label{SUN}
\begin{aligned}
&\left\langle  \Tr\frac{\mu\Phi}{z-\Phi}\right\rangle+\left\langle \sqrt{2}\tilde{Q}^{i}\frac{1}{z-\Phi}Q_{i}\right\rangle - \frac{\sqrt{2}T(z)}{N}\left\langle \tilde{Q}^{i}Q_{i}\right\rangle =2R(z)T(z),\\
&\left\langle  \Tr\frac{\mu\Phi W_{\alpha}W^{\alpha}}{z-\Phi}\right\rangle + \frac{32\sqrt{2}\pi^2}{N}R(z)\left\langle \tilde{Q}^{i}Q_{i}\right\rangle = -32\pi^2 R^{2}(z),\\
&\left\langle \tilde{Q}^{i}\frac{\sqrt{2}\Phi+m_i}{z-\Phi}Q_{i}\right\rangle = N_f R(z).
\end{aligned}
\end{equation}
We can see from the above relations that the anomaly equations for the $SU(N)$ theory with superpotential $\frac{1}{2}\mu\Tr\Phi^2$ are equivalent
to those of the $U(N)$ theory but with a different superpotential ($\frac{1}{2}\mu\Tr\Phi^2-a\mu\Tr\Phi$), where 
\begin{equation}  
a \equiv \frac{\sqrt{2}}{N\mu}\langle\tilde{Q}^{i}Q_{i}\rangle 
\end{equation}   
and $S$ denotes   the gluino condensate. 
Taking this modification into account we obtain from the second equation (see \cite{FDSW}) the relation
\begin{equation}\label{rr}
R(z)=\frac{1}{2}\left(\mu (z-a)-\sqrt{\mu^2(z-a)^2-4S\mu}\right).
\end{equation}
An important point is that  our generating functions are actually defined on a double cover of the z-plane, which we can describe using the matrix 
model curve (a sphere in our case) $$\Sigma : y^2=\mu^{2}(z-a)^2-4\mu S=\mu^2 [(z-a)^2-\tilde{z}],$$ which has a single branch cut. 
The $R(z)$ function actually assumes the above form on the first sheet of this Riemann surface (the one visible semiclassically) 
and for large z it behaves like $R(z)\simeq S/z$, tending to zero for $z\rightarrow\infty$, where the theory can be studied 
semiclassically \cite{FDSW,CSW}. In this limit the cut closes and we recover the invisibility of the second sheet at the 
classical level. On the second sheet the sign of the square root changes and the asymptotic behaviour is $R(z)\simeq W'(z)$. We will exploit this property in section 5.

The $M(z)$ and $T(z)$ generating functions have $N_f$ poles located at the points $z_i\equiv-\frac{m_i}{\sqrt{2}}$ (some on the first and the others on the second sheet of the double cover of the z-plane, depending on the vacuum we are considering) and a pole at infinity. 
Using the above solution for $R(z)$, we can now determine $M(z)$ and $T(z)$ following the derivation in \cite{CSW} (see also eqs. 2.8-2.9 in \cite{UN}). Since our main interest is 
the theory with equal bare masses for the flavors, we will set from now on 
\begin{equation}  z_i=-\frac{m}{\sqrt{2}}   \equiv \eta\;,     \qquad\forall i=1,\dots N_f.  
\end{equation}
 Notice that we are considering the
equal mass limit of the theory with nondegenerate masses
\begin{equation}\label{RTM}
\begin{aligned}
M(z)=&\frac{N_f[R(z)+\frac{\mu}{2}(a-\eta)]+\frac{N_f-2r}{2}\sqrt{\mu^2(a-\eta)^2-4S\mu}}{\sqrt{2}z + m},\\
T(z)=&\frac{\frac{N_f-2r}{2}\sqrt{\mu^2(a-\eta)^2-4S\mu}}{(z+m/\sqrt{2})\sqrt{\mu^2(z-a)^2-4S\mu}} + \frac{N_f/2}{z+m/\sqrt{2}}+\\
&\frac{\mu(N-N_f/2)}{\sqrt{\mu^2(z-a)^2-4S\mu}},
\end{aligned}
\end{equation}
where $r$ is the number of poles located on the first sheet.
We can now explicitly determine the matrix model (Dijkgraaf-Vafa) superpotential \cite{DV,DVI,DVII} (see also \cite{DVIII}) which
encodes the holomorphic data of the theory as in \cite{CSW} (the analogous calculation for the $U(N)$ theory has been done in \cite{UN}). This will be the subject of the next section. 

\subsection{The Dijkgraaf-Vafa superpotential}

The aim of this section is to determine the DV superpotential in the r vacua of the $\mathcal{N}=2$ $SU(N)$ SQCD (for the basic properties of such vacua see e.g.
\cite{APS},\cite{CKM}) softly-broken by a mass term for the adjoint chiral multiplet; such vacua can be described in the "matrix model language'' by putting
 r poles on the first sheet of the double cover of the z-plane and the other $N_f-r$ on the second one (at least for large values of $m$, 
where a semiclassical analysis is reliable). We can follow closely the calculation done in \cite{UN}. The effective superpotential assumes the following form (for quadratic tree level superpotential as in this
case):
\begin{equation}\label{eff}
\begin{aligned}
\W_{DV}=&-\frac{1}{2}N\Pi-\frac{1}{2}\sum_{i,\;r_i=0}\Pi^{0}-\frac{1}{2}\sum_{i,\;r_i=1}\Pi^{1}+\frac{1}{2}(2N-N_f)\W(\Lambda_0)+\\
&\frac{1}{2}\sum_i \W(\eta)-(2N-N_f)\pi iS+S\log\left(\frac{\sqrt{2}^{N_f}\Lambda^{2N-N_f}}{\Lambda_{0}^{2N-N_f}}\right). 
\end{aligned}
\end{equation}
By $\sum_i$ we indicate the sum over flavors and by $\sum_{i,\;r_i=0}$, $\sum_{i,\;r_i=1}$ we mean the sum of the contributions from poles on the second and on the first sheets 
respectively. In the previous formula we have used the notation
\begin{equation}\nonumber
\begin{aligned}
\Pi =& 2\int_{\sqrt{\tilde{z}}+a}^{\Lambda_0}\mu\sqrt{(z-a)^2-4S/\mu}dz=\mu(\Lambda_0-a)^2-2S-2S\log\frac{(\Lambda_0-a)^2 \mu}{S},\\
\Pi^{0} =& -\int_{q}^{\Lambda_0}\mu\sqrt{(z-a)^2-4S/\mu}dz= \frac{-\mu(\Lambda_{0}-a)^{2}}{2}+2S\log(\Lambda_0-a)+\\
& 2S\left[\frac{1}{2}+\frac{\mu(\eta-a)}{4S}\sqrt{(\eta-a)^{2}-\frac{4S}{\mu}}-\log\left(\frac{\eta-a}{2}+\frac{1}{2}\sqrt{(\eta-a)^{2}-\frac{4S}{\mu}}\right)\right],\\
\Pi^{1}=& -\int_{\tilde{q}}^{\Lambda_0}\mu\sqrt{(z-a)^2-4S/\mu}dz= \frac{-\mu(\Lambda_{0}-a)^{2}}{2}+2S\log(\Lambda_0-a)+\\
& 2S\left[\frac{1}{2}-\frac{\mu(\eta-a)}{4S}\sqrt{(\eta-a)^{2}-\frac{4S}{\mu}}-\log\left(\frac{\eta-a}{2}-\frac{1}{2}\sqrt{(\eta-a)^{2}-\frac{4S}{\mu}}\right)\right],\\
\end{aligned}
\end{equation}
where $\Lambda_0$ is a UV cutoff, $q$ and $\tilde{q}$ are the positions of the poles on the first and on the second sheet respectively. Substituting back in (\ref{eff}) we obtain for a r vacuum the result 
\begin{eqnarray}\label{pote}
\W_{DV}&&=S\left[N+\log\left(\frac{\mu^{N}\Lambda^{2N-N_f}}{S^N}\sqrt{2}^{N_f}\right)\right] - N\mu\frac{a^2}{2}\nonumber\\
&&-rS\left[\frac{1}{2}+\frac{\mu\xi}{4S}\sqrt{\xi^2-\frac{4S}{\mu}}-\frac{\mu\xi^2}{4S}-\log\left(\frac{\xi}{2}+\frac{1}{2}\sqrt{\xi^{2}-\frac{4S}{\mu}}\right)\right]\\
&&-(N_{f}-r)S\left[\frac{1}{2}-\frac{\mu\xi}{4S}\sqrt{\xi^2-\frac{4S}{\mu}}-\frac{\mu\xi^2}{4S}-\log\left(\frac{\xi}{2}-\frac{1}{2}\sqrt{\xi^{2}-\frac{4S}{\mu}}\right)\right].\nonumber
\end{eqnarray}
where we have introduced  
\begin{equation}
\xi  \equiv  a-\eta= a+m/\sqrt{2}\;. 
\end{equation}
In the previous formula the parameter $a$ has the meaning of a Lagrange multiplier, so the next step is to set to zero 
\begin{equation}\label{trace}
\frac{\De\W_{DV}}{\De a}=\frac{\mu}{2}\left[(N_{f}-2r)\sqrt{(a-\eta)^{2}-4S/\mu}-(2N-N_f)a-N_{f}\eta\right],
\end{equation}
which enforces $\Tr\Phi=0$, and substitute back in (\ref{pote}). 

The r vacua of our theory are associated to the critical points of the effective superpotential with respect to $S$. From its variation we obtain the equation
\begin{eqnarray}
\frac{\De\W_{DV}}{\De S} &=& r\log\left(\frac{\xi}{2}+\frac{1}{2}\sqrt{\xi^{2}-\frac{4S}{\mu}}\right) + (N_{f}-r)\log\left(\frac{\xi}{2}-\frac{1}{2}\sqrt{\xi^{2}-\frac{4S}{\mu}}\right)\nonumber\\
&& + \log\left(\frac{\mu^{N}\Lambda^{2N-N_f}}{S^N}\sqrt{2}^{N_f}\right)=0\nonumber,
\end{eqnarray}
or (for $r\leq\frac{N_f}{2}$)
\begin{equation}\label{gluino}
(N_{f}-2r)\log\left(\frac{a-\eta}{2}-\frac{1}{2}\sqrt{(a-\eta)^{2}-\frac{4S}{\mu}}\right) + \log\left(\frac{\mu^{N-r}\Lambda^{2N-N_f}}{S^{N-r}}\sqrt{2}^{N_f}\right)=0. 
\end{equation}
From equations (\ref{trace}) and (\ref{gluino}) we can determine $a$ and $S$ and then all the chiral correlators of the theory from (\ref{RTM}) (plugging (\ref{trace}) in
(\ref{RTM}) we can get rid of the square roots in the numerator and rewrite it in the form)
\begin{equation}\label{RTMI}
\begin{aligned}
M(z)=&\frac{\mu Na + N_f R(z)}{\sqrt{2}z + m},\\
T(z)=&\frac{\mu(N-N_f/2)}{\sqrt{\mu^2(z-a)^2-4S\mu}} + \frac{N_f/2}{z+m/\sqrt{2}}-\\
&\frac{\mu Na\left(1-\frac{N_f}{2N}\right)-\frac{N_f}{2\sqrt{2}}\mu m}{(z+m/\sqrt{2})\sqrt{\mu^2(z-a)^2-4S\mu}}.
\end{aligned}
\end{equation}

\subsection{Effective superpotentials and the Konishi anomaly}

The value of the meson and gluino condensates can also be obtained using the technique proposed for $SU(2)$: we extremize (\ref{potM}) and then use the Konishi anomaly equation to determine $S$. Combining the two equations we deduce the relation (no sum over $i$)
\begin{equation}\label{ciao}
S=-M_{i}\frac{\De\W_{NP}}{\De M_i},
\end{equation}
where $M_i$ is a diagonal element of the meson matrix. Using this equation we can write the solution for the meson condensate in the form
$$M_i = \frac{1}{2}\left[\frac{m\mu}{2}+\frac{\Tr M}{N}\pm\sqrt{\left(\frac{m\mu}{2}+\frac{\Tr M}{N}\right)^2-2S\mu}\right],$$
r solutions with the minus sign ($M_{i}^{-}$) and the others with plus ($M_{i}^{+}$) for a r vacuum; the important thing is that it is not necessary to know the precise form of $\W_{NP}$. Notice that taking the sum over $i$ of the above relation we obtain precisely equation (\ref{trace}) since $a=\frac{\sqrt{2}}{N\mu}\Tr M$. We can now use equation (\ref{gluino}) to derive a relation between $S$ and $M$:
\begin{eqnarray}
&&\log\left(\frac{S^N}{\mu^{N}\Lambda^{2N-N_f}\sqrt{2}^{N_f}}\right)= (N_{f}-r)\log\left(\frac{\xi}{2}-\frac{1}{2}\sqrt{\xi^{2}-\frac{4S}{\mu}}\right) + \nonumber\\
&& r\log\left(\frac{\xi}{2}+\frac{1}{2}\sqrt{\xi^{2}-\frac{4S}{\mu}}\right)=N_f\log\frac{S}{\mu} -(N_f -r)\log\left(\frac{\xi}{2}+\frac{1}{2}\sqrt{\xi^{2}-\frac{4S}{\mu}}\right)-\nonumber\\
&& r\log\left(\frac{\xi}{2}-\frac{1}{2}\sqrt{\xi^{2}-\frac{4S}{\mu}}\right)=N_f\log\frac{S}{\mu} -\log\left[ \left(\frac{\sqrt{2}}{\mu}\right)^{N_f}(M_{i}^{-})^{r}(M_{i}^{+})^{N_f -r}\right].\nonumber
\end{eqnarray}
In the last term we recognize the determinant of the meson matrix. We can thus rewrite the previous relation in the form
\begin{equation}\label{olo}
S^{N-N_f}=\frac{\mu^{N}\Lambda^{2N-N_f}}{\deter M}=\frac{\Lambda_{1}^{3N-N_f}}{\deter M}.
\end{equation}
Taking into account (\ref{ciao}) we can deduce from this equation that
\begin{itemize}
\item For $N_f < N$ $\W_{NP}$ is precisely the ADS superpotential given in (\ref{potenziali}).
\item For $N_f = N$ we have the constraint $\deter M=\Lambda_{1}^{2N}$.
\item For $N_f = N+1$ $\W_{NP}$ assumes the form $\deter M/\Lambda_{1}^{2N_f-3}$.
\item For $N_f > N+1$ we find the ``continuation'' of the ADS superpotential (the functional dependence on the fields is the 
same), which can be obtained from the superpotential given in (\ref{potenziali}) by integrating out the dual quarks (this is 
legal when the meson matrix has maximal rank); see \cite{CKM} and especially \cite{EAI,EAII} for a detailed discussion on this point.
\end{itemize}
We see that we can easily recover the nonperturbative part of the superpotential using (\ref{pote}) without having to discuss the various ranges of $N_f$ separately (an analogous result holds for the $U(N)$ theory, as shown in \cite{OO}). The result obtained agrees with (\ref{potenziali}) once the massive fields have been integrated out and the baryons set to zero (we are discussing nonbaryonic vacua).
Such a relation between the DV superpotential and the instanton superpotentials (\ref{potenziali}) for $N_f\leq N$ has been noticed previously in \cite{YD,f=c}.
The derivation there is based on the relation (see e.g. \cite{UN},\cite{CV}) $$\W_{DV\; \text{on-shell}}=\frac{\mu}{2}\langle\Tr\Phi^2\rangle,$$ and on the explicit factorization of the Seiberg-Witten
curve using random matrices \cite{DJ}. Notice that the above relation is easily recovered from what we have done in the previous sections: the r.h.s. can be read off from the $1/z^3$ term of $T(z)$ in (\ref{RTMI}) and is equal to $S(N-\frac{N_f}{2})+\mu ma\frac{N}{2\sqrt{2}}$. The same expression can be obtained from (\ref{pote}), once (\ref{trace}) and (\ref{gluino}) are imposed.

\section{Chiral condensates in the r vacua and pseudo-confining phase}

In this section we use the machinery introduced so far to study the properties of r vacua. Our starting point is the system of equations (\ref{trace}) 
and (\ref{gluino}) and we will concentrate on the even $N_f$ case for simplicity (the other case is similar). The analysis performed so far is valid for 
large $m$, where semiclassical tools are reliable. 

The important point is that we can now let $m$ decrease and follow the r vacua in the nonperturbative region,
comparing with the analysis at small $m$ performed in \cite{CKM} using the SW curve. This process is quite nontrivial and, as we will see, many interesting phenomena emerge. 

We will show that for particular values of $m$ the r vacua merge in a superconformal fixed point and that in many cases for $m$ small enough the r vacua cross the cut 
of the $\mathcal{N}=1$ curve, signaling that a perturbative analysis is no more adequate. This is the basic ingredient which will allow us to understand precisely the correspondence between semiclassical (large $m$) and quantum r vacua.  

\subsection{Coalescence of the r vacua}

Solving (\ref{trace}) and (\ref{gluino}) in the general case is a hard task. However, the equations simplify considerably in the case $r=\frac{N_f}{2}$, leading to the solution
\begin{equation}\label{rmax}
a=\frac{N_f}{2N-N_f}\frac{m}{\sqrt{2}}, \quad S=2^{\frac{N_f}{2N-N_f}}\omega^{k} \mu\Lambda^{2} \quad k=0,\dots,N-\frac{N_f}{2}-1.
\end{equation}
Here $\omega$ is the $(N-\frac{N_f}{2})$-th root of unity, giving the expected number of vacua. Once we have determined these quantities we can calculate all the chiral condensates and determine the position of the vacuum in the $\mathcal{N}=2$ moduli space using equations (\ref{RTM}) and (\ref{curva}). If we now tune appropriately the bare mass of the quarks the poles associated to the matter fields (located at $\eta$)
coalesce with one of the branch points of the $\mathcal{N}=1$ curve. Imposing this condition we find (for $k=0$) 
\begin{equation}\label{mcrit}
m=\pm2^{\frac{6N-2N_f}{4N-2N_f}}\frac{2N-N_f}{2N}\Lambda.
\end{equation}
One can notice that for these particular values of the masses something special happens: the solutions (\ref{rmax}) become solutions of (\ref{trace}) and (\ref{gluino})
for every r! Since the position in the $\mathcal{N}=2$ moduli space is uniquely determined by $a$ and $S$ we find that for every r branch one vacuum coalesces with the
one we have considered so far, giving a superconformal point (in the $\mu=0$ limit) characterized by a higher singularity of the SW curve. This is clear if
we look at the factorization equation \cite{CV}
\begin{equation}\label{fact}
\begin{aligned}
& P_N(z)^2-4\Lambda^{2N-N_f}(z+m)^{N_f}=H^2(z)F(z),\\
& \quad y^2=\mu^2[(z-a)^2-4S/\mu]=Q^2(z)F(z).
\end{aligned}
\end{equation}
Since $S$ is nonzero the $\mathcal{N}=1$ curve is not a square and is divided by $z+m$ if the above condition is satisfied. In a $r=\frac{N_f}{2}$ vacuum $H^2(z)$
contains at least a $(z+m)^{N_f}$ factor so the curve can be rewritten as $(z+m)^{N_f+1}G(z)$. If $N_f=2N-2$ this point coincides with the maximally singular
point of \cite{GST}. As we change the value of $m$ the r vacua separate again. 

\subsection{Transition from the pseudo-confining to the Higgs phase}

Following the discussion in \cite{UN} we will now study the process of passing poles through the cut of the matrix model curve. As pointed out in \cite{CSW}, this fact signals the transition from the Higgs to the pseudo-confining phase. To discuss this issue we set $r=0$, so that equation (\ref{gluino}) becomes
\begin{equation}\label{taglio}
\tilde{\eta}-\tilde{a}=-\omega_{N_f}^{k}\frac{\hat{S}^{\frac{N}{N_f}}}{\sqrt{2}}-\omega_{N_f}^{-k}\hat{S}^{1- \frac{N}{N_f}}\sqrt{2},\quad \hat{S}=\frac{S}{\mu\Lambda^2},\; \tilde{\eta}-\tilde{a}=\frac{(\eta-a)}{\Lambda}.
\end{equation}
On the other hand, from the reduced $\mathcal{N}=1$ curve we deduce that the two branch points linked by the cut are located at $$\tilde{z}=\tilde{a}\pm 2\sqrt{\hat{S}}.$$
These two equations represent the basic ingredient for our analysis (from now on we will omit the tilde and the hat). We consider moving the $N_f$ poles on top of each other 
(since we are considering the equal mass limit) from infinity on the second sheet (since we are considering the $r=0$ case) towards the origin along a line of constant phase $\theta$ on the complex z-plane $$\eta=Re^{i\theta}.$$
With a suitable phase redefinition (also of $\eta$) we can put (\ref{taglio}) in the form $$\eta-a=-\frac{S^{\frac{N}{N_f}}}{\sqrt{2}}-\sqrt{2}S^{1- \frac{N}{N_f}}.$$  

\subsubsection*{$N_f=N$}

Let us start from the simplest case $N_f=N$ in which the previous equation becomes $$a-\eta=\frac{S}{\sqrt{2}}+\sqrt{2},$$ 
and from (\ref{trace}) we obtain the relation $$a=-\frac{S}{\eta}.$$ In order to understand how the cut moves as we change
$\eta$, we can use the above equations to reexpress the position of the branch points as a function of $\eta$. We find
\begin{equation}\label{branch}
z=a\pm 2\sqrt{S}= \sqrt{2}\pm\sqrt{-4\sqrt{2}\eta}.
\end{equation}

From this equation we can see that, as long as we keep the phase of $\eta$ constant, the cut does not rotate and its lenght is proportional to 
$\sqrt{\vert \eta\vert}$. The important point to notice is that the branch cut and the line of constant phase (the dashed line in figure1) always intersect at distance $\sqrt{2}$
from the origin, so the poles can pass through the cut only when $\vert \eta\vert=\sqrt{2}$ (or equivalently when the length of the cut is $4\sqrt{2}$,
as one can easily see from (\ref{branch})). We thus find the following picture: starting from infinity on the second sheet, the poles cross the cut as we reach
$\vert m\vert=2\Lambda$ (notice that this is precisely the value found in (\ref{mcrit}), if we set $N_f=N$) and end up in the first sheet. When we reach the origin (massless case) the cut closes
up. If we start increasing $\vert \eta\vert$, as we pass the critical value seen before, the poles are kicked back to the second sheet (see figure1). We thus learn that
the vacuum is always in the pseudo-confining phase for large values of the mass! 

\begin{figure}
\centering{\includegraphics[width=\textwidth]{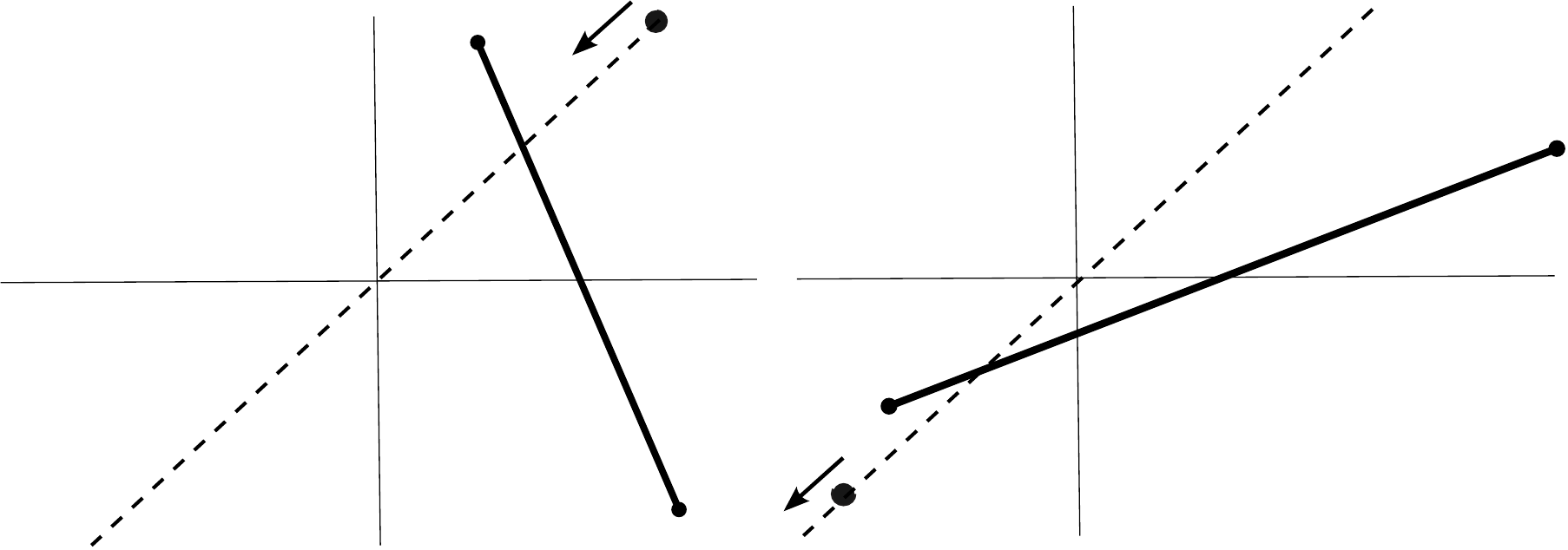}} 
\caption{\emph{(On the left) as the poles located at $\eta$ (represented by the dot on the dashed line) move towards the origin the cut (the thick line in the figure)
shrinks without rotating. (On the right) as the poles move from the origin on the first sheet to infinity they cross the cut (which opens up) and are sent to the second sheet.}}
\end{figure}

\subsubsection*{$N_f\neq N$}

Solving explicitly (\ref{taglio}) in this case is not simple, nonetheless we can deduce the basic features without a detailed calculation. Our expectation
is that for $N_f>N$ it is impossible to cross the cut only once and go to infinity on the first sheet, interpolating from the pseudo-confining to the Higgs
phase (see \cite{CSW} and \cite{UN}); we will now see that this is indeed the case. Let us recall our basic equations: (\ref{gluino})
\begin{equation}\label{asy}
\frac{S^q}{\sqrt{2}}=\frac{a-\eta}{2}\pm\frac{1}{2}\sqrt{(a-\eta)^2-4S},\quad q\equiv\frac{N}{N_f}, 
\end{equation}
where the + or - sign corresponds to a pole on the first or second sheet respectively (as we can read from \ref{gluino}) and (\ref{trace}) 
\begin{equation}
(2q-1)a+\eta=\mp \sqrt{(a-\eta)^2-4S} 
\end{equation}
(- for poles on the first sheet and + for poles on the second one). If we plug the second equation into the first one we obtain
$$\frac{S^q}{\sqrt{2}}=a(1-q)-\eta,$$ and comparing now with the second equation squared we find the relation
$$S^{1-q}=a\frac{q}{\sqrt{2}}.$$ Summing the last two relations we end up with 
$$\frac{S^q}{\sqrt{2}}+\sqrt{2}S^{1-q}=a-\eta.$$ We can deduce now that, for $N\leq N_f<2N$ (or $\frac{1}{2}<q\leq 1$), the asymptotic behaviour for 
large $\vert \eta\vert$ is $$\eta\simeq S^q,\quad a\simeq S^{1-q},$$ and this is incompatible with (\ref{asy}) taken with the + sign ,as we can see 
expanding the square root: the requirement that $R(z)$ vanishes on the first sheet for $z\rightarrow\infty$ (see the discussion after equation (\ref{rr})) leads to the following expansion for large mass 
\begin{equation}\label{conv}
\sqrt{\xi^2-4S}\simeq-\xi+\frac{2S}{\xi}+\dots
\end{equation}

We can conclude that, for $2N>N_f \geq N$, we cannot cross the cut once and go off to infinity on the first sheet. 
Clearly, this argument does not apply in the $N_f<N$ case: from (\ref{taglio}) we cannot conclude that S tends to infinity for large $\vert \eta\vert$ 
and the following asymptotic behaviour is allowed $$S\rightarrow 0, \quad \eta\simeq S^{1-q}.$$ This is compatible with (\ref{asy}), taken with the + sign.

\subsection{Generic r vacua: semiclassical analysis}

The equations for a generic r vacuum are not more difficult than those analized in the previous section so, the methods used there are applicable also
in this more general context. This section is devoted to the analysis of generic r vacua, resulting in a precise understanding of the relation between 
semiclassical and quantum vacua. The first thing one could ask is the following: from the analysis of the equations of motion (reliable in the large $m$
case) we can conclude that $r\leq\min{[N-1,N_f]}$. On the other hand, from what we have done so far, only the bound $r\leq N_f$ seems to be implied. Can
we recover the classical result from the matrix model framework? The answer is positive and we will now show it.

Let us consider a $r\geq N$ vacuum. Equations (\ref{trace}) and (\ref{gluino}) can be recast in the form (written in terms of the dimensionless variables
introduced in (\ref{taglio}))
\begin{equation}\label{key}
\begin{aligned}
&(N_{f}-2r)\sqrt{\xi^{2}-4S}=(2N-N_f)a+N_{f}\eta,\\
&\frac{\xi}{2}+\frac{1}{2}\sqrt{\xi^{2}-4S} = \frac{S^{1+\frac{N-r}{2r-N_f}}}{\sqrt{2}^{\frac{N_f}{2r-N_f}}}. 
\end{aligned} 
\end{equation}
If we now plug the first equation into the second one we obtain
\begin{equation}
\frac{S^{1+\frac{N-r}{2r-N_f}}}{\sqrt{2}^{\frac{3N_f-4r}{2r-N_f}}}=\xi\frac{2r-2N}{2r-N_f}+\eta\frac{2N}{N_f-2r}. 
\end{equation}
Squaring instead the second equation we find
\begin{equation}
\xi=\frac{S^{1+\frac{N-r}{2r-N_f}}}{\sqrt{2}^{\frac{N_f}{2r-N_f}}}+\sqrt{2}^{\frac{N_f}{2r-N_f}} S^{\frac{r-N}{2r-N_f}}. 
\end{equation}
Taking now the limit $\eta\rightarrow\infty$ (large $m$ or semiclassical limit) and recalling that $0\leq\frac{r-N}{2r-N_f}<\frac{1}{2}$ for $N_f<2N$ and 
$r\geq N$, we can deduce from these two equations that $$\xi\rightarrow\infty,\quad\xi\simeq S^{1+\frac{N-r}{2r-N_f}}.$$
As a consequence, we have that the ratio $S/\xi^2$ tends to zero and expanding the square root directly in (\ref{key}) we find, using (\ref{conv}) as before $$\xi\simeq S^{\frac{r-N}{2r-N_f}}.$$
Since we obtain two different asymptotic expansions for $\xi$, we can conclude that equations (\ref{key}) are inconsistent with the asymptotic behaviour of $R(z)$ given before, and we can discard r vacua
with r larger than (or equal to) N. We thus recover the semiclassical result.

We can understand in a similar way the semiclassical behaviour for general r. In the range $N_f-N_c<r<N$ (in the case $N_f>N$), depending on whether r is less or greater than
$\frac{N_f}{2}$, equation (\ref{gluino}) becomes respectively
$$\frac{\xi}{2}-\frac{1}{2}\sqrt{\xi^{2}-4S} = \frac{S^{\frac{N-r}{N_f-2r}}}{\sqrt{2}^{\frac{N_f}{N_f-2r}}},$$ or 
$$\frac{\xi}{2}+\frac{1}{2}\sqrt{\xi^{2}-4S} = \frac{S^{1+\frac{N-r}{2r-N_f}}}{\sqrt{2}^{\frac{N_f}{2r-N_f}}}.$$ We pass from the first to the second set
of vacua crossing the cut of the matrix model curve. In the first case, since $\frac{N-r}{N_f-2r}>\frac{1}{2}$
we have for large $\vert\eta\vert$ the following asymptotic behaviour from (\ref{conv}) $$S\rightarrow\infty,\quad\xi\simeq S^{\frac{N-r}{N_f-2r}}.$$
In the second case instead, from the fact that $\frac{N-r}{2r-N_f}>0$ we find $$S\rightarrow 0,\quad\xi\simeq S^{\frac{r-N}{2r-N_f}}.$$ 

A special role is played by vacua with $r=N_f-N$: in \cite{CKM} it was argued that at the quantum level they are actually part of the baryonic root. The 
argument involves showing that for $m=0$ the SW curve becomes a perfect square, whereas for nonbaryonic vacua it is characterized by two single roots.
We will see in a moment that our formalism allows to recover such a result in a simple way. Equations (\ref{trace}) and (\ref{gluino}) become in this case
\begin{equation}
\begin{aligned}
&\left(\frac{\xi}{2}-\frac{1}{2}\sqrt{\xi^2-4S}\right)^{2N-N_f}=\frac{S^{2N-N_f}}{\sqrt{2}^{N_f}},\\
&\sqrt{\xi^2-4S}=\xi+\frac{2N}{2N-N_f}\eta.
\end{aligned}
\end{equation}
This system can be solved explicitly, leading to  the $2N-N_f$ solutions 
$$S=\sqrt{2}^{\frac{N_f}{2N-N_f}}\frac{N}{2N-N_f}\,\eta\,\omega^k,\quad\xi=-(\sqrt{2}^{\frac{N_f}{2N-N_f}}\omega^k+\frac{N}{2N-N_f}\eta),$$
where $\omega$ is the $2N-N_f$-th root of unity and $k=1,\dots,2N-N_f$ (note that in the special case $N=N_f$ we recover the result of the previous section).

The above solution tells us that $S$ vanishes in the massless case, recovering the result that the SW curve becomes a perfect square in that
limit, as we can see from the factorization equation (\ref{fact}). On the other hand, the vanishing of the gluino condensate signals a singularity in the 
description we are giving, as pointed out in \cite{UN}. This singularity is due to the presence of massless degees of freedom that we are missing; in 
this case they can be identified with the baryons, characteristic of the vacua in the baryonic root. 

The vacua with $r< N_f-N$ have the same asymptotic behaviour as those with $N_f-N<r<\frac{N_f}{2}$. The only difference is that crossing the cut we end 
up with a vacuum characterized by $r>N$, which does not exist semiclassically, as we have seen before. We conclude that it is not possible in this case to cross the cut once and 
then go off to infinity. A special case is given by the $r=0$ vacua discussed in the previous section.
It is anyway worth discussing them because the remaining baryonic vacua fall in this class. If we square equation (\ref{trace})
we find the relation
\begin{equation}\label{first}
S=\left(\frac{N-r}{N_f-2r}\xi+\frac{N}{N_f-2r}\eta\right)\left(\frac{N_f-N-r}{N_f-2r}\xi-\frac{N}{N_f-2r}\eta\right). 
\end{equation}
Plugging this into (\ref{gluino}) (and using (\ref{trace}) again) results in the equation
\begin{equation}
\begin{aligned}
&\sqrt{2}^{N_f}\left(\frac{N_f-N-r}{N_f-2r}\xi-\frac{N}{N_f-2r}\eta\right)^{N_f-2r}=\\
&\left(\frac{N-r}{N_f-2r}\xi+\frac{N}{N_f-2r}\eta\right)^{N-r}\left(\frac{N_f-N-r}{N_f-2r}\xi-\frac{N}{N_f-2r}\eta\right)^{N-r}.\\ 
\end{aligned}
\end{equation}
If $r>N_f-N$, then $N_f-2r<N-r$ and this equation has degree $2N-N_f$, giving the expected degeneracy of nonbaryonic vacua. If instead 
$r<N_f-N$ the previous equation becomes
$$\sqrt{2}^{N_f}\left(\frac{N_f-N-r}{N_f-2r}\xi-\frac{N}{N_f-2r}\eta\right)^{N_f-N-r}=\left(\frac{N-r}{N_f-2r}\xi+\frac{N}{N_f-2r}\eta\right)^{N-r}$$
In this case the degree is $N-r=2N-N_f+(N_f-N-r)$. We thus find more solutions than the $2N-N_f$ associated to nonbaryonic roots;
these are precisely the missing baryonic vacua: in the limit $\eta\rightarrow0$ the above equation gives $N_f-N-r$ zero solutions. On the other
hand, from (\ref{first}) one can easily see that $S$ vanishes as well in this limit and the discussion made for the $r=N_f-N$
applies in this case too. The only difference is that the full $U(N_f)$ flavor symmetry is restored in this class of vacua (the vev
of the meson matrix vanishes in the massless limit).
 
In the theories with $N_f<N$ all the r vacua ($0\leq r\leq N_f$) fall in the class $N_f-N<r<N$ analysed above and also their asymptotic behaviour is the same,
so we do not need to discuss them further.

Let us summarize what we have found in this section:
\begin{itemize}
\item The r vacua exist for $0\leq r\leq\min{[N-1,N_f]}$.
\item For $N_f<N$ there are $2N-N_f$ vacua for every r in the above range.
\item For $r<N_f-N$ (so $N_f>N$) we have $2N-N_f$ nonbaryonic vacua and $N_f-N-r$ baryonic vacua characterized by a restoration 
of the flavor symmetry for every r.
\item For $r=N_f-N$ we have $2N-N_f$ baryonic vacua characterized by dynamical breaking of the flavor symmetry.
\item For $N_f-N<r<N$ we have found $2N-N_f$ nonbaryonic vacua for every r. 
\end{itemize}
In the nonbaryonic vacua the pattern of flavor symmetry breaking is $U(N_f)\rightarrow U(r)\times U(N_f-r)$. If we deform the 
quark masses taking them to be unequal, every r vacuum splits into $\binom{N_f}{r}$ vacua, as shown in \cite{CKM}. 
We thus recover precisely the vacuum counting performed there. 

\subsection{Classical vs quantum $r$ vacua}

Let us now move to the main result of this section. In \cite{CKM,BK} it was noticed that there is a two-to-one correspondence, mapping both r and $N_f-r$ semiclassical vacua to r quantum vacua, which exist only for  $r \le  \tfrac{N_{f}}{2}$.  Making use of the matrix model technique we will be able to understand precisely the origin of this map.

\subsubsection{Interpolating between $r$ and $N_f-r$ classical vacua}

Let us start with a r vacuum with $N_f-N<r\leq\frac{N_f}{2}$ and large $\vert\eta\vert$. If we now let the mass decrease our vacuum enters the nonperturbative region and can cross the cut of the matrix model curve. Depending on whether the vacuum crosses the cut or not, it is characterized
by r or $N_f-r$ poles on the second sheet respectively, and can be described by the (by now familiar) system of equations 
\begin{equation}\label{map}
\begin{aligned}
&\pm(2r-N_{f})\sqrt{\xi^{2}-4S}=(2N-N_f)\xi+2N\eta,\\
&\left(\frac{\xi}{2}\pm\frac{1}{2}\sqrt{\xi^{2}-4S}\right)^{N_f-2r} = \frac{S^{N-r}}{\sqrt{2}^{N_f}}. 
\end{aligned} 
\end{equation}
The sign is plus if it crosses the cut and minus in the other case. Our purpose is now to determine the locus on the mass plane on which our r vacuum crosses the cut. In order to do that we can add to the above system the equation 
\begin{equation}\label{cross}
\eta=a+2t\sqrt{S}\rightarrow\xi^2=4t^2S,\quad t\in[-1,1],
\end{equation}
and try to solve it. If we square the first equation in (\ref{map}), we find the relation 
$$4S=\xi^2-\left(\frac{2N-N_f}{N_f-2r}\xi+\frac{2N}{N_f-2r}\eta\right)^2.$$
Combining this with (\ref{cross}) we can rewrite $\eta$ in terms of $\xi$ and t:
\begin{equation}\label{eta}
\eta=\pm\frac{N_f-2r}{2Nt}\xi\left(\sqrt{t^2-1}\mp t\frac{2N-N_f}{N_f-2r}\right).
\end{equation}
Combining now the second equation in (\ref{map}) with the above relations we find
$$\left(\frac{\xi^2}{4t^2}\right)^{N-r}=\sqrt{2}^{N_f}\left(\frac{\xi}{2}\mp\frac{\sqrt{t^2-1}}{2t}\xi\right)^{N_f-2r}.$$ Notice that $N-r>N_f-2r$ since we are considering the case $r>N_f-N$, so we can simplify the equation and put it in the form (unless $\xi$ vanishes or equivalently $t=0$, but we will see that this is not a problem) $$\xi^{2N-N_f}=\sqrt{2}^{N_f}2^{2N-N_f}t^{2N-N_f}(t\mp\sqrt{t^2-1})^{N_f-2r}.$$
Using now (\ref{eta}) we can finally write the solution to our problem in the form
\begin{equation}\label{fund}
\eta=-\omega^k\frac{2N-N_f}{N}\sqrt{2}^{\frac{N_f}{2N-N_f}}(t\mp q\sqrt{t^2-1})(t\mp\sqrt{t^2-1})^{q},
\end{equation}
where $\omega$ is the $2N-N_f$-th root of unity, $k=1,\dots,2N-N_f$ and with $q$ we have indicated the ratio 
$$q\equiv\frac{N_f-2r}{2N-N_f}.$$ This is positive since $2r<N_f$ and $2N>N_f$ and is smaller than 1 because $r>N_f-N$.

In (\ref{fund}) we have found $4N-2N_f$ solutions: $2N-N_f$ is the number of r vacua and the sign ambiguity doubles that 
quantity. This is due to the fact that crossing the cut the r vacuum becomes a $N_f-r$ vacuum and (\ref{fund}) takes into 
account the contribution from both sets of vacua (notice that a change of sign in (\ref{fund}) can be undone sending 
$q\rightarrow-q$, or equivalently $r\rightarrow N_f-r$). 
\begin{figure}
\centering{\includegraphics[width=.3\textwidth]{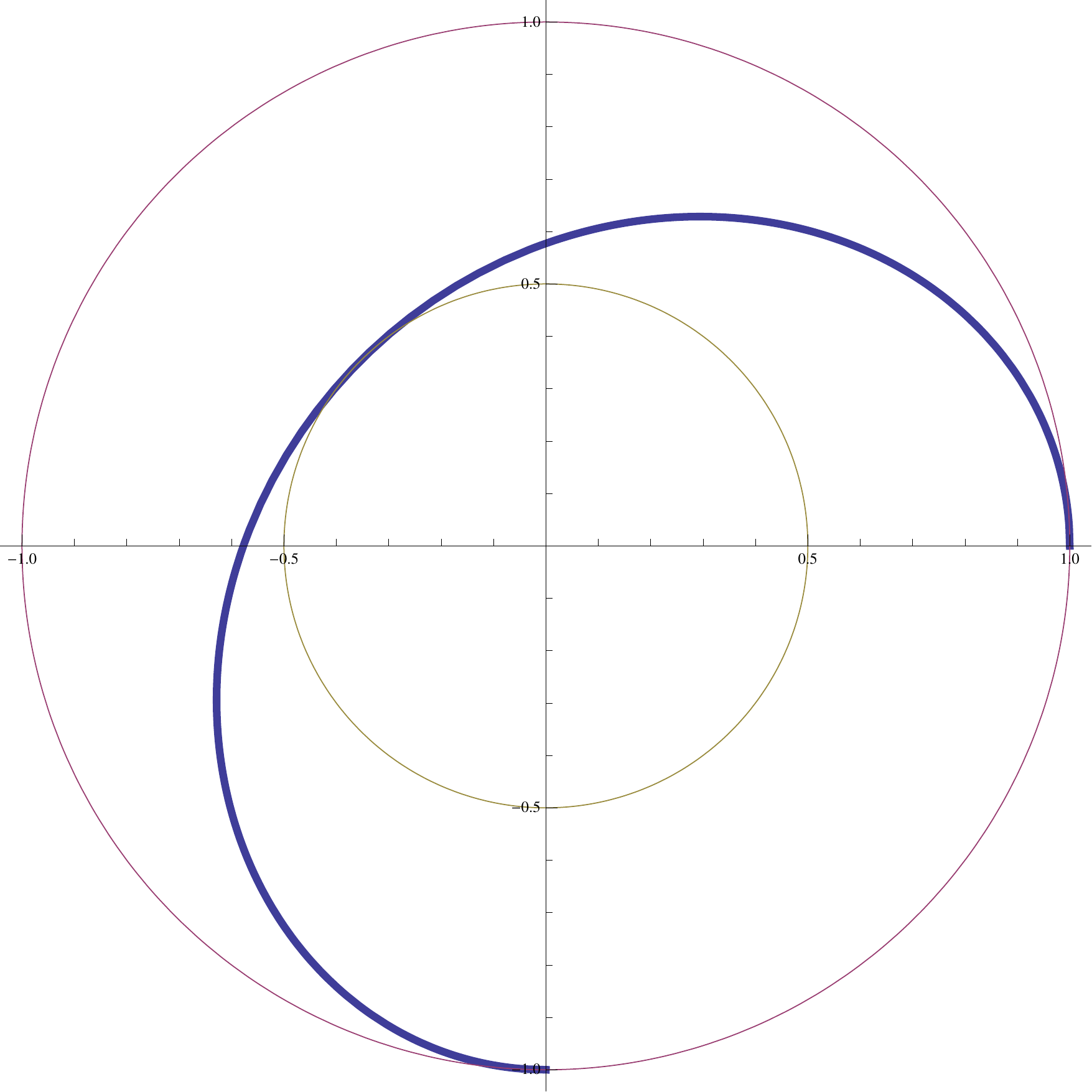}} 
\caption{\emph{We have plotted (\ref{fund}) on the complex plane for $q=\frac{1}{2}$ (the thick blue line). For $t=1$ the phase is 0, for $t=0$ it
is equal to $\frac{3}{4}\pi$ and for $t=-1$ it reaches $\frac{3}{2}\pi$. The line is open (as expected for $q<1$) and the phase of $\eta$ changes by 
$\frac{3}{2}\pi$. $\vert\eta\vert$ attains its minimum for $t=0$ and its maximum for $t=\pm1$.}}
\end{figure}
Focusing now on a specific vacuum (we set $k=0$, the discussion 
is essentially unchanged in the other cases) we can study the locus of points we have just determined: the first property 
is $$\sqrt{2}^{\frac{N_f}{2N-N_f}}\frac{N_f-2r}{2N}\leq\vert\eta\vert\leq\sqrt{2}^{\frac{N_f}{2N-N_f}}{2N-N_f}{N},$$ the 
maximum is attained for $t=\pm1$ and the minimum for $t=0$. This tells us that we can cross the cut and therefore interpolate
 between $r$ and $N_f-r$ vacua only for values of $m$ low enough. Consequently, a semiclassical approach will always suggest 
us that we are dealing with vacua of different kind. Anyway, we are now in the position to compare the semiclassical and 
quantum behaviours: looking at (\ref{fund}) (let's say with the plus sign, the other case can be recovered simply by complex 
coniugation) we can see that, as $t$ goes from 1 to $-1$, the phase of $\eta$ changes by $(1+q)\pi$. The crucial point now 
is that $q<1$ in the range we are discussing, so the curve will not be closed! 

From this analysis we learn that, starting in the semiclassical regime with a r vacuum, depending on how we choose to change 
the value of $\eta$, we can reach the very strongly coupled region $m\simeq0$ either crossing the cut or not; in the second 
case we still have a vacuum with $r$ poles on the first sheet, but in the first one r and $N_f-r$ are interchanged and going 
back to infinity without crossing the cut again we can freely interpolate between the two sets of vacua. Notice that such a
process requires passing in the ``strongly coupled region" of the m-plane, where a fully quantum description is needed. The 
same considerations are also valid in a $r'=N_f-r$ vacuum, apart from the fact that we have to interchange the first case 
with the second one. The result is thus that the $2N-N_f$ solutions of (\ref{map}) 
(considering together the + and - cases) 
can all be obtained starting from a vacuum with $r<\frac{N_f}{2}$ and will coincide with those found starting from a $r'$ 
vacuum. Since $\xi$ and $S$ determine uniquely the location of the vacuum in the $\mathcal{N}=2$ moduli space, we conclude that there 
is no actual distinction between  $r$ and $N_f-r$ vacua at the quantum level (of course when both exist). This nicely 
explains the results found in \cite{CKM,BK} by matching the semi-classical and quantum vacua with the same flavor symmetry 
breaking pattern. 

\subsubsection{Seiberg-Witten curve at classical $r$ vacua}

With the machinery developed in this chapter we can go further and prove that the SW curve factorizes in the same way for $r$ and 
$N_f-r$ classical vacua.

In \cite{APS} it was shown that the vacua which are not lifted 
by the $\mathcal{N}=1$ perturbation $\mu\Tr\Phi^2$ are all the points in the moduli space such that the SW curve factorizes as 
\beq\label{fac} y^2=(x+m)^{2r}Q_{N-r-1}^{2}(x)(x-\alpha)(x-\beta),\quad r\leq\frac{N_f}{2},\eeq and in such a vacuum the effective low energy theory 
includes an abelian $U(1)^{N-r-1}$ sector and a non-abelian one which is an infrared free $U(r)$ gauge theory with $N_f$ 
massless flavors. It is then clear that classical $r$ vacua with $r<N_f/2$ fall in this class. 
In order to determine the form of the SW curve at classical $r$ vacua, our starting point will be  equation (\ref{curva}), relating the SW 
curve and the chiral condensates:
\be P_{N}(z)=z^{N}e^{-\sum_{i}\frac{U_i}{z^i}}|_{+}  +\Lambda^{2N-N_f}\frac{(z+m)^{N_f}}{z^N}e^{\sum_{i}\frac{U_i}{z^i}}|_{+},  \label{SWcurve}  \ee
where $U_i$ are the vacuum expectation values  $U_{i}=\frac{1}{i}\langle\Tr\Phi^i\rangle$ and  the symbol $ ...|_{+} $ indicates that only terms with nonnegative powers of
$z$ are kept ($P_{N}(z)$ is thus a polynomial).    These in turn can be computed from the generalized Konishi anomaly relations found in section 3.3:  
\begin{equation}\begin{aligned} \left\langle\Tr\frac{1}{z-\Phi}\right\rangle=& \sum_{i\geq0}\frac{\langle\Tr\Phi^i\rangle}{z^{i+1}}=
\frac{\frac{N_f-2r}{2}\sqrt{\mu^2(a+m)^2-4S\mu}}{(z+m)\sqrt{\mu^2(z-a)^2-4S\mu}}\\ 
& + \frac{N_f/2}{z+m}+\frac{\mu(N-N_f/2)}{\sqrt{\mu^2(z-a)^2-4S\mu}},   \label{KArel}   \end{aligned}\end{equation}
where $S$ and  $a$ are the gaugino and meson condensates 
$$    S  \equiv  \frac{g^{2}}{32 \pi^{2}} \langle W^{\alpha}  W_{\alpha}  \rangle\;;
\qquad       a \equiv \frac{\sqrt{2}}{N\mu}\langle\tilde{Q}^{i}Q_{i}\rangle\;. $$
These can be determined from the Dijkgraaf-Vafa superpotential. The corresponding equations, determining at once $S$ and $a$ both 
for $r$ and $N_f-r$ classical vacua, are
\footnote{As explained before, the equations obtained by extremizing the DV superpotential can be brought in this form 
only for $r\geq N_f-N$. This will be enough for our purpose of discussing the $r\leftrightarrow N_f-r$ correspondence. 
The equations we have given in the previous sections look  slightly different from the ones given here. This is due to a 
different normalization: if $m'$ and $\Lambda'$ denote the 
parameters used so far the following relations hold: $m=m'/\sqrt{2}$ and $\Lambda^{2N-N_f}=\sqrt{2}^{N_f}(\Lambda')^{2N-N_f}$. 
We changed notation because these are the parameters entering in the SW curve.}
\be\nonumber\left(\frac{(N-r)a}{N_f-2r}-\frac{rm}{N_f-2r}\right)^{N-r}\left(\frac{N_f-N-r}{N_f-2r}a+\frac{(N_f-r)m}{N_f-2r}\right)^{N+r-N_f}=\Lambda^{2N-N_f}\ee
\be\label{eqxS} S=\mu\left(\frac{N-r}{N_f-2r}a-\frac{r}{N_f-2r}m\right)\left(\frac{N_f-N-r}{N_f-2r}a+\frac{N_f-r}{N_f-2r}m\right).\ee
Of the $2N-N_f$ solutions $N-r$ describe classical $r$ vacua and the remaining $N+r-N_f$ correspond to $N_f-r$ vacua. We can 
distinguish the two groups of solutions exploiting the fact that $S$ tends to infinity in the large $m$ limit only for classical $r$ 
vacua with $r<N_f/2$.

Solving these equations in general is very hard and we will not attempt to do this. However, in the massless limit they greatly 
simplify making it possible to check our claim, as we will now see.
In the massless case the equation for $a$ can be easily solved, leading to the $2N-N_f$ solutions $$a=\text{const.}\,  \omega_{2N-N_f}^{k}\Lambda;\quad 
k=1,\dots,2N-N_f,$$ where $\omega_{2N-N_f}$ is the $2N-N_f$-th root of unity.    If we consider two roots $a$ and $a'$ such that 
$a'=\omega_{2N-N_f}^{j}a$ for some integer $j$, we have from (\ref{KArel})
$$ \sum_{i\geq0}\frac{\langle\Tr\Phi^i\rangle(a')}{(z')^{i+1}}=\frac{1}{\omega_{2N-N_f}^{j}}
\sum_{i\geq0}\frac{\langle\Tr\Phi^i\rangle(a)}{z^{i+1}}\Longrightarrow \sum_{i}\frac{U_i(a')}{(z')^i} =  \sum_{i}\frac{U_i(a)}{z^i}\;, 
$$
where we have defined $z'=\omega_{2N-N_f}^{j}z$.  Eq.~(\ref{SWcurve}) tells then that the SW curve factorizes in the 
same way in all $N-r$ vacua of a given $r$, and in all vacua with $r'=N_{f}-r$. Since we know that in all $r$ vacua with 
$r<N_f/2$ the low energy physics can be described as an infrared free SQCD with r colors for any value of $m$ (see the above discussion), 
we know from \cite{APS} that the SW curve factorizes precisely as in (\ref{fac}). This proves our claim.
We also checked the above relations for generic $m$ in the case of $SU(5)$  theory with $N_{f}=6$. We have verified 
by solving (\ref{SWcurve})-(\ref{eqxS}) with Mathematica that both $r=2$ and $r=4$ classical vacua are described by the 
identical singularity of the SW curve
\be   y^{2}=  P_N(z)^2-4\Lambda^{2N-N_f}(z+m)^{N_f} \sim   (z+m)^{4}\;,
\ee
corresponding to the quantum $r=2$ theory of Section~\ref{quantumr}. The expression for $P_{5}(x)$ from equation (\ref{SWcurve}) is very 
lengthy and we will not write it. However, there are two limiting cases worth mentioning: the massless and the semiclassical 
($\Lambda\rightarrow0$) one. In the massless limit we get the following four solutions for $P_{5}(x)$:
\begin{align}
 & x^5-\sqrt{\frac{4}{3}}i\Lambda^2x^3\pm\frac{16}{9}\left(-\frac{1}{3}\right)^{1/4}\Lambda^3x^2,\nonumber\\
 & x^5+\sqrt{\frac{4}{3}}i\Lambda^2x^3\pm\frac{16}{9}\left(-\frac{1}{3}\right)^{1/4}i\Lambda^3x^2.\nonumber
\end{align}
One can easily check that the SW curve satifies the factorization condition (\ref{fac}) with $r=2$. In the $\Lambda\rightarrow0$ limit 
we expect instead to recover the semiclassical result: three $r=2$ classical vacua ($-m$ is a root of $P_{5}(x)$ with multiplicity
two) and one $r=4$ vacuum ($-m$ is a root of $P_{5}(x)$ with multiplicity four). Defining $z=x+m$ we find infact the four solutions
$$\left(z-\frac{5}{3}m\right)^3z^2,\quad\left(z-\frac{5}{3}m\right)^3z^2,\quad\left(z-\frac{5}{3}m\right)^3z^2,\quad (z-5m)z^4.$$
Notice that this result is obtained discarding all subleading terms (higher orders in $\Lambda/m$); in the exact solution all the 
polynomials are divisible just by $z^2$. The point is that the coefficients for the cubic and quadratic terms in $z$ for the fourth 
polynomial are negligible in this limit.

\section{Concluding remarks and discussion}

In the present chapter we have seen how the Dijkgraaf-Vafa superpotential and the generalized Konishi anomaly allow to analyse 
the properties of the vacua of softly-broken $\mathcal{N}=2$ SQCD, even when a semiclassical analysis is not valid. We recover all
the properties derived in previous works and we are able to perform the analysis for generic $m$, interpolating between the quantum and
semiclassical regimes. Some comments are in order regarding the results of section 4.4 and the meaning of the two to one map of
\cite{CKM,BK}. 

As discussed in \cite{BK}, when $r>\frac{N_f}{2}$ the low energy theory is asymptotically free and becomes strongly 
coupled in the infrared; the dynamics in that regime is best described in terms of dual variables. The proposal of \cite{BK} is
that the dual theory has $N_f$ flavors and $N_f-r$ colors (like in Seiberg duality) and is thus infrared free. This is the origin
of the relation between $r$ and $N_f-r$ vacua we have discussed: the infrared dynamics is the same. Notice that this argument requires the presence of $N_f$ 
massless flavors in the low energy theory and this in turn implies the degeneracy of the bare masses for the matter fields. 

In section 3.4 we showed that both $r$ and $N_f-r$ semiclassical vacua correspond to quantum $r$ vacua and proved that the SW curve factorizes in the same way in both cases. This correspondence is related
in our language to the possibility of crossing the cut which is invisible at the classical level. This interpretation is also supported by the
fact that the two sheets are indistinguishable in the small $z$ region, as noticed in \cite{CSW}: for large $z$, where a semiclassical analysis is reliable,
the first sheet is characterized by the vanishing of $R(z)$ and one can characterize each vacuum specifying the number $r$ of poles on the
first sheet. In a neighborhood of the origin instead, where quantum effects cannot be neglected, the only meaningful quantity is the partition $(r,N_f-r)$. However, the relation we found requires
that the flavor symmetry (and thus the degeneracy of the masses) is not modified. One might in fact observe that by passing the
poles through the cut one by one the relation between $r$ and $N_f-r$ vacua does not hold anymore. This is certainly true but does not contradict our result, as we will
argue now. 

Starting from a $r$ vacuum (let us consider for concreteness $r>\frac{N_f}{2}$) at large $m$ (so the vacuum is characterized by r poles on the first sheet), let us change the value of one bare mass (this will be
sufficient to illustrate this point) and tune it to zero without crossing the cut (let us suppose the corresponding pole is located on the first sheet). If we now tune to zero the other $N_f-1$ masses 
crossing the cut (and mantaining their degeneracy) we end up with a vacuum characterized by $r-1$ poles on the second
sheet. If we now let the common mass parameter become large (without crossing the cut) we end up with a $N_f+1-r$ vacuum. The key point
is that we have broken explicitly the flavor symmetry to $SU(N_f-1)$ in the process and the Seiberg-like duality noticed in \cite{BK} relates 
now a theory with $N_f-1$ flavors and $r-1$ colors to another with $N_f-1$ flavors and $N_f-r$ colors. Restoring then the
original flavor symmetry also the gauge symmetry is enhanced (the rank increases by one) and taking the large $m$ limit we recognize the physical properties characterizing a $N_f+1-r$ vacuum.
We thus see that the two descriptions (the one proposed in \cite{BK} and ours) of this process perfectly agree.

Let us notice that all the poles have crossed the cut once apart from the one we ``moved'' at the beginning. The process 
just described is thus equivalent to the following one: we let one pole pass through the cut (keeping the others fixed) and
then go back to its initial position. Then we move all the poles through the cut preserving the full $SU(N_f)$ flavor symmetry.
The first process is precisely the interpolation between Higgs and pseudo-confining phases studied in \cite{CSW} and relates
$r$ with $r-1$ vacua, whereas the second one is the interpolation we described in section 4.4. The mapping arising in this case between
classical and quantum vacua is thus obtained from the one observed in \cite{CKM,BK} and explained in this chapter simply applying
the interpolation process of \cite{CSW} and is perfectly consistent with the picture we are proposing. A more complicated situation in which we move independently several poles can be
interpreted in a similar way. 

A similar issue, namely the relation between classical and quantum vacua in the case $r=N$ for $U(N)$ gauge theories has 
already been studied, using completely different techniques, in \cite{SYI,SYII}.
This case is in a certain sense complementary to our analysis, since in $SU(N)$ theories the $r=N$ vacuum does not exist. This 
vacuum undergoes a crossover transition and the duality
observed in these papers is Seiberg-like as in our case,relating a theory with $N$ colors to a infrared free one with $N_f-N$ colors. This analysis has been extended to other $r$ vacua in \cite{SYIII,SYIV}.

\chapter{Singular points in $\mathcal{N}=2$ SQCD}

\section{Introduction}

This chapter is devoted to the study of singular points (and in particular infrared fixed points) in $\mathcal{N}=2$ SQCD.
It was soon realized that superconformal fixed points are ubiquitous in these models
and their study was initiated in \cite{AD} and \cite{ADSW}. In particular, in the latter reference general properties of conformal theories with
$\mathcal{N}=2$ supersymmetry were derived, such as the fact that mass parameters associated to a non-abelian global symmetry cannot acquire
anomalous dimension. 

A more systematic analysis was initiated in \cite{EH} and especially \cite{EHIY}, in which the authors classified singular
points in $SU(N)$ SQCD with $N_{f}$ flavors. All these papers are based on the idea that the SW curve in a neighbourhood of the singular
point should exhibit scale invariance. Combining this with the requirement that the SW differential has scaling dimension one fixes
the scaling dimensions of all the chiral operators. 

This analysis revealed the existence (for any value of the bare mass $m$ of the flavors) of singular submanifolds in the 
moduli space such that the SW curve factorizes
as $y^2=(x+m)^{2r}Q(x)$ ($2r\leq N_f$). We already came across this formula in chapter 3, while discussing $r$ vacua. We have seen 
that the low energy dynamics is described by a non-abelian $SU(r)$ theory with $N_f$ massless matter fields in the fundamental representation. 
Only in the case $2r=N_f$ we have an interacting fixed point. Tuning $m$ appropriately one can find points in the moduli space where
the curve becomes more singular. In this latter case the approach of \cite{EHIY} leads to anomalous dimensions for the casimirs of the non-abelian flavor group,
contrary to the argument given in \cite{ADSW}. 

More recently the situation has been reanalyzed in \cite{GST}, in which the authors show that this problem can be solved if one
allows for the existence of two scale invariant sectors, weakly coupled by a gauge field. This is analogous to Argyres-Seiberg
duality \cite{AS} (and generalization thereof), apart from the fact that the gauge group appearing in the dual description is infrared free. 
This proposal also removes a possible counterexample to the a-theorem \cite{STI} (see also \cite{KSS,KA}).\\
\indent In \cite{EHIY} the authors analyzed singular points in $USp(2N)$ and $SO(N)$ gauge theories as well. The result of their study is that, as long as
the flavors are massive, the singular points are identical to those of $SU(N)$ SQCD. However, when the masses are set to zero the flavor symmetry enhances
and one finds a different class of fixed points. At that time no tools were available to study them but now, with the techniques of \cite{AS} and
the methods developed by Gaiotto in \cite{G} and Tachikawa in \cite{DT}, the problem can be approached. The scope of this chapter is to make a systematic
analysis of these singular points.\\
\indent The extensive analysis of these models and of their softly broken versions (an incomplete list of references is \cite{APS,CKM,APSII,CKMII})    
revealed that the vacua which are not lifted by the massive perturbation $\mu\Tr\Phi^2$ are generically characterized by a non-abelian gauge symmetry in the infrared. When the flavors are massive the
properties of these vacua are ``universal'' and do not depend on the gauge group (for classical gauge groups).\\ 
\indent In the massless limit this picture does not
change for $SU(N)$ theories, whereas a different phenomenon occurs for $SO$ and $USp$ theories \cite{CKM,CKMII}: only two vacua remain. One is characterized
by the condensation of baryonic-like composite objects and is in a non-abelian Coulomb phase (this vacuum was identified in \cite{APSII}). The second one
(called Chebyshev point in \cite{CKM,CKMII}) arises from the collision of the other vacua and is in general characterized by a strongly interacting low energy theory, which exhibits conformal
invariance in the $\mathcal{N}=2$ limit. In this sense the study of confinement in the softly broken theory and the analysis of singular points
in the parent $\mathcal{N}=2$ theory are linked. As we will see, the study of maximally singular points will allow us to understand the low energy
physics at the Chebyshev point as well. The properties of the theory once the $\mathcal{N}=1$ perturbation has been turned on will be studied in 
chapter 5.\\
\indent This chapter is organized as follows: in section 2 we review the argument given in \cite{GST}, which will be the key ingredient of our analysis, and explain
the properties of vacua relevant for the perturbation to $\mathcal{N}=1$. In section 3 we determine the structure of the maximally
singular point and the Chebyshev point in $USp(2N)$ gauge theory with $2n$ flavors. In section 4 we repeat this analysis for $SO$ gauge theories
and we conclude with a discussion in section 5. As a byproduct we will recover many of the infinite coupling dualities proposed recently.

\section{$SU(N)$ SQCD with $2n$ flavors and r-vacua}

We will now sketch the argument presented in \cite{GST}. 
As is well known, the SW curve and differential for $SU(N)$ SQCD with $N_f=2n$ flavors are
$$y^2=P_{N}^2(x)-4\Lambda^{2N-2n}\prod_{i=1}^{2n}(x+m_i),\quad \lambda=xd\log\frac{P_N-y}{P_N+y},$$ with $P_{N}(x)=x^N-\sum_{k\geq2}u_kx^{N-k}$.
For our purposes, it is convenient to rewrite the curve in the following way (see \cite{GST} for the details):
$$y^2=(x^N-\dots-u_N)(x^N-\dots+(4\Lambda^{N-n}-u_{N-n})x^n-\dots-u_N)-\sum_{k=2}^{2n}c_kx^{2n-k}.$$
Setting all the coulomb branch coordinates $u_i$ and casimirs of the flavor symmetry $c_i$ to zero, we find the maximally singular (EHIY) point, where the SW curve and SW differential become: 
$$y^2=x^{N+n}(x^{N-n}+4\Lambda^{N-n}), \quad \lambda\approx\frac{y}{x^n}dx.$$ 
Requiring that the SW differential has dimension one gives the relation $[y]=1+(n-1)[x]$. If we further impose the scale invariance
of the curve we find the equation $2[y]=(N+n)[x]$. These relations fix the scaling dimensions of $x$ and $y$ and in particular
imply an anomalous dimension for the cubic and higher casimirs of the flavor group ($[c_i]=(2N+i)/(N+1)$). So, when $n$ is at least two, the above analysis
is inconsistent with the general constraints for theories with $\mathcal{N}=2$ superconformal symmetry (e.g. the $c_i$'s should have 
canonical dimension \cite{ADSW}). 

A natural resolution of this inconsistency is to identify subsectors with different scalings of $x$; clearly the $N+n$ colliding branch points will
distribute among the subsectors. The proposal of \cite{GST} is precisely along this line: the authors introduce a particular
scaling limit in which two subsectors emerge: one is a $D_{N-n+2}$ Argyres-Douglas theory \cite{EHIY,CV} (or maximally singular point of $SU(N-n+1)$ gauge
theory with 2 flavors \cite{GST}; see also \cite{BMT,AMT}) and the other can be described as a three punctured 
sphere in the Gaiotto framework \cite{G}, as we will now see.

The first step is to rewrite the curve in a ``6d form'': one defines $t=y/x^{n-1}$ (so the SW differential becomes $\lambda\approx tdx/x$) and writes the curve as
\begin{equation}\label{curvas}
\begin{aligned}
t^2=&(x^{N-n+2}-u_1x^{N-n+1}-\dots-u_{N-n+2}-\dots-\frac{u_N}{x^{n-2}})\\
&\times(x^{N-n}-\dots+(4\Lambda^{N-n}-u_{N-n})-\dots-\frac{u_N}{x^n})-\sum_{k=2}^{2n}c_kx^{2-k}.
\end{aligned}
\end{equation}
Notice the presence of $u_1$, which is a parameter proportional to $\sum_i m_i$ and not a coordinate on the Coulomb branch. This will
be important in later sections.
To account for the presence of two sectors, we now introduce two scales $\epsilon_A,\epsilon_B\ll1$. Notice that with the above rescaling the 
condition $[\lambda]=1$ implies $[t]=1$ in both sectors.
In the A sector ($\vert x\vert\sim\epsilon_A$) we will impose $[x]=1$, in order to satisfy the constraint $[c_k]=k$. This leads to
the relation $c_k\sim O(\epsilon_{A}^{k})$. Consider now the term $4\Lambda^{N-n}x^{N-n+2}$. It is 
clearly negligible for $\vert x\vert\sim\epsilon_A$ (with respect to, e.g. $\sum_{k=2}^{2n}c_kx^{2-k}$), and consequently has to appear in the B sector
($\vert x\vert\sim\epsilon_B$). This implies $t^2\sim\epsilon_{B}^{N-n+2}$, and since $t$ has scaling dimension one in both sectors, we
deduce the relation \begin{equation}\epsilon_{A}^{2}=\epsilon_{B}^{N-n+2}.\end{equation}
Interestingly, the above considerations and the requirement that all the coulomb branch coordinates appear in at least one sector necessarily imply
 $$u_{k}\sim O(\epsilon_{B}^{k}),\quad k=1,\dots,N-n+2;\quad u_{k}\sim O(\epsilon_{A}^{k+n-N})\quad k=N-n+2,\dots,N.$$  

Now it is possible to read from (\ref{curvas}) the curves for the two subsectors just collecting the leading order terms:
\begin{enumerate}
 \item For $\vert x\vert\sim\epsilon_{A}$ we are left with
\begin{equation}
\begin{aligned}
t^2=&-\sum_{k=2}^{2n}c_kx^{2-k}-(u_{N-n+2}+\dots+\frac{u_N}{x^{n-2}})\\
&\times(4\Lambda^{N-n}-\frac{u_{N-n+2}}{x^2}-\dots-\frac{u_N}{x^n}).
\end{aligned} 
\end{equation}
As discussed in \cite{GST}, this is the SW curve (when $n>2$) for the Gaiotto theory obtained compactifying n M5 branes on a sphere with three regular punctures (two are maximal
and one is described by a Young tableau with columns of height $\{n-2,1,1\}$). Its global symmetry group is $SU(2)\times SU(2n)$ and the corresponding
casimirs are $c_i$ and $u_{N-n+2}+c_2/4\Lambda^{N-n}$. 
This is precisely the interacting theory that enters in the S-dual description of $SU(n)$ theory with 2n flavors and its properties have been studied in detail in \cite{CD}.
For $n=3$ this S-duality coincides with Argyres-Seiberg duality and the A sector describes the $E_6$ theory of Minahan and Nemeschansky \cite{MNI}. For $n=2$
the theory becomes free and describes three doublets of $SU(2)$ (the global symmetry is $SU(2)\times SO(6)\simeq SU(4)$).
\item For $\vert x\vert\sim\epsilon_{B}$ we find instead
\begin{equation}
 t^2=4\Lambda^{N-n}(x^{N-n+2}-u_1x^{N-n+1}-\dots-u_{N-n+2})-c_2.
\end{equation}
This is the SW curve for the $D_{N-n+2}$ theory. For $N=n$ this sector is free and describes a doublet of hypermultiplets.
\end{enumerate}
Both theory 1 and theory 2 have an $SU(2)$ flavor symmetry; in our context the diagonal combination has been gauged and the
SW curve for $\epsilon_A<\vert x\vert<\epsilon_B$ describes the tubular region associated to this gauge group 
($t^2=-4\Lambda^{N-n}u_{N-n+2}-c_2$). 

As explained in \cite{GST}, one can now evaluate the beta function for this $SU(2)$ gauge group from the above curve: a closed BPS string
located in the tubular region at constant $\vert x\vert$ describes a W-boson with central charge $a$, whereas a geodesic connecting a branch point at
$\vert x\vert\sim\epsilon_{B}$ and another at $\vert x\vert\sim\epsilon_{A}$ describes a monopole with central charge $a_{D}$
(see \cite{GNM} for a detailed discussion on this point).
\begin{equation*}
\begin{aligned}
 a=&\int_{\vert x\vert=\text{const.}}\lambda=2\pi i\alpha;\quad \alpha^{2}=-4\Lambda^{N-n}u_{N-n+2}-c_2,\\
 a_D=&\int_{\vert x\vert\sim\epsilon_{A}}^{\vert x\vert\sim\epsilon_{B}}\lambda=\alpha\left(\frac{N-n}{N-n+2}\log{\epsilon_{A}}+\text{const.}\right).
\end{aligned} 
\end{equation*}
Using then the relation $\tau=\partial a_{D}/\partial a$ and identifying $\epsilon_{A}$ with the renormalization group scale we obtain
$$\frac{d\tau}{d(\log\epsilon_{A})}=\frac{b_1}{2\pi i}=\frac{1}{2\pi i}\frac{N-n}{N-n+2},$$ where $b_1$ is the one-loop
coefficient of the beta function. We thus learn that this $SU(2)$ group is infrared free. Since the contribution to the beta function from the three punctured
sphere is 3, we can read out the contribution given by the $D_{N-n+2}$ theory: $$b_{D_{N-n+2}}=2\left(1-\frac{1}{N-n+2}\right).$$ Indeed, this matches the
result of \cite{CVC} (the calculation can also be performed using the techniques presented in \cite{ST}).

In view of the breaking to $\mathcal{N}=1$, the relevant points in the moduli space are those such that the SW curve factorizes
as \begin{equation}\label{facto}y^2=(x-\alpha)(x-\beta)Q^2(x).\end{equation} As shown in \cite{APS,CKM}, in the case of equal masses $m_i$, these vacua are labelled by an integer $r$ ($0\leq r\leq N_f/2$),
corresponding to the fact that $Q(x)$ factorizes as $Q(x)=(x+m)^r\tilde{Q}(x)$.When the $\mathcal{N}=1$ perturbation is turned on, for $m\gg\Lambda$ the theory is in the Higgs phase, 
whereas in the limit $m\ll\Lambda$ it becomes confining.  For each value of $r$ there are $2N-N_f$ vacua and the low energy
effective theory is characterized by an abelian sector and a non-abelian one with $U(r)$ gauge group and $N_f$ massless 
matter fields in the fundamental representation. For $r<N_f/2$ the low energy theory is infrared free and admits a lagrangian description.
More interesting is the situation for $r=N_f/2$: in this case the non-abelian sector of the low energy theory is superconformal and
the scaling dimensions of chiral operators can be determined as in \cite{EHIY} (in this case the casimirs have canonical
dimension and there is no need to introduce two sectors).
For generic values of the mass parameter $m$, this is the whole story. However, if one sets $m$ equal to $$m=\omega^{k}_{2N-N_f}\frac{2N-N_f}{N}\Lambda,$$ 
one can show \cite{noi} \footnote{In \cite{noi} the formula for $m$ is different. The discrepancy is simply due to a different 
convention (which is the one adopted in \cite{CSW}): indicating with $\tilde{m}$ and $\tilde{\Lambda}$ the parameters used in those papers, we have 
$m=\tilde{m}/\sqrt{2}$ and $\Lambda^{2N-N_f}=\sqrt{2}^{N_f}\tilde{\Lambda}^{2N-N_f}$.}
that some of the r vacua (one for each value of r) collide and the curve becomes more singular. This signals the transition from Higgs to
confinement phase \cite{noi}. In this limit
$\alpha$ or $\beta$ in equation (\ref{facto}) become equal to $-m$ and the SW curve and differential can be approximated as $$y^2\approx(x+m)^{N_f+1},\quad\lambda\approx\frac{y}{x^n}dx.$$
Here we recognize the EHIY point when $N=n+1$. Indeed, as was argued in \cite{EHIY}, the physics of this singular point is that for the EHIY point of $SU(n+1)$
gauge theory with $2n$ flavors. In this case the B sector is given by the $D_3$ theory.

We thus propose that the low-energy physics at this point is described by:
\begin{itemize}
\item An abelian $U(1)^{N-n-1}$ sector with massless hypermultiplets charged under each $U(1)$ factor.
\item The $D_3$ theory (B sector).
\item The scale invariant theory entering in the S-dual description of $SU(n)$ SQCD with $2n$ flavors (A sector).
\item An infrared free $SU(2)$ gauge multiplet coupled to sectors A and B.
\end{itemize}
This is identical to the proposal made in \cite{GST}, apart from the fact that the abelian sector includes hypermultiplets charged
under the various $U(1)$ factors (one for each $U(1)$). This comes from the requirement that the point we are discussing is not lifted
by the $\mathcal{N}=1$ perturbation \cite{APS,CKM}. This is not the case for the EHIY point discussed in \cite{GST} (apart from
the case $N=n+1$).

\section{$USp(2N)$ SQCD with $2n$ flavors}

Let us turn to $\mathcal{N}=2$ gauge theories with $USp$ gauge group and $N_f$ hypermultiplets in the fundamental representation (we
consider only the equal mass case as before). If the bare mass $m$ for the matter fields is different from zero, the flavour symmetry
is $U(N_f)$ and, as we said in the introduction, one recovers the results found in the previous section; in particular the vacua surviving
the $\mathcal{N}=1$ perturbation have exactly the same structure as the r-vacua of $SU(N)$ SQCD (see \cite{EHIY,ASI,CKM,APSII}) and all the superconformal
points are analogous to those described in the previous section. 

More interesting is the case of massless matter fields: the first main difference is the flavor symmetry, which
is enhanced to $SO(2N_f)$. Moreover, all the r vacua merge into a single superconformal point in this limit \cite{CKM} (we will refer
to it as the Chebyshev point from now on, because its location in the Coulomb branch is determined by Chebyshev polynomials 
\cite{CKM}). As the symmetry enhancement suggests, this fixed point is different from those we have seen so far. 
The purpose of this section is to study the superconformal points of massless $USp(2N)$ SQCD with $N_f=2n$.

The SW curve and SW differential for this model are \cite{ASI}:
\begin{equation}\label{USP}
xy^2=[xP_{N}(x)+2\Lambda^{2N-2n+2}\prod_im_i]^{2}-4\Lambda^{4N-4n+4}\prod_i(x-m_{i}^{2}), 
\end{equation}
\begin{equation}\label{diff}
 \lambda=\frac{\sqrt{x}}{2\pi i}d\log\left(\frac{xP_{N}(x)+2\Lambda^{2N-2n+2}\prod_im_i-\sqrt{x}y}{xP_{N}(x)+2\Lambda^{2N-2n+2}\prod_im_i+\sqrt{x}y}\right).
\end{equation}
where $P_{N}(x)=x^{N}-u_1x^{N-1}-\dots-u_N$ and $m_i$ are the masses for the flavors. Note that in the $SU(N)$ case $u_1$ was a parameter 
while in this case is a coordinate on the Coulomb branch. We can now rewrite the curve as 
$$xy^2=(x^{N+1}-u_1x^{N}-\dots-u_Nx+\tilde{c}_{2n})^{2}-4\Lambda^{4N-4n+4}x^{2n}-\sum_{i}c_{2i}x^{2n-i},$$ where $c_{2i},\tilde{c}_{2n}$
are the $SO(4N)$ casimirs. We can further rewrite it as
$$\begin{aligned}xy^2=&-\sum_{i}c_{2i}x^{2n-i}+(x^{N+1}\dots-u_{N-n+1}x^{n}\dots+\tilde{c}_{2n})\\
&\times(x^{N+1}\dots-(u_{N-n+1}-4\Lambda^{2N-2n+2})x^{n}\dots+\tilde{c}_{2n}),
\end{aligned}$$ where we just redefined $u_{N-n+1}$. If we set to zero all $c_i$ and $u_k$, we find the maximally 
singular point, where the curve and differential become
\begin{equation}\label{scft}
y^2=x^{N+n}(x^{N-n+1}+4\Lambda^{2N-2n+2}),\quad\lambda=\frac{y}{x^{n}}dx. 
\end{equation}
We thus come across the same problem found in the previous section: imposing $[\lambda]=1$ and $[y]=\frac{N+n}{2}[x]$ leads to anomalous
dimensions for the non-abelian casimirs ($c_{2i}=2+(i-1)[x]=2\frac{N-n+1+i}{N-n+2}$). In order to determine the structure of this infrared
fixed point, we can adopt the technique seen before and introduce two different sectors. 

It is now convenient to define $t=y/x^{n-1}$ and rewrite the curve as
\begin{equation}
t^2=(x^{N+2-n}\dots+\frac{\tilde{c}_{2n}}{x^{n-1}})(x^{N+1-n}\dots+\frac{\tilde{c}_{2n}}{x^{n}})-\sum_{i}c_{2i}x^{1-i}. 
\end{equation}
The constraint $[c_{2i}]=2i$ can be satisfied introducing the scale $\epsilon_{A}$ and setting $c_{2i}\sim O(\epsilon_{A}^{2i})$, 
$\vert x\vert\sim\epsilon_{A}^{2}$. This gives $t\sim\epsilon_A$. A second sector emerges as we introduce the scale $\epsilon_{B}$ and set $\vert x\vert\sim\epsilon_{B}^2$. 
The same reasoning adopted in the previous section leads to the relation $t^2\sim x^{N+2-n}$, from which we deduce 
$$\epsilon_{B}^{2N+4-2n}=\epsilon_{A}^2.$$
The Coulomb branch coordinates are then scaled to zero as
$$u_{i}\sim O(\epsilon_{B}^{2i})\; i=1,\dots,N-n+2;\quad u_{N-n+2+i}\sim O(\epsilon_{A}^{2+2i})\; i=0,\dots,n-2.$$ Collecting the leading
terms as before we can now determine the SW curves for the two sectors.

For $\vert x\vert\sim\epsilon_{A}^{2}$ the curve becomes
\begin{equation}\label{ea}\begin{aligned}
t^2=&-\sum_{i}c_{2i}x^{1-i}+\left(u_{N+2-n}+\dots+\frac{\tilde{c}_{2n}}{x^{n-1}}\right)\\
&\times\left(4\Lambda^{2N+2-2n}-\frac{u_{N-n+2}}{x}+\dots+\frac{\tilde{c}_{2n}}{x^{n}}\right). 
\end{aligned}\end{equation}
It has $2n-2$ branch points.

The remaining $N-n+2$ branch points appear in the second sector, for $\vert x\vert\sim\epsilon_{B}^2$. The curve becomes in this case
\begin{equation}\label{eb}
t^2=4\Lambda^{2N+2-2n}(x^{N+2-n}-\dots-u_{N-n+2})-c_2. 
\end{equation}
Let us analyze these two regions:
 
For $\vert x\vert\sim\epsilon_{B}^2$ we recognize the curve we have seen before: this is the SW curve for the $D_{N-n+2}$ theory. The only difference with respect to the
$SU(N)$ case is that, as we noticed before, $u_1$ is a coordinate on the Coulomb branch in the present context. The flavor symmetry
of this theory is thus just $SU(2)$. Two special cases are $N=n$, when the theory becomes free and describes a doublet of $SU(2)$,
and $N=n-1$, when the curve becomes trivial and describes an ``empty'' theory \cite{AS}.

The curve for the region $\vert x\vert\sim\epsilon_{A}^{2}$ is new; it has $SU(2)\times SO(4n)$ flavor symmetry and can be described as the compactification on a three punctured sphere
of the 6d $(2,0)$ $D_n$ theory \cite{DT}, as we will now see, for $n>2$. For $n=2$ it becomes a free theory. 

The SW curve for this class of theories can be written in the form
\begin{equation}\label{sw}
\lambda^{2N}=\sum_{k}\phi_{2k}(z)\lambda^{2N-2k},\quad \lambda=v\frac{dz}{z}. 
\end{equation}
The theory is specified by the singularities on the Riemann surface, which are labelled by Young tableaux (in the case of regular punctures) 
as in the $A_{N}$ case. From the Young tableaux one can read out the pole structure of the various k-differentials and then determine the Coulomb branch coordinates using Riemann-Roch
theorem \footnote{Contrary to the $A_N$ theory, in which this is the general recipe, the $D_N$ theory has a further 
complication: the coefficients one extracts using Riemann-Roch theorem obey in general non-trivial polynomial relations and one must take 
this into account in order to extract the true coordinates on the Coulomb branch (see \cite{CDT} for a detailed analysis of this issue). 
However, this will not be important in the present case.}. The pole structure at each puncture can be determined as follows \cite{DT,CDII,CDT}:
\begin{itemize}
 \item Take the longest even row in the Young tableau which occurs with odd multiplicity (in our case the row of length four) and
remove the last box. Place it at the end of the next available row (such that the result is a Young tableau). Repeat this operation
until it stops (the resulting Young tableau does not contain even rows with odd multiplicity).
\item Number the boxes of the ``corrected'' Young tableau as follows: start with zero in the first box and number the boxes in the
first row with successive integers. When you reach the end of the row, repeat that number in the first box of the following row and continue.
\end{itemize}
The numbers inserted in boxes number $2,4,\dots,2N$ are the orders of the pole of $\phi_{2},\phi_{4},\dots,\phi_{2N}=(\tilde{\phi}_{N})^2$
at the given puncture. As we have seen the algorithm for $A_{N}$ punctures is obtained just neglecting the first step (odd degree
differentials $\phi_{2k+1}$ do not vanish in the $A_{N}$ case, and the corresponding degree of the pole is the integer contained in
boxes number $2k+1$).

\begin{figure}
\centering{\includegraphics[width=.3\textwidth]{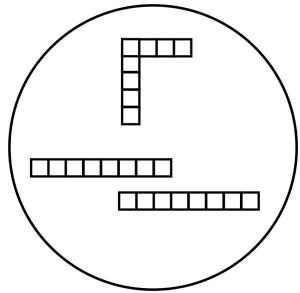}} 
\caption{\label{sdual}\emph{The three punctured sphere that represents the theory entering in the S dual description of $USp(2N)$ with $2N+2$ flavors (in this case $N=3$).}}
\end{figure}

Let us apply the above algorithm to a sphere (depicted in figure\ref{sdual}) with two maximal punctures (labelled by
a Young tableau with a single row of length $2n$) and a third one labelled by a Young tableau (always with $2n$ boxes) with a row of length 4 and the others
of length one (these are all grey punctures in the notation of \cite{DT}). The pole structure at the maximal puncture is $\{1,3,\dots,2n-3;n-1\}$,
whereas the other puncture assigns pole orders $\{1,2,\dots,2;1\}$. The last entry represents the order of the pole of $\tilde{\phi}_{n}$.
The k differentials can thus be written as $$\phi_{2k}=2\frac{u_{2k}z}{(z-1)^2}\left(\frac{dz}{z}\right)^{2k}\; 2k=4,\dots,2n-2;\quad\phi_2=\tilde{\phi}_{n}=0.$$
The SW curve can then be derived just by plugging this result in (\ref{sw}). If we now multiply both sides by $(z-1)^2/v^{2n}$ and
define $y=z-1$, $x=v^2$ we find
$$y^2=2\sum_{k=2}^{n-1}\frac{u_{2k}}{x^k}(y+1)\Longrightarrow \left(y-\sum_{k=2}^{n-1}\frac{u_{2k}}{x^k}\right)^2=\left(\sum_{k=2}^{n-1}\frac{u_{2k}}{x^k}\right)\left(2+\sum_{k=2}^{n-1}\frac{u_{2k}}{x^k}\right).$$
Defining now $(y-\sum_{k=2}^{n-1}\frac{u_{2k}}{x^k})^2=t^2/x^2$ and multiplying both terms by $x^2$ we immediately recognize
(\ref{ea}), with $c_i$ and $u_{N-n+2}$ set to zero. These are the mass parameters associated with the $SO(4n=2N_f)\times SU(2)$
flavor symmetry of the theory. 

Following \cite{GST}, our interpretation is that the infrared physics at the maximally singular point
can be described by the two sectors A ($\vert x\vert\sim\epsilon_A^2$) and B ($\vert x\vert\sim\epsilon_B^2$); both sectors have $SU(2)$ global symmetry and the 
diagonal combination is promoted to a gauge symmetry.

The A sector we have just described (see figure \ref{sdual}) already appeared in \cite{DT} (and for $N=3$ was studied in \cite{CDII}), where it was recognized 
that it enters in the S dual description of the scale invariant $USp(2N)$ theory with $2N+2$ flavors: in the infinite coupling limit the two simple punctures
collide and this three punctured sphere emerges from the collision. Indeed, we can understand this duality using the analysis
of the maximally singular point given above: if we apply the same strategy to the scale invariant case, 
so that the maximally singular point is precisely the origin of the Coulomb branch, we find a B sector which is trivial and thus
the S dual description is given by the A sector, with an $SU(2)$ subgroup of the global symmetry group gauged. The commutant
$SO(2N_f)$ is the flavor group of the original theory. For $N=1$ the theory is $USp(2)\simeq SU(2)$ with 4 flavors, which has $SO(8)$ flavor
symmetry. In this case the A sector becomes free and describes four doublets of $SU(2)$. For $N=2$ the theory is $USp(4)$ with
six fields in the fundamental. This case has been studied by Seiberg and Argyres in \cite{AS}, where it was recognized that the A sector
coincides with the $E_7$ SCFT of Minahan and Nemeschansky \cite{MN} (so in this case there is an enhancement from the naive $SU(2)\times SO(12)$
to $E_7$). Indeed, the scale invariance of the curve requires (for $N=n-1$) that the $SU(2)$ beta function is zero, so the A sector
should give the same contribution as 4 flavors. This obviously works in the $N=1$ case and tells us that the $SU(2)$ central charge
is 8 in the other cases. This is precisely the value found in \cite{CDII} for the puncture with partition $\{2n-3,1,1,1\}$.

We can now determine the beta function of the $SU(2)$ gauge group emerging at the maximally singular point with the same technique adopted in the previous section: 
the curve for $\epsilon_A^2<\vert x\vert<\epsilon_{B}^{2}$ represents a tubular region associated with the $SU(2)$ gauge group. We can 
thus compute $a$ and $a_D$ and then determine the generalized coupling constant $\tau=\partial a_D/\partial a$.
\begin{equation*}
\begin{aligned}
 a=&\int_{\vert x\vert=\text{const.}}\lambda=2\pi i\alpha;\quad \alpha^{2}=-4\Lambda^{2N+2-2n}u_{N-n+2}-c_2,\\
 a_D=&\int_{\vert x\vert\sim\epsilon_{A}^2}^{\vert x\vert\sim\epsilon_{B}^2}\lambda=\alpha\left(\frac{2N-2n+2}{N-n+2}\log{\epsilon_{A}}+\text{const.}\right).
\end{aligned} 
\end{equation*}
Identifying as before $\epsilon_A$ with the energy scale we find
$$b_1=2\left(1-\frac{1}{N-n+2}\right),$$ which is the contribution to the beta function of the $D_{N-n+2}$ theory. Indeed, this is the
expected result, since the contribution from the A sector, as we have just seen, saturates the $SU(2)$ beta function.

The case $n=1$ deserves some comments: the A 
sector becomes trivial and we are left with the $D_{N+1}$ theory, which has $SU(2)$ flavor symmetry and not $SO(4)$! Let us analyze the curve
carefully in this case:
$$xy^2=(x^{N+1}-\dots-u_Nx+\tilde{c}_{2})(x^{N+1}-\dots-(u_N\pm4\Lambda^{2N})x+\tilde{c}_{2})-c_2x-\tilde{c}_2^2.$$
Scaling towards the small $x$ region we find $$y^2=\pm4\Lambda^{2N}(x^{N+1}-\dots-u_Nx)-c_2\pm4\Lambda^{2N}\tilde{c}_2,$$
where the casimir associated to the $SU(2)$ global symmetry is $c_2\pm4\Lambda^{2N}\tilde{c}_2$.
The $\pm$ term reflects the fact that there are two maximally singular vacua, corresponding to $u_N=\pm2\Lambda^{2N}$ (clearly,
this is true also for $n>1$).  The $n=1$ case is special because the two quadratic casimirs of $SO(4)$ enter symmetrically in the scaled
curve and in each one of the two singular points only an $SU(2)$ subgroup acts. Of course, it is well known that this occurs in
the $N=1$ case (i.e. the $USp(2)\simeq SU(2)$ gauge theory with two massless flavors): in this case the two singular points describe
two hypermultiplets which are neutral under an $SU(2)$ subgroup of the $SO(4)$ flavor symmetry group.

We are now in a position to determine the infrared physics at the Chebyshev point. The SW curve and differential are \cite{CKM}
$$y^2\approx x^{2n},\quad \lambda\approx\frac{y}{x^n}dx.$$ This are precisely the curve and differential at the maximally singular
point of $USp(2n)$ theory with $2n$ flavors, in which the B sector describes a doublet of $SU(2)$. We thus propose that the low-energy 
description at the Chebyshev point of $USp(2N)$ theory with $2n$ massless flavors includes:
\begin{itemize}
\item An abelian $U(1)^{N-n}$ sector, with massless particles charged under each $U(1)$ subgroup.
\item The A sector described above, with global symmetry $SU(2)\times SO(4n)$.
\item A third sector consisting of two hypermultiplets, whose symmetry is $SU(2)$. The gauging of the diagonal $SU(2)$ 
couples the last two sectors.
\end{itemize}
Notice that for $n=2$ the A sector becomes free and describes four doublets of $SU(2)$.

\section{$SO(N)$ SQCD}

We can extend this analysis to theories with gauge group $SO(N)$, that exhibit the same phenomena described
in the previous section (coalescence of r vacua and flavor symmetry enhancement in the massless limit). We analyze first theories
with $N$ even and then those with $N$ odd.

\subsection{$SO(2N)$ SQCD with $2n$ flavors}

Let us consider $SO(2N)$ gauge theory with $N_f=2n$ flavors. The theory becomes superconformal for $n=N-1$ and in the massless limit has $USp(4n)$
flavor symmetry. The SW curve and differential are
\begin{equation}\label{SO}
y^2=xP_{N}^{2}(x)-4\Lambda^{4N-4n-4}x^3\prod_i(x-m_{i}^{2}), 
\end{equation}
\begin{equation}\label{dif}
 \lambda=\frac{\sqrt{x}}{2\pi i}d\log\left(\frac{xP_{N}(x)-\sqrt{x}y}{xP_{N}(x)+\sqrt{x}y}\right),
\end{equation}
where $P_N(x)=x^N-\sum_{k=1}^{N-1}u_kx^{N-k}-(u_N)^2$ ($u_1$ is a Coulomb branch coordinate in this case as well). With usual manipulations we can rewrite the curve as
$$\begin{aligned}
y^2=&-\sum_{k=1}^{2n}c_{2k}x^{2n+3-k}+x(x^N-\dots-u_N^2)\\
&\times(x^{N}-\dots+(4\Lambda^{2N-2n-2}-u_{N-n-1})x^{n+1}-\dots-u_N^2).\end{aligned}$$
Here we have simply redefined $u_{N-n-1}+2\Lambda^{2N-2n-2}\rightarrow u_{N-n-1}$. Turning off all the parameters we then get the maximally singular
point: \begin{equation}\label{max}
        y^2=x^{N+n+2}(x^{N-n-1}+4\Lambda^{2N-2n-2}),\quad\lambda=\frac{y}{x^{n+2}}dx.
       \end{equation}
Rescaling the curve as before we obtain ($t=y/x^{n+1}$)
\begin{equation}\label{socurve}
\begin{aligned}
t^2=&-\sum_{k=1}^{2n}c_{2k}x^{1-k}+\left(x^{N-n}-\dots-u_{N-n}-\dots-\frac{u_N^2}{x^{n}}\right)\\
&\times\left(x^{N-n-1}-\dots+(4\Lambda^{2N-2n-2}-u_{N-n-1})-\dots-\frac{u_N^2}{x^{n+1}}\right).
\end{aligned}
\end{equation}
We can now introduce the two sectors imposing $\vert x\vert\sim\epsilon_A^2$ and $\vert x\vert\sim\epsilon_B^2$. Setting $c_{2k}\sim O(\epsilon_{A}^{2k})$ leads to
$t\sim\epsilon_A$ in the A sector and $t\sim\epsilon_{B}^{N-n}$ in the second one, so we deduce $$\epsilon_A=\epsilon_{B}^{N-n}.$$ The same 
argument we gave in sections 2 and 3 then assigns $$u_i\sim O(\epsilon_{B}^{2i})\; i=1,\dots,N-n;\quad u_{N-n+i}\sim O(\epsilon_{A}^{2+2i})\; i<n;\quad u_N\sim O(\epsilon_{A}^{n+1}).$$ 

The SW curve in the B sector is the by now familiar curve for the $D_{N-n}$ theory:
$$t^2=4\Lambda^{2N-2n-2}(x^{N-n}-\dots-u_{N-n})-c_{2}.$$
In the conformal case $N=n+1$ this sector is trivial and describes a doublet of hypermultiplets when $N=n+2$.
The A sector is described by the curve
$$t^2=\left(u_{N-n}-\dots-\frac{u_{N}^{2}}{x^n}\right)\left(4\Lambda^{2N-2n-2}-\frac{u_{N-n}}{x}-\dots-\frac{u_{N}^{2}}{x^{n+1}}\right)-\sum_{k=1}^{2n}c_{2k}x^{1-k}$$ 
This curve has $2n+2$ branch points and contrary to the $USp(2N)$ case the A sector is never free. The global symmetry group is $SU(2)\times USp(4n)$
and, as usual, the $SU(2)$ gauge group is gauged. In the scale invariant case we recover a S-dual description similar to the one for $USp(2N)$: the 
B sector is trivial and we are left with the A sector with an $SU(2)$ subgroup of the flavor symmetry group gauged. Now one can repeat the
calculation of the $SU(2)$ beta function with the same technique adopted in sections 2 and 3; the result is $$b_1=2\left(1-\frac{1}{N-n}\right).$$
This coincides again with the contribution of the $D_{N-n}$ theory, so the contribution from the A sector must saturate the $SU(2)$ beta
function as before.

\begin{figure}
\centering{\includegraphics[width=.3\textwidth]{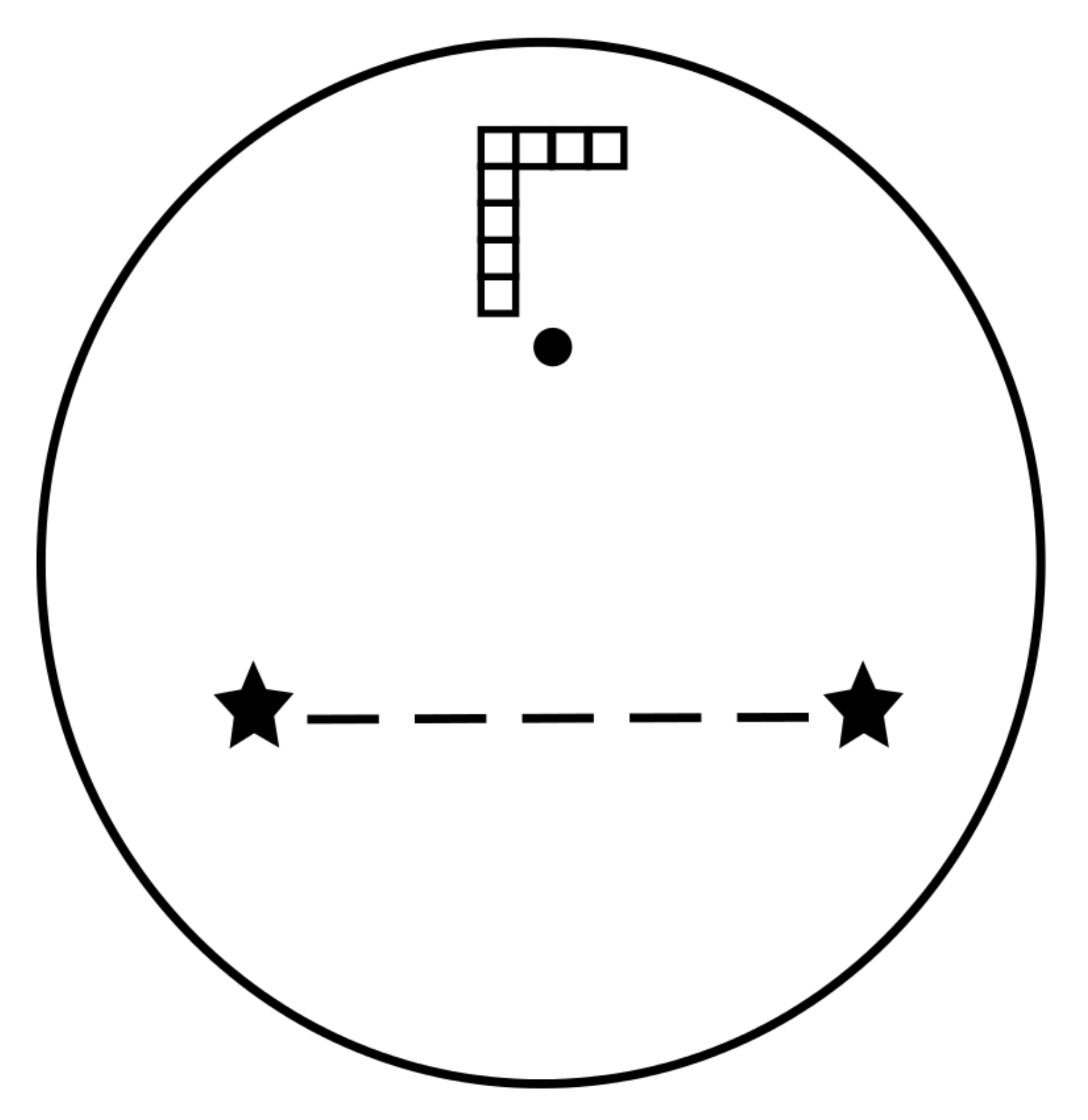}} 
\caption{\label{spdual}\emph{The three punctured sphere associated to the SCFT entering in the S dual description of $SO(2N)$ with $2N-2$ flavors (in this case $N=4$). 
The black dot indicates the D-partition and we have drawn the corresponding Young tableau. The two maximal C-partitions are indicated with $\star$.
To visualize the presence of the twist we draw a dashed line.}}
\end{figure}

Also in this case a description in terms of 6d $(2,0)$ $D_{n+1}$ theory compactified on a three punctured sphere is available (for $n>1$).
The only new ingredient is the presence of black puntures (in the notation of \cite{DT}), or C-partitions in the language of 
\cite{CDII,CDT}. To determine the theory, the simplest way is to notice that the A sector emerges in the dual description (of the 
strong coupling limit) of the scale invariant $SO(2N)$ SQCD. The collision of the simple punctures produces as before the 
D-partition (or grey puncture) described by Young tableau with $2n+2$ boxes, organized in a row of length four and the others of unit length (see figure\ref{spdual}). 
This puncture gives rise to an $SU(2)$ global symmetry group with central charge $k=8$, which is precisely the value needed to saturate the beta function. The 
remaining two punctures are described by a Young tableau with $2n$ boxes and a single row. The pole structure for 
the k-differentials encoded in this puncture has been determined in \cite{DT} and is $\{1,\dots,2n-1;n+1/2\}$ \footnote{The algorithm 
for determining the pole structure for general C-partitions is different with respect to the one described in section 3 and in this paper
we will not need it. The interested reader can find an exhaustive discussion on this point in \cite{CDT}.}. The fractional degree of the pole for 
$\tilde{\phi}_{n+1}$ is not a problem, since we have two such punctures. Turning around one of them we find $\tilde{\phi}_{n+1}\rightarrow-\tilde{\phi}_{n+1}$, 
which is precisely the action of the $\mathbb{Z}_{2}$ outer automorphism of the $D_{n+1}$ Lie algebra \cite{TII} (see figure\ref{spdual}). The k-differentials can thus be written as
$$\phi_{2k}=\frac{u_{2k}z}{(z-1)^2}\left(\frac{dz}{z}\right)^{2k}\; k=2,\dots,n;\quad\tilde{\phi}_{n+1}=\frac{u_{n+1}\sqrt{z}}{z-1}\left(\frac{dz}{z}\right)^{n+1}.$$
Using (\ref{sw}) we find the following SW curve: $$v^{2n+2}=\frac{z}{(z-1)^2}\left(\sum_{k=2}^{n}u_{2k}v^{2n+2-2k}+u_{n+1}^{2}\right).$$ With the same manipulations
described in section 3 we get precisely the curve for the A sector.

The case $n=1$ deserves some comments: the A sector has (at least) $SU(2)\times USp(4)$ global symmetry and turning off all the mass deformations its curve becomes
$$t^2=-\frac{u_{N}^{2}}{x}\left(4\Lambda^{2N-4}-\frac{u_{N}^{2}}{x^{2}}\right).$$ It describes a rank one, scale invariant theory
with a Coulomb branch coordinate of dimension 2. Notice also that the discriminant is proportional to $u_{N}^{6}$. 
The possible scale invariant singularities of rank one curves were classified in \cite{MNI} (see also \cite{AWIII}) and just from these data 
we can associate the above curve with the $D_4$ singularity \footnote{In the present context the A-D-E nomenclature comes from a correspondence between 
the A-D-E affine Lie algebra extended Dynkin diagrams and the pattern of blowups resolving those singularities. 
This should not be confused with the $D_4$ Argyres-Douglas theory we have discussed so far (see e.g. \cite{AWIII}).}.
In \cite{AWII} it was pointed out that this is not enough to identify the A sector: there are different theories associated with 
the same singularity and one should turn on the mass deformations in order to distinguish them. In the case of the $D_4$ singularity 
there are two possibilities: one theory has $SO(8)$ global symmetry and the second one $SU(2)$ (corresponding to $SU(2)$ gauge theory 
with 4 hypermultiplets in the doublet or one in the adjoint). The second possibility can be ruled out since our theory has a larger 
global symmetry group. We can thus identify the A sector with the origin of the Coulomb branch of $SU(2)$ theory with 4 massless flavors.
This theory has central charges $a=23/24$, $c=7/6$ and $SO(8)$ global symmetry with central charge $k_{SO(8)}=4$ (see e.g. \cite{TIII}). 
Here we see once again the phenomenon first described by Argyres and Seiberg in \cite{AS}: $SO(8)$ has a maximal $SU(2)\times USp(4)$ subgroup and by
gauging the $SU(2)$ factor we recover the $USp(4)$ symmetry of the parent gauge theory. 

We can confirm our claim looking at the conformal case, namely $SO(4)$ gauge theory with two flavors: this theory is equivalent to the 
$SU(2)$ gauging of the A sector. We expect the $SU(2)$ beta function to vanish and the value of the central charges $a$ and $c$ 
should match in both descriptions. The $SU(2)$ central charge can be computed
using the formula given in \cite{AS} $$k_{SU(2)}=\mathcal{I}_{SU(2)\hookrightarrow SO(8)}k_{SO(8)},$$ where $\mathcal{I}$ is the 
embedding index introduced in chapter 2. Using for example that the $\bf{8}_{V}$ of $SO(8)$ decomposes as $\bf{8}_{V}=(\bf{3},\bf{1})\oplus(\bf{1},\bf{5})$ under $SU(2)\times USp(4)$,
we obtain \cite{AS} $$\mathcal{I}_{SU(2)\hookrightarrow SO(8)}=\frac{T({\bf3})+5\cdot T({\bf1})}{T(\bf{8}_V)}=2.$$
We thus find that the $SU(2)$ central charge is 8, which is precisely the value needed to saturate the beta function.
Finally, summing the contribution to $a$ and $c$ coming from the $SU(2)$ gauge group and from the A
sector we get precisely the central charges for the $SO(4)$ theory with two flavors, which is nothing but the $SU(2)\times SU(2)$
gauge theory with two hypermultiplets in the $(\bf{2},\bf{2})$.

\subsection{$SO(2N+1)$ SQCD with $2n+1$ flavors}

The above analysis can be repeated for $SO(2N+1)$ gauge theories with odd number of flavors $N_f=2n+1$. The SW curve and differential
are \begin{equation}\label{BN}
y^2=xP_{N}^{2}(x)-4\Lambda^{4N-4n-4}x^2\prod_i(x-m_{i}^{2}), 
\end{equation}
\begin{equation}\label{swdif}
 \lambda=\frac{\sqrt{x}}{2\pi i}d\log\left(\frac{xP_{N}(x)-\sqrt{x}y}{xP_{N}(x)+\sqrt{x}y}\right),
\end{equation}
Redefining $u_{N-n-1}+2\Lambda^{2N-2n-2}\rightarrow u_{N-n-1}$ we can rewrite the curve as
$$\begin{aligned}y^2=&-\sum_{i=1}^{2n+1}c_{2i}x^{2n+3-i}+x(x^{N}-\dots-u_N)\\
&\times(x^{N}-\dots+(4\Lambda^{2N-2n-2}-u_{N-n-1})x^{n+1}-\dots-u_N).\end{aligned}$$
The most singular point can be found setting all $C_{2i}$ and $u_i$ to zero:
$$y^2=x^{N+n+2}(x^{N-n-1}+4\Lambda^{2N-2n-2});\quad\lambda\approx\frac{y}{x^{n+2}}dx.$$
As before, when $N=n+1$ the theory is conformal and this point coincides with the origin of the Coulomb branch. When $N=n+2$ we
recover the Chebyshev point, where the curve degenerates as $y^2\approx x^{N_f+3}$.

We can now set $t=y/x^{n+1}$ and introduce the A and B sectors, in which $\vert x\vert\simeq\epsilon_A^2$
and $\vert x\vert\simeq\epsilon_B^2$ respectively. The same argument given in the previous sections leads us to the relation $\epsilon_A=\epsilon_{B}^{N-n}$
and to the assignment $$u_i\sim O(\epsilon_{B}^{2i})\; i=1,\dots,N-n;\quad u_{N-n+i}\sim O(\epsilon_{A}^{2+2i}).$$ 
The curves describing the theories in the two sectors can now be readily identified: the B sector is the $D_{N-n}$ theory
(when $N=n+2$ it describes a doublet of $SU(2)$ and when $N=n+1$ becomes trivial) and the curve for the A sector is
\begin{equation}\label{eea}\begin{aligned}
t^2=&-\sum_{i=1}^{2n+1}c_{2i}x^{1-i}-\left(u_{N-n}+\dots+\frac{u_N}{x^{n}}\right)\\
&\times\left(4\Lambda^{2N-2-2n}-\frac{u_{N-n}}{x}-\dots-\frac{u_N}{x^{n+1}}\right). 
\end{aligned}\end{equation}
One can see from the above curve that this sector has $SU(2)\times USp(4n+2)$ global symmetry, since the mass parameters precisely 
correspond to the casimirs of this group. To identify the theory, let us start from the $n=1$ case. Turning off
the mass parameters we are left with the curve (setting $4\Lambda^{2N-4}=2$) $$t^2=-\frac{u}{x}\left(2-\frac{u}{x^2}\right).$$ 
To bring it to a more familiar form, it is now convenient to define $y=tx^2$. We then find $$y^2=-ux(2x^2-u);\quad\frac{\partial\lambda}
{\partial u}=\frac{dx}{y}.$$ Making then the change of variables $$y=-\frac{\tilde{y}}{2u};\quad x=-\frac{\tilde{x}}{2u},$$ we recognize the curve
for the $E_7$ SCFT of Minahan and Nemeschansky \cite{MN}: $$\tilde{y}^2=\tilde{x}^3-2u^3\tilde{x};\quad\frac{\partial\lambda}{\partial u}=\frac{d\tilde{x}}{\tilde{y}}.$$ 
The fact that only an $SU(2)\times USp(6)$ subgroup of $E_7$ appears in (\ref{eea}) can be understood in terms of the phenomenon 
mentioned in the previous section \cite{AWII,AW}: there is more than one theory described by the above curve and in order to distinguish them  
one has to turn on the mass deformations. They are characterized by different flavor symmetries and the Minahan-Nemeschansky 
theory is just the one with the largest symmetry group. In \cite{AW} the authors recognized that the $SU(2)\times USp(6)$ theory 
enters in the S-dual description of $SO(5)$ gauge theory with three flavors at the infinite coupling point and called it
submaximal mass deformation of the $E_7$ theory. 

We already encountered the $E_7$ theory in section 3: it is the A sector for $USp(2N)$ SQCD with six flavors. Comparing equations
(\ref{ea}) and (\ref{eea}), we can see that the analogy between the A sectors of these two theories is not limited to this case!
The A sectors of $SO(2N+1)$ SQCD with $N_f=2n+1$ flavors and $USp(2N)$ SQCD with $N_f=2n+4$ flavors are described by the same curve
(once we have set to zero the mass deformations), with Coulomb branch coordinates of the same scaling dimension. However, the 
flavor symmetry groups are different: $USp(4n+2)$ and $SO(4n+8)$ respectively. Here we see some higher rank examples of the 
phenomenon studied in \cite{AWII,AW} (in the language of Argyres and Wittig the first theory represents a submaximal mass 
deformation of the second one). This is not surprising 
since the SW curves and differentials for $SO(2N+1)$ SQCD with $N_f=2N-1$ and $USp(2N)$ SQCD with $N_f=2N+2$ coincide in the
massless case.

\section{Other singular points}

So far we have discussed singular points in $SU(N)$, $USp(2N)$ and $SO(2N)$ SQCD with even number of flavors and $SO(2N+1)$ SQCD 
with odd number of flavors. The most natural question now is: what happens in the other cases? In this section we will try to say something 
about singular points in these models. It is worth anticipating that the two sector structure encountered in the previous sections 
will not appear in this case. We will identify (at least in most cases) the corresponding BPS quivers using the 
technique developed in \cite{CM}. In this paper the authors identify the quiver for an infinite family of SCFTs with ADE flavor 
symmetry. The idea is to start from a 4d theory engineered ``compactifying'' type IIB string theory on the local Calaby-Yau 
hypersurface \begin{equation}\label{CC} W_{G,s}(z,x_1,x_2,x_3)=\Lambda^{b}e^{(s+1)z}+\Lambda^be^{-z}+W_{G}(x_1,x_2,x_3),\end{equation} where $W_{G}$ is the minimal ADE singularity 
of type G. For $s=0$ this gives SYM theory with gauge group G. Using the 2d/4d correspondence one can identify the BPS quiver and then, 
decoupling the G gauge group, one is left with a superconformal theory with (at least) $G$ flavor symmetry which the authors call $D(G,s)$ 
(we will refer to them also as $D_p(G)$ theories, where $p=s+1$). 

In \cite{CMS} we continued the analysis of these models. In particular it was shown that the a and c central charges can be written 
in terms of Lie algebraic invariants. The c central charge is given by the formula 
\be\label{cdp}c=\frac{1}{12}(c_{eff}+\text{rank}B),\ee where $c_{eff}$ coincides with the c central charge of the corresponding 
two dimensional model. For $D_p(G)$ theories it is equal to \be\label{ceff}(p-1)r(G)h(G),\ee where $r(G)$ and $h(G)$ are 
respectively the rank and Coxeter number for $G$. The combination $2a-c$ can be found summing the scaling dimensions of the 
various Coulomb branch operators (see equation (\ref{cariche}) below). These are equal to \be\label{ddp}j+1-\frac{h(G)i}{p},\ee 
where $i=1,\dots,p-1$ and j runs over the set of exponents of $G$ (degree of the Casimirs minus one), subject to the further 
constraint $j>\frac{h(G)i}{p}$. In this section we will argue that some of the models with $s=1$ precisely correspond to the 
singular points we want to study.

\subsubsection{$SU(N)$ SQCD with $N_f$ odd}

Let us start from the $SU(N)$ case. Let us write the SW curve as ($N_f=2n+1$) $$y^2=P_{N}^2(x)-4\Lambda^{2N-2n-1}(x+m)^{2n+1},\quad \lambda=xd{\text log}\left(\frac{P-y}{P+y}\right).$$
The maximally singular point is given choosing $P_N=(x+m)^{n+1}Q(x)$. In the neighbourhood of the singular point the SW curve 
and differential can be approximated as
\begin{equation}\label{SUN1} y^2=x^{2n+1}+(u_{N-n}x^{n}+\dots+u_N)^2,\quad \lambda=\frac{y}{x^{n+1/2}}dx.\end{equation}
This result can be found also sending $\Lambda$ to infinity and scaling accordingly the other parameters.
As long as $N_f<2N-1$, we can find a point of this kind for generic $m$. As noticed in \cite{EHIY},going then to the semiclassical regime ($m\gg\Lambda$) 
we find that $\langle\phi\rangle$ breaks the gauge group to $SU(n+1)$ at the scale $m$. Since we have $2n+1$ flavors this 
theory is still asymptotically free so, this procedure does not tell what is the low-energy theory. 

When $N_f=2N-1$ the formula is slightly different (and we must tune $m$ appropriately):
$$y^2=x^{2N-1}+(u_2x^{N-2}+\dots u_N)^2;\quad\lambda=\frac{y}{x^{N-1/2}}dx.$$
The more compelling reason for splitting the curve in two sectors as we did so far is the constraint on the dimensions of 
the casimirs of the flavor group. This problem does not arise in this case since x has dimension 1. We will thus assume that the 
standard procedure for determining the scaling dimensions of operators is the right option. We will see that this assumption 
passes some nontrivial checks.
For $N_f=3$ we find a familiar theory: it is the $D_4$ Argyres-Douglas theory (one can check this comparing the scaling dimensions 
of the operators), consistently with the enhancement of the flavor symmetry to $SU(3)$ for the $D_4$ theory. For $N_f=5$ we find one 
of the rank two theories studied in \cite{AWV}. The other cases have not been studied so far.

In order to make contact with \cite{CM}, let us start from the $SU(n+1)\times SU(2n+1)$ theory (remember that $N_f=2n+1$) with a 
multiplet in the bifundamental as the only matter field. The SW curve for this model can be written as
$$ \Lambda^bt^{2}+ctP_{n+1}(x)+P_{2n+1}(x)+\frac{\Lambda^b}{t};\quad\lambda_{SW}=\frac{x}{t}dt,$$ where $\Lambda$ is the dynamical 
scale for the $SU(2n+1)$ group and the dynamical scale for the $SU(n+1)$ group ($\Lambda_{SU(n+1)}$) is proportional to $c^{-2}$. To see this drop the term proportional 
to $t^{-1}$ (this is equivalent to turning off the $SU(2n+1)$ gauge coupling). We are then left with $SU(n+1)$ SQCD with $2n+1$ flavors. If we define 
$t'=ct$ the term quadratic in t becomes proportional to $t'^2/c^2$. The coefficient of this term is in turn identifyable with the dynamical scale. 
Sending c to zero thus corresponds to taking the limit $\Lambda_{SU(n+1)}\rightarrow\infty$. We then find precisely the singular point of interest for us.

Sending instead c to zero first we recognize the SW curve for the theory defined by equation (\ref{CC}) in the case $s=1$ (we have set $t=e^z$). Sending 
then $\Lambda$ to zero as explained in \cite{CM} (i.e. dropping the term proportional to $t^{-1}$ as before) we find the $D(SU(2n+1),1)$ theory. The order of the limits is irrelevant and 
both procedures lead to the same geometry so, we identify our singular point with the $D(SU(2n+1),1)$ theory. In the case $n=1$ 
the equivalence with the $D_4$ Argyres-Douglas theory is proved explicitly in \cite{CM}. As a check of our claim we can compute 
the $SU(2n+1)$ flavor central charge (which gives in turn the contribution to the $SU(2n+1)$ beta function of our SCFT) and the 
central charges $(a,c)$. These can be determined using the technique presented in \cite{ST}, once the scaling dimensions of operators 
are known. Reading them from the curve as we described before we find that the contribution to the $SU(2n+1)$ beta function is 
$n+1/2$ and the central charges are $$a=\frac{7n^2+7n}{24};\quad c=\frac{n^2+n}{3}.$$ These values are precisely in agreement 
with those extracted from the BPS quiver (\ref{cdp})-(\ref{ddp}) for the $D(SU(2n+1),1)$ theory.

\subsubsection{The $USp(2N)$ theory with odd number of flavors}

In this section we will concentrate on the massless theory. Also in this case the condition on the casimirs of the flavor group 
is satisfied if we adopt the standard technique to read the dimensions of operators so, we will assume as before that it is correct.

In the case $G=SO(2N)$ equation (\ref{CC}) describes a 4d theory whose SW curve and differential are 
\begin{equation}\label{SAN} \Lambda^bt^{s+1}+\frac{P_{N}(x^2)}{x^2}+\frac{\Lambda^b}{t}=0;\quad\lambda_{SW}=\frac{x}{t}dt,\end{equation}
where $P_{N}(x^2)=x^{2N}+u_2x^{2N-2}+\dots+u_{N}^2$ (see the discussion in \cite{CVC}, section 6.3.1)\footnote{one can identify $X$ in that paper 
with our $x^2$ and $\lambda$ corresponds (modulo a coefficient) to $u_N$.}. For $s=0$ this corresponds to SYM theory with gauge group $SO(2N)$, as we 
remarked above. 

In order to see the connection between these models and the singular points of interest for us let us start from the following model: 
$USp(2k)\times SO(2N)$ gauge theory ($2k=N-1$ if $N$ is odd, $2k=N$ otherwise) with a half-hypermultiplet in the bifundamental and a (massless) 
hypermultiplet in the fundamental of $USp$. The SW curve and differential for this model are \cite{SOP} 
\begin{equation}t^2x^2+ctx^2Q_{k}(x^2)+P_{N}(x^2)+\frac{\Lambda^bx^2}{t}=0;\quad\lambda_{SW}=\frac{x}{t}dt,\end{equation}
where $Q_k$ is a generic monic polynomial of degree k. As in the previous section $\Lambda$ can be identified with the $SO(2N)$ dynamical scale, 
whereas the $USp(2k)$ scale is proportional to $c^{-2}$. If we send $\Lambda$ to zero, thus decoupling the $SO(2N)$ gauge multiplet, we are left with 
$USp(2k)$ SQCD with $N+1$ hypermultiplets in the fundamental. We have been slightly imprecise in the last statement: with the relation given above 
between k and N, when $N$ is even the $USp$ SQCD is asymptotically free and $c^{-2}$ is proportional to the dynamical scale. When N is odd the 
$USp$ theory is scale invariant and c is related to the coupling constant (the flavor symmetry is $SO(2N+2)$ in both cases). If we send c to zero, in 
the first case this corresponds to sending the dynamical scale to infinity (thus scaling towards the singular point); in the second it is equivalent 
to taking the weak-coupling limit for the scale invariant theory. 

\begin{figure}
\centering{\includegraphics[width=\textwidth]{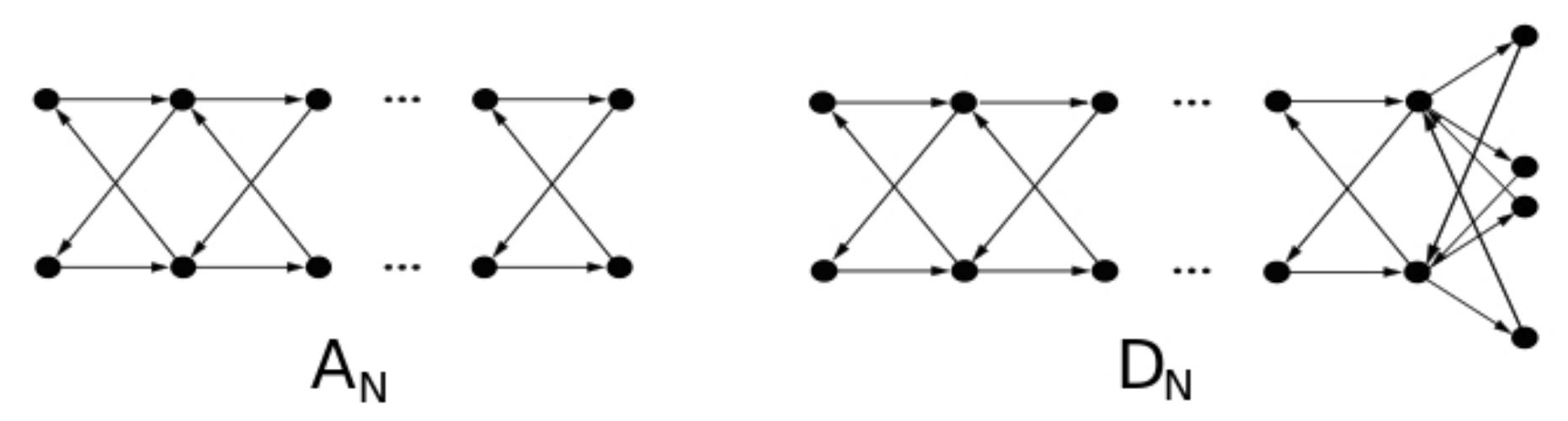}} 
\caption{\label{qqq}\emph{The BPS quivers associated to $D(SU(N+1),1)$ theories (on the left) and to $D(SO(2N),1)$ (on the right). 
Both have $2N$ nodes and describe respectively $SU(\frac{N+1}{2})$ and $USp(N-1)$ SQCD with $N+1$ flavors 
when $N$ is odd. For $N$ even they describe instead the infrared fixed points we have discussed above.}}
\end{figure}

Now we reverse the order of the limits as before. When we set c to zero we recover equation (\ref{SAN}) in the case $s=1$. Turning then off the $SO(2N)$ 
gauge coupling ($\Lambda\rightarrow0$) we are left with the $D(SO(2N),1)$ models of \cite{CM}. For $N$ odd the theory is lagrangian, as we noted above, 
and for N even it describes the maximally singular point of $USp(N)$ SQCD with $N_f=N+1$. We can thus simply read the quiver from \cite{CM}.
Using the technique of \cite{ST} we can easily compute the central charges:
When $N$ is odd we get the scale invariant theory $USp(N-1)$ with $N_f=N+1$, so $$a=\frac{7N^2-5N-2}{48};\quad c=\frac{2N^2-N-1}{12}.$$
For $N$ even we find instead our singular point and the central charges are $$a=\frac{7N^2-5N-10}{48};\quad c=\frac{2N^2-N-2}{12},$$ 
again in agreement with the formulas given above. The BPS quivers associated to $D(SU(N),1)$ and $D(SO(2N),1)$ theories are depicted in Figure \ref{qqq}.

Notice that the above argument allows to extract informations about the maximally singular point of $SO(2n+1)$ SQCD with even number of flavors as well:
it is enough to notice as in the previous section that the SW curves and differentials for $SO(2N+1)$ SQCD with $N_f$ flavors and $USp(2N)$ SQCD 
with $N_f+3$ flavors coincide in the massless case. Notice that in this case the flavor symmetry is ``reduced'' to $USp(2N_f)$.

\subsubsection{The $E_6$ Minahan-Nemeschansky theory}

Notice that $D(SO(8),1)$ coincides with the $E_6$ MN theory: it is a rank one theory with a Coulomb branch coordinate of dimension 
three, the flavor group has rank six and the central charges precisely match those of the $E_6$ theory. The BPS quiver coincides 
as well, as can be easily seen from Figure \ref{qqq}. As a byproduct our observation gives a realization of this theory as an IR fixed point of a lagrangian theory.

In order to see that the SW curves match, we can proceed as follows: in the Gaiotto setting the $E_6$ theory is realized by 
compactifying the $A_2$ six-dimensional theory on a sphere with three maximal punctures (located at let's say $z=0,\lambda,\infty$). 
The SW curve is then $$x^3=-\frac{uz}{(z-\lambda)^2};\quad\lambda_{SW}=\frac{x}{z}dz.$$ If we now take the limit 
$\lambda\rightarrow\infty$, we end up with a two-punctured sphere. The puncture at $z=0$ is unchanged whereas the puncture at 
infinity is now irregular, with a pole of order four for the cubic differential. The curve now becomes $$x^3+uz=0;\quad 
\lambda_{SW}=\frac{x}{z}dz.$$ If we now multiply everything by $x^3$ and set $t=zx^3$ the curve and differential become 
$$x^6+ut=0;\quad\lambda_{SW}=\frac{x}{t}dt.$$ The scaling dimensions of $x$ and $u$ and $t$ are now respectively one, three and 
three. Since both terms appearing in the curve have dimension six we can add a term quadratic in $t$. The complete curve is then 
$$x^6+ut+t^2=0.$$ Setting $u$ to zero we recognize the curve (\ref{SAN}), describing the $D(SO(8),2)$ theory at the conformal point. 

\section{IR fixed points of quiver gauge theories}

In this section we will see that $D(G,s)$ models with $s>1$ correspond to infrared fixed points of linear quiver gauge 
theories (in the cases $G=SU(N),SO(2N)$).

\subsection{$D(SU(N),s)$ theories}

In order to generalize the above argument to the case $s>1$ it is convenient to slightly change perspective as follows:
consider the SW curve for $D(SU(N),1)$ theories $t^2+P_N(x)=0$. The coefficients of the polynomial $P_N$ are the mass parameters 
associated to the $SU(N)$ flavor symmetry of the theory and have canonical dimension, as remarked above. This implies in particular 
that $x$ has scaling dimension one and then, since all the terms appearing in the curve describing a SCFT should have the 
same scaling dimension, we deduce that $t^2$ should have dimension $N$. We can actually deform the above curve adding terms of 
the form $u_ktx^k$. The scaling dimension of the parameters $u_k$ can then be fixed imposing the condition $[u_k]+[t]+k[x]=N$, 
leading to the relation $[u_k]=N/2-k$. If $N$ is even the ``complete'' curve becomes $$t^2+t(x^{N/2}+\dots+u_{0})+P_N(x)=0,$$ 
which is precisely the SW curve for the scale invariant $SU(N/2)$ theory with $N$ flavors. For $N$ odd we get instead 
$$t^2+t(u_{(N-1)/2}x^{(N-1)/2}+\dots+u_0)+P_N(x)=0.$$ Assuming that all the coefficients 
$u_k$ have positive scaling dimension, terms involving higher powers of x necessarily have scaling dimension greater than N and can 
thus be discarded. This curve precisely coincides with (\ref{SUN1}) (modulo a trivial reparametrization) and correctly describes 
the infrared fixed point of $SU(k)$ SQCD ($k\geq\frac{N+1}{2}$) with $N$ flavors studied in the previous section, which coincides 
in turn with $D(SU(N),1)$. 

The above argument can be easily generalized to to the $s>1$ case: the SW curve coming from (\ref{CC}) is $t^p+P_N(x)=0$ ($p=s+1$) 
and we can turn on all possible deformations of the form $u_{ij}t^{p-j}x^{i}$. The only restriction comes from the requirement 
$[u_{ij}]\geq0$. In this way we will relate $D(SU(N),s)$ theories to IR fixed points of linear quivers of unitary gauge groups.
The scaling dimension of $u_{ij}$ can be easily evaluated: $t^p$ has dimension N, so $[t]=\frac{N}{p}$. Imposing 
then the condition $[u_{ij}]+(p-j)[t]+i[x]=N$ we immediately find
\begin{equation}\label{dim}
 [u_{ij}]=\frac{N}{p}j-i.
\end{equation}
We can now readily compute the quantity
$8a-4c$, which gives the effective number of vectormultiplets for the SCFT and can be evaluated using the formula
\begin{equation}\label{cariche}4(2a-c)=\sum_{ij}(2[u_{ij}]-1),\end{equation} where the sum involves all operators whose scaling dimension is larger than one. Using 
(\ref{dim}) this can be rewritten as $$\sum_{j=1}^{p-1}\sum_{i}\left(2\frac{N}{p}j-2i-1\right).$$ We have discarded the contribution 
from $u_{i0}$'s because they are mass parameters associated to the $SU(N)$ symmetry, rather than Coulomb branch operators. Let 
us first evaluate the second summation at fixed j from $i=0$ to $i=\lfloor\frac{N}{p}j\rfloor-1$ (we denote with $\lfloor\;\rfloor$
the integer part): 
$$\sum_{i=0}^{\lfloor\frac{N}{p}j\rfloor-1}\left(2\frac{N}{p}j-2i-1\right)=\left\lfloor\frac{N}{p}j\right\rfloor\left(2\frac{N}{p}j-
\left\lfloor\frac{N}{p}j\right\rfloor\right).$$ Performing now the sum over j does not always lead to the right result for the 
following reason: when $\frac{N}{p}j$ is integer the above summation includes the contribution from a parameter $u_{ij}$ whose 
scaling dimension is one. This will happen whenever N and p are not coprime. 
These should always be regarded as mass parameters, implying the enhancement of the flavor symmetry from the naive $SU(N)$
and their contribution should be discarded in the summation. This can be done simply subtracting $\text{gcd}\{N,p\}-1$ from the 
above formula. The final result is then 
\begin{equation}
8a-4c=\sum_{j=1}^{p-1}\left\lfloor\frac{N}{p}j\right\rfloor\left(2\frac{N}{p}j-
\left\lfloor\frac{N}{p}j\right\rfloor\right)-\text{gcd}\{N,p\}+1. 
\end{equation}
This formula clearly reproduces the expected result for $D(SU(N),s)$ theories, since the scaling dimension of Coulomb branch operators 
are exactly the same (compare (\ref{ddp}) with (\ref{dim})).
We also recover the result found in \cite{CM} that the rank of the flavor symmetry for $D(SU(N),s)$ theories is $N+\text{gcd}\{N,p\}$.

We can now perform a consistency check on the formula for the c central charge derived from the BPS quiver: we have seen that the 
SW curve associated to the $D(SU(N),p-1)$ theory (at the conformal point) and SW differential are $$t^{p}+x^{N}=0;\quad\lambda= 
\frac{x}{t}dt.$$ With the trivial substitution $y=x/t$ the SW differential becomes $ydt$ and we recognize (at least for $p>N$) 
the SW curve for the $(I_{k,N},F)$ theories recently studied by Xie and Zhao in \cite{XIE2}. The parameters $(k,N)$ in that paper correspond to 
$(N,p-N)$ in the present notation. We thus propose to identify $D(SU(k),N+k-1)$ with $(I_{k,N},F)$. This observation gives in 
particular a realization of our models in terms of the 6d $\mathcal{N}=(2,0)$ theory of type $A_{k-1}$ compactified on a sphere 
with two punctures; one irregular of type I (in the language of \cite{XIE2}) and one maximal. It is then natural to propose that 
$D(SO(2N),s)$ and $D(E_N,s)$ theories can be constructed compactifying on a two punctured sphere the 6d theories of type $D_N$ 
and $E_N$ respectively.

The authors of \cite{XIE2} have been able to compute the a and c central charges for $(I_{k,N},F)$ theories when $N$ is a multiple 
of $k$ exploiting the fact that in this case the mirror dual of the $\mathcal{N}=4$ 3d theory obtained compactifying $(I_{k,N},F)$ 
on $S^1$ is lagrangian. By identifying explicitly this theory they can determine the dimension of its Coulomb branch, which in turn 
coincides with the dimension of the Higgs branch of the parent 4d theory. This allows to extract the value of $c-a$ and combining 
this with (\ref{cariche}) one can determine both a and c. They also conjecture a formula for $N$ generic using the results of 
\cite{ST}: in this paper the authors derive a formula (valid for any $\mathcal{N}=2$ SCFT) for a and c which depends on the R-charge 
of the discriminant $\Delta$ of the SW curve (more precisely they consider the R-charge of $B=\Delta^{1/8}$). Using the result 
for a and c derived from the 3d mirror they extract a formula for $R(B)$ and then propose that it is valid for general $k$ and $N$. 
We will now see that our formula for c agrees perfectly with the result obtained using mirror symmetry. At the same time this will 
support our result and confirm the conjecture of \cite{XIE2}. 

Notice first of all that since the SW curve and differential for $D(SU(k),N+k-1)$ and $(I_{k,N},F)$ are the same, the scaling dimension 
of the various operators cannot differ. We thus learn that the rank of the theory (i.e. the dimension of the 
Coulomb branch) and the value of $2a-c$ (see (\ref{cariche})) necessarily coincide as well. It is now convenient to use the 
formula given in \cite{ST} for c: $$c=\frac{1}{3}R(B)+\frac{r}{6},$$ where $r$ is the rank of the theory. This clearly implies that 
a and c coincide for $D(SU(k),N+k-1)$ and $(I_{k,N},F)$, since in principle $R(B)$ can be computed from the curve. However, it is 
hard in general to determine it explicitly as we did in the case $p=2$. In \cite{XIE2} a formula has been proposed using mirror 
duals. We can easily recover this result using the formulas for c and for the Coulomb branch dimension deduced from the BPS 
quiver. Plugging them in the above expression for c we find the answer for generic $p,N$ and any $G=ADE$
\begin{equation}
R(B)=\frac{1}{4}(p-1)r(G)h(G).
\end{equation}
Specializing to the case $G=SU(k)$ (and setting $p=N+k$ as before) we find $$R(B)=\frac{1}{4}k(k-1)(N+k-1),$$ in perfect agreement 
with equation (2.44) of \cite{XIE2}. 

In the above mentioned paper the authors also analyze theories on the sphere with only one irregular 
puncture. In particular they study the so called $(A_N,A_k)$ theories introduced in \cite{CNV} and propose a formula for $R(B)$ also in 
this case. Their argument relies on 3d mirrors as before. Using the formula for the c central charge coming from the BPS quiver 
we are able to confirm their conjecture in this case as well: using as before the relation $$c=\frac{1}{3}R(B)+\frac{r}{6},$$ we find 
$$R(B)=\frac{1}{4}\frac{r(G)r(G')h(G)h(G')}{h(G)+h(G')}$$ for $(G,G')$ theories (see \cite{CNV,CMS}). If we now set $G=A_{N-1}$ and 
$G'=A_{k-1}$ we obtain \begin{equation}R(B)=\frac{N(N-1)k(k-1)}{4(N+k)}.\end{equation} This correctly reproduces equation 
(2.28) of \cite{XIE2}.

I would now like to make some comments about the linear quiver theories associated to $D(SU(N),s)$ SCFTs. As already remarked, 
there is not a unique choice for the rank of the gauge groups and obviously, once a candidate linear quiver has been found, we 
are free to enlarge the rank of the gauge groups since the Coulomb branch of the first theory can be regarded as a submanifold of 
the Coulomb branch of the second. One natural question is then: What is the minimal \footnote{We mean that the sum 
of the ranks of the gauge groups attains the minimum value.} choice? In order to answer this question we must distinguish two cases: 
p greater and smaller than N.

\begin{figure}
\centering{\includegraphics[width=10cm]{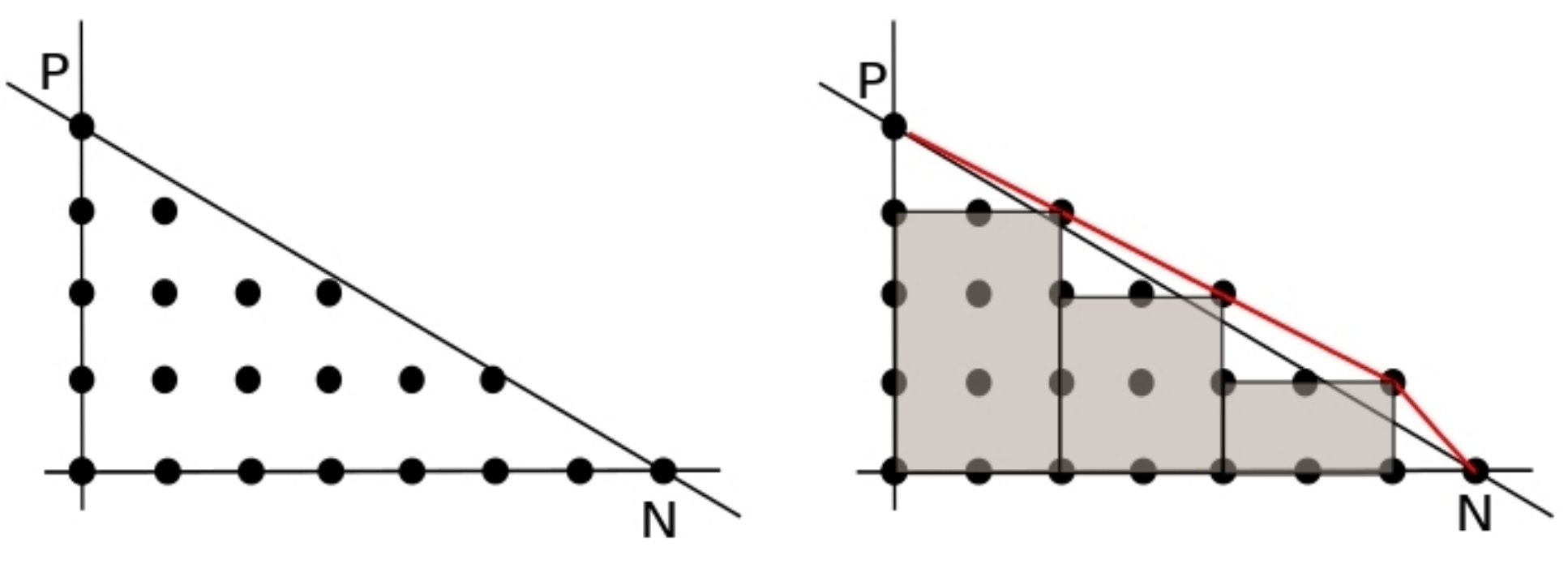}} 
\caption{\label{dotdiagram}\emph{On the left we have the dot diagram associated to $D_{p}(SU(N))$ theory (with $p=4$ and $N=7$). 
The black dots indicate all terms appearing in the SW curve. On the right we have the dot diagram associated to the minimal linear 
quiver having $D_p(SU(N))$ as an infrared fixed point. We can easily see that the linear quiver is $SU(2)-SU(4)-SU(6)-\boxed{7}$. 
Only the $SU(6)$ group is asimptotically free.}}
\end{figure}

For $p<N$ the linear quiver has $p-1$ gauge groups. The first n groups, where $n=N(\text{mod p})$, are 
$$SU\left(\left\lfloor\frac{N+p}{p}\right\rfloor k\right),\quad k\leq n.$$ The remaining $p-n-1$ gauge groups are 
$$SU\left(\left\lfloor\frac{N+p}{p}\right\rfloor n+\left\lfloor\frac{N}{p}\right\rfloor j\right),\quad j\leq p-n-1.$$ 
In order to show this let us draw a diagram on the plane as in Figure \ref{dotdiagram} on the left, in which the term $x^it^j$ 
is represented by a dot located at the point with coordinates $(i,j)$. This is very similar to the Newton polygon used in 
\cite{XIE2}, which is not surprising since we have identified our theories (for $G=A_N$) with the models discussed in that paper. 
It is easy to see that all the terms associated with points lying on the straight line passing through $(N,0)$ and $(0,p)$ have 
dimension $N$ ($i[x]+j[t]=N$), and that the straight lines parallel to it identify lines of constant dimension in the above 
sense. It is then clear that the dots located at points with integer coordinates in the interior of the triangle depicted 
in Figure \ref{dotdiagram} correspond to all the terms entering in the SW curve associated to $D_p(SU(N))$ theory. 

The SW curve of any linear quiver of $SU(N)$ gauge groups has the form $$t^k+\sum_{i,j}u_{ij}t^{k-j}x^{M-i}+x^M=0,$$ and indeed can be 
represented on our diagram. The x-coordinate of the rightmost dot of each row just counts the number of colors of the corresponding 
gauge group. If we connect all rightmost dots as in Figure \ref{dotdiagram} (on the right) we obtain a polygon, and 
the requirement that all the gauge groups in the theory are asymptotically free or conformal is simply equivalent to its 
convexity (just because the number of flavors can be at most twice the number of colors). A given linear quiver will have
$D_p(SU(N))$ as an infrared fixed point only if its associated polygon contains the triangle identified by the straight line 
passing through $(N,0)$ and $(0,p)$ (for our purposes we can assume that $k=p$ and $M=N$) as in the figure. 

The minimal quiver can now be found simply identifying the polygon which satisfies the above requirements and has minimal area. 
In order to see this just consider the grey region in Figure \ref{dotdiagram} on the right. Its area is clearly equal to the rank of the theory 
plus $p-1$ and the area of the whole polygon can be obtained adding the contribution of the various triangles, which is equal 
to $N/2$. It is easy to see that the minimal polygon has exactly four edges (except when $p$ divides $N$) as in the figure and their slope lead to the formula 
given above. Notice that all the groups in the linear quiver but one are conformal. The only asymptotically free group is the one 
associated to the dot at which the two edges meet (see the figure).

If N is a multiple of p the theory is lagrangian and we already know the answer. Notice anyway that our formula works in this case
as well, predicting that the quiver contains the gauge groups $SU(Nk/p)$, with $k\leq p-1$. We thus recover the expected result for 
lagrangian theories. In the limiting case $N=p$, the above rules give a quiver which formally starts with a $SU(1)$ group. The 
corresponding term in the SW curve is $t^{N-1}(x+u_{01})$ and $u_{01}$ has scaling dimension one so, as we have explained above, 
it should be interpreted as a mass parameter. This term in the curve should then be regarded as describing a hypermultiplet in the fundamental 
of the subsequent gauge group, namely $SU(2)$. We thus get the linear quiver $$\boxed{1}-SU(2)-SU(3)-\dots-SU(N-1)-\boxed{N},$$ 
as expected.

For $p>N$ we can just apply a similar argument. In this case it is important to realize that what really matters is that the polygon 
associated to the linear quiver contains all the dots associated to $D_p(SU(N))$. It is not really necessary that it contains 
the whole triangle. In the $p<N$ case these conditions are just equivalent. Since the edges of the polygon are either vertical or 
have slope smaller than one, the minimal polygon is built adding a vertical line and one with slope one.

The minimal linear quiver thus contains a tail which is identical to $D(SU(N),N-1)$ (apart from the doublet of $SU(2)$). The 
remaining part of the quiver is a sequence of $SU(N)$ gauge groups with a hypermultiplet in the bifundamental between neighbouring 
groups: $$SU(2)-SU(3)-\dots-SU(N-1)-SU(N)-\dots-SU(N)-\boxed{N}.$$ The doublet of $SU(2)$ at the beginning is present only if the 
SW curve admits the term $t^{p-k}(x+u_{0k})$, which occurs if and only if p is a multiple of N. In any case the number of gauge 
groups is $p-1-\lfloor p/N\rfloor$. A simple check is in order: for $N=2$ we find a linear quiver of $SU(2)$ groups. For $p=2n-1$ 
there are $n-1$ gauge groups and two doublets at one end. For $p=2n$ the number of gauge groups is the same but we also have an 
extra doublet at the other end of the quiver. Since $D_p(SU(2))$ coincides with the $D_{p}$ AD theory, we precisely 
recover the result found in \cite{XIE}.

\subsection{$D(SO(2N),s)$ theories}

Similar considerations allow to identify $D(SO(2N),s)$ theories with IR fixed points of linear quivers with alternating SO and 
USp gauge groups and half-hypermultiplets in the bifundamental between neighbouring groups. Since the argument is anologous to 
the one given for $SU(N)$, we will be more sketchy.
For $G=D_N$ the curve can be written as $$t^{p}+\frac{P_N(x^2)}{x^2}=0,$$ where $P_N(x^2)=x^{2N}+\dots+u_N^2$. We can then 
add terms of the form $u_{ab}x^{2a}t^{p-b}$. The mass parameters associated to the $SO(2N)$ flavor symmetry will have canonical 
dimension only if $[x]=1$. This implies $$[t]=\frac{2N-2}{p};\quad [u_{ab}]=\frac{2N-2}{p}b-2a.$$ This matches precisely the 
dimension of Coulomb branch operators of the $D(SO(2N),p-1)$ theories (see (\ref{ddp})). Equation (\ref{cariche}) then implies that the above curve reproduces 
the correct effective number of vectormultiplets $2a-c$.

The theory can be lagrangian only if all $u_{ab}$'s have even dimension. This constraint will be satisfied whenever
$(2N-2)/p$ is even, reproducing the expected result. We can determine precisely what the theory is simply collecting all terms 
with $[u_{ab}]=0$. All parameters satisfying this relation are simply combinations of the marginal couplings, rather than operators.
We then find the curve \begin{equation}\label{soso}t^{p}+\sum_{k=1}^{p-1}a_kx^{km}t^{(p-k)}+x^{2N-2}=0,\end{equation} where $pm=2N-2$. In order to make contact with 
\cite{SOP}, it is convenient to multiply everything by $x^2$:
$$x^2t^{p}+\sum_{k=1}^{p-1}a_kx^{km+2}t^{p-k}+x^{2N}=0.$$ This is precisely the curve for a linear 
quiver of alternating SO/USp gauge groups with half-hypermultiplets in the bifundamental. A term $x^{2n}$ corresponds either to 
$USp(2n-2)$ or to $SO(2n)$. If the quiver terminates with a USp gauge group (p even), the term $x^2t^{p}$ indicates the presence 
of one hypermultiplet in the fundamental of the last USp group. If it terminates instead with a SO gauge group (p odd), we only have 
the bifundamentals. We thus find the lagrangian theories (the number inside the boxes indicate the hypermultiplets in the fundamental)
\begin{equation}\begin{aligned}
{\bf p}\; \bf\text{{\bf even}}:\quad &{\boxed N}- USp(2N-2-m) - SO(2N-2m) -\dots-USp(m)-{\boxed 1},\nonumber\\
{\bf p}\; \bf\text{{\bf odd}}: \quad &{\boxed N}- USp(2N-2-m) - SO(2N-2m) -\dots-SO(m+2).\nonumber
\end{aligned}\end{equation}
Since one needs  respectively $2n+2$ or $2n-2$ hypermultiplets in the {\bf 2n} to make a $USp(2n)$ or $SO(2n)$ gauge group conformal, 
it can be easily checked that each node in the above quivers is conformal. These are precisely the lagrangian theories we were 
looking for. Indeed, the number of nodes in the BPS quiver describing these models and their contribution to the $SO(2N)$ beta 
functions match precisely the results found in \cite{CM}.

In order to identify the minimal quivers containing $D(SO(2N),s)$ as an infrared fixed point, it is convenient to go back to formula 
(\ref{soso}). With this normalization the term $x^{2n}$ corresponds either to $USp(2n)$ or to $SO(2n+2)$. We can construct the dot diagram as before, 
inserting for example a dot at the point $(i,j)$ if the curve includes the term $u_{ij}t^jx^{2i}$. The vertices of the 
triangle will then be located at $(0,p)$, $(N-1,0)$ (and obviously $(0,0)$). 
It is easy to see that the condition for UV completeness (as opposed to IR freedom) at all the nodes is again that the associated polygon should be convex.
All the arguments given for $SU(N)$ apply also in this case, so we give directly the result.

For $p<N$ the dots lying on the perimeter of the polygon are associated to terms of the form $t^{p-i}x^{k}$ where k is 
$$2i\left(\left\lfloor\frac{N-1}{p}\right\rfloor+1\right),$$ for $i\leq (N-1)\text{mod} p$ and $$2i\left\lfloor\frac{N-1}{p}\right\rfloor+ 
2\left\lbrace\frac{N-1}{p}\right\rbrace p$$ otherwise (the braces indicate the fractional part). Eploiting now the fact that the quiver contains 
alternating $SO$-$USp$ gauge groups, one can easily reconstruct it explicitly from these data. 

For $p>N-1$ the polygon as a vertical edge and one of slope one as in the $SU(N)$ case. The quiver starts with a tail of the form 
$$\boxed{N}-USp(2N-2)-SO(2N)-USp(2N-2)-\dots$$ and ends in one of the following two ways
\begin{equation}\begin{aligned}
\dots-USp(10)-SO(10)-USp(6)-SO(6)-SU(2)-\boxed{1},\nonumber\\
\dots-USp(12)-SO(12)-USp(8)-SO(8)-USp(4)-SO(4).\nonumber
\end{aligned}\end{equation}
In this case the number of gauge groups in the quiver is $p-\lfloor \frac{p}{N-1}\rfloor$. If this number is odd the first option is the 
correct one, otherwise the quiver terminates as in the second sequence.
One can easily check that all the groups in the tails are conformal, except the group associated to the dot at which the two edges 
meet as for $SU(N)$.

\section{Concluding remarks}

We have made a systematic analysis of singular points in $\mathcal{N}=2$ SQCD with classical gauge groups, focusing on the 
maximally singular points in the moduli space. We have seen that in many cases, in order to satisfy the constraint on the scaling dimensions 
of mass parameters, we are forced to introduce different scale invariant sectors. The introduction of two sectors, which is the simplest
possibility, leads to a unique answer which is consistent with all the strong coupling dualities found recently and allows to 
satisfy the constraints imposed by superconformal invariance. 

We found a common structure for the low-energy description at these points, which is schematically given by:
\begin{itemize}
\item  An abelian sector.
\item The B sector, which is always given either by a $D_N$ theory (with $N>2$) or by a doublet of hypermultiplets. In both cases
the flavor symmetry is $SU(2)$.
\item The A sector, with (at least) $SU(2)\times G$ flavor symmetry, where $G$ is the flavor symmetry of the parent gauge theory. 
This is the only sector that changes as we vary the gauge group and in most cases admits a six-dimensional description.
\item An infrared free $SU(2)$ gauge multiplet coupled to sectors A and B.
\end{itemize}
At the maximally singular point the abelian sector just describes a number of decoupled vector multiplets, as pointed out in \cite{GST},
whereas at points which are not lifted by the $\mathcal{N}=1$ perturbation it includes massless hypermultiplets charged under
each $U(1)$ factor. Chebyshev points in $USp$ and $SO$ gauge theories fall in this second class and are characterized by a free B
sector, which describes two massless hypermultiplets. This is not the case for $SU(N)$ gauge theories. A particularly simple case is given by $USp$ theory with 4 flavors: in this
case the A sector becomes free and it turns out that the Chebyshev point admits a lagrangian description. We will analyze this model in depth in 
the next chapter.

One possible extension of this analysis is to consider more general theories and see whether different structures emerge in the infrared at
the maximally singular points. The superconformal theories $D(G,s)$ and those studied in \cite{XIE} certainly play an important role. In particular,
it would be interesting to understand whether the constraints we imposed always lead to a unique answer. 

This two-sector structure emerges, as we have seen, only for a specific parity of $N_f$. In the other cases we get instead ordinary 
IR fixed points analogous to AD theories. We have seen that we can extract information on the light BPS spectrum in a neighbourhood 
of these points exploiting the analysis of \cite{CM}. From the knowledge of the BPS quiver we can learn a lot about a strongly 
coupled theories (a and c central charges, flavor central charge...) and it would be very interesting to understand how to 
generalize the analysis of \cite{CM}.

Roughly speaking, at the level of the SW curve the construction proposed in \cite{CM} amounts to the following (in the $A_N$ and 
$D_N$ cases): take the curve for the $D_p$ AD theory $x^2=z^p$ (and of course $\lambda=x/zdz$), and replace $x^2$ with $W_G(x)$. 
At the level of the quiver it amounts to considering the triangle tensor product with the $G$ Dynkin diagram but can also be seen 
as a procedure for constructing higher rank theories starting from simpler rank one models (in the language of \cite{G}). This is 
similar in spirit to \cite{GMNN}.

\chapter{Singular SQCD Vacua and Confinement}

\section{Introduction} 

As we have seen, considerable progress is being made in our understanding of the dynamics of supersymmetric non-Abelian gauge 
theories in four dimensions.  A recent remarkable development concerns the better understanding of ${\cal N}=2$  superconformal 
theories (SCFT) discussed in chapte 2. Also many new results on the exact BPS spectra  in the strongly coupled gauge systems are 
now available (see e.g. \cite{CV1,GMN}).  Another venue in which considerable  development has occurred 
is the investigations of soliton vortex and monopoles of non-Abelian type \cite{HT}-\cite{GJK}.   Together, it is quite plausible that these developments help clarifying many issues left still to be elucidated, even after the discovery of Seiberg-Witten solutions of ${\cal N}=2$ gauge theories and the developments which followed.  

The physics of the local $r$ vacua represents a beautiful example of confining vacuum which is  dual Higgs system of non-Abelian 
variety. But even more interesting is the situation in which the singular SCFT's studied in the previous chapter are 
deformed by an ${\cal N}=1$ adjoint scalar mass term $\mu\, \Tr \Phi^{2}$.

The purpose of this chapter is to put together the results found so far in order to study the possible types of strongly-coupled 
gauge systems in confinement phase in the softly broken ${\cal N}=2$  SQCD. We will restrict for simplicity to the cases in which 
the the low-energy theory involves two sectors.
In section 2 we discuss the low energy physics at fixed points in $USp$ $\mathcal{N}=2$ SQCD. These SCFTs become confining when 
we turn on the $\mathcal{N}=1$ perturbation and we discuss the mechanism of confinement occurring in these special cases. Sections 3 
and 4 deal with singular points of $SU(N)$ SQCD, especially those studied in chapter 3. We will see how the two-sector low-energy 
description allows to recover all the results found in chapter 3. Secton 4 is devoted to the analysis of the Tchebyshev point in 
$SO(N)$ SQCD.

\section{Singular points in $USp(2N)$ theory with four flavors \label{uspsection}}

Let us start from the confining vacua of $USp$ SQCD. As we have seen, in \cite{EHIY} it was pointd out that in the massive, equal mass ($m_{i}=m \ne 0$) theory, 
the SCFT vacua occurring in the  $USp(2N)$ theory are the same  $r$ vacua of $SU(N)$ theory, $r=0,1,2,\ldots, N_{f}/2$, exemplifying 
the universality of SCFTs.   In the $m_{i}\to 0$ limit,  however, the $r$ vacua collapse into a singular SCFT  
(``Tchebyshev''  point) \cite{CKM}   with a larger global  symmetry $SO(2N_{f})$.

\subsection{Low-energy effective description}

The analysis given in chapter 4, done following the work Gaiotto, Seiberg and Tachikawa, has 
shown that the relevant SCFT can be analyzed by introducing two different scalings for the scalar VEVs $u_{i}\equiv \langle \Phi^{i} \rangle$ 
(the Coulomb branch coordinates) around the singular point. Our prediction is that the low energy physics at this singular 
point can be described as    
\begin{description}
\item [(i)] $U(1)^{N-n}$ abelian sector, with massless particles charged under each $U(1)$ subgroup.
\item [(ii)]  The  (in general, non-Lagrangian) A sector  with global symmetry $SU(2)\times SO(4n)$.
\item [(iii)] The B sector is free and describes a doubet of hypermultiplets. The flavor symmetry of this system is $SU(2)$.
\item [(iv)] $SU(2)$ gauge fields coupled weakly  to the last two sectors.
\end{description} 

For general $N_{f}$  these involve non-Lagrangian SCFT theories, and it is not easy to analyze the effects of $\mu\, \Tr\, \Phi^{2}$ deformation.  
 In the particular case $n=2$  ($USp(2N)$ theory with $N_{f}=4$), however, the A sector becomes free and describes four doublets 
 of $SU(2)$.  Let us consider the effect of $\mu \, \Phi^{2}$ deformation of this particular  theory focusing on the non-abelian
 sector. The analysis of the $U(1)^{N-n}$ sector was already given in \cite{APS,CKM} and is by now standard.
 
\subsection{Flavor symmetry breaking} 

The superpotential  for a hypermultiplet $Q_{0}$ and four hypermultiplets  $Q_{i}$'s, coupled to $SU(2) \times U(1)$  gauge fields  (only $Q_{0}$ carrying the $U(1)$ charge)   is
\beq\label{vac}  \sqrt{2} Q_{0} A_{D} {\tilde Q}^{0} + \sqrt{2} Q_{0} \phi {\tilde Q}^{0} + \sqrt{2}\sum_{i=1}^{4}  Q_{i} \phi {\tilde Q}^{i}  +  \mu A_{D} \Lambda  + \mu \Tr \phi^{2}\;.   
\eeq
Setting $\phi=\phi_{\alpha}\tau^{\alpha}$, where $\tau^{\alpha}$ are Pauli matrices and $\alpha=1,2,3$, the vacuum equations are 
\beq   \sqrt{2} Q_{0} {\tilde Q}_{0} + \mu \Lambda =0\;;   \label{eQ1}
\eeq
\beq   (\phi + A_{D})  {\tilde Q}_{0}= Q_{0} \, (\phi + A_{D}) =0\;;  \label{eQ2}
\eeq
\beq     \sum_{i=1}^{4} \sqrt{2} Q_{i}^{a} (\tau_{\alpha})_{a}^{b} {\tilde Q}_{b}^{i}  + \sqrt{2} Q_{0}^{a}(\tau_{\alpha})_{a}^{b}
{\tilde Q}^{0}_{b} + \mu \, \phi_{\alpha}=0\;;  \label{eQ3}
\eeq
\beq    \phi \, {\tilde Q}^{i}=  Q_{i}\, \phi =0, \quad \forall i\;.   \label{eQ4}
\eeq
The first says that $Q_{0}\ne 0$ and acting with a gauge transformation we can impose the following condition
\beq  Q_{0}^{1} =  {\tilde Q}_{0\, 1}=  2^{-1/4}\sqrt{-\mu \Lambda} \ne 0; \qquad   Q_{0}^{2} =  {\tilde Q}_{0\, 2}=0.
\eeq
The second equation then implies that $\phi$ is diagonal in this gauge. 

In principle equation (\ref{eQ2}) is satisfied if we impose 
\beq   A_{D} \ne 0,   \qquad \phi=    \left(\begin{array}{cc}a & 0 \\0 & -a\end{array}\right), \quad a=-A_{D}.
\eeq 
But then (\ref{eQ4}) would tell
\beq    Q_{i}= {\tilde Q}^{i} =0. 
\eeq
Plugging now this into equation (\ref{eQ3}) then tells us that the $Q_0$ condensate should be compensated by the $\phi$ field 
alone, implying that $\langle\phi\rangle\sim\Lambda$. However, this solution involves fluctuations of order $\Lambda$ which are 
beyond the validity of our effective theory and should be discarded as they are an artifact of our approximation (see \cite{CKM}).

We thus conclude that the vev of $Q_i$'s cannot vanish for all values of $i$, which in turn imlies (see (\ref{eQ4}))  
\beq   \phi =0, \quad  A_{D}=0\;
\eeq 
and 
\beq     \sum_{i=1}^{4}  Q_{i}^{a} (\tau_{3})_{a}^{b} {\tilde Q}_{b}^{i}  = - Q_{0}^{a}(\tau_{3})_{a}^{b}{\tilde Q}^{0}_{b} =  
 \frac{\mu}{2\sqrt{2}} \Lambda.   
\eeq
By flavor rotation this can be brought to
\beq     Q_{1}^{a}   {\tilde Q}_{b}^{1}  = \frac{\mu}{\sqrt{2}} \Lambda \,  \delta^{a1} \delta^{b1},\qquad  Q_{i}={\tilde Q}_{i}=0, \quad i=2,3,4 
\eeq
or
\beq  Q_{1}^{a}   {\tilde Q}_{b}^{1}  = -\frac{\mu}{\sqrt{2}} \Lambda \,  \delta^{a2} \delta^{b2},\qquad  Q_{i}={\tilde Q}_{i}=0, \quad i=2,3,4. 
\eeq
This means that the flavor symmetry is spontaneously broken as
\beq   SO(8) \to    U(1) \times SO(6) = U(1) \times SU(4). 
\eeq
This is precisely what is  expected from the result known at large $\mu \gg \Lambda$ \cite{CKM}, showing the  consistency of 
the whole picture.

The gauge group is completely broken and the corresponding vectormultiplets become massive. The only massless fields left after 
the breaking are the three doublets $Q_i$, $\tilde{Q}_i$ (i=2,3,4) for a total of twelve chiral multiplets (remember that we have 
only $\mathcal{N}=1$ supersymmetry now). This matches nicely the expected number of Goldstone multiplets coming from the 
breaking of the flavor symmetry ($\text{dim}SO(8)-\text{dim}U(4)=12$). These fields are then the supersymmetric counterpart of 
pions.

\subsubsection{Neither abelian nor non-abelian superconductor}

We find here the first example of a situation we will encounter again in this chapter: the degrees of freedom in the 
effective lagrangian (\ref{vac}) are magnetically charged, so their condensation leads to confinement as should be by now familiar. 
However, this system is different from the one we considered in chapter 1 while discussing confinement in $SU(2)$ SYM: here the 
low energy lagrangian involves a non-abelian gauge theory and the $Q$ doublets should be thought of as non-abelian monopoles. In this 
respect the system seems more similar to $r$ vacua. However, in that case $U(r)$ ``magnetic multiplets'' condense leading both to confinement and to 
chiral symmetry breaking. The resulting confining string is then of non-abelian kind as reviewed in chapter 1. 

Here we find a different behaviour: there is a solitonic vortex coming from the breaking of the $U(1)\times SU(2)$ gauge group, 
but since only one doublet, not all the five of them as would occur in a $r$ vacuum, is charged under $U(1)$ it is not possible 
to exploit the $SU(2)$ factor to construct a non-abelian vortex as in \cite{HT}-\cite{GJK}. For the same reason one can rule out 
the presence of the fractional or semilocal vortices found in \cite{semi}-\cite{frazio}. We are thus forced to conclude that the confining string is a 
standard ANO abelian vortex as in $SU(2)$ SYM and only $Q_0$ can be used to construct it! We can thus say that the $Q_0$ field 
is responsible for confinement and, being uncharged under $SO(8)$, is instead unrelated to (the analog of) chiral symmetry 
breaking which is due to the condensation of $Q_1$. 

We thus find a sort of intermediate situation between the previously known abelian and non-abelian confining systems, in which 
confinement and chiral symmetry breaking are clearly related to each other but are induced by the condensation of different fields.
The feature that usually characterizes abelian confinement in higher rank gauge theories, namely the presence of multiple distinct 
confining strings, is clearly avoided in this system.

\subsection{Adding flavor masses: the counting of vacua}

The strategy adopted in \cite{CKM} to analyze this singularity was to ``resolve'' it, by introducing generic, nearly equal quark masses $m_{i}$ alongside the adjoint scalar mass $\mu$.   
By requiring the factorization property of the Seiberg-Witten curve to be of maximally Abelian type (the criterion for ${\cal N}=1$ supersymmetric vacua), 
this point was found to split into  various $r$ vacua  which are local $SU(r)\times U(1)^{N-r}$ gauge theories, identical to those appearing in the infrared limit of   $SU(N)$  SQCD (the universality of the infrared fixed points).
\be    \binom{N_{f}}{0} + \binom{N_{f}}{2} + \ldots \binom{N_{f}}{N_{f}} =  2^{N_{f}-1}    \label{one}
\ee
whereas the other vacuum splits into odd $r$ vacua, with the total multiplicity
\be    \binom{N_{f}}{1} + \binom{N_{f}}{3} + \ldots \binom{N_{f}}{N_{f}-1} =  2^{N_{f}-1}\;.  \label{two}
\ee
Due to the exact  ${\mathbbm Z}_{2N+2-N_{f}}$ symmetry of the massless theory,  the singular (EHIY) point actually appears $2N+2-N_{f}$ times, and the 
number of the vacua for generic $\mu, m_{i}$ is given \footnote{For even $N_{f}$ we are considering, the $(N+1-N_{f}/2)$-th element of   ${\mathbbm Z}_{2N+2-N_{f}}$  
exchanges the two Chebyshev vacua \cite{CKM}, so that the number of the vacua is  $ (2N+2-N_{f})\, 2^{N_{f}-1}$  and not  $ (2N+2-N_{f})\, 2^{N_{f}}$. }  by  $ (2N+2-N_{f})\, 2^{N_{f}-1}$.     

In order to match these predictions, let us turn on the mass parameters for the flavors as well 
\beq\label{vacmass}\begin{aligned}   &\sqrt{2} \, Q_{0} A_{D} {\tilde Q}^{0} +  \sqrt{2} \, Q_{0} \phi {\tilde Q}^{0} + \sum_{i=1}^{4}  \sqrt{2} \, Q_{i} \phi {\tilde Q}^{i}
+  \sum_{i=1}^{4}   m_{i}\, Q_{i} {\tilde Q}^{i}\\  
&+  \mu A_{D} \Lambda  +  {\mu}\,  \Tr \phi^{2}.    
\end{aligned}\eeq
For equal and nonvanishing masses the system has $SU(4)\times U(1)$ flavor symmetry and in the massless limit the symmetry gets 
enhanced to $SO(8)$, as seen before.    


The vacuum equations are:   
\beq    \sqrt{2} \, Q_{0} {\tilde Q}_{0} + \mu \Lambda =0\;;   \label{eq1}
\eeq
\beq   (\sqrt{2} \, \phi + A_{D})  {\tilde Q}_{0}= Q_{0} \, (\sqrt{2} \, \phi + A_{D}) =0\;;  \label{eq2}
\eeq
\beq  \sqrt{2} \,\,\left[\, \frac{1}{2}  \sum_{i=1}^{4}  Q_{i}^{a}   {\tilde Q}_{b}^{i} -  \frac{1}{4}  \delta_{b}^{a}  Q_{i}{\tilde Q}^{i} +  \frac{1}{2} Q_{0}^{a}{\tilde Q}^{0}_{b}- \frac{1}{4} \delta^{a}_{b}  Q_{0}{\tilde Q}^{0} \, \right] + \mu \, \phi^{a}_{b}=0\;;  \label{eq3}
\eeq
\beq    (\sqrt{2} \,\phi +m_{i} )\, {\tilde Q}^{i}=  Q_{i}\, (\sqrt{2} \,\phi + m_{i})  =0, \qquad \forall i\;.   \label{eq4}
\eeq
The first tells that $Q_{0}\ne 0$.  By gauge choice
\beq  Q_{0} =  {\tilde Q}_{0}= \left(\begin{array}{c}   2^{-1/4}\sqrt{-\mu \Lambda}  \\ 0 \end{array}\right)
\eeq
so that 
\beq    \frac{1}{2} Q_{0}^{a}{\tilde Q}^{0}_{b}- \frac{1}{4}  (Q_{0}{\tilde Q}^{0}) \, \delta^{a}_{b} = \frac{(-\mu \Lambda)}{4 \sqrt{2}} \, \tau^{3}\;.  
\eeq
The second equation can be satisfied by adjusting $A_{D}$.

As in the $m_{i}=0$  case  discussed previously, we must discard  the  solution 
\beq  \phi=  a\, \tau^{3}, \qquad      a=  \frac{\Lambda}{4}, \qquad Q_{i} ={\tilde Q}_{i}=0, \quad \forall i\;
\eeq
as it involves a fluctuation ($\sim\Lambda$) far beyond the validity of the effective action.

The true solutions can be found by having one of $Q_{i}$'s canceling the contributions of   $Q_{0}$ and $\phi$ in   Eq~(\ref{eq3}).  Which of $Q_{i}$ is nonvanishing is related to the
value of $\phi$  through Eq~(\ref{eq4}).   For instance, four  solutions can be found by choosing  ($i=1,2,3,4$)
\beq   a=  -\frac{m_{i}}{\sqrt{2}}, \qquad   Q_{i}= {\tilde Q}_{i} = \left(\begin{array}{c}f_i \\0\end{array}\right) ;\qquad  Q_{j}= {\tilde Q}_{j}=0, \quad  j \ne i\;, 
\eeq
such that  
\beq   f_{i}^{2}=    \frac{\mu \Lambda - 4\, a }{\sqrt{2}} =   \mu ( \frac{\Lambda}{\sqrt{2}} + 2 m_{i})\;.    
\eeq

There are four more solutions of the form,  ($i=1,2,3,4$)
\beq   a= + 
\frac{m_{i}}{\sqrt{2}}, \qquad   Q_{i}= {\tilde Q}_{ i} = \left(\begin{array}{c}0 \\g_i\end{array}\right);\qquad  Q_{j}= {\tilde Q}_{j}=0, \quad  j \ne i\;,   
\label{sol11}
\eeq
and 
\beq   g_{i}^{2}=    \frac{-\mu \Lambda + 4\, a }{\sqrt{2}} = -   \mu (\frac{ \Lambda}{\sqrt{2}}  -  2 m_{i})\;.   
\label{sol22}   \eeq
Note that the solutions (\ref{sol11}) and (\ref{sol22}) are unrelated by any $SU(2)$ gauge transformation. 
In all, we have found $2^{3}=8$ solutions consistently with  Eq.~(\ref{one}).

In the equal mass limit the $8$ solutions group into two  set of four nearby vacua, obviously connected by the $SU(4)$. 
So these look like the $4+4 =8$,      two  $r=1$ vacua,     from one of the Chebyshev vacua, see Eq.~(\ref{two}).  The other Chebyshev vacuum should give $1 + 6 + 1=8$ vacua, corresponding to  $r=0,2$ vacua. Where are they?

A possible solution is that in the other Chebyshev vacuum  the superpotential has a similar form as (\ref{vacmass}) but with $Q_{i}$'s carrying different flavor charges.   
The $SU(4)$ symmetry of the equal mass theory may be represented as  $SO(6)$:
\beq\label{vacmassBis}\begin{aligned}  &\sqrt{2} \,Q_{0} A_{D} {\tilde Q}^{0} +  \sqrt{2} \, Q_{0} \phi {\tilde Q}^{0} + \sum_{i=1}^{4}  \sqrt{2} \,Q_{i} \phi {\tilde Q}^{i}  
+     \sum_{i=1}^{4}   {\tilde m}_{i}\, Q_{i} {\tilde Q}^{i}\\  
&+  \mu A_{D} \Lambda  +  {\mu}\,  \Tr \phi^{2},
\end{aligned}\eeq
where      
\bea   {\tilde m}_{1} =    \frac{1}{4}  (m_{1}+m_{2}- m_{3}-m_{4})\;;  \nonumber \\   
{\tilde m}_{2}=    \frac{1}{4}  (m_{1}-m_{2}+ m_{3}-m_{4})\;;  \nonumber \\ 
  {\tilde m}_{3}=    \frac{1}{4}  (m_{1}-m_{2}- m_{3}+m_{4})\;;\nonumber \\ 
    {\tilde m}_{4}=  \frac{1}{4}  (m_{1}+m_{2} + m_{3} + m_{4})\;.    \label{masses}
\eea
The correct realization of the underlying symmetry in various cases is not obvious, so let us check them  all.  
\begin{description}
\item[(i)]  In the equal mass limit, $m_{i}=m_{0}$,  
\beq       {\tilde m}_{4}=m_{0}, \qquad    {\tilde m}_{2}= {\tilde m}_{3}= {\tilde m}_{4}=0\;, 
\eeq
so  the symmetry is
\beq   U(1)\times SO(6) = U(1)\times SU(4)\;.
\eeq
  Clearly in the $m_{i}=0$ limit the symmetry is enhanced to $SO(8)$. 
  \item[(ii)]   $m_{1}=m_{2}$,    $m_{3}$, $m_{4}$ generic.  
  In this case    ${\tilde m}_{2}= -  {\tilde m}_{3}$ and  ${\tilde m}_{4}$ and  ${\tilde m}_{1}$ are generic, so  the symmetry is   $U(1)\times U(1) \times U(2)$, as in the underlying theory;
  \item[(iii)] $m_{1}=m_{2} \ne 0$,    $m_{3}=m_{4}=0$.    In this case,  ${\tilde m}_{4}=  {\tilde m}_{1}\ne 0$ and ${\tilde m}_{2}=  {\tilde m}_{3} = 0$, so
  obviously the symmetry is $U(2)\times SO(4)$ both in the UV and in  (\ref{vacmassBis}).  
  
  \item[(iv)] $m_{1}=m_{2}=m_{3}\ne 0$,   $m_{4}$ generic.  In this case,  ${\tilde m}_{1}=  {\tilde m}_{2} =  -  {\tilde m}_{3}  \ne 0$,     ${\tilde m}_{4}$ generic. 
  Again the symmetry is $U(3)\times U(1)$ both at the UV and IR.

  \item[(v)]  $m_{1}=m_{2} \ne 0$  and   $m_{3}=m_{4} \ne 0$ but $m_{1}\ne m_{3}$.  In this case  $ {\tilde m}_{2} =  {\tilde m}_{3} = 0$ and  
  $ {\tilde m}_{4}$  and  ${\tilde m}_{1}$ generic.  The flavor symmetry is 
  \beq    SO(4) \times  U(1) \times U(1) =  SU(2)\times SU(2) \times  U(1) \times U(1)\;; 
  \eeq
  this is equal to the symmetry 
  \beq    (SU(2) \times U(1)) \times   (SU(2)\times U(1))
  \eeq
  of the underlying theory.
  \item[(vi)]    $m_{1}\ne 0$,  $m_{2}=m_{3}=m_{4}=0.$  In this case  ${\tilde m}_{1}={\tilde m}_{2}={\tilde m}_{3}={\tilde m}_{4}\ne 0$.
  The symmetry is $U(1)\times SO(6)$ in the UV, and  $U(4)$ in the   infrared. 
   \item[(vii)]    $m_{1}\ne 0$,  $m_{2}\ne 0,$  $ m_{1} \ne m_{2}$,  $m_{3}=m_{4}=0.$  In this case  ${\tilde m}_{1}={\tilde m}_{4}$   ${\tilde m}_{2}={\tilde m}_{3}\ne   {\tilde m}_{1}$.
  The symmetry is $U(1)^{2}\times SO(4)$ in the UV, and  $U(2)\times U(2)$ in the   infrared. 

\end{description}
The cases of masses equal except sign, e.g.,   $m_{1}= - m_{2}$,  are similar. 

Thus in all cases Eq.(\ref{vacmassBis}) has the correct symmetry properties as the underlying theory.  The vacuum solutions which follow from  it are similar to those
found from   Eq.(\ref{vacmass}),  with simple replacement, 
\beq   m_{i} \to {\tilde m}_{i}
\eeq
so  there are  $8$ of them.   The interpretation and their positions in the quantum moduli space (QMS) are different, however.    In the equal mass limit, $m_{i}\to m_{0}$,   The two solutions with 
\beq  a = - \frac{{\tilde m}_{4}}{\sqrt{2}}, \qquad {\rm or} \qquad    a =  \frac{{\tilde m}_{4}}{\sqrt{2}}, 
\eeq
can be regarded as  two  $r=0$ vacua. Note that as $|f_{1}| \ne |g_{1}|$ they correspond to distinct points of the moduli space.  
On the other hand, in the other six vacua  $a=0$ always and     $|f_{i}| = |g_{i}| $, these six solutions correspond to the same point of  
the moduli space: they may be associated with the $r=2$ (sextet) vacuum.

\noindent{\bf Remarks:} ~~ The assumption that the $Q_{i}$ fields have different mass assignment in the two Chebyshev vacua, 
as in Eq.~(\ref{vacmass}) and Eq.~(\ref{vacmassBis}) (with (\ref{masses})), is indeed mainly motivated by the fact that the two 
Chebyshev points in QMS are known to behave differently under the mass perturbation \cite{CKM}.  One of the points splits into 
(for nearly equal masses $m_{i}\ne 0$) two nearby groups of $4+4$ vacua  (see Eq.~(\ref{two})) corresponding to two $r=1$ vacua, 
whereas the other is resolved into three groups of $1+1+6$ vacua, which correspond to two $r=0$ and one $r=2$ vacua 
(Eq.~(\ref{one})).  The analysis of this section shows that these properties are precisely reproduced
by our low energy effective action.  
The flavor charges (\ref{masses}) suggest that $Q$'s  are really non-Abelian magnetic monopoles, as semiclassically magnetic 
monopoles appear in the spinor representations of $SO(2N_{f})$.     

\section{Singular points in $SU(N)$ SQCD}

In this section we study two types of singular points in $SU(N)$ SQCD (with an even number   $N_f=2n$ of  flavors) which are relevant for the breaking to $\mathcal{N}=1$. 
In the first case, which we have studied in chapter 3, the singular point arises from the collision of different $r$ vacua. Dynamical flavor symmetry breaking does 
not occur in this case. The second class of singular points arise in a ``degeneration limit'' of $r=n$ vacua, in 
which the SW curve becomes more singular. Vacua with different $r$ are not involved in this case and the pattern of flavor symmetry 
breaking remains to be $U(N_f)\rightarrow U(n)\times U(n)$.

\subsubsection{Colliding $r$-vacua of  the  $SU(N)$ SQCD  \label{collision}  }

The singular point analyzed in chapter 3 arises when quark mass parameters are fine-tuned to a particular value of the order of $\Lambda$,
\be   m=m^{*}\equiv   \omega^{k}\, \frac{2 N - N_{f}}{N}\, \Lambda\;, \quad (k=1,\ldots,  2N-N_{f}, \;\;  \omega^{2N- N_{f}}=1).   \label{collidingvacua}
\ee
 All the $r$-vacua with $r=0,1,\ldots, \tfrac{N_{f}}{2}$ (more precisely, one representative from each $r$ vacua) coalesce to form a single vacuum \footnote{In \cite{Bolo}
 an analogous phenomenon was studied, but by using an appropriate ${\cal N}=1$ superpotential $W(\Phi)$ and selecting particular vacua.}
 \be     y^{2}\sim   (x+m^{*})^{N_{f}+1}.
\ee
This corresponds to the SCFT of the highest criticality \cite{EHIY} for  
\be   N= n+1 \; \qquad (N_{F}=2n).  \label{collapse}
\ee   
 
Exploiting the analysis of \cite{GST} it is easy to see that the low energy physics at the singular point of interest for us can be described as
\begin{description}
\item [(i)] A $U(1)^{N-n-1}$ abelian sector, with massless particles charged under each $U(1)$ subgroup.
\item [(ii)]  The  (in general, non-Lagrangian) A sector  with global symmetry $SU(2)\times SU(N_f)$.
\item [(iii)] The B sector is the most singular superconformal point of $SU(2)$ theory with two flavors (or the $D_3$ Argyres-Douglas
theory), with $SU(2)$ flavor symmetry.
\item [(iv)] $SU(2)$ gauge fields coupled to the last two sectors.
\end{description} 
The presence of the $\mu \Phi^{2}$ term, breaking $SU_{R}(2)$ explicitly, is expected to generate nonvanishing gaugino condensate through anomaly, and induce the symmetry breaking 
\be   {\mathbbm Z}_{2N-N_{f}} \to   {\mathbbm Z}_{2}\;. 
\ee
We are not able to deduce such a result directly  with the  GST dual description. 
However, the analysis of chapter 3 shows that the meson  and gaugino condensates are of the form (see Eq.~(\ref{eqxS}))
\be    \langle\tilde{Q}^{i}Q_{i}\rangle \sim \, \mu \Lambda\,, \quad ({\rm indep. ~ of} \,\,  i)\;;  \qquad  \langle  \lambda \lambda  \rangle  \sim \mu \Lambda \ne 0\;. 
\ee
The $SU(N_{f})\times U(1)$ symmetry remains unbroken, in contrast to what happens (for generic $m$)  in single $r$ vacua, (\ref{symbr}). 

\subsubsection{Higher order singularity  \label{non collision}}

The vacuum arising from the collision of r vacua is not the only higher singular point in softly broken $\mathcal{N}=2$ $SU(N)$ SQCD.
In order to illustrate this point let us consider the simplest example, namely $SU(4)$ theory with $N_f=4$: the SW curve is
$$y^{2}=(x^4-u_2x^2-u_3x-u_4)^{2}-4\Lambda^{4}(x+m)^4.$$ In this case the $r=2$ vacuum can be found easily and the curve assumes the form
$$y^2=(x+m)^{4}(x-m)^{2}(x-m-2\Lambda)(x-m+2\Lambda).$$ From here we easily see that when $m=\pm\Lambda$ the curve can be approximated as
$y^2\approx(x+m)^5$ and we recover the case studied before. On the other hand, in the limit $m=0$ the curve becomes more singular and 
reduces to $y^2\approx x^6$. Of course, this singular point exists in the general case, as long as $n<N-1$ (this was already noticed in \cite{CKM}):
in a $r=\frac{N_f}{2}$ vacuum the SW curve assumes the form $$y^2=(x+m)^{N_f}Q_{N-n-1}^{2}(x)(x-\alpha)(x-\beta).$$ The roots 
of $Q_{N-n-1}$ have multiplicity one and are located at (see \cite{CKM}) 
\be   x=\frac{N_f}{2N-N_f}m+2\Lambda\cos\left(\frac{2k\pi}{2N-N_f}\right);\quad k=1,\dots,N-\frac{N_f}{2}-1.   \label{tcheb}\ee
When the bare mass is chosen in such a way that $-m$ coincides with one of these roots the SW curve can be approximated as $y^2\approx(x+m)^{N_f+2}$.

From the analysis performed in \cite{GST} we can conclude that the low energy physics at this singular point can be described as follows:
\begin{description}
\item [(i)] A $U(1)^{N-n-2}$ abelian sector, with massless particles charged under each $U(1)$ subgroup.
\item [(ii)]  The A sector  with global symmetry $SU(2)\times SU(N_f)$ described in the previous section.
\item [(iii)] The B sector is the most singular point of $SU(3)$ theory with two flavors (or the $D_4$ Argyres-Douglas
theory), with $SU(2)\times U(1)$ flavor symmetry \footnote{Actually, it was recently shown in \cite{AMT} that in this case the 
flavor symmetry enhances to $SU(3)$. However, an $SU(2)$ subgroup is gauged and the manifest flavor symmetry is the commutant 
of $SU(2)$ inside $SU(3)$, which is $U(1)$.}.
\item [(iv)] $SU(2)$ gauge fields coupled to the last two sectors.
\end{description} 

From the analysis performed in chapter 3 it is easy to see that the pattern of symmetry breaking (once the superpotential for the adjoint field is
 turned on) is $U(N_f)\rightarrow U(N_f/2)\times 
U(N_f/2)$, the same as an $r=N_f/2$ vacuum. The main difference with respect to the previous case is that this vacuum does 
not arise from the coalescence of different r vacua.

\section{Breaking to $\mathcal{N}=1$ in the singular vacua}

In this section we want to test our proposal for the low-energy effective description at the singular points by reproducing the 
correct pattern of flavor symmetry breaking occurring once the $\mu\Tr\Phi^2$ perturbation is turned on.

As we have seen, for generic $m$ the most singular point in the moduli space is the $r=N_f/2$ vacuum \footnote{As is well known, 
the curve can become even more singular. However, such points are not relevant in view of the breaking to $\mathcal{N}=1$.} (in 
which the SW curve can be approximated as $y^2\approx(x+m)^{N_f}$). Its low-energy physics 
involves a scale invariant sector with $SU(N_f/2)$ gauge group, whose coupling constant depends on $m$. For 
special values $m^{*}$ of order $\Lambda$ (or zero) the SW curve degenerates further ($y^2\approx(x+m)^{N_f+1}$ or $y^2\approx(x+m)^{N_f+2}$,
as we have seen before ). It is easy to see that, as we approach these critical values the coupling constant of the $SU(N_f/2)$ 
theory diverges. In this limit it is convenient to adopt the Argyres-Seiberg dual description \cite{AS} in which an
 $SU(2)$ gauge group emerges, coupled to an hypermultiplet in the doublet (B sector) and to a strongly coupled interacting sector 
 which coincides precisely with the A sector introduced above.
 
In order to describe the low energy physics at the most singular point in a neighbourhood of the critical values $m^{*}$, it is 
thus convenient to introduce these two sectors. As we approach the critical value $m^{*}$ the curve becomes more singular and the B sector, 
which is free in the $r$ vacuum, becomes interacting ($D_3$ or $D_4$ Argyres-Douglas theory for the two classes of 
singular points we are interested in). In the process the A sector is just a spectator.


Finding the effective low energy description at these singular points once the $\mu\Tr\Phi^2$ term is turned on is in general very hard. However, 
in the $N_f=4$ case the problem is greatly simplified, since the A sector is free. For $m$ close to $m^*$ the low energy physics in the 
$r=2$ vacuum admits a lagrangian description analogous to (\ref{vac}). The only difference is that the A sector describes three 
doublets of $SU(2)$ instead of four. Imposing the F-term equations we find as before a non vanishing condensate for the $Q_{0}$ and $Q_1$ fields
 (we use the same notation as in (\ref{vac})),
reproducing the correct pattern of symmetry breaking: $$U(1)\times SO(6)\rightarrow U(1)\times U(1)\times SO(4)\simeq U(2)\times U(2).$$
Clearly, if the condensate for $Q_0$ vanishes, the one for $Q_1$ vanishes as well, restoring the full $U(N_f)$ flavor symmetry of 
the theory. The $Q_0$ condensate can be determined focusing on the B sector only, which is the most singular point in $SU(2)$ or 
$SU(3)$ theory with two flavors in the cases of interest for us. The problem is thus reduced to computing the abelian condensates 
in $SU(2)$ or $SU(3)$ theories (the calculation is confined in the appendix).
The result is that the $Q_0$ condensate vanishes in the $SU(2)$ case but not in the $SU(3)$ one, reproducing the expected pattern of 
flavor symmetry breaking.

As we have seen the effective theory is non lagrangian in this case. However, in order to check the counting of vacua in the 
mass-deformed theory, we can exploit the fact that we know much about the interacting B sectors of these two models: they can be 
realized as IR fixed points of simple lagrangian theories. We will precisely use this idea in the rest of the section.

\subsection{Colliding $r$ vacua  of the  $SU(N)$, $N_{f}=4$ Theory   \label{collidingrvacua}}  

We have seen that the effective GST dual is made of 
the A sector describing the three doublets of free hypermultiplets and the B sector,  which is the most singular SCFT of the $SU(2)$, $N_{f}=2 $ theory 
\beq     D_{3} - SU(2) -  {\boxed 3}  \;.         \label{GSTD3}
\eeq
In order to see the effect of the ${\cal N}=1$ perturbation ${\mu \Phi^{2}}$   in this vacuum,  let us replace the B sector 
(the $D_{3}$ theory) by  a new $SU(2)$ theory and a bifundamental 
field $P$,
\beq     SU(2) \, {\stackrel {P}{-}} \,  SU(2) -  {\boxed 3}  \;.    \label{replace}
\eeq
The superpotential has the form, 
$$\sum_{i=1}^{3} \sqrt{2}  Q_{i}  \Phi {\tilde Q}^{i}  + \sum_{i=1}^{3}  {\tilde m}_{i}  \,  Q_{i} {\tilde Q}^{i} +      \mu \Phi^{2} +  \sqrt{2}   P  \Phi {\tilde P}
+   \sqrt{2}  {\tilde P} {\chi}     P +  \mu  \, \chi^{2}  +   m^{'} \,  {\tilde P}   P  ,    
$$
where $P=P_{a}^{\alpha}$ and $\chi$ is the adjoint scalar of the new $SU(2)$ gauge multiplet.   The new $SU(2)$ intereactions are  asymptotically free and become strong in the infrared. 
As the  $SU(2)$  of  GST is weakly coupled,  the dynamics of the new $SU(2) $ is not affected by it.  
Let us recall that the $D_{3}$ singular SCFT arises in the $SU(2)$ theory with two flavors (with the same bare mass $m^{'}$) as 
the result of collision of a doublet singularity (at $u = m^{' \, 2}$)  with another, singlet vacuum. This occurs when $m^{'}$ coincides 
with the dynamical scale $\Lambda^{'}$.

If we perturb the $D_{3}$ singularity, setting  
\beq  m'  \simeq  \pm \Lambda^{'},  \eeq
but not exactly, the AD point splits as mentioned before in two vacua. Let us analize the resulting systems. The physics of the doublet singularity can be described as follows.  The $P$ system dynamically Abelianizes and gives rise to a superpotential, 
\beq \label{su3eff}\begin{aligned} &\sum_{i=1}^{3} \sqrt{2}  Q_{i}  \Phi {\tilde Q}^{i}  + \sum_{i=1}^{3}  {\tilde m}_{i}  \,  Q_{i} {\tilde Q}^{i}+   \mu \Phi^{2}   +  \sqrt{2}   M  \Phi {\tilde M}\\
&+   \sqrt{2}  {\tilde M}  A_{\chi}   M  +  \mu  \, A_\chi  \Lambda^{'},   
\end{aligned}\eeq
where a doublet of  $M$ represent light Abelian monopoles.  The mass parameters are assumed to have the form \cite{KT}
\bea  {\tilde m}_{1}&=&    \frac{1}{4}  (m_{1}+m_{2}- m_{3}-m_{4})\;;  \nonumber \\
   {\tilde m}_{2}&=&    \frac{1}{4}  (m_{1}-m_{2}+ m_{3}-m_{4})\;;  \nonumber \\
   {\tilde m}_{3} &=&    \frac{1}{4}  (m_{1}-m_{2}- m_{3}+m_{4})\;,      \label{massessu4}  \eea
as in (\ref{masses}) but without ${\tilde m}_{4}$, in terms of the bare quark masses of the underlying $SU(3)$ theory.

Again the flavor symmetry in various cases works out correctly (in all cases a $U(1)$ in the infrared comes from $M$): 
\begin{description}
\item[(i)]  In the equal mass limit, $m_{i}=m_{0}$   ($i=1,2,3, 4$)\;,
\beq          {\tilde m}_{1}= {\tilde m}_{2}= {\tilde m}_{3}=0\;, 
\eeq
so  the symmetry is
\beq   U(1)\times SO(6) = U(1)\times SU(4)\;. 
\eeq
 Note that in contrast to the $USp(2N)$ theory, the symmetry is not enhanced and remains to be $SU(4)\times U(1)$  in the $m_{i}=0$ (so  ${\tilde m}_{i}=0$)  limit 
  \item[(ii)]   $m_{1}=m_{2}$,    $m_{3}$, $m_{4}$ generic.  
  In this case    ${\tilde m}_{2}= -  {\tilde m}_{3}$ and    ${\tilde m}_{1}$ is  generic, so  the symmetry is   $U(1)\times U(1) \times SU(2)$, as in the underlying theory;
  \item[(iii)] $m_{1}=m_{2} \ne 0$,    $m_{3}=m_{4}=0$.    In this case,  $  {\tilde m}_{1}\ne 0$ and ${\tilde m}_{2}=  {\tilde m}_{3} = 0$, so
 the symmetry is $U(2)\times U(2)$  in the UV while  $U(1)\times U(1) \times SO(4)$  in  (\ref{vacmassBis}), which is the same. 
  
  \item[(iv)] $m_{1}=m_{2}=m_{3}\ne 0$,   $m_{4}$ generic.  In this case,  ${\tilde m}_{1}=  {\tilde m}_{2} =  -  {\tilde m}_{3}  \ne 0$. 
  Again the symmetry is $U(3)\times U(1)$ both in the UV and IR.

  \item[(v)]  $m_{1}=m_{2} \ne 0$  and   $m_{3}=m_{4} \ne 0$ but $m_{1}\ne m_{3}$.  In this case  $ {\tilde m}_{2} =  {\tilde m}_{3} = 0$ and  
  $ {\tilde m}_{4}$  and  ${\tilde m}_{1}$ generic.  The flavor symmetry is 
  \beq    SO(4) \times  U(1) \times U(1) =  SU(2)\times SU(2) \times  U(1) \times U(1)\;; 
  \eeq
  this is equal to the symmetry 
  \beq    SU(2) \times U(1) \times   SU(2)\times U(1)
  \eeq
  of the underlying theory.
  \item[(vi)]    $m_{1}\ne 0$,  $m_{2}=m_{3}=m_{4}=0.$  In this case  ${\tilde m}_{1}={\tilde m}_{2}={\tilde m}_{3}\ne 0$.
  The symmetry is $U(1)\times U(3)$ both in the UV and IR.
   \item[(vii)]    $m_{1}\ne 0$,  $m_{2}\ne 0,$  $ m_{1} \ne m_{2}$,  $m_{3}=m_{4}=0.$  In this case  ${\tilde m}_{1} \ne 0$,    ${\tilde m}_{2}={\tilde m}_{3}\ne   {\tilde m}_{1}$.
  The symmetry is $U(1)^{2}\times U(2)$ in the UV, and  $U(1)\times U(1) \times U(2)$ in the   infrared. 

\end{description}

Note that the flavor symmetry in various cases is not the same in $USp(2N)$  and  $SU(N)$ theories.  When  at least two  masses are zero, the symmetry is larger in the $USp(2N)$ theory, 
so the exact matching of the flavor symmetry in the UV and in the IR   is  quite nontrivial.  

The vacuum equations are now  
\beq    M {\tilde M} + \mu \Lambda^{\prime} =0\;;   \label{eq1bis}
\eeq
\beq   (\phi + A_{\chi})  {\tilde M}=M \,(\phi + A_{\chi})  =0\;;  \label{eq2bis}
\eeq
\beq  \sqrt{2} \,\left[\,  \frac{1}{2}  \sum_{i=1}^{3}  Q_{i}^{a}   {\tilde Q}_{b}^{i} -  \frac{1}{4}  \delta_{b}^{a}  (Q_{i}{\tilde Q}^{i})  +  \frac{1}{2}M^{a}{\tilde M}_{b}- \frac{1}{4} \delta^{a}_{b} M {\tilde M}\,\right]+ \mu \, \phi^{a}_{b}=0\;;  \label{eq3bis}
\eeq
\beq    (\phi + {\tilde m}_{i} )\, {\tilde Q}^{i}=  Q_{i}\, (\phi + {\tilde m}_{i})  =0, \quad \forall i\;.   \label{eq4bis}
\eeq
The first says that $M \ne 0$.  By gauge choice
\beq  M =  {\tilde M}= \left(\begin{array}{c}  2^{-1/4}\sqrt{-\mu \Lambda^{\prime}}  \\ 0 \end{array}\right)   \label{gchoice}
\eeq
so that 
\beq    \frac{1}{2} M^{a}{\tilde M}_{b}- \frac{1}{4}  (M{\tilde M}) \, \delta^{a}_{b} = \frac{(-\mu \Lambda^{'})}{4\sqrt{2}} \, \tau^{3}\;.  
\eeq
The second equation   is satisfied by adjusting $A_{\chi}$, whichever valur $\phi$ takes. The solution of the third equation (\ref{eq3bis}) without $Q$ vevs is not acceptable as it implies a
large ($O(\Lambda')$)  vev for $\phi.$   We are led to conclude that  ($i=1,2,3$): 
\beq   a=  -{\tilde m}_{i}, \qquad   Q_{i}= {\tilde Q}_{i} = \left(\begin{array}{c}h_i \\0\end{array}\right) ;\qquad  Q_{j}= {\tilde Q}_{j}=0, \quad  j \ne i \;.  
\eeq
such that  
\beq   h_{i}^{2}=    \frac{\mu \Lambda'}{\sqrt{2}}   +  2\, {\tilde m}_{i} \, \mu 
\eeq
or 
\beq   a=  {\tilde m}_{i}, \qquad   Q_{i}= {\tilde Q}_{ i} = \left(\begin{array}{c}0 \\  k_i\end{array}\right);\qquad  Q_{j}= {\tilde Q}_{j}=0, \quad  j \ne i  \;,   
\label{sol111}
\eeq
and 
\beq   k_{i}^{2}=   - \frac{\mu \Lambda' }{\sqrt{2}}  +   2\, {\tilde m}_{i} \, \mu \;.    
\label{sol222}   \eeq
These  give six vacua (corresponding to the $r=2$ vacua).  Where are other, $r=0,1$ vacua?

Now in the singlet vacuum the low energy physics of the new $SU(2)$ theory is
an Abelian gauge theory with a single monopole, $N$,  thus our effective superpotential is similar with (\ref{su3eff}) but with $N$ field having no coupling to the weak GST $SU(2)$
gauge fields: 
\beq       \sum_{i=1}^{3} \sqrt{2}  Q_{i}  \Phi {\tilde Q}^{i}  + \sum_{i=1}^{3}  {\tilde m}_{i}  \,  Q_{i} {\tilde Q}^{i}+   \mu \Phi^{2}   
+   \sqrt{2}  {\tilde N}  A   N  +  \mu  \, A  \Lambda^{'}+   m^{'} \,  {\tilde N}  N.   \label{su3effBis}
\eeq
Now the $U(1)$ part gets higgsed as usual,  and the GST $SU(2)$ gauge theory becomes asymptotically free, having ${\tilde N}_{f}=3$  hypermultiplets with small masses ${\tilde m}_{i}$. 
The infrared limit of this theory is well known: there is one vacuum with four nearby singularities and one singlet vacuum \cite{SWII}. In the quadruple vacuum,  the mass perturbation  ${\tilde m}_{i}$
give  four nearby vacua, the light hypermultiplets have masses, in the respective vacua \cite{KT},
\beq   {\hat m}_{1}=   {\tilde m}_{1}+  {\tilde m}_{2}+ {\tilde m}_{3} = \frac{1}{4} ( 3 m_{1}- m_{2}-m_{3}- m_{4})\;;
\eeq
\beq   {\hat m}_{2}=   -{\tilde m}_{1}-  {\tilde m}_{2}+ {\tilde m}_{3} = \frac{1}{4} ( 3 m_{4}- m_{1}-m_{2}- m_{3})\;;
\eeq
\beq   {\hat m}_{3}=  - {\tilde m}_{1}+  {\tilde m}_{2} -{\tilde m}_{3} = \frac{1}{4} ( 3 m_{3}- m_{1}-m_{2}- m_{4})\;;
\eeq
\beq   {\hat m}_{4}=   {\tilde m}_{1} -  {\tilde m}_{2} - {\tilde m}_{3} = \frac{1}{4} ( 3 m_{2}- m_{1}-m_{3}- m_{4})\;:
\eeq
they correctly represent the physics of the $r=1$ vacuum where the light hypermultiplets appear in the ${\underline 4}$ of the underlying $SU(N_{f})= SU(4)$ group.

Finally, in the singlet vacuum of the new $SU(2)$,  ${\tilde N}_{f}=3$ theory, the light hypermultiplet is a singlet of the flavor group, so it is also a singlet of the original 
$SU(4)$ flavor group.  


 \subsection{Singular $r=2$ vacua of the $SU(N)$, $N_{f}=4$ Theory \label{singularsu(4)}} 
 
In the case of the higher singularity of $SU(4)$, $N_{f}=4$ theory the GST  dual description is  
\beq      D_{4} - SU(2)  -   {\boxed 3}  \label{noncollidr}
\eeq
where $D_{4}$ is the most singular SCFT of  $SU(3)$ theory with ${\tilde N}_{f}=2$ flavors and  $ {\boxed 3} $  represents 
three doublets as before.
 
We therefore replace the above with another system 
\beq     SU(3) \, {\stackrel {B}{-}} \,  SU(2) -  {\boxed 3}  
\eeq
with a bifundamental field $B^{\alpha}_{a}$ carrying both $SU(3)$ and $SU(2)$ charges, that is, 
\beq\label{particular}\begin{aligned} &\sum_{i=1}^{3} \sqrt{2}  Q_{i}  \Phi {\tilde Q}^{i}  + \sum_{i=1}^{3}  {\tilde m}_{i}  \,  Q_{i} {\tilde Q}^{i} +      \mu \Phi^{2} +  \sqrt{2}   B  \Phi {\tilde B}\\
& + \sqrt{2}  {\tilde B} {\chi}     B   +  \mu  \, \chi^{2}  +   m^{''} \,  {\tilde B}   B.    
\end{aligned}\eeq
where $\chi$  is the adjoint scalar of the new $SU(3)$ and ${\tilde m}_{i}$ are given by  (\ref{massessu4}). For simplicity we 
have set the mass parameters for $\Phi$ and $\chi$ to be equal.  ${\tilde m}_{i}=0$, $i=1,2,3$ in the equal mass limit of the underlying theory,  $m_{i}=m_{0}$, $i=1,\ldots, 4$,   
so the system has the correct flavor symmetry,  $SO(6)\times U(1)= SU(4)\times U(1)$.   The flavor symmetry in various cases of unequal masses works out as in Subsection~\ref{collidingrvacua}.  

The $SU_{GST}(2)$  interactions are weak and do not affect significantly the $SU(3)$ gauge interactions.  Actually,  in order 
to study the system (\ref{noncollidr}), we must  focus our attention to one 
particular $SU(3)$  vacuum  (i.e., $D_{4}$ SCFT).   $D_{4}$ SCFT  appears as the $r=\frac{N_{f}}{2} =1$  vacuum of the new 
$SU(3)$, ${\tilde N}_{f}=2$ theory in the limit  $m^{''} \to 0$ (see the appendix).   In contrast to the case discussed 
in the previous subsection, the $r=1$ vacuum does not collide with the $r=0$ vacuum.

The $SU(3)$, ${\tilde N}_{F}=2$ theory is asymptotically free and becomes strongly coupled
in the infrared. For $m^{''}\neq0$ the low energy dynamics at the $r=1$ vacuum is described by a $U(1)^{2}$ theory \cite{SWII}:  two types of  
massless monopole hypermultiplets 
$M$ and $N$  appear, each carrying one of the local $U(1)$ charges, and one of them  ($M$) is a doublet of the  flavor 
$SU({\tilde N}_{F})=SU_{GST}(2)$.
Therefore the low-energy effective superpotential is given by 
 \bea   &&  \sum_{i=1}^{3} \sqrt{2}  Q_{i}  \Phi {\tilde Q}^{i}  + \sum_{i=1}^{3}  {\tilde m}_{i}  \,  Q_{i} {\tilde Q}^{i} +      \mu \Phi^{2} +  \sqrt{2}   M  \Phi {\tilde M}
+   \sqrt{2}  {\tilde M} A_{\chi}  M  +  \mu  \, A_{\chi}  \Lambda^{'}    \nonumber \\
&&   +   m^{''} \,  {\tilde M}   M +\sqrt{2}  {\tilde N} A  N  +  \mu  \, A  \Lambda^{'}    \label{abelian}
\eea
where $M$ is now a doublet of Abelian monopoles and $N$ is the Abelian monopole, singlet of the flavor $SU({\tilde N}_{F})=SU(2)$. The vacuum equations are  
\beq    \sqrt{2} \, M {\tilde M} + \mu \Lambda^{\prime} =0\;;  
\eeq
\beq   ( \sqrt{2} \phi + A_{\chi}+ m^{''}  )  {\tilde M}=M \,( \sqrt{2} \phi + A_{\chi} +  m^{''} )  =0\;;  
\eeq
\beq  \sqrt{2} \left[\, \frac{1}{2}  \sum_{i=1}^{3}  Q_{i}^{a}   {\tilde Q}_{b}^{i} -  \frac{1}{4}  \delta_{b}^{a}  (Q_{i}{\tilde Q}^{i})  +  \frac{1}{2}M^{a}{\tilde M}_{b}- \frac{1}{4} \delta^{a}_{b} M {\tilde M}\,\right]  + \mu \, \phi^{a}_{b}=0\;;  
\eeq
\beq    ( \sqrt{2} \phi + {\tilde m}_{i} )\, {\tilde Q}^{i}=  Q_{i}\, ( \sqrt{2} \phi + {\tilde m}_{i})  =0, \quad \forall i\;.   
\eeq
The solution of these equations are given by Eqs. (\ref{gchoice})-(\ref{sol222})  with the replacement ${\tilde m}_{i} \to m_{i}$. 
We therefore find six vacua, corresponding to the $r=2$  vacua of the underlying theory.
  
The degeneracy of vacua can also  be determined integrating out the $SU(3)$ $\psi$ field and adding the ADS superpotential. 
This procedure will inevitably produce the whole set of vacua of the model, including all vacua of the $SU(3)$ theory, whereas 
we are interested only in one of the  $r={\tfrac{{\tilde N}_{f}}{2}}=  1$ vacua, since we are interested in the $D_4$ sector. Our strategy will be to make the computation in the general case and then discard all the 
unwanted solutions. Integrating out $\psi$ the effective superpotential becomes 
\beq\nonumber \begin{aligned}
\mathcal{W}=&\sum_{i=1}^{3} \sqrt{2}  Q_{i}  \Phi {\tilde Q}^{i}  + \sum_{i=1}^{3}  {\tilde m}_{i} Q_{i} {\tilde Q}^{i} +\mu \Phi^{2}
+m\Tr M+\frac{\mu^3\Lambda^4}{\det M}+\Tr(\Phi M)\\
&-\frac{1}{2\mu}\left(\Tr M^2-\frac{(\Tr M)^2}{3}\right),
\end{aligned}\eeq where $M_{ab}$ is the meson field $\tilde{B}_{a\alpha}B_{b}^{\alpha}$. The meson matrix can be supposed diagonal and we 
will parametrize it as $M=aI+2b\tau_3$, so
$$\Tr(\Phi M)=\Phi_3b,\; \det M=a^2-b^2,\;\Tr M=2a,\; \Tr M^2=2(a^2+b^2)\;. $$ 
The superpotential then becomes
\beq\begin{aligned}
\mathcal{W}=&\sum_{i=1}^{3} \sqrt{2}  Q_{i}  \Phi {\tilde Q}^{i}  + \sum_{i=1}^{3}  {\tilde m}_{i} Q_{i} {\tilde Q}^{i} +\mu \Phi^{2}
+\Phi_3b+2ma\\
&+\frac{\mu^3\Lambda^4}{a^2-b^2}+\frac{4a^2}{6\mu}-\frac{a^2+b^2}{\mu}.
\end{aligned}\eeq Modulo a gauge choice we can diagonalize the $\Phi$ field, so that $\Phi_3$ is the only nonvanishing component. We thus find 
the following F-term equations:
\beq2m-\frac{2a}{3\mu}-\frac{2\mu^3\Lambda^4}{(a^2-b^2)^2}a=0\;,\eeq
\beq\Phi_3-\frac{2b}{\mu}+\frac{2\mu^3\Lambda^4}{(a^2-b^2)^2}b=0\;,\eeq
\beq(\sqrt{2}\Phi+m_i)Q_i=0,\quad b+\mu\Phi_3+\frac{\sqrt{2}\tilde{Q^i}Q_i\vert_{3}}{2}=0\;,\eeq
where $\tilde{Q}Q\vert_3$ is the component proportional to $\tau_3$. If the vev of $Q$ is nonzero, then $\Phi_3=\pm\sqrt{2}m_i$ 
(as before, only one $Q_i$ can have vev and one of the two components must vanish). Since there are six possibilities we will get 
the six vacua we were looking for. The last equation then tells that 
$$\tilde{Q^i}Q_i=\pm 2\mu m_i-b\;.$$
 The first two equations imply that $a^2-b^2\propto\mu^2\Lambda^2$.
If the vev of $Q$ is zero we have two possibilities: b nonzero and $\mu\Phi_3=-b$. But then we get $b\sim\mu\Lambda$ which in turn 
implies that $\Phi_3\sim\Lambda$ and this solution should be discarded. The other possibility is 
$b=\Phi_3=0$ and then from the first equation $a\sim\mu\Lambda$. We get four solutions which precisely correspond to the $r=0$ 
vacua of the $SU(3)$ theory. Since we are not interested in these vacua we simply discard them. 

The solutions corresponding to the 
$r=2$ vacua of our theory are correctly characterized by a nonvanishing $Q$ condensate and thus the pattern of flavor symmetry breaking 
is the expected $$U(4)\rightarrow U(2)\times U(2)\;.$$
Indeed we can proceed as before and choosing $m\neq0$ the singular point describing the $D_4$ theory evolves into a $r=1$ vacuum 
whose massless spectrum includes two vector multiplets that we will denote $A$ and $B$, two hypermultiplets charged under e.g.,  $A$ and forming a doublet of the $SU(2)$ flavor symmetry of the theory and a third hypermultiplet charged under $B$ which is 
a singlet of the flavor symmetry. The effective superpotential is thus
\beq\nonumber\begin{aligned}
\mathcal{W}=&\sum_{i=1}^{3} \sqrt{2}  Q_{i}  \Phi {\tilde Q}^{i}  + \sum_{i=1}^{3}  {\tilde m}_{i} Q_{i} {\tilde Q}^{i} +\mu \Phi^{2}
+ \sqrt{2}\tilde{M}{\phi}M+ \sqrt{2}\tilde{M}{A}M\\
&+\sqrt{2}\tilde{R}{B}R+\mu_1\Lambda A +\mu_2\Lambda B. 
\end{aligned}\eeq
This correctly describes the physics of the perturbed $r=2$ vacuum of $SU(4)$ SQCD with four flavors, leading to six vacua. The computation 
proceeds as in the previous sections.

\section{Singular points of $SO(N)$ SQCD \label{SO(N)}}

In chapter 4 Tchebyshev points of $SO(N)$ theories were analyzed as well. The outcome was the by now familiar two-sector structure (for $SO(2N)$ SQCD 
with even $N_f$ or $SO(2N+1)$ SQCD with $N_f$ odd): one hypermultiplet charged under an Abelian gauge group and a SCFT which can be described in terms of the 6d $D_N$ theory compactified 
on a three-punctured sphere \cite{DT}. These two sectors are coupled as before through an infrared free $SU(2)$ vector multiplet. The analysis of the 
$\mathcal{N}=1$ breaking in the general case is still out of reach but we can analyze in detail a couple 
of examples with low number of flavors, since as expected the superconformal sector simplifies enough to make the problem approachable. We will study 
the cases $N_f=1,2$ which already involve a nontrivial structure hard to guess without performing the analysis in the above mentioned paper.

\subsection{$SO(2N+1)$ theory with one flavor}

The SW curve at the Chebyshev point  of $SO(2N+1)$ SQCD  with one flavor becomes $y^2=x^4$. The superconformal sector entering the GST description 
becomes free in this case and describes one hypermultiplet in the adjoint of $SU(2)$, thus saturating its beta function. Notice that starting in the 
UV from a theory with a single matter field in the vector representation, we end up with an infrared effective description involving an $SU(2)$ 
gauge group coupled to matter fields in different representations! The expectation from the semiclassical analysis \cite{CKMII}  is that the $SU(2)$ flavor symmetry 
is dynamically broken to $U(1)$ when the mass term $\mu\Tr\Phi^2$ is turned on. Furthermore, if we give mass to the flavor (or to the hyper in the 
adjoint in the effective description) we expect to get two vacua  ($2^{N_{f}}=2$). Since our infrared effective theory admits a Lagrangian description, all these properties 
should be reproduced by the equations of motion. We will now check that this is case.

In the $\mathcal{N}=1$ language we describe the hypermultiplet in the adjoint using two chiral multiplets $X_1$ and $X_2$. The superpotential is
\beq\label{SO1}\begin{aligned}\mathcal{W}=&\sqrt{2} \tilde{Q} A_{D} Q +  \sqrt{2} \tilde{Q} \Phi Q +  \mu A_{D} \Lambda  +  {\mu}\,  \Tr \Phi^{2}\\
&+ \sqrt{2}i\Tr(\Phi[X_1,X_2])+ m\Tr(X_1X_2).\end{aligned}\eeq 
The variation with respect to $A_D$ tells that $Q$ has a nonvanishing vev. By gauge choice we can 
set $Q_2=0$ and then the variation with respect to $\tilde{Q}$ implies that $\Phi$ is diagonal. The variation with respect to the fields in the adjoint 
give the equations (we write them as $X=X^{a}\tau_{a}$):
\beq\label{phi}\sqrt{2} \tilde{Q} \tau_{a} Q + \mu\Phi_{a}+\sqrt{2}i\Tr(\tau_{a}[X_1,X_2])=0\;,\eeq
\beq\label{x1}\frac{m}{2}X_{2}^{a} + \sqrt{2}i\Tr(\Phi[\tau^{a},X_2])=0\;,\eeq
\beq\label{x2}\frac{m}{2}X_{1}^{a} - \sqrt{2}i\Tr(\Phi[\tau^{a},X_1])=0\,.\eeq
Since $\Phi_1=\Phi_2=0$, equations (\ref{x1},\ref{x2})  imply that $X_{1}^{3}=X_{2}^{3}=0$ (it suffices to note that $\Tr(\tau_3[\tau_3,\cdotp])=0$).
The nontrivial equations become then
 $$\begin{aligned}
&\frac{m}{2}X_{2}^{i}-\frac{\sqrt{2}}{2}\epsilon_{ij3}\Phi_3X_{2}^{j}=0,\quad \frac{m}{2}X_{1}^{i}+\frac{\sqrt{2}}{2}\epsilon_{ij3}\Phi_3X_{1}^{j}=0,\\
&\mu\Phi_3-\frac{\mu\Lambda}{2}-\frac{\sqrt{2}}{2}\epsilon_{ij3}X_{1}^{i}X_{2}^{j}=0.
\end{aligned}$$
Notice that the above equations imply that none of the unknowns can vanish (if one of the $X_i$'s vanish we would have $\Phi_3\sim\Lambda$, which we 
must discard as explained in the previous sections). Setting $X_1^{i=1}\equiv a$, $X_1^{i=2}\equiv b$, the second equation leads to the system 
$$\frac{m}{2}a+\frac{\Phi_3}{\sqrt{2}}b=0,\quad \frac{m}{2}b-\frac{\Phi_3}{\sqrt{2}}a=0\;.$$
 Writing $a$ in terms of $b$ and $\Phi_3$ using the first 
relation and substituting in the second we directly get $\Phi_3^2=-m^2/2$ and thus $a=\pm ib$. Using an analogous argument the third equation leads 
to $d=\pm ic$, where $X_2^{i=1}\equiv c$, $X_2^{i=2}\equiv d$. Notice that if we choose e.g. $a=+ib$ we are forced to set $d=+ic$ and not $-ic$, 
otherwise the term $\epsilon_{ij3}X_{1}^{i}X_{2}^{j}$ would vanish. Since the D-term equations imply that $\vert b\vert=\vert c\vert$, we get two solutions 
as expected. In the massless limit $m=0$, both the gauge and flavor symmetries are broken by the vevs of $Q$ and $X_i$'s. However, a diagonal combination 
of the gauge and flavor Abelian subgroups leaves the vevs invariant. We thus recover the expected $U(1)$ flavor symmetry.

\subsection{$SO(2N)$ theory with two flavors  \label{SO(N)two}    }

The SW curve at the Chebyshev points in these theories becomes $y^2=x^6$  \cite{CKMII}. In \cite{Simone} it was found that the superconformal sector does not become 
free in this case but turns out to be a well known Lagrangian SCFT: $SU(2)$ SQCD with four flavors. Symbolically the system can be represented as 
\beq      {\boxed 1}\, {\stackrel {Q}{-}}  \,  SU(2) \,{\stackrel {M_{i}}{-}}  \,SU(2)\;.
\eeq
An $SU(2)$  subgroup of the $SO(8)$ flavor symmetry is gauged 
in the present context, leaving the commutant $USp(4)$ ungauged, matching the UV flavor symmetry of the theory. The $USp(4)$ symmetry tells us that 
the low energy theory at the Chebyshev point is a $SU(2)\times SU(2)$ gauge theory with two hypermultiplets in the bifundamental and one doublet 
charged under only one of the $SU(2)$ factors. The semiclassical analysis predicts that the flavor symmetry is dynamically broken to $U(2)$ when 
we turn on the mass term for the chiral multiplets in the adjoint, while keeping the bifundamentals massless \cite{CKMII}. If we give mass to the bifundamentals as well, 
we expect to find $2^{N_f}=4$ vacua. We shall now see that the equations of motion for the infrared effective theory reproduce all these features.

We indicate in $\mathcal{N}=1$ notation the hypermultiplets in the bifundamental with $M_1$, $\tilde{M}_1$, $M_2$, $\tilde{M}_2$ and the chiral 
multiplets in the adjoint with $\Phi$ and $\Psi$. The superpotential will then be (the sum over $i=1,2$ for the bifundamentals is implied)
\beq\label{so2}\begin{aligned}
\mathcal{W}=&\sqrt{2} \tilde{Q} A_{D} Q +  \sqrt{2} \tilde{Q} \Phi Q 
+\sqrt{2}\Tr(\tilde{M}_i\Phi M^i)+\sqrt{2}\Tr(M^i\Psi \tilde{M}_i)\\
&+ m_i\Tr(\tilde{M}_iM^i) +  \mu A_{D} \Lambda  +  {\mu} \Tr \Phi^{2} + {\nu} \Tr \Psi^{2},\end{aligned}
\eeq where $\mu$ and $\nu$ are of the same order.
As usual, the variation with respect to $A_D$ implies that the doublet $Q$ has non vanishing vev. We can then use the gauge freedom to set $Q_2$ 
to zero and to diagonalize $\Psi$. The equation for $\tilde{Q}$ will then imply that $\Phi$ is diagonal too. The equations coming from the variation 
of $\Phi$ and $\Psi$ are then 
\beq\label{phibis}\mu\phi_3-\frac{ \mu \Lambda}{2}+\sqrt{2}\Tr(\tilde{M}_i\tau_3 M^i)=0,\quad \sqrt{2}\Tr(\tilde{M}_i\tau_{1,2} M^i)=0\;,   \label{this}   \eeq
\beq\label{psibis}\nu\Psi_3+\sqrt{2}\Tr(M^i\tau_3 \tilde{M}_i)=0,\quad \sqrt{2}\Tr(M^i\tau_{1,2} \tilde{M}_i)=0\;.    \label{that}  \eeq 
Since this will play a role later in the derivation, we would like to draw the reader's attention to the fact that we require 
the vevs of both $\Phi$ and $\psi$ to be much smaller than $\Lambda$. This in particular implies that $\Tr(\tilde{M}_i\tau_3 M^i)$ 
and $\Tr(M^i\tau_3 \tilde{M}_i)$ cannot be of the same order (the first should be much larger than the second in order to compensate 
the term proportional to $\mu\Lambda$ in (\ref{phibis})). The variation with respect to 
the bifundamental fields gives $$\sqrt{2}\Phi M_i + m_iM_i + \sqrt{2}M_i\Psi=0,\quad \sqrt{2}\Psi\tilde{M}_i+m_i\tilde{M}_i+\sqrt{2}\tilde{M}_i\Phi=0\;.$$ 
It is convenient to rewrite these two equations in matrix form:
\beq\label{matr}\begin{aligned}
\left(\begin{array}{cc}
(m_i+\frac{\Phi_3}{\sqrt{2}}+\frac{\Psi_3}{\sqrt{2}})a_i & (m_i+\frac{\Phi_3}{\sqrt{2}}-\frac{\Psi_3}{\sqrt{2}})b_i\\
(m_i-\frac{\Phi_3}{\sqrt{2}}+\frac{\Psi_3}{\sqrt{2}})c_i & (m_i-\frac{\Phi_3}{\sqrt{2}}-\frac{\Psi_3}{\sqrt{2}})d_i\\
\end{array}\right)&=0\;,\\
\left(\begin{array}{cc}
(m_i+\frac{\Phi_3}{\sqrt{2}}+\frac{\Psi_3}{\sqrt{2}})e_i & (m_i-\frac{\Phi_3}{\sqrt{2}}+\frac{\Psi_3}{\sqrt{2}})f_i\\
(m_i+\frac{\Phi_3}{\sqrt{2}}-\frac{\Psi_3}{\sqrt{2}})g_i & (m_i-\frac{\Phi_3}{\sqrt{2}}-\frac{\Psi_3}{\sqrt{2}})h_i\\
\end{array}\right)&=0\;,\end{aligned}
\eeq 
where we have set (we will take into account D-terms later) 
$$ M_i=\left(\begin{array}{cc}
a_i & b_i\\
c_i & d_i\\
\end{array}\right),\qquad\tilde{M}_i=\left(\begin{array}{cc}
e_i & f_i\\
g_i & h_i\\
\end{array}\right),\quad i=1,2 \;.$$
It is clear that for $m_1$ and $m_2$ generic, equation (\ref{matr}) requires that some entries of $M_i$ and $\tilde{M}_i$ vanish. 

One can check that there are no solutions if only one of the bifundamentals (for instance,  $M_1$) is to have nonvanishing vev. This can be shown 
as follows: it is easy to see that at least two entries (both for $M_1$ and $\tilde{M}_1$) must be zero. Taking into account 
the rightmost equations in (\ref{phibis}) and (\ref{psibis}), one can easily check that actually at most one entry can be different from 
zero, but then $\Tr(\tilde{M}_i\tau_3 M^i)$ and $\Tr(M^i\tau_3 \tilde{M}_i)$ differ at most by a sign and are thus of the same 
order and this is in conflict with the observation we made before. 

We are then forced to let both $M_1$ and $M_2$ be nontrivial. This can be achieved by imposing e.g. the equations 
\beq\label{solu}m_1+\frac{\Phi_3}{\sqrt{2}}+\frac{\Psi_3}{\sqrt{2}}=0,\qquad m_2+\frac{\Phi_3}{\sqrt{2}}-\frac{\Psi_3}{\sqrt{2}}=0\;,  \label{choice}  \eeq 
which determine both $\Phi_3$ and $\Psi_3$ in terms of the mass parameters.
Clearly both $M_1$ and $M_2$ can have only one nonvanishing  entry.  

Let us now count the number of possible solutions: one naively 
has four possible choices for $M_1$; in two cases the matrix is diagonal and in the other two offdiagonal. Actually, the action 
of the $SU(2)$ element   
\beq    T =  \left(\begin{array}{cc}0 & 1 \\-1 & 0\end{array}\right) 
\eeq
of the second $SU(2)$ gauge factor interchanges these two 
possibilities and we may assume, e.g., that $M_1$ is diagonal.  There are two solutions according to which diagonal element is chosen to be nonvanishing.  

The other two solutions (in which $M_{1}$ is offdiagonal) are gauge equivalent to these and should not 
be considered distinct \footnote{We cannot act with an analogous subgroup of the other gauge factor as 
it is already broken by the $Q$ vev.}.      Having $M_{1}$ of diagonal form, equation (\ref{matr}) then implies that $M_2$ is offdiagonal and we have two possible 
choices. If we now take into account the D-term condition  we find that (modulo a phase) $\tilde{M}_1=M_1^{\dagger}$ and 
$\tilde{M}_2=M_2^{\dagger}$. Each one of the four possible choices lead then to a single solution once equations (\ref{phibis}), (\ref{psibis}) 
are taken into account. We thus find    four solutions as anticipated. One of the solutions, corresponding to the choice, (\ref{choice}), takes the form, 
\beq   M_{1}=  \left(\begin{array}{cc}  a_{1}  & 0 \\0 & 0 \end{array}\right)\;; \qquad M_{2} = \left(\begin{array}{cc}0 & 0 \\  g_{2}  & 0\end{array}\right)\;,
\eeq
where  $a_{1}$ and $g_{2}$ are determined by Eqs.~(\ref{this}), (\ref{that}), (\ref{choice}) and are of the order of $O(\mu \Lambda, \mu\, m_{i})$.

In the massless limit the $\Phi$ and $\Psi$ vevs go to zero. The $Q$ condensate breaks the first $SU(2)$ gauge symmetry factor and 
the vev of the bifundamentals breaks the second $SU(2)$ gauge factor. The $USp(4)$ flavor symmetry is broken as well;  however, 
there is a diagonal combination of the global $SU(2)$ gauge transformations (coming from the second gauge factor) and (an $SU(2)$ 
subgroup of) flavor transformations which acts trivially on our solution of the field equations and thus remains  unbroken. Furthermore, 
the second Cartan generator of the flavor symmetry group of the theory can combine with the Cartan of the first $SU(2)$ gauge group 
to give the generator of a $U(1)$ group which is unbroken. The color-flavor locking mechanism thus leads to the 
$U(2)$  unbroken global symmetry, which is the correct unbroken symmetry expected from the analysis made at large $\mu$ \cite{CKMII}.  

\section{Discussion} 

The fate of an ${\cal N}=2$ SCFT  upon deformation by  ${\cal N}=1$, adjoint mass perturbation, $\mu  \Phi^{2}$, can be 
of  several different types.   A nontrivial ${\cal N}=2$ SCFT  in the UV  might smoothly flow into an ${\cal N}=1$ SCFT in the infrared (see \cite{TachiWecht} for some
 beautiful observations).   An infrared fixed-point  SCFT in an ${\cal N}=2$ theory might get lifted upon $\mu  \Phi^{2}$ deformation,
  as in the case of the original AD point in the pure  ${\cal N}=2$, $SU(3)$ theory. 
  
 Infrared fixed-point ${\cal N}=2$ SCFT's might also be brought into confinement phase, as shown in the original  Seiberg-Witten work \cite{SWI,SWII}, in 
 the case of local $r$ vacua \cite{CKM}, or  in the cases of singular SCFT's   discussed in the present paper.    What distinguishes these systems is the 
 presence of $U(1)$ factors in the effective gauge symmetry.  More precisely the property required is the nontrivial fundamental group, 
 \beq   \pi_{1} (G_{eff})  \ne {\bf 1}\;
 \eeq
 where $G_{eff}$ is the low-energy gauge group,  and that  all the $U(1)$ factors are broken upon $\mu \, \Phi^{2}$ perturbation. If the underlying gauge group is simply connected, the 
 vortices of the low-energy theory should not exist in the full theory. If the low-energy theory is magnetic, then the condensation leading to the breaking
 $G_{eff} \to {\bf 1}$  implies confinement of color. 
 
 What was not known in earlier studies  \cite{CKM,CKMII,Konishi:2005qt}  is what happens in the singular Chebyshev vacua 
 (EHIY points), and if the system would  be brought into confinement phase, which kind of confinement phase it would be.  Since such a  system apparently involved
 (infinitely) strongly-coupled, relatively nonlocal monopoles and dyons,  it was not at all evident whether or not the standard (weakly-coupled) dual Higgs picture worked.   
 
The checks made in this chapter have been primarily aimed at ascertaining that one is indeed correctly  describing
 the infrared physics of these SCFT's of highest criticality, {\it  deformed by a small  $\mu \Phi^{2}$ perturbation}, in terms of the GST duals \cite{GST}.  
 Once such a test is done, one can safely discuss the infrared physics in the limit of singular  SCFT, directly.
 
Let us take the example of the theory  discussed in Section \ref{uspsection}.  
In the case of $USp(2N)$,  $N_{f}=4$ theory, the GST dual is 
 \beq       {\boxed 1} -   SU(2) - {\boxed 4}\;.
 \eeq
  The effect of $\mu \, \Phi^{2}$ deformation of this particular theory can then be analyzed straightforwardly  in the massless 
  theory by using the superpotential,  
\beq     \sqrt{2} \, Q_{0} A_{D} {\tilde Q}^{0} +  \sqrt{2} \, Q_{0} \phi {\tilde Q}^{0} + \sum_{i=1}^{4}  \sqrt{2} \, Q_{i} \phi {\tilde Q}^{i}  +  \mu A_{D} \Lambda  +  {\mu}\, \Tr \phi^{2}\;.
\eeq
The solution of the equations of motion are:
\beq  Q_{0} =  {\tilde Q}_{0}= \left(\begin{array}{c}   2^{-1/4}\sqrt{-\mu \Lambda}  \\ 0 \end{array}\right)
\eeq
\beq   \phi =0, \quad  A_{D}=0\;.  \label{sol2}
\eeq 
The contribution from $Q_{i}$'s must then cancel that of   $Q_{0}$ in  Eq.~(\ref{eq3}).   
By flavor rotation  the nonzero VEV can be attributed to $Q_{1}, {\tilde Q}^{1}$, i.e., either of the form
\beq     (Q_{1})^{1} =  ({\tilde Q}^{1})_{1} =2^{-1/4} \sqrt{\mu \Lambda} \;,    \qquad  Q_{i}={\tilde Q}_{i}=0, \quad i=2,3,4.    \label{sol3a}
\eeq
or  
\beq     (Q_{1})^{2} =  ({\tilde Q}^{1})_{2}   = 2^{-1/4}  \sqrt{-  \mu \Lambda} \;,    \qquad  Q_{i}={\tilde Q}_{i}=0, \quad i=2,3,4. \label{sol3b}
\eeq
 The $U(1)$ gauge symmetry is broken by the $Q_{0}$ condensation:  an ANO vortex is formed. As the gauge group of the underlying theory is  simply connected,
such a low-energy vortex must end. The quarks are  confined. The flavor symmetry breaking 
\beq 
  SO(8) \to   U(1) \times  SO(6) =   U(1) \times SU(4) = U(4),  \label{right}
\eeq
is induced by the condensation of  $Q_{1}$, which does not carry the $U(1)$ gauge charge.  The pattern of the symmetry breaking agrees with that found at large $\mu$  \cite{CKM}.\\ 
\indent The vortex is made of the $Q_{0}$ field and the effective Abelian gauge
field. The most interesting feature of this system is that there is no dynamical Abelianization, i.e., the effective low-energy gauge group is $SU(2)\times U(1)$.  
The confining string is unique and does not lead to the doubling of the meson spectrum. The global symmetry breaking of the 
low-energy effective theory is the right one (\ref{right}), but the vacuum is not color-flavor locked. The confining string 
is of Abelian type, and is not a non-Abelian vortex as the one appearing in an $r$ vacuum.  
These facts clearly  distinguish the confining system found here both from the standard Abelian dual superconductor type 
systems and from the non-Abelian dual Higgs system found in the $r$-vacua of SQCD.   The dynamical symmetry breaking and 
confinement are linked to each other (the former is induced by the $Q$ condensates, which in turn, is triggered by the 
$Q_{0}$ condensation which is the order parameter of confinement), but not described by one and the same condensate.\\  
\indent The $SO(N)$ systems discussed in Section \ref{SO(N)} present other examples of confining vacua, with similar properties. In the 
$SO(2N+1)$ theory with one flavor the low energy description involves fields in different representations of $SU(2)$: one in 
the fundamental and one in the adjoint. As before an abelian confining string made of the field in the dublet is produced, 
whereas flavor symmetry breaking is induced by the field in the adjoint. In the $SO(2N)$ theory with two flavors one of the 
sectors in the GST description is interacting, as opposed to the two cases just discussed. However, this SCFT turns out to be a 
well-known lagrangian theory ($SU(2)$ with four flavors) which can be studied with standard techniques. In this case we come 
across the flavor-locking mechanism: the bifundamental fields break both gauge and flavor symmetry but the diagonal combination 
of the two is unbroken, thus reproducing the expected global symmetry of the theory. Also in this case the vortex is abelian and 
the two bifundamentals are not related to its formation.\\
\indent We conclude with a brief comment on the nature of the GST variables. The mass assignment such as in Eq.~(\ref{masses}) which 
reproduces correctly the flavor symmetry property of the underlying theory, is a clear sign of the magnetic monopole nature of 
the low-energy matter content. Their condensation therefore implies confinement of the color-electric charges. Nevertheless, 
the way they realize the dynamical flavor symmetry breaking and confinement appears to present various new features as compared 
to the straightforward dual superconductor picture of confinement, abelian or non-abelian, and seems to urge a better 
understanding of the new confinement phases.

\section*{Appendix}
\addcontentsline{toc}{section}{Appendix}

\subsection*{Monopole condensates in $SU(2)$ and $SU(3)$ $N_f=2$ theories}\label{23}

The SW curve for the $SU(2)$ theory with two flavors can be written as $$y^2=(x^2-u)^2-4\Lambda^2(x+m)^2,$$ and if we set $u=m^2$ it 
degenerates to \beq\label{22f} y^2=(x+m)^2(x-m-2\Lambda)(x-m+2\Lambda).\eeq The low energy physics at this point is described by an Abelian $U(1)$ theory with 
two massless electrons and when we turn on the $\mathcal{N}=1$ deformation $\mu\Tr\Phi^2$, the corresponding effective action includes the superpotential 
$$\sqrt{2}\tilde{Q}_1A Q_1+\sqrt{2}\tilde{Q}_2A Q_2+\mu U; \quad U\equiv\langle\Tr\Phi^2\rangle.$$ The equations of motion thus impose the constraint 
$$\langle\tilde{Q}_1Q_1\rangle+\langle\tilde{Q}_2Q_2\rangle=-\frac{\mu}{\sqrt{2}}\frac{\partial U}{\partial A}.$$ In order to compute the 
condensate we now have to evaluate $\partial U/\partial A$. This can be done noticing that $$\frac{\partial U}{\partial A}^{-1}= 
\frac{\partial A}{\partial U}=\int_{\gamma}\frac{\partial\lambda}{\partial U},$$ where the contour $\gamma$ is a small circle surrounding the point 
$x=-m$. We can now exploit the fact that the SW differential for $SU(N)$ SQCD satisfies the relation \cite{AF} $$\frac{\partial\lambda}{\partial U}=\frac{dx}{y}
x^{N-2}.$$ From (\ref{22f}) we then obtain $$\frac{\partial U}{\partial A}\propto\sqrt{(\Lambda+m)(\Lambda-m)}.$$ This quantity vanishes for $m=\pm\Lambda$,
which are precisely the values such that the SW curve degenerates further and we encounter the $D_3$ Argyres-Douglas point. This shows that the $Q_0$ 
condensate (in the notation of \ref{vac}) vanishes at this point. 

The computation for $SU(3)$ is similar: the SW curve in this case is $$y^2=(x^3-Ux-V)^2-4\Lambda^4(x+m)^2$$ and setting 
$U=2\Lambda^2+\frac{3}{4}m^2$, $V=2m\Lambda^2-\frac{m^3}{4}$ we reach the $r=1$ vacuum, the point we are looking for. 
The SW curve at this point factorizes as \beq\label{32f}y^2=(x+m)^2(x-\frac{m}{2})^2(x^2-mx+\frac{m^2}{4}-4\Lambda^2),\eeq
and the low energy effective action describes an Abelian $U(1)^2$ theory with two massless hypermultiplets charged under 
one $U(1)$ factor and another hypermultiplet charged under the second one. In the $m\rightarrow0$ limit the curve degenerates 
further and we find the maximally singular point. In order to find the $Q_0$ condensate we have to evaluate as before
$$\frac{\partial A}{\partial U}=\int_{\gamma}\frac{\partial\lambda}{\partial U},$$ where the contour $\gamma$ is again a circle around the point 
$x=-m$. The crucial difference with respect to the $SU(2)$ case is the fact that now $$\frac{\partial\lambda}{\partial U}=\frac{xdx}{y},$$
leading to the relation $$\frac{\partial A}{\partial U}\propto\left(\sqrt{4\Lambda^2-\frac{9}{4}m^2}\right)^{-1}.$$ This quantity 
remains finite in the $m\rightarrow0$ limit (notice that $\partial A/\partial V$ diverges instead). 
The computation of $\partial U/\partial A$, the quantity we are interested in, is slightly more delicate with respect to the $SU(2)$ case, since the 
Coulomb branch has now complex dimension two and the $A$ cycle is a function of both $U$ and $V$. It is convenient to introduce the homology cycle $B$, which 
satisfies the equation
$$\frac{\partial B}{\partial U}=\int_{\gamma'}\frac{\partial\lambda}{\partial U},\quad \frac{\partial B}{\partial V}=\int_{\gamma'}\frac{\partial\lambda}{\partial V},$$ 
where $\gamma'$  is a loop around the point $x=\frac{m}{2}$. We can take $A$ and $B$ as a basis of ``electric'' cycles. From the above formulas it is 
clear that $\partial A/\partial U$ and $\partial B/\partial U$ are both finite in the massless limit, whereas $\partial A/\partial V$ and $\partial B/\partial V$
are both proportional to $\sim\frac{1}{m}$ for small $m$. Considering now the equations
$$\frac{\partial A}{\partial U}\frac{\partial U}{\partial A}+\frac{\partial A}{\partial V}\frac{\partial V}{\partial A}=1;\quad 
\frac{\partial B}{\partial U}\frac{\partial U}{\partial A}+\frac{\partial B}{\partial V}\frac{\partial V}{\partial A}=0,$$
we can easily see that they cannot be satisfied if $\partial U/\partial A$ vanishes. This guarantees that the $Q_0$ condensate 
does not vanish. A similar computation of condensates using the SW curve has been performed in \cite{SYIV,SYnew}.

\chapter*{Concluding remarks}
\addcontentsline{toc}{chapter}{Concluding remarks}

Despite the existence of magnetic monopoles has been proposed more than eighty years ago \cite{Dirac,Dirac2}, they have not found 
yet their place in our understanding of nature. These particles often appear in spontaneously broken gauge theories and play a leading role in the 
dynamics of supersymmetric gauge theories. In this thesis we 
have studied how the dynamics of magnetic monopoles induces confinement and chiral symmetry breaking, when we perturb an infrared 
fixed point in $\mathcal{N}=2$ SQCD breaking softly extended supersymmetry.

A proper understanding of a conformal field theory appearing as an infrared fixed-point is in many cases of physics of fundamental importance, as it reveals the collective behavior 
of the underlying degrees of freedom, which determines the long-distance physics of the system. Critical phenomena and phase transitions are typical situations in which 
such a consideration plays the central role. A closely related problem is that of quark confinement: even though the UV behavior 
of the quark and gluon degrees of freedom is well understood (asymptotic freedom), the collective behavior of color in the 
infrared is still covered in mystery. 

The idea that at a certain mass scale the system dynamically Abelianizes and produces Abelian monopoles of the dual $U(1)^{2}$ 
theory,  and that its dynamical Higgsing induces confinement \cite{TP}, is yet to be demonstrated.  An interesting alternative 
possibility is that the system does not completely Abelianize, with non-Abelian monopoles of the gauge breaking
$SU(3)\rightarrow SU(2)\times U(1)$ acting as the effective dual degrees of freedom.  As the $u$ and $d$ quarks are light, it 
is possible that the QCD vacuum is in an $SU(2)$ color-flavor locked phase, in which confinement and chiral symmetry breaking 
occur simultaneously, via  the condensation of the  non-Abelian monopoles carrying $u$, $d$ flavor charges  \cite{Ken}. 
This would alleviate the problem associated with the dynamical Abelianization:  the problem of too-many Nambu-Goldstone bosons
and of the doubling of the meson spectrum.  

At the same time, however, it introduces a new difficulty. In contrast to what happens in the $r$-vacua of the softly broken ${\cal N}=2$ 
supersymmetric QCD, it is likely that the interactions among non-Abelian monopoles associated with the above pattern of gauge 
symmetry breaking are asymptotically free and become strong at low energies,  as the sign flip of the beta function is rather 
difficult in non-supersymmetric QCD.
It is possible that ultimately one must accept the idea that the color magnetic degrees of freedom of QCD are strongly coupled 
and that confinement and dynamical chiral symmetry breaking are described in a way subtler than a straightforward dual 
superconductivity picture.  

From this point of view the physics of singular points in ${\cal N}=2$ supersymmetric QCD with $SU(N)$, $USp(2N)$ and $SO(N)$ 
gauge groups that we have studied represent a class of models of considerable interest.  These SCFT's occur at particular points 
of the vacuum moduli and/or for special choices of the bare quark mass parameters.  A straightforward interpretation of the 
points of the highest criticality would involve monopoles and dyons in an infinite-coupling regime, therefore making their 
physical interpretation a highly nontrivial task \cite{Konishi:2005qt}. 
 
Exploiting the recent progress in the understanding of $\mathcal{N}=2$ theories in four dimensions, especially superconformal 
ones, we have been able to derive a more tractable description of the low-energy physics at these singular points. This was 
done extending the analysis for $SU(N)$ SQCD of Gaiotto, Seiberg and Tachikawa (GST) \cite{GST}. In the particular cases of $USp(2N)$ 
theory with four flavors, $SO(2N+1)$ theory with one flavor and $SO(2N)$ theory with two flavors our description simplifies 
considerably providing a peaceful local lagrangian description, from which one can then extract all the desired information just 
analyzing the equations of motion. We recovered the condensation of magnetic monopoles (confinement), the pattern of dynamical 
symmetry breaking and the multiplicity of vacua in the mass deformed theory, finding perfect agreement with the semiclassical 
analysis and the study of the SW curve presented in \cite{CKM,CKMII}.

The analysis of singular points in $SU(N)$ SQCD is subtler because the GST description in this case always involves a strongly 
interacting sector. However, in the case $N_f=4$ we can exploit the fact that the SCFTs entering in our dual description arise 
as infrared fixed points of lagrangian theories. As long as we are interested in determining the pattern of flavor symmetry 
breaking and the multiplicity of vacua we can indeed focus on these lagrangian theories and work out the desired information 
applying standard techniques. Our findings are in perfect agreement with \cite{CKM} and with the results of chapter 3.

As a byproduct we find an unconventional mechanism of confinement, in which confining strings of abelian kind are accompanied by 
non-abelian monopoles. This peculiar scenario, which is roughly speaking half-way between the abelian and non-abelian superconductor 
pictures, allows to bypass the usual problems mentioned above.

For generic number of flavors the problem becomes considerably more involved, since our description now includes nonlagrangian 
superconformal sectors which have no known realization as infrared fixed points of more standard theories. The analysis of 
these cases is still beyond the tools developed here and it is quite unlikely that a local description can ever be found in this 
case. However, the understanding of confinement and dynamical symmetry breaking in supersymmetric models requires only the 
evaluation of chiral condensates, which do not require a detailed knowledge of the dynamics. Presumably in the near future new 
techniques will be developed, allowing to complete the present analysis.

The most serious problem in extending this kind of analysis to models with less supersymmetry is the absence of two scales: what 
makes it possible to understand confinement and chiral symmetry breaking in softly broken $\mathcal{N}=2$ theories is the fact 
that the scale at which these two phenomena occur is much smaller than $\Lambda_{\mathcal{N}=2}$, the scale at which our low-energy 
description breaks down. This allows to address such questions with standard field theory techniques in the framework of the 
infrared effective theory. In YM or $\mathcal{N}=1$ SYM theory this structure is not present, making the problem considerably 
more difficult. Indeed, we can flow to $\mathcal{N}=1$ SYM starting from the $\mathcal{N}=2$ theory just decoupling the chiral 
multiplet in the adjoint representation. However, the scale of validity of the SW effective action goes to zero in the decoupling 
limit, signalling that the knowledge of the IR of $\mathcal{N}=2$ SYM is not enough.

In the analysis presented in chapter 5 we found several examples of solutions to the F-term equations of the low-energy effective 
lagrangian which are not compatible with the pattern of flavor symmetry breaking predicted semiclassically. However, they all 
involve VEVs for the scalar field in the $\mathcal{N}=2$ vectormultiplet which are of order $\Lambda$.
Indeed, they are solutions of our effective theory but have nothing to do with the SQCD models we want to study:
as we have seen they must be discarded and this is confirmed by the couting of supersymmetric vacua. A situation of this kind 
would be unavoidable in a theory with a single scale, raising the question whether an effective low-energy description can be 
useful to address these problems in QCD.


\end{document}